\documentclass[openany, 11pt, reqno]{amsbook}

\usepackage{amsmath,amssymb,amsfonts}
\usepackage{algorithm}
\usepackage{algpseudocode}
\usepackage{algorithmicx}

\usepackage{graphicx}
\usepackage{booktabs}
\usepackage{tikz}

\usepackage[roman]{complexity}
\usepackage{imakeidx}

\usepackage{chngcntr} 
\counterwithin{figure}{chapter} 

\usepackage{times}

\usepackage{tcolorbox}

\usepackage[normalem]{ulem} 

\numberwithin{table}{chapter}

\makeindex[columns=2, name= authors, title=Author Index]
\makeindex[columns=2, name=subject, title=Subject Index]

\newtheorem{theorem}{Theorem}[chapter]
\newtheorem{lemma}[theorem]{Lemma}

\newtheorem{fact}[theorem]{Fact}
\newtheorem{corollary}[theorem]{Corollary}
\newtheorem{claim}[theorem]{Claim}
\newtheorem{proposition}[theorem]{Proposition}

\newtheorem{notenum}{Theorem}
\newtheorem{note}[notenum]{Note}

\theoremstyle{definition}
\newtheorem{definition}[theorem]{Definition}
\newtheorem{example}[theorem]{Example}

\theoremstyle{remark}
\newtheorem{remark}[theorem]{Remark}

\numberwithin{section}{chapter}
\numberwithin{equation}{chapter}

\usepackage{etoolbox}

\makeatletter
\renewcommand{\listofalgorithms}{\@starttoc{loa}{List of Algorithms}}
\let\l@algorithm=\l@figure
\patchcmd{\@tocline}
  {\hfil}
  {\leaders\hbox{\,.\,}\hfil}
  {}{}
\newcommand\numberprefix{}
\def\numberline#1{\hb@xt@\@tempdima{\numberprefix #1\hfil}}
\def\l@figure#1#2{\@tocline{0}{3pt plus2pt}{0pt}{5pc}{}%
  {\renewcommand\numberprefix{Figure~}#1}{#2}}
\def\l@table#1#2{\@tocline{0}{3pt plus2pt}{0pt}{5pc}{}%
  {\renewcommand\numberprefix{Table~}#1}{#2}}
\makeatother

\newcommand{\SD}{\mbox{\rm SIZE-DEPTH}}
\newcommand{\FSD}{\mbox{\rm FSIZE-DEPTH}}
\newcommand{\MAJ}{\mbox{\rm MAJ}}
\renewcommand{\NSPACE}{\mbox{\rm NSPACE}}
\newcommand{\coNSPACE}{\mbox{\rm co-NSPACE}}

\newcommand{\Lredn}{\leq^{\mbox{{\rm L}}}_m}

\newcommand{\LTredn}{\leq^{\mbox{{\rm L}}}_{\mbox{{\rm T}}}}
\newcommand{\boldNL}{\mbox{\rm \bf NL}}
\renewcommand{\L}{\mbox{\rm L}}
\newcommand{\NLH}{\mbox{\rm NLH}}
\newcommand{\LModLpoly}{\mbox{\ensuremath{\L^{\ModL}{\rm /poly}}}}
\renewcommand{\coNL}{\mbox{\rm co-NL}}

\newcommand{\boldUL}{\mbox{\rm \bf UL}}

\newcommand{\coUL}{\mbox{\rm co-UL}}

\newcommand{\ULredn}{\mbox{{\rm UL}}}
\newcommand{\parityL}{\mbox{\ensuremath{\oplus\L}}}

\newcommand{\NCone}{\ensuremath{\NC^1}}
\newcommand{\NCtwo}{\ensuremath{\NC^2}}
\newcommand{\BPNCtwo}{\mbox{\rm BP.NC$^2$}}

\newcommand{\TCzero}{\ensuremath{\TC^0}}
\newcommand{\TCone}{\ensuremath{\TC^1}}

\newcommand{\ACzero}{\mbox{\rm AC$^0$}}

\newcommand{\sharpP}{\mbox{$\sharp$\rm P}}
\renewcommand{\ModkP}{\mbox{\rm Mod$_k$P}}
\newcommand{\ModpP}{\mbox{\rm Mod$_p$P}}
\newcommand{\sharpL}{\mbox{$\sharp$\rm L}}
\newcommand{\sharpLH}{\mbox{$\sharp$\rm LH}}
\newcommand{\GapLH}{\mbox{\rm GapLH}}
\newcommand{\ModLH}{\mbox{\rm ModLH}}
\newcommand{\boldsharpL}{\mbox{$\textrm{\bf $\mathbf{\sharp}$}$\rm \bf L}}
\newcommand{\boldGapL}{\mbox{\rm \bf GapL}}
\newcommand{\ModpL}{\mbox{\rm Mod$_p$L}}
\newcommand{\ModpLH}{\mbox{\rm Mod$_p$LH}}
\newcommand{\ModkLH}{\mbox{\rm Mod$_k$LH}}
\newcommand{\ModiL}{\mbox{\rm Mod$_i$L}}
\newcommand{\ModjL}{\mbox{\rm Mod$_j$L}}
\newcommand{\ModjkL}{\mbox{\rm Mod$_{jk}$L}}

\newcommand{\ModpeL}{\mbox{\rm Mod$_{p^e}$L}}
\newcommand{\ModpeplusL}{\mbox{\rm Mod$_{p^{e+1}}$L}}

\newcommand{\ModpieiL}{\mbox{\rm Mod$_{p_i^{e_i}}$L}}
\newcommand{\ModponetopmL}{\mbox{\rm Mod$_{p_1p_2\cdots p_m}$L}}
\newcommand{\CeqL}{\mbox{\rm C$_=$L}}

\newcommand{\boldCeqL}{\mbox{\rm \bf C{\bf $_\mathbf{=}$}L}}
\renewcommand{\PL}{\mbox{\rm PL}}
\newcommand{\PLH}{\mbox{\rm PLH}}
\newcommand{\boldPL}{\mbox{\rm \bf PL}}
\newcommand{\coPL}{\mbox{\rm co-PL}}
\newcommand{\coCeqL}{\mbox{\rm co-C$_=$L}}

\newcommand{\ModL}{\mbox{\rm ModL}}
\newcommand{\boldModL}{\mbox{\rm \bf ModL}}
\newcommand{\ModLredn}{\mbox{{\rm ModL}}}
\newcommand{\FL}{\mbox{\rm FL}}
\newcommand{\FLModL}{\mbox{\ensuremath{\FL ^{\ModL}}}}
\newcommand{\FLGapL}{\mbox{\ensuremath{\FL ^{\mbox{\GapL}}}}}
\newcommand{\LModL}{\mbox{\ensuremath{\L ^{\ModL}}}}
\newcommand{\LGapL}{\mbox{\ensuremath{\L ^{\mbox{\GapL}}}}}
\newcommand{\LsharpL}{\mbox{\ensuremath{\L ^{\mbox{\sharpL}}}}}
\newcommand{\LCeqL}{\mbox{\ensuremath{\L ^{\mbox{\CeqL}}}}}

\renewcommand{\mod}{\mbox{\textrm mod~}}
\newcommand{\Z}{\mathbb{Z}}
\renewcommand{\R}{\mathbb{R}}
\newcommand{\Q}{\mathbb{Q}}
\newcommand{\N}{\mathbb{N}}

\newcommand{\F}{\mathcal{F}}
\newcommand{\x}{\mbox{\bf x}}

\renewcommand{\b}{\mbox{\bf b}}

\newcommand{\acc}{\mbox{\it acc}}
\newcommand{\rej}{\mbox{\it rej}}
\newcommand{\gap}{\mbox{\it gap}}
\newcommand{\ctt}{\mbox{{\rm ctt}}}
\newcommand{\dtt}{\mbox{{\rm dtt}}}
\newcommand{\trt}{\mbox{{\rm tt}}}
\newcommand{\LDAG}{\mbox{\rm SLDAG}}
\newcommand{\DSTCON}{\mbox{\rm DSTCON}}
\newcommand{\LDAGSTCON}{\mbox{\rm SLDAGSTCON}}

\newcommand{\FUL}{\mbox{\rm FUL}}
\newcommand{\ssum}{\mbox{{\ensuremath{\Sigma}}}}

\renewcommand{\poly}{\mbox{\rm poly}}

\newcommand{\NLpoly}{\mbox{\ensuremath{\NL{\rm /poly}}}}
\newcommand{\ULpoly}{\mbox{\ensuremath{\UL{\rm /poly}}}}

\begin{document}
\setlength{\parskip}{0cm}
\raggedbottom
\frontmatter

\title{\textbf{THE COMPLEXITY OF LOGARITHMIC\\ SPACE BOUNDED COUNTING CLASSES}}
\author{\bf \scalebox{1}{T. C. VIJAYARAGHAVAN}}
\address{\begin{center}
{\bf (SECOND EDITION)}
\end{center}}

\subjclass[2010]{\emph{Primary:} {68Q10, 68Q15}\vspace{175mm}}


\date{\today}

\begin{abstract}
In this monograph, we study complexity classes that are defined using $O(\log n)$-space bounded non-deterministic Turing machines. We prove salient results of Computational Complexity in this topic such as the Immerman-Szelepcsenyi Theorem, the Isolating Lemma, theorems of Meena Mahajan and V. Vinay on the determinant and many consequences of these very important results. The manuscript is intended to be a comprehensive textbook  on the topic of The Complexity of Logarithmic Space Bounded Counting Classes.\vspace{140mm}\newline\newline\newline\newline\newline
\copyright T. C. Vijayaraghavan. This work is subject to copyright. All rights are reserved by the author of the work.
\end{abstract}

\maketitle

\cleardoublepage
\setcounter{page}{5}

\tableofcontents


\chapter*{Preface to the Second Edition}
The second edition of this monograph was brought about due to a pressing need to remove the incorrect result that $\NL =\CeqL$, which I claimed to be true using a buggy proof in the first edition of this monograph. I am extremely grateful to Jacobo Toran for pointing out this error even before the first edition was published. I however did not understand the argument given by Jacobo Toran all by myself because of keeping sight on other portions of this book. However after publishing the first edition and discussing with Eric Allender who was very patient in explaining the error in my proof, in which I claimed that $\CeqL\subseteq\NL$ and with a bit more diligence on my part, I understood that the result I have claimed is incorrect. Consequences of assuming the statement which are incorrect in various chapters have also been taken care of and corrected to the best of my knowledge.

Eric Allender gave invaluable comments about many other portions of the first edition, especially in pointing out in Chapter 2, the result $\NL /\poly =(\UL\cap\coUL )/\poly$ requires some more refinement that polynomially many weight functions are required as an advice string to isolate a directed path from $s$ to $t$ in the input directed graph, helped me gain clarity and proper understanding of the results on this topic. I am extremely grateful to Eric Allender for careful proof reading of major portions of the second chapter of this book. I once again thank Jacobo Toran and Eric Allender for proof reading many other portions of this book and sending me their valuable comments.

Shortly after publishing the first edition, I also felt the need to give more precision to the introductory material contained in Chapter 1 of this monograph. In Chapter 1, we have included some more basic material starting from definitions of deterministic and non-deterministic Turing machines as in \cite{Koz1997}, refined the notion of a configuration of a space bounded Turing machine by defining mini configuration and succinct configuration, and to state explicitly and justify observations concerning definitions of these two types of configurations. Following this in Section 1.2 of Chapter 1, we have stated certain very useful complexity upper bound results from \cite{CDL2001} based on Boolean circuits and the Chinese remainder representation. Following this, we have included an useful subsection on logarithmic space bounded computation.

Certain important typographical errors in various results or algorithms such as in line 26 of the algorithm of Theorem \ref{chap2-NLcomplement} which existed in the first edition have been corrected. In page 8 of Chapter 1, I have stated that we usually assume that the computation binary tree of a non-deterministic Turing machine $M$ is a complete binary tree. However, I have clarified in page 29 of Chapter 2 that this is not always the case; that is, the computation tree of a non-deterministic Turing machine is not always a complete binary tree since this assumption seems to be unrealistic in the context of some logarithmic space bounded counting classes, especially $\GapL$. We carry forward this word of caution in many places in following chapters of this monograph. Sections 2.3 and 2.4 of Chapter 2 have been thoroughly revised.

Chapter 3 remains almost the same as in the first edition except that a theorem of Kummer and its corollary, the Prime Number Theorem and the Chinese Remainder Theorem which are useful for showing closure properties of $\ModpL$ and $\ModL$ have been explicitly stated. Chapter 4 also remains almost the same.

Chapter 5 has been thoroughly revised with most notable changes being consequences of dropping the incorrect result that $\CeqL\subseteq\NL$. The list of complete problems given in the second edition now separately includes problems that are logspace many-one complete for $\CeqL$. Table 5.1 has been revised and Section 5.9 does not deal with the $\CeqL$ hierarchy or Boolean circuits that have $\CeqL$ oracle gates. Chapter 5 ends with the Notes section which contains Figure 5.1 that shows the landscape of important logarithmic space bounded counting classes. This figure has also been carefully redrawn.

Chapter 6 has undergone major revision in terms of the manner in which we have introduced basic concepts on permutations in Section 6.1. We have introduced certain terminologies such as orbicycle, $2$-orbicycle, and orbicycle cover. I felt that using the same term ``cycle" in the context of permutations and in the context of directed graphs is obviously confusing. To solve this problem, I decided to introduce the notion of a cycle in a permutation and call it as an orbicycle. The notion of a $2$-cycle (also known as a transposition in the first edition) is called as just $2$-orbicycle here in the second edition. An even permutation and an odd permutation are defined based on the parity of the number of $2$-orbicycles. Some more examples are included in Section 6.1. The notion of the sign of a permutation is defined in Definition 6.19 and it is interestingly justified in Theorem 6.21. The notion of an orbicycle cover is related to a cycle cover of a directed graph in Definitions 6.38-6.39 and in Lemma 6.40 in Section 6.2. Rest of Section 6.2 remains the same. Section 6.3 discusses complexity upper bound results on problems in Linear Algebra which follow as a consequence of the logspace many-one completeness of computing the determinant of integer matrices for $\GapL$. The complexity upper bound results have been revised following us noting that $\CeqL\subseteq\NL$ has been incorrectly claimed to be true in the first edition. 

Exercise problems of all chapters have been carefully looked into and changes have been made to correct any error which might have been caused due to incorrect statements in the first edition. Notes of all chapters have been read carefully. Any error which remains in this manuscript is due to me, solely.

It is interesting to note that even though this topic of logarithmic space bounded counting classes is a branch of computational complexity and structural complexity, it borrows notions, ideas and results from many other areas of Mathematics. 
The second edition of this book includes a small Appendix A on Mathematical prerequisites which contains separate sections for Number Theory, Asymptotic notation and Basics of Algebra \& notation.

I am once again indebted to V. Arvind for his guidance and continuous encouragement in helping me prepare this manuscript. Any useful comments regarding this book may be sent to {\tt tcv.tclsbc@gmail.com}.

\noindent {Chennai,}{\rightline{\it T. C. Vijayaraghavan,\hspace{13mm}}}
\noindent{India.}{\rightline{\it March 2026.\hspace{8mm}}}

\chapter*{Preface to the First Edition}
It is a fact that the study of complexity classes has grown enormously and an innumerable number of results have been proved on almost every complexity class that has been defined. A counting class is defined as any complexity class whose definition is based on a function of the number of accepting computation paths and/or the number of rejecting computation paths of a non-deterministic Turing machine. By restricting the space used by a non-deterministic Turing machine to be $O(\log n)$, where $n$ is the size of the input, we can define many logarithmic space bounded counting complexity classes. The first and fundamental logarithmic space bounded counting complexity class that one can easily define is Non-deterministic Logarithmic space ($\NL$). The definition of any other logarithmic space bounded counting complexity class is based on $\NL$.

The purpose of this research monograph is to serve as a textbook for teaching the topic of logarithmic space bounded counting complexity classes and almost all the salient results on them. In this monograph, we introduce the beautiful and sophisticated theory of logarithmic space bounded counting complexity classes. In Chapter 1, we briefly introduce the Turing machine model and the Boolean circuit model of computation. In Chapters 2 to 5 of this monograph, we define logarithmic space bounded counting complexity classes and we prove structural properties of these complexity classes. In particular we study some important results on a number of complexity classes whose definitions are based on Turing machines and which are contained between the circuit complexity classes $\NCone$ and $\NCtwo$. The complexity classes of interest to us are $\NL$, $\sharpL$, $\GapL$, $\CeqL$, $\UL$, $\ModpL$, $\ModkL$, $\ModL$ and $\PL$, where $k,p\in\N$, $k\geq 2$ and $p$ is a prime. Results we prove in Chapters 2 to 5 are diverse in the ideas and techniques involved such as the following:
\begin{itemize}
\item[$\bf\star$] the non-deterministic counting method invented to prove $\NL=\coNL$ which is the Immerman-Szelepcsenyi Theorem in the logarithmic space setting and some of its useful consequences which is to show that $\L^{\mbox{\NL}}=\NL$ \xout{and $\NL=\CeqL$} in Chapter 2,
\item[$\bf\star$] using the Isolating Lemma to show that $\NLpoly =(\UL\cap\coUL ){\rm /poly}$ in Chapter 2,
\item[$\bf\star$] by proving closure properties of $\sharpL$ and $\GapL$, and using results from elementary number theory we prove many interesting closure properties of $\ModpL$ and a characterization of $\ModkL$ in Chapter 3, where $k,p\in\N$, $k\geq 2$ and $p$ is a prime,
\item[$\bf\star$] the double inductive counting method used to prove a combinatorial closure property of $\sharpL$ under the assumption that $\NL =\UL$ in Chapter 2 and its implications for $\ModL$ in Chapter 3, and
\item[$\bf\star$] using polynomials to approximate the sign of a $\GapL$ function and its applications to show closure properties of $\PL$ in Chapter 4 such as the closure of $\PL$ under logspace Turing reductions.
\end{itemize}
In Chapter 5, we list a set of problems which are complete for logarithmic space bounded counting classes under logspace many-one reductions, all of which are based on the results that we have discussed in Chapters 2 to 4. In Chapter 5, we also define hierarchies of logarithmic space bounded counting complexity classes and complexity classes based on Turing reductions that involve Boolean circuits containing oracle gates for various logarithmic space bounded counting complexity classes. We show that these two notions coincide for all logarithmic space bounded counting complexity classes and as a consequence of the results shown in Chapters 2 to 4, some of the hierarchies also collapse to their logarithmic space bounded counting complexity class itself. In Chapter 6 of this monograph, we deal exclusively with one of the very important and useful theorems and its consequences on logarithmic space bounded counting classes that computing the determinant of integer matrices is logspace many-one complete for $\GapL$. We state and explain two very beautiful and deep theorems of M. Mahajan and V. Vinay on the determinant and also show many applications of the logspace many-one completeness of the determinant for $\GapL$ to classify the complexity of linear algebraic problems using $\GapL$ and other logarithmic space bounded counting complexity classes.

Since counting classes are defined based on non-deterministic Turing machines, invariably every theorem statement that intends to prove a property of a logarithmic space bounded counting class which we have covered in this monograph has at least one non-deterministic algorithm which has been either made explicit or explained in an easy to understand manner. We refrain from giving explicit algorithms for logspace many-one reductions since they are routine algorithms computed by deterministic Turing machines. We do not claim uniqueness over the order in which the chapters of this textbook is written since many theorems or statements proved in this monograph may have various proofs depending upon how we introduce this topic and order the results.

As a pre-requisite to understand this monograph, we assume that the student is familiar with computation using Turing machines and has undertaken a basic course on the Theory of Computation in which a proper introduction to complexity classes, reductions, notions of hardness and completeness, and oracles have been given.

I am extremely grateful to my doctoral advisor V. Arvind for the continuous encouragement and support he has given to me in pursuing research ever since I joined the Institute of Mathematical Sciences, C.I.T. Campus, Taramani, Chennai-600113 as a research scholar in Theoretical Computer Science. His invaluable guidance, ever encouraging words and timely help have rescued me from tough situations and helped me shape this research monograph. I am extremely grateful to Jacobo Toran for a gentle proof reading of the results shown in this manuscript. His valuable comments and suggestions have significantly improved the presentation of the results shown in this monograph. I also thank Meena Mahajan for some useful discussions pertaining to the result that computing the determinant of integer matrices is logspace many-one complete for $\GapL$.
 
I am extremely thankful to the higher officials of the Vels Institute of Science, Technology \& Advanced Studies (VISTAS), Pallavaram, Chennai-600117 for their encouragement given to me in preparing this monograph. I also thank my colleagues in the Department of Computer Science and Information Technology, School of Computing Sciences, VISTAS and my friends for many enjoyable discussions while I was preparing this monograph.

I am extremely indebted to my mother for always being prompt and never late in cooking and providing me food everyday and in taking care and showing attention pertaining to any issue of mine if I lacked confidence and direction in tackling it. Any useful comments regarding this book may be sent by email to the email address {\tt ~tcv.tclsbc@gmail.com}.
\vspace{12pt}

\noindent {Chennai,}{\rightline{\it T. C. Vijayaraghavan.\hspace{16mm}}}
\noindent{India.}
\mainmatter

\chapter{Introduction}\label{chap01-introduction}

\section{The Turing machine model of computation}\label{TM-model-of-computation}
In this monograph, every Turing machine that we deal with has a two-way semi-infinite read-only input tape and a two-way semi-infinite read-write work tape. The input tape and the work tape have separate tape heads. Also the input tape and the work tape are assumed to be two-way to mean that corresponding tape heads can move in both left and right directions. These two tapes are assumed to be semi-infinite to mean that there are only finitely many tape cells on the left-hand side and both these tapes have infinitely many cells on the right-hand side.

Using the input tape head, the Turing machine can only read symbols in the input tape which are from the input alphabet $\Sigma$. Also using the work tape head, the Turing machine can read and write symbols from the output alphabet $\Gamma$ on the work tape. We assume that both the input tape and the work tape have a de-limiter ($\vdash$) in the leftmost tape cell which should prevent tape heads from moving any further to the left. The input tape is assumed to have a de-limiter ($\dashv$) in the tape cell which is to the right of the tape cell containing the last symbol of the input string $x$. The input tape head is assumed not to move to the right of the right most or the last symbol of $x$, because it is prevented by the de-limiter ($\dashv$). We follow the standard assumption that the $\vdash, \dashv$ and the blank symbol $\sqcup\not\in\Sigma$, where $\Sigma$ is the input alphabet. In every Turing machine, it is assumed that $\Sigma\cup\{\sqcup ,\vdash ,\dashv\}\subseteq\Gamma$.

We also assume that the accepting state ($q_{acc}$) and the rejecting state ($q_{rej}$) of our Turing machine are unique among its set of states. Any Turing machine decides to accept or reject an input string depending on whether it enters its accepting state or rejecting state. The Turing machine stops its computation when it enters the unique accepting state or the unique rejecting state.\textcolor{white}{\index[subject]{accepting state}\index[subject]{rejecting state}} Given $L\subseteq \Sigma^*$, we say that a Turing machine accepts $L$ if and only if $M$ accepts exactly those strings in $L$.

\begin{figure}[h!]
\begin{center}
\includegraphics[scale=0.92]{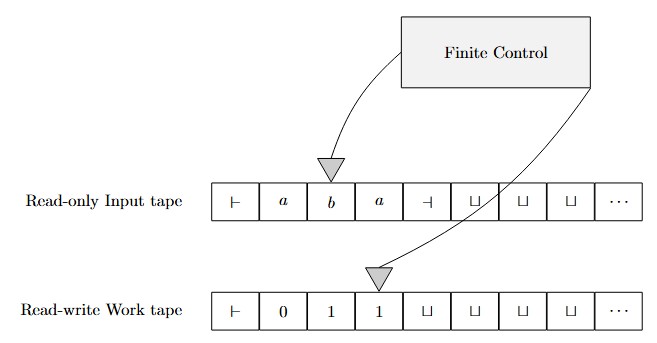}
\caption{A Turing machine.}
\end{center}
\end{figure}

\begin{definition}\label{chap1-defn-DTM}
We formally define a deterministic Turing machine as $M=(Q,\Sigma ,\Gamma ,\vdash ,\dashv ,\sqcup ,\delta ,q_0,F)$, where \textcolor{white}{\index[subject]{deterministic Turing machine}}
\begin{enumerate}
    \item $Q$ denotes the set of states of $M$,
    \item $\Sigma$ is the input alphabet,
    \item $\vdash$ and $\dashv$ are de-limiters on the left-hand side and on the right-hand side of the input string on the input tape; $\vdash$ is the de-limiter on the left-hand side of the work tape,
    \item $\sqcup$ is the blank symbol,
    \item $\Gamma =\Sigma\cup\{ \vdash ,\dashv ,\sqcup\}$ denotes the output alphabet,    
    \item $q_0\in Q$ is the initial state,
    \item $F=\{ q_{acc},q_{rej}\}\subseteq Q$ is the set of final states of the Turing machine, where $q_{acc}$ denotes the unique accepting state and $q_{rej}$ denotes the unique rejecting state, and
    \item $\delta$ is the transition function of $M$, defined as $\delta :\{Q\setminus F\}\times\Gamma\times\Gamma\rightarrow Q\times \Gamma\times\{ L,R\}\times \Gamma \times \{ L,R\}$.
\end{enumerate}
\end{definition}
\begin{definition}\label{chap1-defn-NDTM}
We formally define a non-deterministic Turing machine as $M=(Q,\Sigma ,\Gamma ,\vdash ,\dashv ,\sqcup ,\delta ,q_0,F)$, where \textcolor{white}{\index[subject]{non-deterministic Turing machine}}
\begin{enumerate}
    \item $Q$ denotes the set of states of $M$,
    \item $\Sigma$ is the input alphabet,
    \item $\vdash$ and $\dashv$ are de-limiters on the left-hand side and on the right-hand side of the input string on the input tape; $\vdash$ is the de-limiter on the left-hand side of the work tape,
    \item $\sqcup$ is the blank symbol,
    \item $\Gamma =\Sigma\cup\{ \vdash ,\dashv ,\sqcup\}$ denotes the output alphabet,    
    \item $q_0\in Q$ is the initial state,
    \item $F=\{ q_{acc},q_{rej}\}\subseteq Q$ is the set of final states of the Turing machine, where $q_{acc}$ denotes the unique accepting state and $q_{rej}$ denotes the unique rejecting state, and
    \item $\delta$ is the transition function of $M$, defined as $\delta :\{Q\setminus F\}\times\Gamma\times\Gamma\rightarrow 2^{Q\times \Gamma\times\{ L,R\}\times \Gamma \times \{ L,R\}}$, where $2^{Q\times \Gamma\times\{ L,R\}\times \Gamma \times \{ L,R\}}$ denotes the set of all subsets of $Q\times \Gamma\times\{ L,R\}\times \Gamma \times \{ L,R\}$.
\end{enumerate}
\end{definition}
The state of a Turing machine along with the transition function is collectively called as the finite control of the Turing machine. Since any Turing machine does no further computation after entering its accepting state or its rejecting state both these states are called as \emph{halting states}.\textcolor{white}{\index[subject]{halting states}}

\begin{definition}
We say that a Turing machine $M$ is an $O(S(n))$-space bounded Turing machine if the number of tape cells of the work tape that are being read or written by the tape head of the work tape of $M$ when it is given an input of length $n$ is $O(S(n))$, where $S(n)\in\Omega (\log n)$.\textcolor{white}{\index[subject]{$O(S(n))$-space bounded Turing machine}}
\end{definition}
\begin{definition}\label{chap1-defn-L}
We define the complexity class $\L$ to be the class of all languages that are accepted by a $O(\log n)$-space bounded determinsitic Turing machine.\textcolor{white}{\index[subject]{L}}
\end{definition}

\subsection{Configuration}\label{sec-configuration}\textcolor{white}{\index[subject]{configuration}\index[subject]{mini configuration}\index[subject]{mini initial configuration}}
Given a Turing machine $M$, the configuration of $M$ is a snapshot of the computation in progress. It precisely gives information about the Turing machine which includes the state of the Turing machine, contents of the input tape and the work tape. In this monograph, since we deal exclusively with space bounded computation and any Turing machine is $O(S(n))$-space bounded, where $S(n)\in\Omega (\log n)$, we define the mini configuration of $M$ first as follows.
\begin{definition}({\bf mini configuration})\label{chap1-defn-miniconfiguration}
Let $\Sigma$ be the input alphabet, $\Gamma$ be the output alphabet and let $M$ be a Turing machine. Given an input string $x\in\Sigma^*$, we define the mini configuration of $M$ on $x$ as the tuple $(q,s_1,p_i,s_2s_3,t_1,p_w,t_2t_3)$, where $q\in Q$ denotes the state of $M$, $s_1,s_2,s_3\in\Sigma$, and $t_1,t_2,t_3\in\Omega$. The input tape head scans the $p_i^{th}$ cell after the left-most cell which is the first cell containing the de-limiter $\vdash$ of the input tape. The cell which is scanned has the symbol $s_2$ with the symbol $s_1$ in the preceding cell and the symbol $s_3$ in the next cell of the input tape. Here $p_i$ is a positive integer which is $\leq n$, where $n$ is the size of the input. This configuration also indicates that the work tape head scans the $p_w^{th}$ cell of the work tape after the left most cell which is the first cell of the work tape containing the de-limiter $\vdash$. This cell has the symbol $t_2$ with the symbol $t_1$ in the preceding cell and $t_3$ in the next cell of the work tape. Similar to $p_i$, here $p_w$ is a positive integer which is $\leq k\cdot S(n)$, where $k$ is a constant, $O(S(n))$ is the space bound of the Turing machine and $n$ is the size of the input. 
\end{definition}
\begin{definition}({\bf mini initial configuration})\label{chap1-defn-initialminiconfiguration}
Let $\Sigma$ be the input alphabet. Given an input string $x=s_1s_2\cdots s_n\in\Sigma^*$ to a Turing machine $M$, the mini initial configuration of $M$ is $(q_0,\vdash ,1,s_1s_2,\vdash ,1,\sqcup \sqcup)$.
\end{definition}
\begin{proposition}\label{chap1-prop-miniconfspacereqd}
Let $S(n)\in \Omega (\log n)$. The amount of space required to store a mini configuration of a $O(S(n))$-space bounded Turing machine is $O(S(n))$.
\end{proposition}
\begin{proof}
It follows from Definitions \ref{chap1-defn-DTM}, \ref{chap1-defn-NDTM} and \ref{chap1-defn-miniconfiguration} that the amount of space required to store a mini configuration of $O(S(n))$-space bounded Turing machine is $max(O(\log n),O(S(n))=O(S(n))$. 
\end{proof}
Depending on the transition function of the Turing machine, it changes its state, tape heads of the input tape and the work tape move in exactly one of the either directions by exactly one tape cell and only the contents of the work tape cell scanned by its tape head are modified. As a result, the Turing machine will change from the present mini configuration to another mini configuration based on its state and the transition function.
\begin{proposition}\label{chap1-prop-config-count}
Let $S(n)\in \Omega (\log n)$. The number of mini configurations of a $O(S(n))$-space bounded Turing machine is upper bounded by $O(n\cdot S(n))$.
\end{proposition}
\begin{proof}
From Definitions \ref{chap1-defn-DTM}, \ref{chap1-defn-NDTM} and \ref{chap1-defn-miniconfiguration} it is easy to see that the number of mini configurations that can exist is upper bounded by $k\cdot |Q|\cdot |\Gamma |^3\cdot n\cdot |\Gamma |^3\cdot S(n)=k\cdot |Q|\cdot |\Gamma |^6\cdot n\cdot S(n)$, where $S(n)\in\Omega (\log n)$, $n$ is the length of the input and $k>0$ is a constant. This is because the state of the Turing machine contributes to the factor $|Q|$ in the product, the symbols of the input tape $s_1,s_2,s_3\in\Gamma$ and symbols $t_1,t_2,t_3\in\Gamma$ of the work tape contribute to factor $|\Gamma |^6$ in the product, and the position of the tape heads on the respective tapes contributes to a factor of $k\cdot n\cdot S(n)$ to the product, where $k>0$ is a constant.
\end{proof}

Let $S(n)\in\Omega (\log n)$. For any non-deterministic Turing machine $M$ that uses at most $O(S(n))$ space, since there exists an upper bound on the number of mini configurations of $M$ on any given input $x\in\Sigma^*$, we assume without loss of generality that $M$ halts on all inputs either in the unique accepting state or in the unique rejecting state. We also assume without loss of generality that when a Turing machine has arrived at a decision regarding the membership of the input string $x=s_1s_2\cdots s_n$ in a language $L$, before it enters the accepting state or the rejecting state it does a clean up of the work tape contents. In other words, it replaces the contents of the work tape with the symbol $\sqcup$ in all cells by entering a special state called as the \emph{clean-up state} and brings its tape heads to the left most cell of the input tape and the work tape. Following this, our Turing machine enters its unique accepting state ($q_{acc}$) or its unique rejecting state ($q_{rej}$).\textcolor{white}{\index[subject]{clean-up state}}
\begin{definition}({\bf mini accepting and mini rejecting configuration})\label{chap1-defn-acceptrejectminiconfig}
Let $\Sigma$ be the input alphabet of a Turing machine $M$. The mini accepting configuration of $M$ is $(q_{acc},\vdash ,1,s_1s_2,\vdash ,1,\sqcup\sqcup )$ and the mini rejecting configuration of $M$ is $(q_{rej},\vdash , 1,s_1s_2,\vdash ,1,\sqcup\sqcup )$, where $s_1,s_2\in\Sigma$.\textcolor{white}{\index[subject]{mini accepting configuration}\index[subject]{mini rejecting configuration}}
\end{definition}

\begin{tcolorbox}[colback=gray!35!white,colframe=white]
{\it\textbf{Note that unlike the configuration of a polynomial time bounded Turing machine introduced in standard texts, we do not include the entire contents of the input tape or the work tape as part of the mini configuration of our space bounded Turing machine.}}
\end{tcolorbox}

Since we modify only the contents of the work tape of a Turing machine, let us try to relax this stringent condition by including entire contents of only the work tape in the configuration of a $O(S(n))$-space bounded Turing machine which we call as a \emph{succinct configuration}, where $S(n)\in\Omega (\log n)$.
\begin{definition}({\bf succinct configuration})\label{chap1-defn-configuration}\textcolor{white}{\index[subject]{succinct configuration}}
Let $\Sigma$ be the input alphabet and $\Gamma$ be the output alphabet of a Turing machine $M$. Given an input $x=s_1s_2\cdots s_n\in\Sigma^*$, we define a succinct configuration of $M$ as $(q,s_1,p_i,s_2s_3,x_{w1},p_w,x_{w2})$, where $q\in Q$ is a state of $M$, $s_1,s_2,s_3\in\Sigma$ and $1\leq p_i\leq n$ is a positive integer, and $n$ is the size of the input. The input tape head scans the $p_i^{th}$ cell after the left-most cell which is the first cell containing the de-limiter $\vdash$ of the input tape. The cell which is scanned has the symbol $s_2$ in it with $s_1$ in the preceding cell and $s_3$ in the next cell of the input tape. The work tape of the succinct configuration contains the string $x_{w1}x_{w2}\in\Gamma^*$ and the work tape head scans the $p_w^{th}$ cell after the left most cell of the work tape. The left most cell of the work tape is the first cell of the work tape containing the de-limiter $\vdash$ and the symbol of the work tape scanned by $M$ is the first symbol of $x_{w2}$. Similar to $p_i$, here $1\leq p_w\leq k\cdot S(n)$ is a positive integer, where $k$ is a constant, $O(S(n))$ is the space bound on the Turing machine and $n$ is the size of the input.
\end{definition}
\begin{definition}({\bf succinct initial configuration})\label{chap1-defn-initialconfiguration}\textcolor{white}{\index[subject]{succinct initial configuration}}
Let $\Sigma$ be the input alphabet and let $S(n)\in\Omega (\log n)$. Given an input string $x=s_1s_2\cdots s_n\in\Sigma^*$ to a Turing machine $M$, the succinct initial configuration of $M$ is $(q_0,\vdash ,1,s_1s_2,\vdash ,1,\underbrace{\sqcup\cdots \sqcup}_{O(S(n))})$.
\end{definition}
\begin{definition}({\bf succinct accepting and succinct rejecting configuration})\label{chap1-defn-acceptrejectconfig}\textcolor{white}{\index[subject]{succinct accepting configuration}\index[subject]{succinct rejecting configuration}}
Let $\Sigma$ be the input alphabet, and let $\Gamma$ be the output alphabet of a Turing machine $M$. The succinct accepting configuration of $M$ and the succinct rejecting configuration of $M$ are $(q_{acc},\vdash ,1,s_1s_2,\vdash ,1,\underbrace{\sqcup\cdots\sqcup}_{O(S(n))})$ and $(q_{rej},\vdash , 1,s_1s_2,\vdash ,1,\underbrace{\sqcup\cdots\sqcup}_{O(S(n))})$ respectively, where $s_1,s_2\in\Sigma$.
\end{definition}

As a result of Definition \ref{chap1-defn-configuration}, we infer the following which is similar to Propositions \ref{chap1-prop-miniconfspacereqd} and \ref{chap1-prop-config-count}.
\begin{proposition}\label{chap1-prop-config-size-count-2}
Let $\Sigma$ be the input alphabet, $\Gamma$ be the output alphabet of a $O(S(n))$-space bounded Turing machine $M$, where $S(n)\in \Omega (\log n)$. $M$ requires at most $O(S(n))$ space to store its succinct configuration on the work tape and the number of succinct configurations is upper bounded by $O(n\cdot |\Gamma|^{S(n)})=O(|\Gamma|^{S(n)})=|\Gamma |^{O(S(n))}$. Clearly the number of mini configurations of $M$ is less that or equal to the number of succinct configurations of $M$.
\end{proposition}
\begin{tcolorbox}[colback=gray!35!white,colframe=white]
{\it\textbf{From now onwards, in all chapters to follow, we assume that we uniformly deal either with mini configurations or succinct configurations of $O(\log n)$-space bounded Turing machines, and not both of them. We however call both a mini configuration or a succinct configuration as just a {\bf \emph{``configuration"}}}}.
\end{tcolorbox}
As result of Propositions \ref{chap1-prop-miniconfspacereqd}, \ref{chap1-prop-config-count} and \ref{chap1-prop-config-size-count-2}, for the purpose of results shown in this monograph, irrespective of whether we consider a mini configuration or a succinct configuration, we note the following observation.
\begin{theorem}\label{chap1-thm-config-size-count-NL}
Let $\Sigma$ be an input alphabet and let $M$ be a deterministic or a non-deterministic Turing machine that accepts a language $L\subseteq\Sigma^*$. The size of a mini configuration or a succinct configuration of $M$ is $O(\log n)$ and the number of mini configurations or succinct configurations of $M$ is upper bounded by a polynomial $n^k$ on any given input $x\in\Sigma^*$, where $n=|x|$ and $k>0$ is a constant.
\end{theorem}
Once our Turing machine enters its accepting state or rejecting state, it halts and does no further computation. Since the Turing machine halts and does no further computation after it enters the accepting state or the rejecting state, we say that the accepting and rejecting configurations are {\bf\emph{halting configurations}}\textcolor{white}{\index[subject]{halting configuration}}.
\subsection{Computation tree}\label{chap1-sec-comptree}\textcolor{white}{\index[subject]{computation tree}}
Let $S(n)\in\Omega (\log n)$. It is easy to observe that the computation of a $O(S(n))$-space bounded non-deterministic Turing machine $M$ on a given input is in fact a tree which we call the computation tree of $M$. Since there is an upper bound on the number of configurations of $M$, the running time of $M$ is at most $|\Gamma |^{O(S(n))}$. Since the running time of $M$ can be taken to be the length of the longest computation path of $M$, its computation tree is finite and it contains finite number of nodes and edges.
\begin{fact}
Let $S(n)\in\Omega(\log n)$ and let $M$ be $O(S(n))$-space bounded non-determinsitic Turing machine. Given any input $x\in\Sigma^*$, in the computation of $M$ on input $x$ we assume that no configuration gets repeated. Since otherwise there will exist a path in the computation tree of $M$ on $x$ which extends infinitely long and this contradicts the fact that the computation tree contains finitely many nodes and edges.    
\end{fact}
\begin{proposition}
    Let $\Sigma$ be the input alphabet. Let $M$ be a non-deterministic Turing machine. Given an input $x\in\Sigma^*$, let $M$ be in a configuration $c$ during its computation. The number of configurations to which $M$ can move from $c$ is a constant.
\end{proposition}
\begin{proof}
Given a mini configuration $c$ of $M$, it is easy to note that the number of mini configurations which we can obtain in one step from $c$ based on the transition function of $M$ is upper bounded by $|Q|\cdot 2\cdot |\Gamma |\cdot 2\cdot |\Gamma |^2=4\cdot |Q|\cdot |\Gamma |^3$ which is a constant. 

Since we define a succinct configuration in Definition \ref{chap1-defn-configuration} as an extension of a mini configuration, in one transition on a given input, $M$ can move from the present succinct configuration to at most $4\cdot |Q|\cdot |\Gamma |^3$ succinct configurations, which is also a constant.
\end{proof}
It is also possible to restrict the possibility of moving from one configuration to more than $2$ configurations in a transition, to less than or equal to $2$ configurations by modifying the definition of the transition function of $M$. We can do this by introducing new states in $Q$, new symbols in the output alphabet $\Gamma$ and new transitions in $\delta$.

For example, in the Turing machine $M$, if we have the transition $\delta (q,s,t)=\{(q_1,s,R,t_1,R),(q_2,s,R,t_2,R),(q_3,s,R,t_3,R)\}$ then we introduce new states $p_1,p_1'\in Q$ and a new symbol $\hat{t}\in\Omega$ and modify transitions of $M$ such that,
\begin{itemize}
    \item $\delta (q,s,t)=\{(q_1,s,R,t_1,R),(p_1,s,R,\hat{t},R)\}$, \item $\delta (p_1,s',t')=\{(p_1',s',L,t',L)\}$
    \item $\delta (p_1',s,\hat{t})=\{(q_2,s,R,t_2,R),(q_3,s,R,t_3,R)\}$
\end{itemize}

Therefore, treating any configuration as a node $u$ in the computation tree of $M$ such that $u$ has $k$ configurations $v_1,\ldots ,v_k$ as children connected by directed edges $(u,v_i)$, where $k$ is a constant and $1\leq i\leq k$, it is possible to convert this $k$-ary tree of depth $1$ into a binary tree with $u$ as the root by introducing new intermediate vertices to the $k$-ary tree such that $v_1,\ldots ,v_k$ are the leaf nodes of this binary tree with $u$ as the root. The depth of this binary tree is at least $\log k$ and it is at most $k$ and its size is also a constant. 

As a result, it is possible to assume that whenever $M$ makes a non-deterministic choice in its computation, the number of branches in the computation tree is $2$. We therefore obtain the following.
\begin{proposition}
Let $M$ be a non-determinsitic Turing machine. The computation tree of $M$ is a binary tree. It contains finitely many nodes and edges. Every node of the computation binary tree of $M$ denotes a non-deterministic choice of $M$. Each node is either an internal node which has exactly $2$ children or the node is a leaf.
\end{proposition}
\begin{fact}
If $M$ is a non-deterministic Turing machine that requires at most $O(\log n)$ space, where $n$ is the size of the input, then the depth of its computation binary tree of $M$ denotes the running time of $M$. The length of a computation path from the root to a leaf is upper bounded by the number of configurations of $M$ which is a polynomial in $n$. The number of nodes in the computation tree of $M$ is exponential in $n$.
\end{fact}
\begin{figure}[h!]
\begin{center}
\includegraphics[scale=1.0]{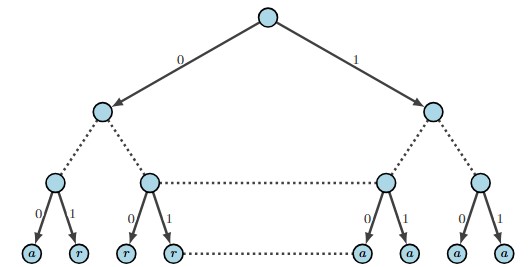}
\caption{This figure is the computation tree of a non-deterministic Turing machine $M$, where $0$ and $1$ on the edges denote the two non-deterministic choices using which $M$ branches at every stage on its computation path. Leaf nodes are marked as ``$a$" or ``$r$" denoting that the computation path ends in an accepting configuration or in a rejecting configuration respectively. Also all computation paths are of the same length.}
\end{center}
\end{figure}

In the computation binary tree of a $O(S(n))$-space bounded non-deterministic Turing machine, we usually assume that all computation paths of $M$ on a given input have the same length. That is, the computation proceeds for the same number of steps along any computation path in the computation tree of a non-deterministic Turing machine, whether it is an accepting computation path or a rejecting computation path. Since in any computation binary tree that we consider, there are no internal nodes which have only one child, if we assume that the length of a directed path from the root to any leaf are equal then we obtain that the tree is in fact a complete binary tree. Note that the converse of this statement is also true: in a complete binary tree the length of a directed path from the root to any leaf is the same. As a result, we usually assume that the computation binary tree of a non-deterministic Turing machine $M$ is a complete binary tree.

The topic of logarithmic space bounded counting complexity classes is dependent on studying the computation binary tree of $\NL$-Turing machines. On a given input $x\in\Sigma^*$, much more than following the computation of a $\NL$-Turing machine $M$ along a computation path, we prove many interesting properties of logarithmic space bounded counting complexity classes by considering an aerial view of the entire computation tree of $M$ on $x$ and designing non-deterministic $O(\log n)$ space algorithms based on observing the number of accepting computation paths and the number of rejecting computation paths of $M$ on $x$.
\subsection{Computing functions}\label{functionoracleaccess}
Let $g:\Sigma^*\rightarrow(\Gamma\setminus\{\sqcup ,\vdash ,\dashv\})^*$. Given $x\in\Sigma^*$, in order to output $g(x)$, we assume that the Turing machine has a separate two-way semi-infinite write-only output tape which has a separate output tape head that can move in both left and right directions, and which is used by the Turing machine to write $g(x)$ as the output on the output tape. As in the case of the work tape, we assume that the output tape has the de-limiter symbol $\vdash$ in the first cell of the output tape. The transition function $\delta$ is defined such that the tape head of the output tape does not move to the left of this first cell of the output cell. The definition of the configuration of any such Turing machine or the notion of its computation tree is the same as a Turing machine that does not have any output tape as described in Sections \ref{sec-configuration} and \ref{chap1-sec-comptree}.

Alternately, we may also consider computing $g(x)$ as a language in a symbol-wise manner as follows. We say that $g(x)$ is computable by $O(S(n))$-space bounded Turing machine $M$, if given an input $x\in\Sigma^*$, we check if the $i^{th}$ symbol of $g(x)$ is $a\in\Gamma\setminus\{\sqcup\}$ by submitting the triple $(x,i,a)$ as an input to $M$ and finding out if $M$ halts in the accepting configuration or in the rejecting configuration in its computation, where $1\leq i\leq |g(x)|$. Therefore by submitting various input instances $(x,i,a)$ to $M$ for different values of $a\in\Gamma$, it is possible to find out the correct value of the $i^{th}$ symbol of $g(x)$, given $x\in\Sigma^*$. Clearly there is no need for any output tape in these Turing machines.

\subsection{Oracle Turing machines}\label{chap1-sec-oracleturingmachine}\textcolor{white}{\index[subject]{oracle Turing machine}}
If $M^A$ is a $O(S(n)$-space bounded Turing machine that has access to an oracle $A$ then $M^A$ has a separate semi-infinite two-way write-only oracle tape using which it can submit queries to the oracle $A$. As in the case of the input tape and the work tape of $M^A$, the oracle tape of $M^A$ is assumed to be semi-infinite, which means that there are only finitely many tape cells to the left and that this tape also has infinitely many cells on the right side. The oracle tape of $M^A$ has a separate two-way tape head, which means that it can move in both left and right directions. We assume that the oracle tape also has a de-limiter ($\vdash$) in a cell to the left of the first cell to prevent the oracle tape head from moving to the left of the first cell of the oracle tape. Also $M^A$ has three exclusive states $q_{QUERY},q_{YES}$ and $q_{NO}$. When $M^A$ completes writing an oracle query string $y\in\Sigma^*$ in the oracle tape of $A$, it puts the de-limiter ($\dashv$) in the last cell (the $|y|+1$ cell) and enters the $q_{QUERY}$ state. In the next instant the oracle $A$ erases the contents of the oracle tape, brings the oracle tape head to the left most cell of the oracle tape and changes the state of $M^A$ from $q_{QUERY}$ to either $q_{YES}$ or $q_{NO}$ depending on whether $y\in A$ or $y\not\in A$ respectively. A configuration of an oracle Turing machine is denoted by the tuple $(q,s_1,p_i,s_2s_3,t_1,p_w,t_2t_3,p_{or})$, where $q,s_1,s_2,s_3,t_1,t_2,t_3,p_i,p_w$ are as in the definition of a configuration of a Turing machine. Here, $p_{or}$ is a positive number that indicates the position of the oracle tape head on the oracle tape. It is upper bounded by $|\Gamma|^{O(S(n))}$. Clearly the amount of space required to store a configuration of a $O(S(n))$-space bounded oracle Turing machine is also $O(S(n))$, where $S(n)\in\Omega (\log n)$. Therefore if $M^A$ is a $O(S(n))$ space bounded non-deterministic Turing machine then the number of configurations of $M^A$ is also upper bounded by $|\Gamma|^{O(S(n))}$. From this we infer that the length of any query to the oracle $A$ and the number of queries that $M^A$ can submit to the oracle $A$ are both at most $|\Gamma|^{O(S(n))}$.

\subsection{Functions as oracles to Turing machines}\label{functionoracleaccess2}
Let $\Sigma$ denote the input alphabet and let $\Gamma =\Sigma \cup \{\sqcup ,\vdash ,\dashv, \#\}$ be the output alphabet. Instead of a language $A$, if the Turing machine $M^f$ has access to a function $f:\Sigma^*\rightarrow \Sigma^*$ as an oracle then $M^f$ first computes an upper bound on the length of any query string that it submits to the oracle $f$. It also computes an upper bound on the length of the value of the function $f$. Given an input $x$ of length $n$, if $M^f$ needs to submit a query to the oracle $f$, it computes the query string $y$ bit-by-bit and writes it on the oracle tape. $M^f$ actually submits the string $(y\# j\# b\# )$ to the oracle $f$ and enters the $q_{QUERY}$ state, where $1\leq j\leq m$ denotes the index of the string $f(y)$ that $M^f$ wants to verify if it is $b$.  In the next instant the oracle $A$ erases the contents of the oracle tape, brings the oracle tape head to the left most cell of the oracle tape and changes the state of $M^A$ from $q_{QUERY}$ to either $q_{YES}$ or $q_{NO}$ depending on whether $f(y)_j=b$ or not.

\subsection{Ruzzo-Simon-Tompa oracle access mechanism}\label{chap1-NOTM}
Let us consider a non-deterministic oracle Turing machine $M^A$ with access to a language $A$ as an oracle. We then follow the \emph{Ruzzo-Simon-Tompa oracle access mechanism} in\textcolor{white}{\index[subject]{Ruzzo-Simon-Tompa oracle access mechanism}} which we assume that $M^A$ always submits its queries to oracle $A$ in a deterministic manner only. This means that all the queries of $M^A$ are submitted to $A$ after doing some computation and even before the first non-deterministic choice has been made in the computation on the given input. All the algorithmic results involving non-deterministic oracle Turing machines in this monograph follow the Ruzzo-Simon-Tompa oracle access mechanism.

\section[Boolean circuit model of computation]{Boolean circuit model of computation}\label{chap1:defnBooleancktsLSBCC}
\begin{definition}\label{chap1-bddfaninbasis-defn-booleanbasis}
We define $\mathcal{B}_0=\{\neg ,(\wedge ^2),(\vee ^2)\}$ as the standard bounded fan-in basis of Boolean gates.\textcolor{white}{\index[subject]{standard bounded fan-in basis, $\mathcal{B}_0$}}
\end{definition}
\begin{definition}\label{chap1-sdbddfaninbasis-defn}
Let $\mathcal{B}_0$ be a standard bounded fan-in basis, and let $s,d:\N\rightarrow\N$. We define the complexity class, $\SD_{\mathcal{B}_0}(s,d)$ as the class of all sets $A\subseteq \{ 0,1\}^*$ for which there is a circuit family $(C_n)_{n\in\N}$ over the basis $\mathcal{B}_0$ of size $O(s)$ and depth $O(d)$ that accepts $A$.\textcolor{white}{\index[subject]{$\SD_{\mathcal{B}_0}(s,d)$}}
\end{definition}
\begin{definition}\label{chap1-unbfaninbasis-defn-booleanbasis}
We define $\mathcal{B}_1=\{\neg ,(\wedge ^n)_{n\in\N},(\vee ^n)_{n\in\N}\}$ as the standard unbounded fan-in basis of Boolean gates.\textcolor{white}{\index[subject]{standard unbounded fan-in basis, $\mathcal{B}_1$}}
\end{definition}
\begin{definition}\label{chap1-sdunbfaninbasis-defn}
Let $\mathcal{B}_1$ be a standard unbounded fan-in basis, and let $s,d:\N\rightarrow\N$. We define the complexity class, $\SD_{\mathcal{B}_1}(s,d)$ as the class of all sets $A\subseteq \{ 0,1\}^*$ for which there is a circuit family $(C_n)_{n\in\N}$ over the basis $\mathcal{B}_1$ of size $O(s)$ and depth $O(d)$ that accepts $A$.\textcolor{white}{\index[subject]{$\SD_{\mathcal{B}_1}(s,d)$}}
\end{definition}

\begin{definition}\label{chap1-admissibleencoding}
Let $(C_n)_{n\in\N}$ be a circuit family over a basis $\mathcal{B}$. An \emph{admissible encoding scheme} of the circuit family is defined as follows: First fix an arbitrary numbering of the elements of $\mathcal{B}$. Second, for every $n$, fix a numbering of the gates of $C_n$ with the following properties:\textcolor{white}{\index[subject]{admissible encoding scheme}}
\begin{itemize}
    \item in $C_n$ the input gates are numbered $0,\ldots ,n-1$,
    \item if $C_n$ has $m$ output gates then these are numbered $n,\ldots ,n+m-1$,
    \item let $s(n)$ be the size of $C_n$. There is a polynomial $p(n)$ such that, for every $n$, the highest number of a gate in $C_n$ is bounded by $p(s(n))$.
\end{itemize}
The encoding of a gate $v$ in $C_n$ is now given by a tuple $\langle g,b,g_1,\ldots ,g_k\rangle$, where $g$ is the gate number assigned to $v$, $b$ is the number of $v$ in the basis $\mathcal{B}$, and $g_1,\ldots ,g_k$ are the numbers of the predecessor gates of $v$ in the order of the edge numbering of $C_n$. Fix an arbitrary order of the gates of $C_n$. Let $v_1,v_2\ldots ,v_s$ be the gates of $C_n$ in that order. Let $\overline{v_1},\overline{v_2},\ldots ,\overline{v_s}$ be their encodings. The admissible encoding of $C_n$, denoted by $\overline{C_n}$, is defined as $\langle \overline{v_1},\dots ,\overline{v_s}\rangle$.
\end{definition}
We recall the definition of $\L$ from Definition \ref{chap1-defn-L}.
\begin{definition}\label{chap1-luniformcircuitfamily}
A circuit family $(C_n)_{n\in\N}$ of size $p(n)$ is $\L$-uniform (also called as \emph{logspace-uniform}) if there is an admissible encoding scheme $\overline{C_n}$ of $C_n$ such that the map $1^n\rightarrow \overline{C_n}$ is computable by a deterministic Turing machine using space $O(\log n)$, for all $n\geq 1$, where $p(n)$ is a polynomial in $n$ and $n$ is the size of the input.\textcolor{white}{\index[subject]{$\L$-uniform}index[subject]{logspace uniform}}
\end{definition}

\begin{definition}\label{chap1-aczero-defn}
Let ${\rm U_L}$-$\ACzero$ denote $\L$-uniform $\ACzero$. The complexity class ${\rm U_L}$-$\ACzero=\L$-uniform $\SD_{\mathcal{B}_1}(p(n),O(1))$, where $p(n)$ is a polynomial in $n$, and $n$ is the size of the input.\textcolor{white}{\index[subject]{$\L$-uniform $\ACzero$}\index[subject]{${\rm U_L}$-$\ACzero$}}
\end{definition}

\begin{definition}\label{chap1-ncone-defn}
Let ${\rm U_L}$-$\NCone$ denote $\L$-uniform $\NCone$. The complexity class ${\rm U_L}$-$\NCone =\L$-uniform $\SD_{\mathcal{B}_0}(p(n),O(\log$ $n))$, where $p(n)$ is a polynomial in $n$, and $n$ is the size of the input.\textcolor{white}{\index[subject]{$\L$-uniform $\NCone$}\index[subject]{${\rm U_L}$-$\NCone$}}
\end{definition}

As we have it in Definition \ref{chap1-unbfaninbasis-defn-booleanbasis}, let $\mathcal{B}_1$ denote the standard unbounded fan-in basis of Boolean gates. In this definition, along with the standard Boolean operations, it is easy to include oracle gates that compute a function. An oracle gate that is of special interest to us is the $\MAJ$ gate (majority gate) defined as follows:\newline
\noindent{\bf MAJ}\textcolor{white}{\index[subject]{MAJ}}\newline
INPUT: $n$ bits $a_{n-1},\ldots ,a_0$.\newline
QUESTION: Are at least half of the $a_i$ one?

\begin{definition}\label{chap1-TCzero-defn}
Let ${\rm U_L}$-$\TCzero$ denote $\L$-uniform $\TCzero$. The complexity class ${\rm U_L}$-$\TCzero=\L$-uniform $\SD_{\mathcal{B}_1\cup {\rm MAJ}}(n^{O(1)},O(1))$.\textcolor{white}{\index[subject]{$\L$-uniform $\TCzero$}\index[subject]{${\rm U_L}$-$\TCzero$}}
\end{definition}
\begin{definition}\label{chap1-TCone-defn}
Let ${\rm U_L}$-$\TCone$ denote $\L$-uniform $\TCone$. The complexity class ${\rm U_L}$-$\TCone=\L$-uniform $\SD_{\mathcal{B}_1\cup {\rm MAJ}}(n^{O(1)},\log n)$.\textcolor{white}{\index[subject]{$\L$-uniform $\TCone$}\index[subject]{${\rm U_L}$-$\TCone$}}
\end{definition}

A \emph{Chinese remainder representation} (CRR) of an integer $x\geq 0$ is based on a set $m_1,\ldots ,m_n$ of pairwise coprime integers. The set $m_1,\ldots ,m_n$ is called the CRR base and each $m_i$ is called a modulus. We will denote this system by CRR($M$). Let $M=m_1\cdots m_n$. By the Chinese Remainder Theorem, every integer $0\leq x\leq M$ is uniquely represented by its CRR namely $(x_1,\ldots ,x_n)$, where $0\leq x_i<m_i$ and $x_i\equiv x(\mod m_i)$.\textcolor{white}{\index[subject]{Chinese remainder representation}}

\begin{theorem}\label{chap1-crrtheorem}
Let CRR($M$) denote the CRR system based on the $n$ consecutive primes $3=m_1<\cdots <m_n$ and let $M=m_1\cdots m_n$.
\begin{enumerate}
    \item Generating and to output the first $n$ primes is in ${\rm U_L}$-$\NCone$.
    \item Converting an integer $x<2^n$ in binary notation to its CRR in CRR($M$) is in ${\rm U_L}$-$\NCone$.
    \item Converting an integer $x<2^n$ which is given in its CRR as CRR($M$) to its binary notation is in ${\rm U_L}$-$\NCone$.
\end{enumerate}
\end{theorem}
\begin{corollary}\label{chap1-cor-crrtheorem}
Given a $1^n$ and $O(\log n)$ bit integer $m>0$ as input, where $n=sizeof(m)$, we can determine if $m$ is a prime in ${\rm U_L}$-$\NCone$.
\end{corollary}

We also need the following useful results on the complexity of certain basic operations using Boolean circuits.
\begin{theorem}\label{chap1-booleancircuittheorem}
\begin{enumerate}
\item Let $\Sigma=\{ 0,1\}$ be the input alphabet. Given a string $x\in\Sigma^*$ in the unary notation as the input, the problem of computing the length of $x$ and to output it in binary notation is in ${\rm U_L}$-$\NCone$.
\item Let $\Sigma=\{ 0,1\}$ be the input alphabet. Given a string $x\in\Sigma^*$, the problem of outputting the reverse  of the input $x$ is in ${\rm U_L}$-$\NCone$.
\item Let $\Sigma=\{ 0,1\}$ be the input alphabet. Let $x,y$ be integers given in binary notation as input. Here we assume that if $x=x_0x_1\cdots x_{n-1}$ is the binary representation of $x$, then $x_0$ is the least significant bit (LSB) and $x_{n-1}$ is the most significant bit (MSB) of $x$. The MSB denotes the sign of $x$, where $0$ denotes that $x$ is negative and $1$ denotes that $x$ is positive. Computing the sum of $x$ and $y$ in binary notation, and the product of $x$ and $y$ in binary notation is in ${\rm U_L}$-$\NCone$.
\item Let $\Sigma=\{ 0,1\}$ be the input alphabet. Let $x$ and $y$ be positive integers. Determining if $x<y$ is in ${\rm U_L}$-$\NCone$.
\item Let $\Sigma=\{ 0,1\}$ be the input alphabet and let $x,y\in\Sigma^*$ be integers given as input such that $y\neq 0$. Computing the quotient of $\lfloor x/y\rfloor$ is in ${\rm U_L}$-$\NCone$.
\item Let $\Sigma=\{ 0,1\}$ be the input alphabet and let $x,y\in\Sigma^*$ be integers given as input such that $y>0$. Computing $x (\mod y)$ is in ${\rm U_L}$-$\NCone$.
\end{enumerate}
\end{theorem}

\subsection{Logarithmic space bounded computation}\label{chap1-sec-logspacecomputation}
A standard result in computational complexity is that ${\rm U_L}$-$\NCone\subseteq\L$ which relates the Boolean circuit model of computation with the Turing machine model. As a result all basic operations that we have listed above in Theorems \ref{chap1-crrtheorem} and \ref{chap1-booleancircuittheorem} are computable in $O(\log n)$ space using deterministic Turing machines, where $n$ is the size of the input. We therefore get the following observation.
\begin{proposition}\label{chap1-prop-chooseinteger}
Let $n$ be a positive integer given as input in the unary notation. It is possible to non-deterministically choose an integer $1\leq k\leq n$ and also output $k$ in binary notation using $O(\log n)$-space.
\end{proposition}
\begin{proof}
Given a positive integer $n$ in the unary notation $1^n$ as input, it follows from Theorem \ref{chap1-booleancircuittheorem} that we can compute $n$ in binary notation using space $O(\log n)$ since the length of $1^n$ is computable in ${\rm U_L}$-$\NCone$ which is contained in $\L$.

Initially let $x$ denote the empty string. Iteratively, we non-deterministically choose $0$ or $1$ and append it to the string $x$ and check in every iteration if $x\leq n$. Since there exists some iteration in which we are bound to get $x>n$ then we output the string $x$ obtained in the previous iteration as the integer $k$ and stop. Clearly this requires at most $O(\log n)$ space.
\end{proof}

\begin{definition}\label{chap1-aczeroredn-defn}
Let $\Sigma$ be the input alphabet and $L_1,L_2\in\Sigma^*$. We say that $L_1$ is ${\rm U_L}$-$\ACzero$ many-one reducible to $L_2$, denoted by $L_1\leq^{{\rm U_L-}\ACzero}_m L_2$, if there exists a function $f:\Sigma^*\rightarrow\Sigma^*$ such that given any input string $x\in\Sigma^*$, it is possible to compute each symbol of $f(x)$ in ${\rm U_L}$-$\ACzero$ and we have $f(x)\in L_2$ if and only if $x\in L_1$. The function $f$ is called as ${\rm U_L}$-$\ACzero$ many-one reduction from $L_1$ to $L_2$.\textcolor{white}{\index[subject]{${\rm U_L}$-$\ACzero$ many-one reducible}\index[subject]{$\leq^{{\rm U_L-}\ACzero}_m$}\index[subject]{${\rm U_L}$-$\ACzero$ many-one reduction}}
\end{definition}

We once again recall the definition of $\L$ from Definition \ref{chap1-defn-L}.
\begin{definition}\label{chap1-defn-fl}
Let $\Sigma$ be the input alphabet. We define the complexity class $\FL$ to be the class of all functions $f:\Sigma^*\rightarrow\Sigma^*$ such that given an input $x\in\Sigma^*$, each symbol of $f(x)$ is computable by an $O(\log n)$-space bounded deterministic Turing machine, where $n=|x|$.\textcolor{white}{\index[subject]{$\FL$}}
\end{definition}
We observe that the composition of two functions in $\FL$ is a function in $\FL$.
\begin{proposition}\label{chap1-prop-flsimulatefl}
Let $\Sigma$ be the input alphabet and let $f_1,f_2:\Sigma^*\rightarrow\Sigma^*$. For any $x\in\Sigma^*$, we have $f_2(f_1(x))\in\FL$. In other words, $\FL\circ\FL =\FL$.
\end{proposition}
\begin{proof}
Let $M_i$ be the $O(\log n)$-space bounded deterministic Turing machine that computes $f_i(x)$ on any input string $x\in\Sigma^*$, where $i=1,2$, and $n=|x|$. Given an input string $x\in\Sigma^*$, let $M$ be a deterministic Turing machine that first simulates $M_1$ on input $x$ and hence computes $f_1(x)$ in a symbol-wise manner such that it stores at most $O(\log n)$ symbols of $f_1(x)$ at any instant on the work tape, where $n=|x|$. The Turing machine $M$ simulates $M_2$ on the input $f_1(x)$ after computing the required $O(\log n)$ symbols of $f_1(x)$. If $M$ requires any symbol of $f_1(x)$ for its simulation of $M_2$ which is not stored in the work tape then it stores its present configuration on the work tape and starts the simulation of $M_1$ on input $x$ to compute the required symbol. Computation of $M$ on the input $x$ proceeds in this manner and we finally obtain $f_2(f_1(x))$ in the work tape. Since this entire computation requires $O(\log n)$ space and $M$ is deterministic we can compute $f_2(f_1(x))$ in $\FL$.
\end{proof}

\begin{definition}\label{lredndefn}
Let $\Sigma$ be the input alphabet and $L_1,L_2\in\Sigma^*$. We say that $L_1$ is logspace many-one reducible to $L_2$, denoted by $L_1\Lredn L_2$, if there exists a function $f:\Sigma^*\rightarrow\Sigma^*$, which is computable by a deterministic $O(\log n)$ space bounded deterministic Turing machine, such that given any input string $x\in\Sigma^*$, we have $f(x)\in L_2$ if and only if $x\in L_1$, where $n=|x|$. The function $f$ is called as logspace many-one reduction from $L_1$ to $L_2$.\textcolor{white}{\index[subject]{$\Lredn$}\index[subject]{logspace many-one reducible}\index[subject]{logspace many-one reduction}}
\end{definition}

\begin{note}\label{lredn-note}
\begin{enumerate}
\item\label{lredn-note1} We note that in the above definition, if $L_1\Lredn L_2$ using a logspace computable function $f$ then given any input $x\in\Sigma^*$ we have $x\in L_1$ if and only if $f(x)\in L_2$. Therefore $x\in\overline{L_1}$ if and only if $f(x)\in\overline{L_2}$. In other words, $\overline{L_1}\Lredn\overline{L_2}$.
\item\label{lredn-note2} Also it is easy to show that if $L_1$ is logspace many-one reducible to $L_2$ and $L_2$ is in a complexity class $\mathcal{C}$ where $\mathcal{C}$ contains the complexity class $\L$ of all languages that can be decided in deterministic logarithmic space then $L_1$ is in $\mathcal{C}$.
\item If $L_1\Lredn L_2$ and $L_2\Lredn L_3$ then $L_1\Lredn L_3$.
\end{enumerate}
\end{note}

\begin{definition}
Let $\Sigma$ be the input alphabet and let $\mathcal{C}$ be a complexity class. $L\subseteq \Sigma^*$ is logspace many-one hard for $\mathcal{C}$ if for any $L'\in\mathcal{C}$, we have $L'\Lredn L$.\textcolor{white}{\index[subject]{logspace many-one hard}}
\end{definition}

\begin{definition}
Let $\Sigma$ be the input alphabet and let $\mathcal{C}$ be a complexity class. $L\subseteq \Sigma^*$ is logspace many-one complete for $\mathcal{C}$ if $L\in\mathcal{C}$ and $L$ is logspace many-one hard for $\mathcal{C}$.\textcolor{white}{\index[subject]{logspace many-one complete}}
\end{definition}

\begin{definition}\label{L-Turing-reduction}
Let $\Sigma$ be the input alphabet and $L_1,L_2\in\Sigma^*$. We say that $L_1$ is logspace Turing reducible to $L_2$, denoted by $L_1\LTredn L_2$, if there exists a $O(\log n)$ space bounded Turing machine $M^{L_2}$ that has $L_2$ as an oracle such that given any input string $x\in\Sigma^*$, we have $M^{L_2}(x)$ accepts if and only if $x\in L_1$, where $n=|x|$.\textcolor{white}{\index[subject]{$\LTredn$}\index[subject]{logspace Turing reducible}}
\end{definition}
\begin{definition}\label{L-oracle-class}
Let $\Sigma $ be the input alphabet and let $\mathcal{C}$ be a complexity class. We define $\L^{\mathcal{C}}$ as the class of all languages $L\subseteq\Sigma^*$ that is accepted by a $O(\log n)$ space bounded deterministic Turing machine that has oracle access to a language $A\in\mathcal{C}$, where $n$ is the size of the input and $A\subseteq\Sigma^*$.\textcolor{white}{\index[subject]{L$^\mathcal{C}$, logspace Turing reducible to the complexity class $\mathcal{C}$}}
\end{definition}
It is easy to note that if $L_1$ is logspace Turing reducible to $L_2$ and $L_2$ is in a complexity class $\mathcal{C}$, where $\mathcal{C}$ contains the complexity class $\L$ of all languages that can be decided in deterministic logarithmic space then $L_1$ is in $\L^{\mathcal{C}}$.

\begin{definition}\label{reductionbetweenfunctions}
Let $\Sigma$ be the input alphabet. Given functions $f,g:\Sigma^*\rightarrow\Z^+$, we say that $f$ is logspace many-one reducible to $g$, denoted by $f\Lredn g$\textcolor{white}{\index[subject]{$\Lredn$}},  if there exists a function $h:\Sigma^*\rightarrow\Sigma^*$ computable in $O(\log n)$ space, where $n$ is the length of the input, such that $f(x)=g(h(x))$ for every $x\in \Sigma^*$.
\end{definition}

\section*{Notes}
The Turing machine model of computation explained in Section \ref{TM-model-of-computation} is from the excellent textbook by Dexter Kozen \cite[Lectures 28-29]{Koz1997} and it is as introduced in any standard textbook on the Theory of Computation such as \cite{HU1979, LP1998, Sip2013}. The Ruzzo-Simon-Tompa oracle access mechanism for non-deterministic oracle Turing machines stated in Section \ref{chap1-NOTM} has been conceptualized and formulated by Walter L. Ruzzo, Janos Simon and Martin Tompa in \cite{RST1984}.\textcolor{white}{\index[authors]{Ruzzo, Walter L.}\index[authors]{Simon, Janos}\index[authors]{Tompa, Martin}\index[authors]{Borodin, Alan}\index[authors]{Beame, Paul}\index[authors]{Hoover, James}\index[authors]{Kozen, Dexter}\index[authors]{Cook, Stephen A.}}

Background material including definitions and results needed to understand the results of Section \ref{chap1:defnBooleancktsLSBCC} is from \cite[Chapter 1]{Voll1999} and \cite{Weg1987}. Definitions \ref{chap1-bddfaninbasis-defn-booleanbasis} to \ref{chap1-luniformcircuitfamily} are due to Heribert Vollmer and they are from \cite[Chapter 1, pp. 10 \& Chapter 2, pp. 43-44, 47]{Voll1999}. In \cite[Lemma 2.25]{Voll1999} it is shown that in order to define a uniform family of circuits, the concept of admissible encoding scheme and the concept of ``\emph{direct connection language}" are equivalent in terms of complexity. In particular it is stated in \cite[pp. 51]{Voll1999} that if the uniform circuit family ${\{ C_n\}}_{n\geq 1}$ is of size polynomial in the size of the input then, ${\{ C_n\}}_{n\geq 1}$ is logspace-uniform if and only if the admissible encoding scheme is computable from $1^n$ in space $O(\log n)$ if and only if the direct connection language is in $\L$. We need the $\MAJ$ function as an oracle gate along with the standard unbounded fan-in basis of Boolean gates in Boolean circuits to define ${\rm U_L}$-$\TCzero$ in Definition \ref{chap1-TCzero-defn} and ${\rm U_L}$-$\TCone$ in Definition \ref{chap1-TCone-defn}.

The concept of Chinese remainder representation and Theorem \ref{chap1-crrtheorem}, due to A. Chiu, G. Davida and B. Litow \cite{CDL2001}, are based on the Chinese Remainder Theorem. Next, Theorem \ref{chap1-booleancircuittheorem} is based on results shown in \cite{BCH1986,CDL2001,HAB2002}. In fact, the first log-depth, polynomial size uniform circuit family for computing the quotient upon dividing an integer by a non-zero integer is demonstrated in the seminal paper of Paul Beame, Stephen A. Cook and James Hoover \cite{BCH1986}. This complexity upper bound is been improved to DLOGTIME-uniform $\TCzero$ by William Hesse, Eric Allender and David A. Mix Barrington in \cite{HAB2002} and their proof uses descriptive complexity. The well known result that ${\rm U_L}$-$\NCone\subseteq\L$ is due to Alan Borodin \cite{Bor1977}.\textcolor{white}{\index[subject]{DLOGTIME-uniform $\TCzero$}\index[authors]{Chiu, Andrew}\index[authors]{Davida, George}\index[authors]{Litow, Bruce}\index[authors]{Hesse, William}\index[authors]{Barrington, David A. Mix}\index[authors]{Allender, Eric}\index[authors]{Vollmer, Heribert}}

We refer to \cite[Lectures 29-33]{Koz1997} for the notion of simulating one Turing machine by another and for the definition of a reduction.

\chapter{Counting in Non-deterministic Logarithmic Space}\label{chap2}
\section{Non-deterministic Logarithmic Space: $\boldNL$}
\begin{definition}\label{NLdefn}
Let $\Sigma $ be the input alphabet. We define the complexity class Non-deterministic Logarithmic Space, $\NL=\{ L\subseteq \Sigma ^*|$ there exists a $O(\log n)$ space bounded non-deterministic Turing machine $M$ such that $L=L(M)\}$.\textcolor{white}{\index[subject]{$\NL$}}
\end{definition}

\begin{definition}\label{compdefinition}
Let $\Sigma $ be the input alphabet. For a language $L\subseteq\Sigma^*$, the complement of $L$ is $\overline{L}=\Sigma^*-L$. Given a complexity class $\mathcal{C}$, we define the complement of $\mathcal{C}$ as $\mbox{co-}\mathcal{C}=\{ \overline{L}\subseteq \Sigma ^*|L\in\mathcal{C}\}$.\textcolor{white}{\index[subject]{co-$\mathcal{C}$, complement of the complexity class $\mathcal{C}$}\index[subject]{$\overline{L}$, complement of a language $L$}}
\end{definition}

\begin{definition}\label{coNLdefn}
Let $\Sigma $ be the input alphabet. For a language $L\in\Sigma^*$, the complement of the language $L$ is $\overline{L}=\Sigma^*-L$. We define the complexity class co-Non-deterministic Logarithmic Space, $\coNL=\{ \overline{L}\subseteq \Sigma ^*|$ there exists a $O(\log n)$ space bounded non-deterministic Turing machine $M$ such that $L=L(M)\}$.\textcolor{white}{\index[subject]{$\coNL$}}
\end{definition}

\subsection{$\boldNL$ and the Directed $\bf{st}$-Connectivity Problem (DSTCON)}
\begin{definition}\label{dstcon-defn}
Let $G=(V,E)$ be a directed graph given in terms of its adjacency matrix, and let $s,t\in V$. We define ${\rm DSTCON}=\{(G,s,t)|\exists${\rm ~a~ directed~ path~ from~}$s${\rm ~to~}$t$~{\rm in~ }$G\}$.\textcolor{white}{\index[subject]{DSTCON}}
\end{definition}
\begin{tcolorbox}[colback=gray!35!white,colframe=white]
{\bf\emph{In this chapter, any matrix $A\in\R^{n\times n}$ is synonymous with the adjacency matrix of a directed graph $G$ such that the weight of the directed edge $(i,j)$ in $G$ is equal to the entry $A(i,j)$ of the matrix $A$, where $1\leq i,j\leq n$. If $A(i,j)=0$ then the directed edge $(i,j)$ does not exist in $G$.}}
\end{tcolorbox}
\begin{lemma}\label{DSTCONisNLhard}
{\rm DSTCON} is logspace many-one hard for $\NL$.
\end{lemma}
\begin{proof}
Let $\Sigma$ be the input alphabet and let $L\subseteq \Sigma^*$. Assume that $L\in \NL$ and let $M$ be a non-deterministic Turing machine that accepts $L$ using space at most $O(\log n)$ where $n$ is the size of the input. We show that $L\Lredn {\rm DSTCON}$. Given an input string $x\in \Sigma^*$, we consider the set of all configurations of $M$ on an input of length $n=|x|$. It follows from Theorem \ref{chap1-thm-config-size-count-NL} that the number of configurations of $M$ on input of length $n$ is at most $n^k$, for some constant $k>0$. We now define a directed graph $G=(V,E)$ where the set of all vertices $V$ is the set of all configurations of $M$ and there exists an edge from a vertex $u$ to another vertex $v$ in $G$ if the tape head upon reading the symbol it points to in configuration $u$ results in the configuration $v$ of $M$. Since $|V|\leq n^k$ and $M$ is $O(\log n)$ space bounded, any configuration of $M$ is also $O(\log n)$ space bounded, and so given a configuration $u$ of $M$ on $x$ it is always possible to find the neighbors of $u$ in $G$ in $O(\log n)$ space.
Also given $x\in \Sigma^*$ we have $x\in L$ if and only if there exists a directed path from the vertex which is the initial configuration of $M$ on input $x$ to the vertex which is the accepting configuration of $M$ on input $x$. Clearly, given the description of $M$ and an input $x\in\Sigma^*$, we can output the adjacency matrix of the directed graph $G$ and vertices $s$ and $t$, which denote the initial and final configurations of $M$ on $x$ respectively, using $O(\log n)$ space, where $n=|x|$. This shows that $L\Lredn {\rm DSTCON}$ which implies that ${\rm DSTCON}$ is hard for $\NL$ under $\Lredn$ reductions.
\end{proof}
\begin{lemma}\label{DSTCONisinNL}
${\rm DSTCON}\in\NL$.
\end{lemma}
\begin{proof}
Let $G=(V,E)$ be a directed graph which is represented by its adjacency matrix and let $s,t\in V$ be given as input along with $G$. Let us consider the following algorithm which can be implemented by a non-deterministic Turing machine.
\begin{algorithm}[H]
\caption{Directed-Graph-Accessibility\newline {\bf Input:} $(G,s,t)$, where $G=(V,E)$ is a directed graph, and $s,t\in V$.\newline {\bf Output:} \emph{accept} if $\exists$ a directed path from $s$ to $t$ in $G$. Otherwise \emph{reject.}\newline {\bf Complexity:} $\NL$.}\label{algo:directed-graph-accessibility}
\begin{algorithmic}[1]
\State Let $n\leftarrow |V|$.
\State Let $m\leftarrow |E|$.
\State Let $u\leftarrow s$.
\For {$counter\leftarrow 0$ to $(n-1)$}
	\State {Non-deterministically choose a vertex $v\in V$ such that $(u,v)\in E$.}
	\If {$\not\exists v\in V$ such that $(u,v)\in E$}
		reject the input.
	\EndIf
	\If {$v=t$}
		\State {output there exists a directed path from $s$ to $t$ in $G$.}
        \State {accept the input.}
	\EndIf
	\State {$u\leftarrow v$.}
	\State {$counter\leftarrow counter+1$.}
\EndFor
\State {reject the input and stop.}
\end{algorithmic}
\end{algorithm}
The non-deterministic Turing machine that implements the Directed-Graph-Accessibility $(G,s,t)$ algorithm needs at most $O(\log n)$ space to store vertices $u,v\in V$, to store variables $n,m$ and to update the variable $counter$. In addition it looks into the adjacency matrix of the directed graph given as input to find and non-deterministically choose vertex $v$ which is the head of the directed edge $(u,v)$ in each iteration. Clearly this also needs at most $O(\log n)$ space.

Now, if there exists a directed path from $s$ to $t$ then any such path is of length less than or equal to $(n-1)$ and it is possible to non-deterministically guess the vertices along a directed path from $s$ to $t$. Thus, whenever there exists a directed path there exists a computation path of the computation tree of this non-deterministic Turing machine that accepts. Conversely, if there does not exist any path from $s$ to $t$ in $G$ then no computation path of this non-deterministic Turing machine ends in an accepting state. Therefore $\rm DSTCON\in\NL$.
\end{proof}

\begin{theorem}\label{DSTCONisNLcomplete}
{\rm DSTCON} is logspace many-one complete for $\NL$.
\end{theorem}
\begin{proof}
The result follows from Lemma \ref{DSTCONisinNL} and Lemma \ref{DSTCONisNLhard}.
\end{proof}

\begin{definition}
We define $\LDAG =\{ G|G=(V,E)$ is a simple layered directed acyclic graph which is represented by its adjacency matrix, and which does not have any self-loops on vertices or directed cycles or parallel edges between vertices. Also vertices in $G$ are arranged as a square matrix such that there are $n$ rows and every row has $n$ vertices. Any edge in this graph is from a vertex in the $i^{th}$ row to a vertex in the $(i+1)^{st}$ row, where $1\leq i\leq (n-1)$ and $n\geq 2\}$.\textcolor{white}{\index[subject]{SLDAG}\index[subject]{simple layered directed acyclic graph}}
\end{definition}
\begin{figure}[h!]
\begin{center}\includegraphics[scale=1.15]{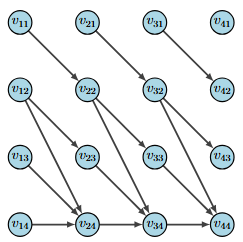}\end{center}
\caption{\it This figure is a directed graph $G$ which is an instance of $\LDAG$. Assuming that vertices $s=v_{11}$ and $t=v_{44}$ we also infer that this instance $(G,s,t)$ is in $\LDAGSTCON$.}
\end{figure}
\begin{definition}\label{sldagstcon}
Let $G=(V,E)\in \LDAG$ be represented by its adjacency matrix, and let $s,t\in V$ such that $s$ is a vertex in the first row and $t$ is a vertex in the last row of $G$. We define $\LDAGSTCON =\{ (G,s,t)|\exists$ a directed path from $s$ to $t$ in $G\}$.\textcolor{white}{\index[subject]{SLDAGSTCON}\index[subject]{simple layered directed acyclic graph $st$-connectivity}}
\end{definition}

\begin{theorem}\label{sldagstconcomplete}
$\LDAGSTCON$ is logspace many-one complete for $\NL$.
\end{theorem}
\begin{proof}
We know from Theorem \ref{DSTCONisNLcomplete} that the $st$-connectivity problem for directed graphs is complete for $\NL$ under logspace many-one reductions.
\begin{figure}[h!]
$\bf G$:\includegraphics[scale=1]{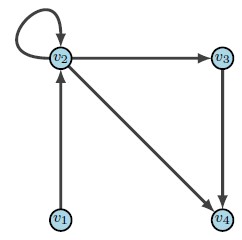}\hspace{1.5cm}$\bf{G'}$:\includegraphics[scale=1]{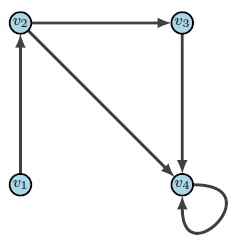}\newline\newline$\bf{G''}$:\includegraphics[scale=1.2]{SLDAGSTCON_instance.png}
\caption{\it This figure explains the reduction from a directed graph $G$ with vertices $s_1=v_1$ to the vertex $t_1=v_4$ to a ``yes" instance $G''$ of $\LDAGSTCON$ where $s=v_{11}$ and $t=v_{44}$.}\label{chap2-figure-rednsldagstcon}
\end{figure}
Therefore the $st$-connectivity problem for directed graphs which does not have any self-loops on vertices or parallel edges between vertices is also complete for $\NL$ under logspace many-one reductions. In other words, the $st$-connectivity problem for simple directed graphs is complete for $\NL$ under logspace many-one reductions. Let $G=(V,E)$ be a simple directed graph given in terms of its adjacency matrix, and let $s,t\in V$. Also let $n=|V|$. Given $G$, we obtain the directed graph $G'=(V',E')$, where $V'=V$ and $E'=E\cup {(t,t)}$. We now reduce $G'$ to $G''=(V'',E'')\in \LDAG$. In $G''$, the vertex set $V''$ has $n^2$ vertices arranged as a $n\times n$ matrix obtained by creating $n$ copies of $V'$. As an example refer to Figure \ref{chap2-figure-rednsldagstcon} in which we obtain an instance of $\LDAG$ from the directed graph given as input. For the sake of convenience, we assume in this example that vertices in each row of the instance of $\LDAG$ are arranged vertically.

Now let the vertex $s$ in the first row be denoted by $(1,s)$ and the vertex $t$ in the last row be denoted by $(n,t)$. A $(i,j)$ vertex in $G''$ is the $j^{th}$ vertex in the $i^{th}$ row of $G''$, where $1\leq i,j\leq n$. Here $E''=\{ ((i,j_1),(i+1,j_2))|(j_1,j_2)\in E'{\rm ,~where~}1\leq i\leq n-1{\rm ~and~} 1\leq j_1,j_2\leq n\}$. Now it is easy to note that there exists a directed path from $s$ to $t$ in $G$ if and only if there exists a directed path from $(1,s)$ to $(n,t)$ in $G''$. Since we can obtain the adjacency matrix of $G''$ from the adjacency matrix of $G$ using a logspace many-one reduction, the result follows.
\end{proof}

\begin{definition}\label{sharpL}
Let $\Sigma$ be the input alphabet. The complexity class $\sharpL$\textcolor{white}{\index[subject]{$\sharpL$, pronounced as sharpL}} is defined to be the class of functions $f:\Sigma^{*}\rightarrow\Z^{+}$ such that there exists a $\NL$-Turing machine $M$ for which we have $f(x)=\acc _M(x)$ where $\acc _M(x)$ denotes the number of accepting computation paths of $M$ on input $x\in\Sigma^{*}$.\textcolor{white}{\index[subject]{$\acc_M(x)$}\index[subject]{$\sharpL$, pronounced as ``sharpL"}}
\end{definition}
\begin{remark}
If a language $L\subseteq\Sigma^*$ is in $\NL$ then there exists a $\NL$-Turing machine $M$ that accepts $L$. Therefore, given any input $x\in\Sigma^*$ we have $x\in L$ if and only if there exists a $y\in\Sigma^*$ which is of length at most $p(n)$ such that $M(x,y)$ \emph{``accepts"}, where $p(n)$ is a polynomial in $n=|x|$. As a result, the problem of counting the number of witnesses $y$ such that $M(x,y)$ \emph{``accepts"} is in $\sharpL$.
\end{remark}
We recall the notion of reduction between functions stated in Definition \ref{reductionbetweenfunctions}.
\begin{definition}
A function $f:\Sigma^*\rightarrow\Z^+$ is logspace many-one hard for $\sharpL$ if every function $g\in\sharpL$ is logspace many-one reducible to $f$.
\end{definition}
\begin{definition}
A function $f:\Sigma^*\rightarrow\Z^+$ is logspace many-one complete for $\sharpL$ if $f\in\sharpL$ and $f$ is logspace many-one hard for $\sharpL$.
\end{definition}

\begin{proposition}\label{dstconsharplcomplete}
Let $(G,s,t)$ be an input instance of $\DSTCON$. Counting the number of directed paths from $s$ to $t$ in $G$, denoted by $\sharp${\rm DSTCON}, is logspace many-one complete for $\sharpL$.\textcolor{white}{\index[subject]{$\sharp$DSTCON}}
\end{proposition}
\begin{proposition}\label{sldagcountsharplcomplete}
Let $(G,s,t)$ be an input instance of $\LDAGSTCON$. Counting the number of directed paths from $s$ to $t$ in $G$, denoted by $\sharp${\rm SLDAGSTCON}, is logspace many-one complete for $\sharpL$.\textcolor{white}{\index[subject]{$\sharp$SLDAGSTCON}}
\end{proposition}
\begin{proposition}\label{parsicomplete}
Let $\Sigma$ be the input alphabet and $f\in \sharpL$ using a $\NL$-Turing machine $M$ such that $f(x)=\acc_M(x)$ on any input $x\in \Sigma^*$.
\begin{enumerate}
\item For any input string $x\in \Sigma^*$, $g(x)=(G,s,t)$, where $g$ is the canonical logspace many-one reduction used to show that $\DSTCON$ is $\NL$-hard and $(G,s,t)$ is an input instance of $\DSTCON$, we have $f(x)$ is equal to the number of directed paths from $s$ to $t$ in $G$,
\item For any input string $x\in \Sigma^*$, $g(x)=(G,s,t)$, where $g$ is the canonical logspace many-one reduction used to show that $\LDAGSTCON$ is $\NL$-hard and $(G,s,t)$ is an input instance of $\LDAGSTCON$, we have $f(x)$ is equal to the number of directed paths from $s$ to $t$ in $G$
\end{enumerate}
\end{proposition}

\section{The Immerman-Szelepcsenyi Theorem}\label{chap02-immermanszelepcsenyitheorem}
\subsection{The Immerman-Szelepcsenyi Theorem in logarithmic space}
\begin{theorem}\label{chap2-NLcomplement}
{$\NL =\coNL$}\textcolor{white}{\index[subject]{Immerman-Szelepcsenyi Theorem}}
\end{theorem}
\begin{proof}
Let $G=(V,E)$ be an input instance of $\LDAG$ which is given in terms of its adjacency matrix and let $s,t\in V$ such that $s$ is a vertex in the first row and $t$ is a vertex in the last row of $G$. We know from Theorem \ref{sldagstconcomplete} that deciding if $(G,s,t)\in\LDAGSTCON$ is logspace many-one complete for $\NL$. Therefore, to prove this theorem, it suffices to prove that the problem of deciding if there does not exist any path from $s$ to $t$ in $G$ is also in $\NL$. In other words we show that $\overline{\LDAGSTCON}\in\NL$.\textcolor{white}{\index[subject]{$\overline{\LDAGSTCON}$}}
\begin{algorithm}[H]
\caption{Immerman-Szelepcsenyi-in-Logspace. \newline {\bf Input:} $(G,s,t)$, where $G=(V,E)$ is an input instance of $\LDAG$ and $s,t\in V$.\newline {\bf Output:} \emph{accept} if $\not\exists$ a directed path from $s$ to $t$ in $G$. Otherwise \emph{reject.} \newline {\bf Complexity:} $\NL$.}\label{algo:IS-Logspace}
\begin{algorithmic}[1]
\State Let $\alpha\leftarrow 1$.
\For {$i\leftarrow 1$ to $(n-1)$}
	\State Let $\beta\leftarrow 1$.
	\State Let $d\leftarrow 0$.
	\State $flag\leftarrow 0$.
	\For {each node $v$ in layer $(i+1)$ in $G$}    
		\For {each node $u$ in layer $i$ in $G$}
            \State {Non-deterministically either perform or skip \{step $9$ to step $16$\}.}
				\State {Non-deterministically follow a path of length $\leq (i-1)$ from $s$.}
					\If {this path does not end at $u$}
						\State reject and stop.
					\EndIf
                \State $d\leftarrow d+1$.                    
            \If {$(u,v)$ is an edge of $G$ and $flag=0$} 										
				\State $\beta\leftarrow \beta+1$ and $flag\leftarrow 1$.
			\EndIf
		\EndFor
		\If {$d\neq \alpha$}
			\State reject and stop.
		\EndIf
        \State $flag\leftarrow 0$.
	\EndFor
    \State $\alpha\leftarrow \beta$.
\EndFor
\State Let $d\leftarrow 0$.
\For {each node $u$ in layer $n$ in $G$}
	\State {Non-deterministically either perform or skip \{step 27 to step 34\}.}
	\State {Non-deterministically follow a path of length $(n-1)$ from $s$.}
	\If {this path does not end at $u$}
		\State reject and stop.
	\EndIf
	\If{$u=t$}
		\State reject and stop.
	\EndIf
	\State $d\leftarrow d+1$.    
\EndFor
\algstore{firstpart}
\end{algorithmic}
\end{algorithm}
\clearpage
\begin{algorithm}[H]
\setcounter{algorithm}{1}
\renewcommand{\addcontentsline}[3]{}
\caption{Immerman-Szelepcsenyi-in-Logspace (continued)}\label{algo:IS-Logspace-continued}
\begin{algorithmic}[1]
\algrestore{firstpart}
\If{$d\neq \alpha$}
	\State reject and stop.
\Else
	\State accept and stop.
\EndIf
\end{algorithmic}
\end{algorithm}
Let us consider our algorithm Immerman-Szelepcsenyi-in-Logspace($G,s,t$). We show that this algorithm can be implemented by a $\NL$-Turing machine $M$. As a result $M$ must have at least one accepting computation path if and only if there does not exist any directed path from $s$ to $t$ in $G$.

Let us assume that we know $\alpha$ which is the number of nodes in layer $n$ of $G$ that are reachable from $s$ in $G$. We assume that $\alpha$ is provided as an input to $M$ and we show in steps 24 to 40 of the above algorithm Immerman-Szelepcsenyi-in-Logspace($G,s,t$), that it is possible to use $\alpha$ to prove that $\overline{\LDAGSTCON}\in\NL$ using a method which we call as {\bf \emph{non-deterministic counting}}. Later we prove that $M$ also computes $\alpha$ in steps 1 to 23 using the same non-deterministic counting method.\textcolor{white}{\index[subject]{non-deterministic counting}}

In steps 24 to 40, given $G,s,t$ and $\alpha$, the Turing machine $M$ operates as follows. One by one $M$ goes through all the $n$ nodes in layer $n$ of $G$ and non-deterministically guesses whether each one is reachable from $s$. Whenever a node $u$ is guessed to be reachable, $M$ attempts to verify this guess by guessing a path of length $n$ or less from $s$ to $u$. If this computation path fails to verify this guess, it rejects. The Turing machine $M$ counts the number of nodes that have been verified to be reachable. When a path has gone through all of $G$'s nodes in layer $n$, it checks that the number of nodes that are verified to be reachable from $s$ equals $\alpha$, the number of nodes that are actually reachable from $s$, and rejects if not. In other words, if $M$ non-deterministically chooses exactly $\alpha$ nodes reachable from $s$, not including $t$, and proves that each is reachable from $s$ by guessing the path, $M$ knows that the remaining nodes including $t$ are not reachable and therefore it can accept.

Next we show in steps 1 to 23 described in the above algorithm Immerman-Szelepcsenyi-in-Logspace($G,s,t$), how to calculate $\alpha$: the number of nodes that are reachable from $s$ in layer $n$ in $G$. We describe a non-deterministic logspace procedure whereby at least one computation path has the correct value of $\alpha$ in which it accepts and all the other paths reject.

Let $A_0=\{s\}$. For each $i$ from $1$ to $(n-1)$, we define $A_i$ to be the collection of all  nodes in layer $(i+1)$ that are reachable from $s$ by a directed path of length $i$. So $A_{n-1}$ is the set of all nodes in layer $n$ that are reachable from $s$ by a directed path of length $(n-1)$. Let $\beta=|A_{i-1}|$. In steps 6 to 22 of the Immerman-Szelepcsenyi-in-Logspace($G,s,t$) algorithm we show how to calculate $|A_i|$ from $|A_{i-1}|$. Repeated application of this procedure yields the desired value of $\alpha$.

We calculate $|A_i|$ from $|A_{i-1}|$ using an idea similar to the one presented earlier in this proof. In the algorithm we go through all the nodes of $G$ in layer $(i+1)$ and determine whether each is a member of $A_i$ and also count members to find $|A_i|$.

To determine whether a node $v$ in layer $(i+1)$ is in $A_i$, we use the innermost {\bf for} loop starting in line 7, to go through all the nodes of $G$ in layer $i$ and guess whether each node is in $A_{i-1}$. Each positive guess is verified by a path of length $(i-1)$ from $s$. For each node $u$ verified to be in $A_{i-1}$, the algorithm tests whether $(u,v)$ is an edge of $G$. If it is an edge, $v$ is in $A_i$. The variable $\beta$ is used to count the number of vertices which are in $A_i$. We use the $flag$ variable to prevent the vertex $v$ from being counted more than once. Additionally the number of nodes verified to be in $A_{i-1}$ is counted. At the completion of the innermost {\bf for} loop, if the total number of nodes verified to be in $A_{i-1}$ is not $\alpha$ then entire $A_{i-1}$ has not been found, so this computation path rejects. If the count equals $\alpha$ and $v$ has not been shown to be in $A_i$, we conclude that $v$ is not in $A_i$. Then we go to the next vertex $v$ in layer $(i+1)$ in the outer loop. Since we use constant number of variables each of which takes values from $0$ to $n$ in the this algorithm, it is easy to note that the Turing machine $M$ that implements it is a $\NL$-Turing machine. As a result we have shown that $\overline{\LDAGSTCON}\in\NL$ which implies $\coNL\subseteq\NL$ from which the result follows.
\end{proof}

We recall the notions of logspace Turing reducibility from Definition \ref{L-Turing-reduction} and the definition of $\L^{\mathcal{C}}$ from Definition \ref{L-oracle-class}, where $\mathcal{C}$ is a complexity class.
\begin{theorem}\label{nlclosureunderTuring}
$\L^{\mbox{\NL}}=\NL$.\textcolor{white}{\index[subject]{$\L^{\mbox{\NL}}$, complexity class of languages logspace Turing reducible to $\NL$}}
\end{theorem}
\begin{proof}
It is trivial to note that $\NL\subseteq\L^{\rm\mbox{\NL}}$. To prove our result we have to show that $\L^{\mbox{\NL}}\subseteq\NL$. Let $L'\in\L^{\mbox{\NL}}$. Therefore there exists a $O(\log n)$-space bounded deterministic Turing machine $M^A$, that has access to a language $A\in\NL$ as an oracle such that given any input string $x\in\Sigma^*$, $M^A$ correctly decides if $x\in L'$. Since $M$ requires at most $O(\log n)$ space on any input of size $n$, we infer that the number of queries that $M$ submits to the oracle $A$ is a polynomial in the size of the input. Let $M_A$ be the $\NL$-Turing machine that decides if an input string is in $A$ or not. Due to Theorem \ref{chap2-NLcomplement} it follows that $\overline{A}$ is also in $\NL$. Let $M_{\overline{A}}$ be the $\NL$-Turing machine that correctly decides if any input string is in $\overline{A}$. Now let $\Sigma$ be the input alphabet and let $x\in\Sigma^*$ be the input string. Let us consider the following algorithm implemented by a non-deterministic Turing machine $N$.
\begin{algorithm}[H]
\caption{Closure-Logspace-Turing-for-$\NL$. \newline {\bf Input:} $x\in \Sigma^*$.\newline {\bf Output:} \emph{accepts} if $x\in L$, where $L\in\L^{\mbox{\NL}}$. Otherwise \emph{reject.} \newline {\bf Complexity:} $\NL$.}\label{algo:nllogspaceTuringclosed}
\begin{algorithmic}[1]
\While{$N$ has not reached any of its halting configurations}
	\State $N$ simulates $M$ on input $x$ until a query $y$ is generated.
	\State$N$ simulates $M_A$ on input $y$.
	\If{$M_A$ accepts $y$}
		\State $N$ assumes that the reply of the oracle to the query $y$ is ``YES".
\algstore{firstpart:lnl}
\end{algorithmic}
\end{algorithm}
\clearpage
\begin{algorithm}[H]
\setcounter{algorithm}{2}
\renewcommand{\addcontentsline}[3]{}
\caption{Closure-Logspace-Turing-for-$\NL$ (continued)}\label{algo:nllogspaceTuringclosed-contd}
\begin{algorithmic}[1]
\algrestore{firstpart:lnl}
	\Else
		\State $N$ simulates $M_{\overline{A}}$ on $y$
        \If{$M_{\overline{A}}$ accepts $y$}
            \State $N$ assumes that the reply of the oracle to the query $y$ is ``NO".
		\Else
			\State $N$ rejects the input $x$ and stops.
		\EndIf
	\EndIf
	\State $N$ continues to simulate $M$ on $x$.
\EndWhile
\end{algorithmic}
\end{algorithm}
We now show that $N$ correctly decides if $x\in L$. It is easy to note that if for any oracle query string $y$, either $M_A$ or $M_{\overline{A}}$ can accept $y$ and not both of them can accept $y$. Also if $M_A$ accepts $y$ then $y\in A$. Similarly if $M_{\overline{A}}$ accepts $y$ then $y\in\overline{A}$. As a result when either of these two $\NL$-Turing machines accept then this can be taken to be the output of the oracle and $N$ will proceed with its simulation of $M$. On the contrary if neither $M_A$ nor $M_{\overline{A}}$ accept since it follows from Theorem \ref{chap2-NLcomplement} that $\NL$ is closed under complement, we know that there exists at least one accepting computation path in exactly one of these Turing machines and we did not simluate both the Turing machines along any such path. Therefore $N$ also rejects $x$ and stops the computation. Since $N$ simulates only $\NL$-Turing machines, it shows that $N$ is also a $\NL$-Turing machine. This shows that $\L^{\mbox{\NL}}\subseteq\NL$ from which we get that $\L^{\mbox{\NL}}=\NL$.
\end{proof}

\begin{corollary}\label{nlunionintersection}
$\NL$ is closed under union and intersection.
\end{corollary}

\begin{definition}\label{NL-oracle-class}
Let $\Sigma $ be the input alphabet and let $\mathcal{C}$ be a complexity class. We define $\NL^{\mathcal{C}}$ to be the class of all languages $L\subseteq\Sigma^*$ that is accepted by a $O(\log n)$ space bounded non-deterministic Turing machine $M$ that has oracle access to a language $A\in\mathcal{C}$, where $n$ is the size of the input and $A\subseteq\Sigma^*$. Here we assume that $M$ submits queries to the oracle $A$ according to the Ruzzo-Simon-Tompa oracle access mechanism.\textcolor{white}{\index[subject]{$\NL^{\mathcal{C}}$, non-deterministic logspace Turing reducible to the complexity class $\mathcal{C}$}}
\end{definition}

\begin{theorem}\label{chap2-NLhierarchycollapses}
$\NL^{\mbox{\NL}}=\NL$.\textcolor{white}{\index[subject]{$\NL^{\rm{\mbox{\NL}}}$, non-deterministic logspace Turing reducible to $\NL$}}
\end{theorem}
\begin{proof}
It is trivial to note that $\NL\subseteq\NL^{\rm{\mbox{\NL}}}$. Therefore we have to prove that $\NL^{\mbox{\NL}}\subseteq\NL$. Let $L\in\NL^{\mbox{\NL}}$. Therefore there exists a $\NL$-Turing machine $M^A$ that has access to a language $A\in\NL$ as an oracle such that given any input string $x\in\Sigma^*$, $M^A$ decides if $x\in L$ correctly. It follows from Section \ref{chap1-NOTM} that our $\NL$-Turing machine $M^A$ follows the Ruzzo-Simon-Tompa oracle access mechanism to submit queries to the oracle $A$.  Due to Proposition \ref{chap1-thm-config-size-count-NL}, since the number of configurations of a $\NL$-Turing machine is at most a polynomial in the size of the input, we that the number of queries that are submitted to the oracle $A$ is also a polynomial in the size of the input. Let $M_A$ be the $\NL$-Turing machine that decides if an input string is in $A$ or not. Since we have shown in Theorem \ref{chap2-NLcomplement} that $\NL$ is closed under complement, it follows that $\overline{A}$ is also in $\NL$. Let $M_{\overline{A}}$ be the $\NL$-Turing machine that correctly decides if any input string is in $\overline{A}$. Now let $\Sigma$ be the input alphabet and let $x\in\Sigma^*$ be the input string. Let us once again consider the Algorithm \ref{algo:nllogspaceTuringclosed}, Closure-Logspace-Turing-for-$\NL$, in Theorem \ref{nlclosureunderTuring} implemented by a non-deterministic Turing machine $N$.

We now show that $N$ correctly decides if $x\in L$. It is easy to note that if for any oracle query string $y$, either $M_A$ or $M_{\overline{A}}$ can accept $y$. Also if $M_A$ accepts $y$ then $y\in A$. Similarly if $M_{\overline{A}}$ accepts $y$ then $y\in\overline{A}$. As a result when either of these two $\NL$-Turing machines accept then this can be taken to be the output of the oracle and $N$ will proceed with its simulation of $M$. On the contrary if neither $M_A$ nor $M_{\overline{A}}$ accept since $\NL$ is closed under complement we know that there exists at least one accepting computation path in exactly one of these Turing machines and we did not simluate both the Turing machines along any such path. Therefore $N$ also rejects $x$ and stops the computation. Since $N$ only simulates $\NL$-Turing machines it shows that $N$ is also a $\NL$-Turing machine. This shows that $\NL^{\mbox{\NL}}=\NL$.
\end{proof}

\begin{theorem}\label{chap2-2SATisinNL}
Let {\rm 2SAT}=$\{\phi|\phi$ {~is~a~Boolean~formula~in~the~}2-CNF{~such~that} {$\phi$ ~is~satisfiable\}}. $\overline{\rm 2SAT}~{is~in~\NL}$ and ${\rm 2SAT}~{is~also~in~\NL.}$\textcolor{white}{\index[subject]{2SAT}}\textcolor{white}{\index[subject]{2-CNF}}\textcolor{white}{\index[subject]{$\overline{\rm 2SAT}$, the language of all unsatisfiable 2-CNF formulae}}
\end{theorem}
\begin{proof}
We want to show that the language of all unsatisfiable Boolean formulae in the 2-CNF, denoted by $\overline {\rm 2SAT}$, is in $\NL$. Let $\phi=(\phi _1\wedge\phi _2\wedge\cdots\wedge\phi_m)$ be in 2-CNF. We know that each $\phi _i$ is a conjunction of exactly two literals, where $1\leq i\leq m$.

Now given a 2-CNF Boolean formula $\phi$ we define a directed graph $G=(V,E)$ as follows. For each variable $x$ that occurs in $\phi$ we define two vertices in $G$. More precisely, we define a vertex for the variable $x$ and a vertex for $\neg x$. Clearly if there are $n$ variables in $\phi$ then there are $2n$ vertices in $G$. If $\phi _i=(\neg x\vee y)$ is a clause then we include directed edges $(x,y)$ and $(\neg y,\neg x)$ in $E$ of $G$. In other words, if we have a clause of $\phi$ to be $(\neg x\vee y)$ which is $(x\Rightarrow y)$ then we include the directed edge $(x,y)$ in $E$. By commutativity, since $(\neg x\vee y)$ is also $(y\vee \neg x)$ which is $(\neg y\Rightarrow\neg x)$ we also include the edge $(\neg y,\neg x)$ in $E$ of $G$. Clearly if there are $m$ clauses in $\phi$ then there are $2m$ directed edges in $G$.
\begin{figure}[h!]
\begin{center}
\includegraphics[scale=1.2]{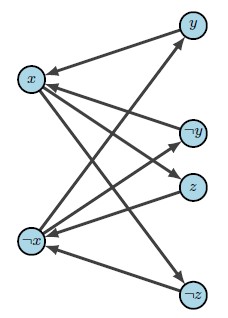}
\end{center}
\caption{\emph{This figure is the directed graph $G$ obtained in the logspace many-one reduction from $\overline{\mbox{\rm 2SAT}}$ to {\rm DSTCON}. We are given the Boolean formula $\phi =(x\vee y)\wedge (x\vee \neg y)\wedge (\neg x\vee z)\wedge (\neg x\vee \neg z)$ as input and we obtain the directed graph $G$ in this figure as the output of the reduction. Clearly, there are $4$ clauses in $\phi$, $8$ vertices and $8$ edges in $G$. It is easy to see that $\phi$ is not satisfiable and there is a directed path in fact, from every literal to its negation in $G$.}}
\end{figure}
Also it is easy to observe that there exists a directed path from a vertex $\alpha$ to a vertex $\beta$ in $G$ if and only if there exists a directed path from $\neg \beta$ to $\neg \alpha$ in $G$. The 2-CNF Boolean formula $\phi$ is the conjunction of implications of the form $(\alpha\Rightarrow\beta )$, where $\alpha$ and $\beta$ are literals in the clauses of $\phi$. Given the conjunction of two clauses $(\alpha\Rightarrow\beta )\wedge (\beta\Rightarrow\gamma )$ we can simplify this conjunction and derive the clause $(\alpha\Rightarrow\gamma )$ from them. Using this we infer that the 2-CNF Boolean formula $\phi$ is unsatisfiable if and only if we can obtain a contradiction which is that there exists at least one variable $x$ such that we can simplify the conjunction of clauses in $\phi$ to obtain the clause $(x\Rightarrow \neg x)$ and also that $(\neg x\Rightarrow x)$. However this is equivalent to the existence of edges in $E$ of $G$ such that there is a directed path from $x$ to $\neg x$ and a directed path from $\neg x$ to $x$.

Now given a 2-CNF Boolean formula $\phi$ such that $n$ is the size of $\phi$, a $O(\log n)$-space bounded deterministic Turing machine can output the adjacency matrix of the directed graph $G$ described above. We then submit a query to the $\NL$ oracle to determine if there exists a directed path from a variable $x$ to $\neg x$ and also from $\neg x$ to $x$ in $G$ for all the variables $x$ that occur in $\phi$. If for some variable $x$ we get the oracle reply to be ``YES" for the queries submitted then we accept $\phi$ and output that $\phi$ is unsatisfiable. Otherwise we reject the input $\phi$ and output that $\phi$ is satisfiable. Now it follows from Theorem \ref{nlclosureunderTuring} that $\overline{\rm 2SAT}$ is in $\NL$. Using Theorem \ref{chap2-NLcomplement} it follows that ${\rm 2SAT}\in\NL$.
\end{proof}

\begin{note}
In the proof of Theorem \ref{chap2-2SATisinNL} we first show that $\overline{\rm 2SAT}\in\NL$. Given a subformula $\psi$ of a Boolean formula $\phi$ which is in 2-CNF, we say that $\psi$ is a contradiction if $\psi$ is not satisfiable. We say that $\psi$ does not have any redundant clause if there does not exist any clause $\psi '$ in $\psi$ such that $\psi$ is unsatisfiable even after deleting $\psi '$ from $\psi$. Any directed path from a variable $x$ to $\neg x$ and therefore a directed path from $\neg x$ to $x$ is a proof or a witness that the input 2-CNF formula $\phi$ is not satisfiable. This is equivalent to stating that a there exists a subformula $\psi$ of $\phi$ which is a contradiction and which is free from any redundant clauses and that it is a proof that $\phi$ is not satisfiable. As a result counting the number of directed paths from a variable $x$ to $\neg x$ is in $\sharpL$. Equivalently counting the number of subformulae of $\phi$, each of which is a contradiction and which do not have any redundant clauses among them, is in $\sharpL$.
\end{note}

\begin{theorem}\label{chap2-2SATisNLhard}
{$\overline{\rm 2SAT}$} {\it is logspace many-one hard for $\NL$.}
\end{theorem}
\begin{proof}
We prove this statement by showing that ${\rm DSTCON}\Lredn\overline{\rm 2SAT}$. Let $G=(V,E)$ be a directed graph and let $(G,s,t)$ be the input instance of DSTCON. Let $n=|V|$. We construct a 2-CNF Boolean formula $\phi$ from $(G,s,t)$ such that there exists a directed path from $s$ to $t$ in $G$ if and only if $\phi$ is not satisfiable. Let $x$ be the Boolean variable corresponding to the vertex $s$ in $G$ and $\neg x$ be the literal corresponding to the vertex $t$. For each of the remaining vertices $v_i$ in $G$, let there be a Boolean variable $v_i$. For each edge $(u,v)\in E$ let us include the clause $(u\Rightarrow v)$ in $\phi$. In other words, for each edge $(u,v)\in E$ we include the clause $(\neg u\vee v)$ in $\phi$. Apart from these clauses we also include the conjunction of the clause $(x)$ in $\phi$. Since the clause $(x)$ is not a disjunction of two distinct literals we introduce a new variable $y$ and include the conjunction of the following 2-CNF Boolean subformula $(x\vee y)\wedge (x\vee \neg y)$ in $\phi$. Clearly a $O(\log n)$-space bounded deterministic Turing machine can output $\phi$ given $(G,s,t)$ as input.

\begin{figure}[h!]
\begin{center}
\includegraphics[scale=1.2]{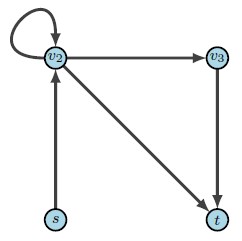}
\end{center}
\caption{\emph{This figure is a directed graph $G$ in the reduction from {\rm DSTCON} to $\overline{\mbox{\rm 2SAT}}$. We therefore obtain the Boolean formula $\phi =(\neg x\vee v_2)\wedge (\neg v_2\vee v_2)\wedge (\neg v_2\vee v_3)\wedge (\neg v_2\vee \neg x)\wedge (\neg v_3\vee \neg x)\wedge (x\vee y)\wedge (x\vee \neg y)$. Since there exists a directed path from $s$ to $t$ in $G$ it is easy to see that $\phi$ is not satisfiable.}}
\end{figure}
Now consider the case when there exists a directed path from $s$ to $t$ in $G$. Let the vertices that form the directed path be the following in sequence: $s,v_i,v_j,\ldots ,$ $v_k,t$, where $1\leq i,j,k\leq n$. We claim that $\phi$ is not satisfiable. To prove this claim consider the case when $x=False$. Then $\phi$ is not satisfiable since we have clauses $(x\vee y)\wedge (x\vee \neg y)$ in $\phi$. For the case when $x=True$ let us consider the 2-CNF Boolean subformula of $\phi$ formed by the clauses due to the edges in a directed path from $s$ to $t$ in $G$: $(\neg x\vee v_i)\wedge (\neg v_i\vee v_j)\wedge\cdots\wedge (\neg v_k\vee \neg x)$. It is easy to see that the Boolean subformula $(\neg x\vee v_i)\wedge (\neg v_i\vee v_j)\wedge\cdots\wedge (\neg v_k\vee \neg x)$ is not satisfiable and so $\phi$ is also not satisfiable.

Conversely, assume that there does not exist any directed path from $s$ to $t$ in $G$. Let $A$ be the subset of vertices in $V$ that are reachable from $s$ in $G$, $B$ be the subset of vertices in $V$ from which $t$ is reachable and $C=V\setminus (A\cup B)$. Since there is no directed path from $s$ to $t$ in $G$ it is easy to observe that the subsets of $V$ of vertices in $A,B$ and $C$ are pairwise disjoint and there are no edges from $A$ to $B\cup C$ and from $A\cup C$ to $B$ in $G$.
\begin{figure}[h!]
\begin{center}
\includegraphics[scale=1.2]{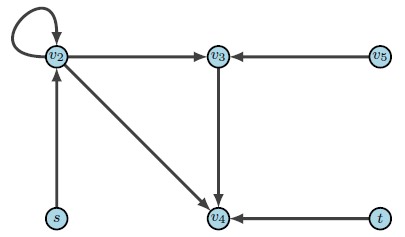}
\end{center}
\caption{\emph{This figure is a directed graph $G$ in the reduction from {\rm DSTCON} to $\overline{\mbox{\rm 2SAT}}$. We therefore obtain the Boolean formula $\phi =(\neg x\vee v_2)\wedge (\neg v_2\vee v_2)\wedge (\neg v_2\vee v_3)\wedge (\neg v_2\vee v_4)\wedge (\neg v_3\vee v_4)\wedge (\neg v_5\vee v_3)\wedge (\neg x\vee v_4)\wedge (x\vee y)\wedge (x\vee \neg y)$. Here $A=\{ x,v_2,v_3,v_4\}, B=\{ t\}$ and $C=\{ v_5\}$. Clearly, $A\cap B=B\cap C=A\cap C=\emptyset$. Also $\not\exists$ any directed edge from a vertex in $A$ to $B\cup C$ and from a vertex in $A\cup C$ to $B$. Since there does not exist any directed path from $s$ to $t$ in $G$ it is easy to see that $\phi$ is satisfiable: we can assign the truth value {\rm True} to every variable in $A\cup C$ including the variable $x$ and the truth value {\rm False} to the literal $\neg x$ corresponding to the vertex $t$. We assign either the truth value {\rm True} or {\rm False} to the variable $y$.}}
\end{figure}
Now let us assign the truth value $T$ to every variable in $A\cup C$ including the variable $x$ and the truth value $F$ to every variable in $B$ including the literal $\overline{x}$. It is easy to verify that this truth value assignment satisfies $\phi$. This shows that ${\rm DSTCON}\Lredn\overline{\rm 2SAT}$.
\end{proof}

\begin{theorem}\label{2SATisNLcomplete}
{\rm 2SAT} {\it is logspace many-one complete for $\NL$.}
\end{theorem}
\begin{proof}
This result follows from Theorem \ref{chap2-2SATisinNL}, Theorem \ref{chap2-2SATisNLhard} and Theorem \ref{chap2-NLcomplement}.
\end{proof}

{\bf\underline{CONJECTURE}}: Counting the number of satisfying assignments of a 2SAT formula, denoted by $\sharp$2SAT, is logspace many-one complete for $\sharpL$.\textcolor{white}{\index[subject]{$\sharp$2SAT}}

\subsection{Non-deterministic space bounded complexity classes above $\boldNL$}
\begin{definition}
Let $\Sigma$ be the input alphabet and $S(n)\in\Omega (\log n)$. We define the complexity class $\NSPACE (S(n))=\{ L\subseteq \Sigma^*|\exists$ a $O(S(n))$ space bounded non-deterministic Turing machine M such that $L=L(M)\}$.\textcolor{white}{\index[subject]{$\NSPACE (S(n))$}}
\end{definition}

\begin{definition}
Let $\Sigma$ be the input alphabet and $S(n)\in\Omega (\log n)$. We define the complexity class $\coNSPACE (S(n))=\{ \overline{L}\subseteq \Sigma^*|L\in\NSPACE (S(n))\}$.\textcolor{white}{\index[subject]{$\coNSPACE (S(n))$}}
\end{definition}

\begin{theorem}\label{ISTheorem}
{\rm\bf(The Immerman-Szelepcsenyi Theorem)}$\NSPACE(S(n))$ is closed under complement, where $S(n)\in\Omega(\log n)$. In other words, $\NSPACE$ $ (S(n))$ $=$$\coNSPACE(S(n))$, where $S(n)\in\Omega(\log n)$.\textcolor{white}{\index[subject]{Immerman-Szelepcsenyi Theorem}}
\end{theorem}
\begin{proof}
Let $\Sigma$ be the input alphabet and let $L\subseteq \Sigma^*$ such that $L\in\NSPACE$ $(S(n))$. Let $M$ be the $\NSPACE(S(n))$ Turing machine that accepts $L$. As in the proof of Lemma \ref{DSTCONisNLhard} we observe that given an input $x\in\Sigma^*$, all the configurations of $M(x)$ is computable by a deterministic Turing machine using space at most $O(S(n))$ and therefore the adjacency matrix of the configuration graph of $M(x)$, which is a directed graph,  can also be output by a $O(S(n))$ space bounded deterministic Turing machine. As in Theorem \ref{sldagstconcomplete} we observe that the directed $st$-connectivity problem for the simple layered directed acyclic graph obtained from the configuration graph $G$ of a $\NSPACE (S(n))$ Turing machine $M(x)$ is a canonical complete problem for $\NSPACE (S(n))$. Now following the algorithm in the proof of Theorem \ref{chap2-NLcomplement} we are able to decide if there does not exist a directed path from a vertex $s$ in layer $1$ of the above graph $G$ to the vertex $t$ in layer $n$ of $G$ in $\NSPACE (S(n))$. This shows that $\coNSPACE (S(n))\subseteq \NSPACE (S(n))$ from which our theorem follows.\textcolor{white}{\index[subject]{NSPACE$(S(n))$}}\textcolor{white}{\index[subject]{co-NSPACE$(S(n))$}}
\end{proof}

\section{Logarithmic Space Bounded Counting classes}
In this section we assume without loss of generality that the input alphabet $\Sigma=\{ 0,1\}$. We informally say that a complexity class $\mathcal{C}$ is a logarithmic space bounded counting class if the criterion to decide if an input string $x\in\Sigma^*$ is in a language $L\in\mathcal{C}$ is based on the number of accepting computation paths and/or the number of rejecting computation paths of a $O(\log n)$ space bounded non-deterministic Turing machine $M$. {\bf It follows from this informal definition that $\NL$ is the first and fundamental logarithmic space bounded counting class.}\textcolor{white}{\index[subject]{rejecting computation path}}
We recall the defintion of $\sharpL$ from Definition \ref{sharpL}.
\begin{proposition}\label{chap2-prop-countcomppaths}
Let $\Sigma$ be the input alphabet and let $M$ be a $O(\log n)$-space bounded non-deterministic Turing machine. Counting the number of computation paths of $M$ on input $x$ is in $\sharpL$, where $x\in\Sigma^*$.
\end{proposition}
\begin{proof}
We define a $O(\log n)$-space bounded non-deterministic Turing machine $M'$ which simulates $M$ on input $x\in\Sigma^*$ such that $M'$ accepts $x$ along all computation paths. In other words, given any input $x\in\Sigma^*$, if $M(x)$ accepts then $M'(x)$ also accepts. Otherwise if $M(x)$ rejects then $M'(x)$ still accepts. As a result $L(M')=\Sigma^*$. Also the number of computation paths of $M'$ on $x$ is equal to $acc_{M'}(x)=acc_M(x)+rej_M(x)$, which is the number of computation paths of $M$ on $x$, where $x\in\Sigma^*$.
\end{proof}

\begin{definition}\label{GapL}
Let $\Sigma$ be the input alphabet. The complexity class $\GapL$ is defined to be the class of functions $f:\Sigma^{*}\rightarrow\Z$ such that there exists a $\NL$-Turing machine $M$ for which we have $f(x)=\acc _M(x)-\rej _M(x)$ where $\acc _M(x)$ and $\rej _M(x)$ denote the number of accepting computation paths and the number of rejecting computation paths of $M$ on any input $x\in\Sigma^{*}$ respectively. We also denote $(\acc _M(x)-\rej _M(x))$ by $\gap _M(x)$.\textcolor{white}{\index[subject]{$\GapL$}}\textcolor{white}{\index[subject]{$rej_M(x)$}} \textcolor{white}{\index[subject]{$gap_M(x)$}}
\end{definition}
\begin{tcolorbox}[colback=gray!35!white,colframe=white]
{\bf\textit{In Section \ref{chap1-sec-comptree}, we have assumed that the computation binary tree of an $O(S(n))$-space bounded nondeterministic Turing machine is a complete binary tree. That is, all computation paths in the computation binary tree of an $O(S(n))$-space bounded nondeterministic Turing machine have the same length. However, note that under this assumption any $f\in\GapL$ will never be an odd integer on any input $x\in\Sigma^*$, where $\Sigma$ is the input alphabet, since the number of leaves in the computation tree is an integer which is a power of $2$. This suggests an unrealistic computation tree. So to overcome this problem, we assume that in many places such as when considering $\GapL$ functions that the computation binary tree for $L\in\NL$ contains computation paths which are not all of the same length. In other words, the computation binary tree of a non-deterministic Turing machine on any given input is not necessarily a complete binary tree.}}
\end{tcolorbox}

\begin{definition}\label{pldefinition}
A language $L$ belongs to the logarithmic space bounded counting class $\PL$ if there exists a $\GapL$ function $f$ such that $x\in L$ if and only if $f(x)>0$.\textcolor{white}{\index[subject]{$\PL$}}
\end{definition}

\begin{definition}\label{ceqldefinition}
A language $L$ belongs to the logarithmic space bounded counting class $\CeqL$ if there exists a $\GapL$ function $f$ such that $x\in L$ if and only if $f(x)=0$.\textcolor{white}{\index[subject]{$\CeqL$}}
\end{definition}

We recall the definition of $\FL$ from Definition \ref{chap1-defn-fl}. We need the following result.
\begin{proposition}\emph{(Folklore)}\label{sharplfolklore}
\begin{enumerate}
\item Let $\Sigma$ be the input alphabet and $f:\Sigma^{*}\rightarrow\Z$ such that $f(x)\geq 0,~\forall x\in\Sigma^*$ and $f\in \FL$. Then $f\in\sharpL$.
\item $\sharpL$ is closed under addition.
\item $\sharpL$ is closed under multiplication.
\item $\GapL$ is closed under addition.
\item $\GapL$ is closed under multiplication.
\end{enumerate}
\end{proposition}
\begin{proof}
\begin{enumerate}
\item Let $f$ be a $\FL$ function that takes non-negative values on all inputs. Given an input $x\in\Sigma^*$, without loss of generality, we assume that $f(x)\in\Z^+$ is represented in binary notation.

Let us define a $\NL$-Turing machine $M$, which on input $x$ first computes $f(x)$, and thereby finds the size of $f(x)$. If $f(x)=0$ then $M$ rejects $x$ and stops. Otherwise, let $l$ denote the size of $f(x)$. It is easy to see that $l\in O(\log n)$, where $n=|x|$. Given $f(x)$, we denote the $i^{th}$ bit of $f(x)$ by $f(x)_i$, where $1\leq i\leq l$. Let $f(x)=f(x)_lf(x)_{l-1}\cdots f(x)_1$. Here $f(x)_l$ denotes the most significant bit of $f(x)$ and $f(x)_1$ denotes the least significant bit of $f(x)$. Since we assume that $f(x)\neq 0$, we have $f(x)_l=1$ always.

After finding $l$, let us define $M$ as follows: $M$ starts by sequentially choosing exactly $l$ bits from $\{ 0,1\}$ in a non-deterministic fashion. For $1\leq i\leq l$, after choosing the $i^{th}$ bit, $M$ computes $f(x)_{l-i+1}$. If $f(x)_{l-i+1}=1$ and the $i^{th}$ bit chosen by $M$ is $0$, then $M$ accepts the input $x$ along all of its computation paths obtained by choosing the remaining $(l-i)$ bits and stops. Instead, if $f(x)_{l-i+1}$ equals the $i^{th}$ bit chosen by $M$, then $M$ continues to iterate the above step of choosing the $(i+1)^{st}$ bit non-deterministically as long as $(i+1)\leq l$. However, if $f(x)_{l-i+1}=0$ and the $i^{th}$ bit chosen by $M$ is $1$ then $M$ rejects the input $x$ along all of its computation paths obtained by choosing the remaining $(l-i)$ bits and stops.

It is easy to see that $M$ accepts $x$ if and only if any of the computation paths which $M$ has guessed is among the first $f(x)$ lexicographically least computation paths in the computation tree of $M$ which has depth $l$ and $2^l$ leaves. Clearly $\acc_M(x)=f(x)$ whenever $f(x)\geq 0, \forall x\in\Sigma^*$.
\item Let $f_1,f_2\in\sharpL$. Since we can define a $\NL$-Turing machine which first branches once and simulates the $\NL$-Turing machine corresponding to $f_1$ along one branch and the $\NL$-Turing machine corresponding to $f_2$ along the other branch, we get the result that $(f_1+f_2)\in\sharpL$.
\item Let $f_1,f_2\in\sharpL$. We define a $\NL$-Turing machine $M$ which simulates the $\NL$-Turing machine corresponding to $f_1$ and then at the end of all of its accepting computation paths it simulates the $\NL$-Turing machine corresponding to $f_2$. At the end of the rejecting computation paths of the $\NL$-Turing machine corresponding to $f_1$, $M$ also rejects the input. It is the easy to see that $(f_1f_2)\in\sharpL$.
\item Proof is identical to (2).
\item  Let $f_1,f_2\in\GapL$. We define a $\NL$-Turing machine $M$ which simulates the $\NL$-Turing machine corresponding to $f_1$ and then at the end of all of its computation paths it simulates the $\NL$-Turing machine corresponding to $f_2$. $M$ accepts if and only if the $\NL$-Turing machine corresponding to $f_2$ accepts. It is the easy to see that $(f_1f_2)\in\GapL$.
\end{enumerate}
\end{proof}
\begin{lemma}\label{nlpropertyplusone}
Let $\Sigma$ be the input alphabet and $L\subseteq\Sigma^*$ be a language in $\NL$. $\exists f\in\sharpL$ such that for any given input $x\in\Sigma^*$, we have $x\in L$ if and only if $f(x)> N$, where $2\times N$ is the number of computation paths of the $\NL$-Turing machine $M$ of $f$.
\end{lemma}
\begin{proof}
Let $M'$ be the $\NL$-Turing machine that accepts $L$. For any given input $x\in\Sigma^*$, we know that $x\in L$ if and only if $acc_{M'}(x)>0$. Let $M''$ be the $O(\log n)$-space bounded non-deterministic Turing machine as shown in Proposition \ref{chap2-prop-countcomppaths} which accepts $\Sigma^*$ such that, given any input $x\in\Sigma^*$, $M''$ simulates $M'$ and finally accepts $x$ and never rejects any input $x$. It is easy to see that the number of accepting computation paths of $M''(x)$, denoted by $acc_{M''}(x)$, is equal to the number of computation paths of $M'$ on $x$.

Given an input $x\in\Sigma^*$, let $M$ be the non-deterministic Turing machine that non-deterministically decides to either simulate $M'$ on input $x$ along the left branch or it decides to simulate $M''$ on input $x$ on the right branch of its computation tree. It is easy to see that, given input $x\in\Sigma^*$, the number of computation paths of $M$ on $x$ is equal to $2\times N$, where $N$ is the number of computation paths of $M'$ on $x$. Also if $x\in L$ then we have $acc_M(x)> N$. Otherwise if $x\not\in L$ then $acc_M(x)=N$. We take $f\in\sharpL$, as the $\sharpL$ function which is the number of accepting computation paths of $M$ on input $x\in\Sigma^*$.
\end{proof}
\begin{theorem}\label{nlpl}
$\NL\subseteq\PL$.
\end{theorem}
\begin{proof}
Follows from Definition \ref{pldefinition} and Lemma \ref{nlpropertyplusone}.
\end{proof}

\begin{definition}
The symmetric difference of two sets $A$ and $B$ is the set of elements that belong to exactly one of $A$ and $B$. The symmetric difference of two sets $A$ and $B$ is denoted by $A\triangle B$. In other words, $A\triangle B=(\overline{A}\cap B)\cup (A\cap\overline{B})$.\textcolor{white}{\index[subject]{$A\triangle B$}}
\end{definition}
\begin{definition}
Let $\Sigma$ be the input alphabet. We define $(\NL\triangle\NL)=\{ L\subseteq\Sigma^*|\exists L_1,L_2\in\NL$ and $L=L_1\triangle L_2$, where $L_1,L_2\subseteq\Sigma^*\}$.\textcolor{white}{\index[subject]{$\NL\triangle\NL$}}
\end{definition}

\begin{theorem}\label{nldeltanllnl}
$(\NL\triangle\NL)\subseteq\L^{\mbox{\NL}}$.
\end{theorem}
\begin{proof}
Given two sets $A$ and $B$ it follows from the definition that $(A\triangle B)=(\overline {A}\cap B)\cup (A\cap \overline{B})$. Now, let $A,B\in\NL$. Due to Theorem \ref{chap2-NLcomplement} and Corollary \ref{nlunionintersection}, it follows that $(A\triangle B)\in\NL$. Clearly, to determine if an input string $x\in\Sigma^*$ is in $(A\triangle B)$ a logarithmic space bounded deterministic Turing machine that has access to a $\NL$ oracle shall make one query and output if $x$ is in $(A\triangle B)$ or not. This shows that $(\NL\triangle\NL)\subseteq\L^{\mbox{\NL}}$.
\end{proof}

\begin{theorem}
$\NL =(\NL\triangle\NL)$.
\end{theorem}
\begin{proof}
Let $\Sigma$ be the input alphabet and let $L\subseteq\Sigma^*$ be a non-trivial language such that $L\in\NL$. It is easy to see that $L=L\triangle\emptyset$, where $\emptyset$ denotes the empty set. We therefore get $\NL\subseteq\NL\triangle\NL$. Converse follows from Theorem \ref{nldeltanllnl} and the closure of $\NL$ under $\leq ^{\rm \L}_{\rm T}$ reductions, which we have shown in Theorem \ref{nlclosureunderTuring}.
\end{proof}

\begin{lemma}\label{chap2-lemm-power2paths}
Let $\Sigma$ be the input alphabet, $L\subseteq\Sigma^*$, $L\in\NL$ and let $M$ be a $\NL$-Turing machine such that $L(M)=L$. There exists a $\NL$-Turing machine $M'$ such that $L(M')=L$ and, on any input $x\in\Sigma^*$, the computation tree of $M'$ is a complete binary tree such that it has has exactly $2^{n^k}$ computation paths and $acc_{M'}(x)=acc_M(x)$, where $n=|x|$ and $k>0$.
\end{lemma}
\begin{proof}
Let $M$ be the $\NL$-Turing machine such that the number of configurations of $M$ on input $x\in\Sigma^*$ is $\leq n^c$, where $c>0$ is a constant. Without loss of generality, we assume that the computation tree of $M$ on $x$ is a binary tree which has finite number of nodes and edges. Let us define another $\NL$-Turing machine $M'$ which on input $x$ does the following:
\begin{enumerate}
    \item simulate $M$ on $x$, and 
    \item start a (deterministic) counter $ctr$ with its initial value $n^k$, where $k>c>0$, and decrement $ctr$ by $1$ in each step until $ctr>0$.
\end{enumerate}
In the simulation of $M$ by $M'$, each configuration we encounter has either $0$ or $1$ or $2$ successors.
\begin{itemize}
    \item if the present configuration has $2$ successors, then $M'$ also has two successor configurations and it continues to simulate $M$ on $x$,
    \item if the present configuration has $1$ successor, then $M'$ is defined such that it has $2$ successor configurations among which one successor configuration continues to simulate $M$ on $x$, and the other is the root of a subtree which rejects along all of its computation paths until $ctr>0$.
    \item if the present configuration has $0$ successors then it is a halting configuration of $M$ on $x$. If the halting configuration is a rejecting configuration, then $M'$ forms a rejecting subtree, as done above by decrementing $ctr$ by $1$ in each step until $ctr>0$, with this node as the root. Otherwise if the halting configuration is an accepting configuration, then $M'$ will generate a complete subtree, as done above by decrementing $ctr$ by $1$ in each step until $ctr>0$, wherein the lexicographically least computation path accepts and remaining computation paths of this subtree reject the input.
\end{itemize}
It is easy to see that the computation tree of $M'$ is a complete binary tree and $acc_{M'}(x)=acc_M(x)$.
\end{proof}

\begin{lemma}\label{nlcoceql}
$\NL\subseteq\coCeqL$.\textcolor{white}{\index[subject]{$\coCeqL$}}
\end{lemma}
\begin{proof}
Let $L\subseteq \Sigma^*$ such that $L\in\NL$ and let $M'$ be the $\NL$-Turing machine which accepts $L$ as in Lemma \ref{chap2-lemm-power2paths}. Let $f\in\sharpL$ such that on any input $x\in\Sigma^*$ we have $f(x)=acc_{M'}(x),~\forall x\in\Sigma^*$. For any input $x\in\Sigma^*$ such that $n=|x|$, we assume without loss of generality that all the computation paths of the computation tree of $M'$ on input $x$ have length $n^k$, where $k>0$ is a constant and which clearly depends on the $O(\log n)$ space used by $M'(x)$ in any of its computation paths.
\begin{algorithm}[H]
\caption{$\NL$-contained-in-$\coCeqL$ $\newline$ {\bf Input:} $x\in\Sigma^*$\newline {\bf Output:} \emph{accept} if $x\in L$, where $L\in\NL$. Otherwise \emph{reject}. \newline {\bf Complexity:} $\coCeqL$.}\label{algo:nl-contained-coceql}
\begin{algorithmic}[1]
\State Non-deterministically choose one of the following two options. \State Simulate $M(x)$, or
\State Form a computation tree having depth $n^k$ and accept the input $x$ on all of these computation paths
\end{algorithmic}
\end{algorithm}
Let us consider the algorithm given above which can be implemented by a $\NL$-Turing machine, say $M''$, on a given input $x\in\Sigma^*$, where $n=|x|$. On any given input $x\in\Sigma^*$, if $x\in L$ then $f(x)\geq 1$. As a result $\gap_{M''}(x)=\acc_{M''}(x)-\rej_{M''}(x)>0$. However if $x\not\in L$ then all the computation paths of $M'$ on input $x$ end in the rejecting configuration and therefore $acc_{M''}(x)=rej_{M''}(x)$ which clearly implies $gap_{M''}(x)=0$. This shows that $L\in\coCeqL$ from which we get $\NL\subseteq\coCeqL$.
\end{proof}
\begin{theorem}\label{nlceql}
$\NL\subseteq\CeqL$.
\end{theorem}
\begin{proof}
From Lemma \ref{nlcoceql} we infer that $\coNL\subseteq \CeqL$. Also as a consequence of Theorem \ref{chap2-NLcomplement} we obtain that $\NL\subseteq\CeqL$.
\end{proof}
\begin{proposition}
$\CeqL$ is closed under union.
\end{proposition}
\begin{proof}
It follows from Proposition \ref{sharplfolklore} due to the closure of $\GapL$ under multiplication.
\end{proof}
\subsection{Properties of $\boldsharpL$ and $\boldGapL$ functions}\label{salientGapL}
We recall the definition of $\sharpL$ and $\GapL$ from Definitions \ref{sharpL} and \ref{GapL}
\begin{lemma}\label{computationpathsupperbound}
Let $\Sigma$ be the input alphabet. If $f(x)$ is a $\sharpL$ or $\GapL$ function then there exists a polynomial $p(n)$ such that the absolute value of $f(x)$ is bounded above by $2^{p(n)}$ for every $x\in\Sigma^*$, where $n=|x|$.
\end{lemma}
\begin{proof}
Either $f(x)=acc _M(x)$ or $\gap _M(x)$ for some $\NL$-Turing machine $M$. Let $s(n)\in O(\log n)$ be an upper bound the amount of space that $M$ uses on input strings of length $n$. It is easy to note that the running time of $M(x)$ is at most $p(|x|)=2^{s(|x|)}$, which is a polynomial in $|x|$. As a result the number of distinct computation paths of $M$ on any input $x$ is upper bounded by $2^{p(|x|)}$ from which it follows that the number of  accepting computation paths and rejecting computation paths of $M$ on any input $x$ is at most $2^{p(|x|)}$ and this proves our claim.
\end{proof}

\begin{proposition}\label{gaplcomplement}
If $f\in\GapL$ then $-f\in\GapL$.
\end{proposition}

\begin{theorem}\label{chap06-gaplsharpl}
Let $\Sigma$ be the input alphabet. For every $\NL$-Turing machine $M$, there is a $\NL$-Turing machine $N$ such that $\gap_N(x)=\acc_M(x)$, on any input $x\in\Sigma^*$.
\end{theorem}
\begin{proof}
Given an input $x$, let $\overline{M}(x)$ denote the $\NL$-Turing machine obtained by reversing the decision of computation paths of $M(x)$. In other words, $\overline{M}(x)$ rejects every accepting computation path of $M(x)$ and accepts every rejecting computing path of $M(x)$. Our $\NL$-Turing machine $N$ guesses a path $p$ of $M(x)$. If $p$ is accepting, $N$ accepts. Otherwise, $N$ branches once, accepting on one branch and rejecting on the other. We have for all $x$, $\gap_N(x)=\acc_N(x)-\rej_{N}(x)=\acc_N(x)-\acc_{\overline{M}}(x)=(\acc_M(x)+\acc_{\overline{M}}(x))-\acc_{\overline{M}}(x)=\acc_M(x).$ Note that all the computation paths of $N$ are not necessarily having the same length.
\end{proof}
\begin{corollary}\label{corollary-gaplsharpl}
$\sharpL\subset\GapL$.
\end{corollary}
\begin{proof}
Follows from Theorem \ref{chap06-gaplsharpl} and the fact that $\GapL$ contains functions that take negative values on many inputs.
\end{proof}

Note that a $\sharpL$ function can take only nonnegative values and so this class cannot capture all the functions in $\FL$ that evaluate to integer values on any given input. However, it does capture those functions in $\FL$ that only take nonnegative integer values on all inputs as shown in Proposition \ref{sharplfolklore}. The class $\GapL$ includes any function in $\FL$ that takes integer values on any input without any restriction. For the rest of this chapter, we assume without loss of generality that, any function $f\in \FL$ which we consider always takes integer values on any input.
\begin{theorem}\label{FLsharpLGapL}
Every $\FL$ function $f$ is in $\GapL$.
\end{theorem}
\begin{proof}
Let $f\in\FL$. Define a $\NL$-Turing machine $N$ which on input $x$ first computes $f(x)$. For any $f(x)$, since the absolute value of $f(x)$ (denoted by $|f(x)|$) is greater than $0$, it follows from Proposition \ref{sharplfolklore} that there exists a $\NL$-Turing machine $N''$ such that $acc_{N''}(x)=|f(x)|$ on any input $x\in\Sigma^*$. Using Corollary \ref{corollary-gaplsharpl} we get that there exists a $\NL$-Turing machine $N'$ such that $\gap_{N'}(x)=acc_{N''}(x)=|f(x)|$. Therefore, after computing $f(x)$, if $f(x)>0$ then $N$ simulates $N'$ on input $x$. However, if $f(x)<0$, then $N$ simulates $N'$ on input $x$ and rejects at the end of the computation paths in which $N'$ accepts and accepts at the end of the computation paths in which $N'$ rejects. This shows that $\gap_N(x)=f(x)$.
\end{proof}

\begin{theorem}\label{sharpLGapLconnect}
Let $\Sigma$ be the input alphabet. If $h\in\GapL$, then there exists $f,g\in\sharpL$ such that $h(x)=f(x)-g(x)$ on any input $x\in\Sigma^*$. Conversely, if $f,g\in\sharpL$, then there exists $h\in\GapL$ such that $h(x)=f(x)-g(x)$ on any input $x\in\Sigma^*$.  In other words, we can write every function in $\GapL$ as the difference of two functions in $\sharpL$. Equivalently, we say that $\GapL =\sharpL -\sharpL$.
\end{theorem}
\begin{proof}
Let $h\in\GapL$ and let $M$ denote the $\NL$-Turing machine corresponding to $h$. Given a $\NL$-Turing machine $M$ and an input $x$, let $\overline{M}(x)$ denote the $\NL$-Turing machine obtained by reversing the decision of computation paths of $M(x)$. In other words, $\overline{M}(x)$ rejects every accepting computation path of $M(x)$ and accepts every rejecting computing path of $M(x)$. For any $M$ we have $\gap_M(x)=\acc_M(x)-\acc_{\overline{M}}(x)$ by definition, so if $h\in\GapL$ then $h$ is the difference of two $\sharpL$ functions.

Conversely, let $f,g\in\sharpL$. Using Corollary \ref{corollary-gaplsharpl} we get that $f,g\in\GapL$. It follows from Theorem \ref{chap06-gaplsharpl} that there exists $\NL$-Turing machines $M$ and $N$ such that $gap_M(x)=f(x)$ and $gap_N(x)=g(x)$ on any input $x\in\Sigma^*$. We use Proposition \ref{gaplcomplement} and Lemma \ref{sharplfolklore} and obtain that $(f(x)-g(x))=(gap_M(x)-gap_N(x))\in\GapL$, on any input $x\in\Sigma^*$.
\end{proof}

\begin{theorem}\label{gaplsharplflconnect}
Let $\Sigma$ be the input alphabet. If $h\in\GapL$, then there exists $f\in\sharpL$ and $g\in\FL$ such that $h(x)=f(x)-g(x)$ on any input $x\in\Sigma^*$. Conversely, if $f\in\sharpL$ and $g\in\FL$, then there exists $h\in\GapL$ such that $h(x)=f(x)-g(x)$ on any input $x\in\Sigma^*$.  In other words, we can write every function in $\GapL$ as the difference of a function in $\sharpL$ and a function in $\FL$. Equivalently, we say that $\GapL =\sharpL -\FL$.
\end{theorem}
\begin{proof}
Let $h\in\GapL$. We know from Theorem \ref{sharpLGapLconnect} that there exists $f,g\in\sharpL$ such that, on any given input $x\in\Sigma^*$, we have $h(x)=f(x)-g(x)$. Let $M_f$ and $M_g$ be $\NL$-Turing machines corresponding to $f$ and $g$ respectively. As shown in Lemma \ref{chap2-lemm-power2paths}, we assume that the computation tree of both $f$ and $g$ are complete binary trees such that both $M_f$ and $M_g$ have exactly $2^{n^k}$ computation paths on any input of length $n$, where $k>0$ is a constant. Let $M_{\overline{g}}$ be the $\NL$-Turing machine which reverses the decision of a computation path of $M_g$. In other words, along any computation path, $M_{\overline{g}}$ accepts if and only if $M_g$ rejects. Let us consider a Turing machine $M$ which non-deterministically branches once and simulates $M_f$ along the left branch and $M_{\overline{g}}$ along the right branch. Then, given any input $x\in\Sigma^*$, we have
\[\begin{array}{lll}
(f(x)-g(x)) & = & acc_{M_f}(x)-acc_{M_g}(x)\\
 & = & acc_{M_f}(x)+acc_{M_{\overline{g}}}(x)-2^{n^k}\\
 & = & acc_{M}(x)-2^{n^k}.
 \end{array}
\]
Since a $\NL$-Turing machine can output $2^{n^k}$ in binary notation when it is given an input $x\in\Sigma^*$ such that $n=|x|$, we get $h=(f-g)\in\sharpL -\FL$.

Conversely, let $f\in\sharpL$ and let $g\in\FL$. We know from Theorem \ref{chap06-gaplsharpl} that there exists a $\NL$-Turing machine $N_1$ such that $gap_{N_1}(x)=f(x)$, for any $x\in\Sigma^*$. We also know from Theorem \ref{FLsharpLGapL} that $g\in\GapL$. Therefore there exists a $\NL$-Turing machine $N_2$ such that $gap_{N_2}(x)=g(x)$, for any $x\in\Sigma^*$. It is now easy to see using Propositions \ref{gaplcomplement} and \ref{sharplfolklore} that $(f-g)\in\GapL$.
\end{proof}

We first recall the notion of an oracle Turing machine from Section \ref{chap1-sec-oracleturingmachine}. We also recall the notion of having a function as an oracle from Section \ref{functionoracleaccess2}. Let us also recall the definition of logspace Turing reducibility using functions from Definitions \ref{L-Turing-reduction} and \ref{L-oracle-class}.
\begin{note}\label{note-function-oracle-definition}
\begin{enumerate}
\item In Definition \ref{L-Turing-reduction}, if we replace $L_2$ by a function $f\in\mathcal{F}$, where $\mathcal{F}$ is a complexity class of functions such as $\sharpL$ then $L_1\LTredn f$ and we say that $M^f$ accepts $L_1$. In this context of having a function $f$ such as $\sharp$DSTCON as an oracle, we note that to obtain the value of the oracle function when it is given an oracle query string as input, the $O(\log n)$-space bounded deterministic Turing machine has to submit at least one and at most $p(n)$ many queries, where $p(n)$ is a polynomial in $n$, and $n$ is the size of the input.
\item In Definition \ref{L-oracle-class} if we replace $\mathcal{C}$ by $\mathcal{F}$, where $\mathcal{F}$ is a complexity class of functions such as $\sharpL$ then $\L ^{\mathcal{F}}$ is the complexity class of all languages that is accepted by a $O(\log n)$ space bounded deterministic Turing machine that has access to a function $f\in\mathcal{F}$ as an oracle.
\end{enumerate}
\end{note}
\begin{definition}
We define $\FL^{\sharpL}$ to be the complexity class of all functions that are logspace Turing reducible to the complexity class of functions in $\sharpL$.\textcolor{white}{\index[subject]{$\FL^{\sharpL}$}}
\end{definition}
\begin{lemma}\label{gaplequivsharpl}
$\GapL\subseteq\FL^{\sharpL}$.
\end{lemma}
\begin{proof}
We know from Proposition \ref{dstconsharplcomplete} that $\sharp${\rm DSTCON} is logspace many-one complete for $\sharpL$. Let $\Sigma$ be the input alphabet. It follows from Proposition \ref{parsicomplete} that if $L\in\NL$ such that $M$ is the $\NL$-Turing machine which decides if an input $x\in\Sigma^*$ is in $L$, then there exists an instance $(G,s,t)$ of DSTCON which is obtained using the logspace many-one reduction shown in Theorem \ref{DSTCONisNLhard} that $acc_M(x)$ is equal to the number of directed paths from $s$ to $t$ in $G$.

Let $f\in\GapL$ and let $M$ be the $\NL$-Turing machine such that given an input $x\in\Sigma^*$, we have $f(x)=acc_M(x)-rej_M(x)$. Let $n$ be the size of the input $x$. Let $\overline{M}$ be the $\NL$-Turing machine which is obtained from $M$ by reversing the decision of the computation paths of $M(x)$. In other words, $\overline{M}(x)$ rejects every accepting computation path of $M(x)$ and accepts every rejecting computation path of $M(x)$. As a result, $acc_{\overline{M}}(x)=rej_M(x)$. Since $\sharp$DSTCON is logspace many-one complete for $\sharpL$, using a $O(\log n)$-space bounded determinsitic Turing machine $M'$, we can obtain $2$ instances $(G_1,s_1,t_1)$ and $(G_2,s_2,t_2)$ of DSTCON from $x$ such that $acc_M(x)$ is equal to the number of directed paths from $s_1$ to $t_1$ in $G_1$ and $acc_{\overline{M}}(x)$ is equal to the number of directed paths from $s_2$ to $t_2$ in $G_2$. $M'$ can therefore use these $2$ instances of DSTCON and submit $p(n)$ many oracle queries to find $acc_M(x)$ and $acc_{\overline{M}}(x)=rej_M(x)$ in a bit-wise manner, where $p(n)$ is a polynomial in $n$. After finding these values, $M'$ computes $f(x)=acc_M(x)-rej_M(x)$ and outputs it in the output tape of $M'$.
\end{proof}
\begin{proposition}\label{lsharplequalsllsharpl}
$\L^{\mbox{\LsharpL}}$ $=\LsharpL$.\textcolor{white}{\index[subject]{$\L^{\mbox{\LsharpL}}$}\index[subject]{$\mbox{\LsharpL}$}}
\end{proposition}
\begin{proof}
It is trivial to show that $\LsharpL\subseteq \L^{\mbox{\LsharpL}}$. We have to therefore show that $\L^{\mbox{\LsharpL}}\subseteq\LsharpL$. Let $L\in\L^{\mbox{\LsharpL}}$ such that there exists a deterministic $O(\log n)$ space bounded oracle Turing machine $N^{L'}$ which correctly decides if any input string $x\in\Sigma^*$ is in $L$, where $L'\in{\mbox{\LsharpL}}$ is the oracle for $N$. Let $M'$ denote a deterministic oracle Turing machine which correctly decides if an input string $y'\in L'$.

Let $M$ be a deterministic $O(\log n)$ space bounded Turing machine which simulates $N^{L'}$ on the given input string $x\in\Sigma^*$ till $N^{L'}$ generates an oracle query $y'$ which is to be submitted to the oracle $L'$. In other words, $M$ continues its simulation of $N^{L'}$ and stops its simulation in the step before it enters the $q_{QUERY}$ state. Now $M$ stores the present configuration of $N^{L'}$ which includes its present state, position of the tape head of the input tape, the contents of the work tape, and the position of the tape head of the work tape. Clearly $M$ requires at most $O(\log n)$ space to store this information. $M$ then simulates $M'$ on the query $y'$ with access to a function in $\sharpL$ as the oracle. Based on this simulation of $M'$ by $M$, it is possible to correctly decide if the oracle query string $y'$ is in $L'$. Therefore after correctly deciding if $y'$ is in the oracle, $M$ restores itself to the state of $N^{L'}$ which it had saved and continues with its simulation repeating these steps whenever a oracle query string is generated. As a result it correctly decides if the input string $x\in L$ or not. Since $M$ simulates a $\LsharpL$ computation and uses at most $O(\log n)$ space we get that $L\in\LsharpL$.
\end{proof}
\begin{proposition}\label{lgaplequalsllgapl}
$\L^{\mbox{\LGapL}}$ $=\LGapL$.\textcolor{white}{\index[subject]{$\L^{\mbox{\LGapL}}$}\index[subject]{$\mbox{\LGapL}$}}
\end{proposition}
\begin{theorem}\label{lgaplequalslsharpl}
$\L^{\mbox{\GapL}}=\L^{\mbox{\sharpL}}$.
\end{theorem}
\begin{proof}
It follows from Corollary \ref{corollary-gaplsharpl} that $\sharpL\subset\GapL$. As a result $\L^{\mbox{\sharpL}}\subseteq\L^{\mbox{\GapL}}$. To prove the other way inclusion, we have already shown that $\GapL\subseteq\L^{\mbox{\sharpL}}$ in Lemma \ref{gaplequivsharpl}. So using Proposition \ref{lsharplequalsllsharpl}, we get $\L^{\mbox{\GapL}}\subseteq\L^{\mbox{\sharpL}}$. We therefore get 
$\L^{\mbox{\GapL}}=\L^{\mbox{\sharpL}}$.
\end{proof}

\begin{lemma}\label{composeFLsharpLGapL}
Let $f$ be a $\FL$ function and $g$ be a $\sharpL$ or $\GapL$ function. Then $g(f(x))$ is a $\sharpL$ or $\GapL$ function, respectively.
\end{lemma}
\begin{proof}
Let $M$ be a $\NL$-Turing machine that defines a $\sharpL$ function $g$. Define $N$ to be a $\NL$-Turing machine that on input $x$ simulates $M(f(x))$. Then $\acc_N(x)$ is exactly $g(f(x))$. The proof for $\GapL$ is identical.
\end{proof}

We refer to Definition \ref{appendix-defn-choosefunction} and Theorem \ref{appendix-thm-choosefunction} in the Appendix A for ${n\choose k}$, where $n,k\in\Z^+$.
\begin{theorem}\label{sharplchooseconstant}
If $f\in\sharpL$ and $k\in\Z^+$ is a constant, then $g(x)={{f(x)}\choose k}\in\sharpL$.
\end{theorem}
\begin{proof}
Let $M$ be a $\NL$-Turing machine which has $f(x)$ accepting computation paths on any input $x\in\Sigma^*$. We now define a $\NL$-Turing machine $N$ such that the number of accepting computation paths of $N$ on input $x\in\Sigma^*$ is ${{f(x)}\choose k}$.

Without loss of generality, assume that $k>0$. Otherwise, if $k=0$ the $\NL$-Turing machine $N$ is defined such that it accepts the input $x\in\Sigma^*$ at the end of the lexicographically least computation path and rejects on all the other computation paths. Now let $k>0$. $N$ starts to simulate $M$ on the given input $x\in\Sigma^*$, and it first stores the initial configuration of $M$ on the work tape. Clearly, this requires $O(\log n)$ space, where $n=|x|$. $N$ maintains a counter which is initialized to $1$. $N$ will store constantly many configurations of $M$ on its work tape and the counter will keep track of how many configurations have been stored.

Now every time when $N$ has to make a non-deterministic choice, in an iterative manner $N$ cycles through all the configurations stored on its work tape and chooses one of the following three possibilities:

(a) replace the configuration which it is at present considering by its left successor configuration

(b) replace the configuration which it is at present considering by its right successor configuration

(c) replace the configuration by its left successor configuration and add the right successor configuration to the list of configurations stored on the work tape, and increment the counter by $1$.

In the above non-determinisitic algorithm, $N$ rejects $x$ if the counter exceeds $k$, or if any of the stored configurations are rejecting configurations of $M$ on $x$, or if all the configurations are accepting; however the counter is less than $k$. $N$ accepts $x$ if all the configurations are accepting and the counter is exactly $k$. Clearly, the space used by $N$ is $O(k\log n)=O(\log n)$ since $k$ is a  constant, where $n=|x|$. Also it is clear that $N$ chooses $k$ many distinct computation paths from the computation of $M(x)$. As a result, the number of accepting computation paths of $N$ on input $x\in\Sigma^*$ is ${{f(x)}\choose k}$.
\end{proof}
Let us recall Proposition \ref{sharplfolklore}.
\begin{theorem}\label{gaplsharplpolynomialaddition}
\begin{enumerate}
\item\label{sharpladdition} Let $\Sigma$ be the input alphabet, $f\in\sharpL$ and $p(n)$ be a polynomial. Let $h:\Sigma^*\rightarrow \Z$ be defined for all $x\in\Sigma^*$ by $h(x)=\Sigma_{1\leq i\leq p(|x|)}f(\langle x,i\rangle)$. $h\in\sharpL$.
\item\label{sharplmultiply} Let $\Sigma$ be the input alphabet, $f\in\sharpL$ and $p(n)$ be a polynomial. Let $h:\Sigma^*\rightarrow \Z$ be defined for all $x\in\Sigma^*$ by $h(x)=\prod_{1\leq i\leq p(|x|)}f(\langle x,i\rangle)$. $h\in\sharpL$.
\item\label{gapladdition} Let $\Sigma$ be the input alphabet, $f\in\GapL$ and $p(n)$ be a polynomial. Let $h:\Sigma^*\rightarrow \Z$ be defined for all $x\in\Sigma^*$ by $h(x)=\Sigma_{1\leq i\leq p(|x|)}f(\langle x,i\rangle)$. $h\in\GapL$.
\end{enumerate}
\end{theorem}
\begin{proof}
\begin{enumerate}
\item Let $f(\langle x,i\rangle)=\acc_M(\langle x,i\rangle)$ for some $\NL$-Turing machine $M$. Define $N$ to be the $\NL$-Turing machine that on input $x\in\Sigma^*$, first guesses a positive integer $i$ between $1$ and $p(|x|)$. If the guessed integer $i$ is greater than $p(|x|)$ then $N$ rejects and stops.

Let us assume that $1\leq i\leq p(|x|)$. $N$ simulates a computation path of $M$ on the input $\langle x,i\rangle$. $N$ accepts the input $\langle x,i\rangle$ if $M$ accepts and rejects the input otherwise. Then, for every $x\in\Sigma^*$, we get $\acc_N(x)=h(x)$.
\item Let $f(\langle x,i\rangle)=\acc_M(\langle x,i\rangle)$ for some $\NL$-Turing machine $M$. Define $N$ to be the $\NL$-Turing machine that on input $x\in\Sigma^*$ simulates a computation path of $M$ on the input $\langle x,i\rangle$, for all $1\leq i\leq p(|x|)$, in sequence one after the other. If for some $i$ in the simulation of $M$ on $\langle x,i\rangle$ by $N$, the computation path rejects then $N$ also rejects and stops. Otherwise $N$ continues to simulate $M$ on $\langle x,i\rangle$. Then for every $x\in\Sigma^*$ we get $acc_N(x)=h(x)$.
\item Let $f(\langle x,i\rangle)=\gap_M(\langle x,i\rangle)$ for some $\NL$-Turing machine $M$. Define $N$ to be the $\NL$-Turing machine which on input $x\in\Sigma^*$, first guesses a positive integer $i$ between $1$ and $p(|x|)$. If the guessed integer $i$ is greater than $p(|x|)$ then $N$ branches exactly once more and accepts along one branch and rejects along the other branch.

Let us assume that $1\leq i\leq p(|x|)$. $N$ simulates a computation path of $M$ on the input $\langle x,i\rangle$. $N$ accepts the input $\langle x,i\rangle$ if $M$ accepts and rejects the input otherwise. Then, for every $x\in\Sigma^*$, we get $\gap_N(x)=h(x)$.
\end{enumerate}
\end{proof}

\begin{theorem}\label{gaplpolynomialmultiplication}
Let $\Sigma$ be the input alphabet, $f\in\GapL$ and $p(n)$ be a polynomial. Let $h:\Sigma^*\rightarrow \Z$ be defined for all $x\in\Sigma^*$ by $h(x)=\prod _{1\leq i\leq p(|x|)}f(\langle x,i\rangle)$. $h\in\GapL$.
\end{theorem}
\begin{proof}
Let $f=\gap_M$ for some $\NL$-Turing machine $M$. For each $x\in\Sigma^*$ and $i$ such that $1\leq i\leq p(|x|)$, let $S(x,i)$ denote the set of all computation paths of $x$ on input $\langle x,i\rangle$ and, furthermore, for each $\pi\in S(x,i)$, define $\alpha (x,i,\pi )=1$ if $\pi$ is an accepting computation path and it is $-1$ otherwise. Then, for each $x\in\Sigma^*$, $h(x)=\Sigma_{\pi_1\in S(x,1)}\cdots \Sigma_{\pi_{p(|x|)}\in S(x,p(|x|))}$$\alpha (x,1,\pi_1)\cdots $ $\alpha (x,p(|x|$$),\pi_{p(|x|)})$.

Define $N$ to be the $\NL$-Turing machine that on input $x\in\Sigma^*$ behaves as follows: $N$ non-deterministically guesses and simulates a path of $M$ on input $\langle x,i\rangle$, for all $1\leq i\leq p(|x|)$. In the course of doing this, $N$ computes the parity of the number of values of $i$, $1\leq i\leq p(|x|)$, such that $M$ on input $\langle x,i\rangle$ rejects. When all the simulations have been completed, $N$ accepts $x$ on its current computation path if and only if this value is even. Note that, for every $x\in\Sigma^*$, on the path of $N$ on input $x$ corresponding to the guesses $(\pi _1,\pi _2,\ldots ,\pi _{p(|x|)})$, the product $\alpha (x,1,\pi _1)\cdots \alpha (x,p(|x|),\pi _{p(|x|)})$ is $1$ if and only if $N$ accepts at the end of this computation path and that the product is $-1$ if and only if $N$ rejects at the end of this computation path. Thus for every $x\in\Sigma^*$, we have $\gap_N(x)=h(x)$. Since $p$ is a polynomial and $M$ is a $\NL$-Turing machine, $N$ is also a $\NL$-Turing machine. Thus $h\in\GapL$.
\end{proof}
\section{The Isolating Lemma}\label{Isolating-Lemma}
Let $A = \{a_1,a_2,\ldots ,a_m\}$ be a set of elements and let $\F$ be a collection of non-empty subsets of $A$. Let $W$ be a finite number of consecutive integers, which are called as integer weights, and assume that we shall assign integer weights to elements of $A$ using a function $f:A\rightarrow W$. Define the weight of a subset $S$ of $A$, denoted by $f(S)$, to be the sum of the weights of the elements in the subset $S$.\textcolor{white}{\index[subject]{Isolating Lemma}}

We say that a weight function $f$\emph{ is good for }$\F$ if there is exactly one minimum-weight set in $\F$ with respect to $f$ and we say that $f$\emph{ is bad for }$\F$ otherwise.
\begin{theorem}{\textrm{\bf (The Isolating Lemma)}}\label{Isolating}
Let $A$ be a finite set of cardinality $m$ and let $\F$ be a family of non-empty subsets of $A$. Let $r>2m$ and let $f$ be a function $f:A\rightarrow\{1,\ldots ,r\}$ that assigns a value independently and uniformly at random to each element in $a_i\in A$ from the set $\{ 1,\ldots ,r\}$, where $1\leq i\leq m$. Then,
\[
\Pr_{f}[f~is~good~for~\F ]\geq\frac{1}{2}.
\]
\end{theorem}
\begin{proof}
Let us define the $MinWeight_f(\F )=\min \{f(S)|S\in\F\}$ and let $MinWeightSet_f(\F )=\{S\in\F |f(S)=MinWeight_f(\F )\}$. Recall from the definition of good and bad functions stated above that any weight function $f$ is bad for $\F$ if and only if $|MinWeightSet_f(\F )|\geq 2$.

For $x\in A$, we say that minimum-weight sets of $\F$ with respect to $f$ are ambiguous about the inclusion of $x$ if there exists at least one pair of sets $S,S'\in MinWeightSet_f(\F )$ such that $x\in (S\setminus S')\cup (S'\setminus S)$. In other words, we say that minimum-weight sets of $\F$ with respect to $f$ are ambiguous about the inclusion of $x$ if there exists at least one pair of sets $S,S'\in MinWeightSet_f(\F )$ such that $x$ is in exactly one of $S$ or in $S'$, and that $x$ is not in both $S$ and in $S'$. Conversely, minimum-weight sets of $\F$ with respect to $f$ are unambiguous about the inclusion of $x$, if $x$ is either in the intersection of all minimum-weight sets or $x$ is not in any minimum-weight set of $\F$. It is easy to see that $f$ is bad for $\F$ if and only if there exists some $x\in A$ such that minimum weight sets of $\F$ with respect to $f$ are ambiguous about the inclusion of $x$. Since any weight function $f$ assigns only positive weights to elements of $A$, it is also easy to see that, if $f$ is a bad weight function then $\exists S,S'\in MinWeightSet_f(\F )$ such that $S\neq S'$ and $|(S\setminus S')\cup (S'\setminus S)|\geq 2$. As a result, if $f$ is a bad weight function, then there exists an element $y\in A$ such that $y\neq x$, $x\in S$, $y\in S'$, and, minimum-weight sets of $\F$ with respect to $f$ are ambiguous about the inclusion of $y$ also.  In fact, we will have at least $2\leq k\leq min(m,{|MinWeightSet_f(\F)|\choose 2})$ such elements in $A$ such that minimum-weight sets of $\F$ with respect to $f$ are ambiguous about the inclusion of each of these $k$ elements.

Let us assume that a function $f$ is bad for $\F$. Therefore $|MinWeightSet_f(\F )|$ $\geq 2$, and minimum weight sets of $\F$ with respect to $f$ are ambiguous about the inclusion of $x$, for some $x\in A$. Let us fix this $x\in A$, and let us not bother about the number of ambiguous elements of $A$ with respect to $f$. Let us define $f':A\rightarrow \{ 1,\ldots, r\}$ suitably so that $MinWeight_{f'}(\F )=MinWeight_{f}(\F )$. In defining $f'$, we first ensure that $MinWeightSet_{f'}$ $(\F )=\{ S\in MinWeightSet_f($$\F$ $ )|x\not\in S\}$.

Let $B=\cup_{\{T|T~does~not~contain~x, ~T\in {MinWeightSet_f(\F)}\}}T$. It is easy to see that $|B|\geq 2$. First, determine if there exists $z'\not\in B$ such that $z'\neq x$. If there does not exist any such $z'$ then we arbitrarily delete a pair of elements from $B$, say $z_1,z_2$, and we do not assign any weight to these two elements immediately as we do for other elements in $B$: if $z\in B$ then let $f'(z)=f(z)$. We define $f'$ for elements $z''\not\in B$, where $z''\neq x$, or the pair of elements $z_1,z_2$ that might exist as mentioned above such that $f'(z'')\neq f(z'')$, or $\{ f'(z_1)\neq f(z_1)$ and $f'(z_2)\neq f(z_2)\}$. This is to ensure that we obtain a function $f'$ which is not $f$ itself. Before we assign weights for $x$ in $f'$, let us consider $\F '$ which is the collection of subsets of $A$ in $\F$ such that if $x$ is in a set $S\in\F$ then we delete $x$ from $S$. It is important to note that by defining $f'$ in this manner we obtain $MinWeight_{f'}(\F ')=MinWeight_f(\F ')=MinWeight_f(\F )$. Let us finally assign a weight $f'(x)\neq f(x)$ for $x$ and and let us consider $\F$. 

{\emph{\textbf{Claim.}}} Minimum weight sets of $\F$ with respect to $f'$ are unambiguous about the inclusion of $x$.

{\emph{\textbf{Proof of Claim.}}}
To prove this, let $\delta (x)=f'(x)-f(x)$. Suppose that $\delta (x)>0$. For any $S\subseteq A$, if $x\in S$ then $f'(S)=f(S)+\delta (x)$ and $f'(S)=f(S)$ otherwise. This implies that $MinWeight_f(\F )=MinWeight_{f'}(\F )$ and $MinWeightSet_{f'}$ $(\F )=\{ S\in MinWeightSet_f($$\F$ $ )|x\not\in S\}$. Therefore, if $\delta (x)>0$ then there does not exist any minimum-weight set of $\F$ with respect to $f'$ that contains $x$. Next, suppose that $\delta (x)<0$. Then, for all $S\in \F$, we have $f'(S)=f(S)-|\delta (x)|$ if $x\in S$ and $f'(S)=f(S)$ otherwise. This implies that $MinWeight_f(\F )=MinWeight_{f'}(\F )-|\delta (x)|$ and $MinWeightSet_{f'}(\F )=\{ S\in MinWeightSet_f$ $(\F )|x\in S\}$. Therefore, if $\delta (x)<0$ then all minimum-weight sets of $\F$ with respect to $f'$ contain $x$. This proves our claim.

Therefore, if $f'(x)=f(x)$ then $f'$ is a bad weight function and minimum weight sets of $\F$ with respect to $f'$ are ambiguous about the inclusion of $x$. Given $\F$, we need not have every $x\in A$ as a possible candidate for being ambiguous since some $x$ can be in every subset of $A$ in $\F$. However if we have obtained one element $x\in A$ as ambiguous with respect to a bad weight function $f$, then the number of bad weight functions $f'$ that we can obtain from $f$, for each of which $x$ is ambiguous, is $\leq r^{m-1}$. Since there can be at most $m$ choices for $x$, the number of bad weight functions that can exist is at most $mr^{m-1}$. As a result, the proportion of weight functions $f$ such that $f$ is bad for $\F$ is $\leq\frac{mr^{m-1}}{r^m}< \frac{1}{2}$. Therefore, the proportion of weight functions $f$ that are good for $\F$ is $\geq\frac{1}{2}$ and this proves our theorem.
\end{proof}

\begin{corollary}\label{chap2-coro-family-weight-Isolating}
Let $A=\{ a_1,\ldots ,a_m\}$ be a finite set of cardinality $m$ and let $\F _1,\F _2,\ldots ,\F _n$ be a collection of families of non-empty subsets of $A$, where $n<m^k$ for some $k\geq 0$. Let $r>2nm$ and let $f$ be a function $f:A\rightarrow \{ 1,\ldots ,r\}$, which assigns a value independently and uniformly at random to each element $a_i\in A$ from the set $\{ 1,\ldots ,r\}$, where $1\leq i\leq m$. Then,
\[
\Pr_{f}[f~is~good~for~each~\F_i]\geq\frac{1}{2},
\]
where $1\leq i\leq n$.
\end{corollary}
\begin{proof}
If we have only one family $\F$, then we have shown in Theorem \ref{Isolating} that
\[
\Pr_{f}[f~is~bad~for~\F ]\leq\left(\frac{mr^{m-1}}{r^m}\right).
\]
If there are $n$ families $\F_1,\ldots ,\F_n$, then
\[
\Pr_{f}[f~is~bad~for~at~least~one~\F_j,~where~1\leq j\leq n ]\leq\frac{nmr^{m-1}}{r^m}<\frac{1}{2}.
\]
As a result, we get
\[
\Pr_{f}[f~is~good~for~all~\F_j,~where~1\leq j\leq n]> \left( 1-\frac{nmr^{m-1}}{r^m}\right)\geq\frac{1}{2}.
\]
\end{proof}

\begin{corollary}\label{chap3-coro-nlpolyulpolytailormade}
Let $A=\{ a_1,\ldots ,a_m\}$ be a finite set of cardinality $m$ and let $\F _1,\F _2,\ldots ,\F _n$ be a collection of families of non-empty subsets of $A$, where $n<m^k$, for some $k\geq 0$. Let $r>4nm$. There is a collection of weight functions $f_1,\ldots f_{m}$ such that, each $f_i:A\rightarrow\{ 1,\ldots ,r\}$ and for every collection $\F_1,\ldots ,\F_n$ of families of non-empty subsets of $A$, there exists some $f_i$ such that $f_i$ is good for all $\F_j$, where $1\leq i\leq m$ and $1\leq j\leq n$.
\end{corollary}
\begin{proof}
Since we have chosen $r>4nm$, it follows from Corollary \ref{chap2-coro-family-weight-Isolating} that
\[
\Pr_{f}[f~is~good~\forall\F_j,~where~1\leq j\leq n]\geq \left(\frac{3}{4}\right).
\]
We first note that there are $2^m$ possible subsets of $A$. Among these subsets, let us consider $S\subseteq A$ such that $a_i\in S$ if and only if $a_i$ is a member of at least one subset of $A$ in some family $\F_j$, where $1\leq i\leq m$ and $1\leq j\leq n$.
As a result considering $S$ and the given a collection of families $\F _1,\F _2,\ldots ,\F _n$ of non-empty subsets of $A$, if $f_1,\ldots f_{m}$ are random weight functions then,
\[
\Pr_{f}[\not\exists ~1\leq j\leq n:~(f_j~is~not~good~\forall \F_j,~where~1\leq j\leq n)]< \left(\frac{1}{4^{m}}\right).
\]
Since there $2^m$ possible subsets of $A$, given any $S\subseteq A$ from whose elements we get families $\F_1,\ldots ,\F_n$ we get
\[
\Pr_{f}[\not\exists ~1\leq j\leq n:~(f_j~is~not~good~\forall \F_j,~where~1\leq j\leq n)]< \left(\frac{2^m}{4^m}\right)<1.
\]
Therefore, there exists a collection of weight functions $f_1,\ldots ,f_m$ such that for every collection of subsets $\F_1,\ldots ,\F_n$ of $A$, there exists some $f_i$ such that $f_i$ is good for all $\F_j$, where $1\leq i\leq m$ and $1\leq j\leq n$.
\end{proof}

\begin{corollary}\label{family-Isolating}
Let $A$ be a finite set of cardinality $m$ and let $\F _1,\F _2,\ldots ,\F _n$ be a collection of families of non-empty subsets of $A$, where $n<m^k$ for some $k\geq 0$. Also let $r>\max (m^4,n^4)$. There exists a function $f:A\rightarrow\{1,\ldots ,r\}$ such that for all families $\F _1,\F _2,\ldots ,\F _n$ of non-empty subsets of $A$, we have $f$ is good for every $\F _i$, where $n< m^k$, for some $k\geq 0$, and $1\leq i\leq n$.
\end{corollary}
\begin{proof}
Form a $0,1$ matrix $M$ of dimension $n\times r^m$ with $n$ rows and $r^m$ columns, where the $i^{th}$ row of $M$ corresponds to the family $\F _i$ and the $j^{th}$ column corresponds to the function $f_j$ in the lexicographically increasing order, for $1\leq i\leq n$ and $1\leq j\leq r^m$. (Here we assume that every function $f$ is considered as a vector, and given two functions $f$ and $g$ we say that $f$ is lexicographically smaller than $g$, denoted by $f\prec g$, if $f$ is less than $g$ when compared component-wise as vectors). We put $M_{ij}=1$ if the function $f_j$ is good for the family $\F _i$ and $M_{ij}=0$ otherwise. For a given $1\leq i\leq n$, by the choice of $r$ and from the proof of Theorem \ref{Isolating}, it follows that
\[
{\Pr_f}\left[f~is~good~for~\F _i\right]>\left (1-\frac{1}{m^k}\right ),
\]
where $f$ is a function chosen independently and uniformly at random. For sufficiently large $m$, it is easy to see that there exists a column in $M$ that contains only $1$. The function corresponding to this column satisfies the condition stated in this theorem.
\end{proof}

\subsection{Min-unique graphs and Unambiguous Logarithmic space, $\boldUL$}\label{min-unique-graphs}
\begin{definition}\label{minunique}
A min-unique graph is a weighted directed graph with positive weights associated with each edge where for every pair of vertices $u,v$, if there is a path from $u$ to $v$, then there exists a unique minimum weight path from $u$ to $v$. Here, the weight of a path is the sum of the weights on its edges.\textcolor{white}{\index[subject]{min-unique graph}} Any such weight function on the edges of the min-unique graph $G$ is defined as a min-unique weight function. If the maximum of the positive integer weights assigned by the weight function are less than or equal to a polynomial in the size of the directed graph then we say that the weight function is polynomially bounded.\textcolor{white}{\index[subject]{min-unique weight function}}

Let $f$ be a function that assigns positive weights to the edges of the directed graph. If only for some vertex $s$ the minimum weight directed path from $s$ to every other vertex reachable from $s$ is unique, then the weight function $f$ is min-unique with respect to $s$.
\end{definition}

We define the complexity class unambiguous logarithmic space, $\UL$.
\begin{definition}\label{UL}
Let $\Sigma$ be the input alphabet. We say that $L\subseteq\Sigma ^{*}$ is a language in the complexity class $\UL$ if there exists a function $f\in \sharpL$ such that for any input $x\in\Sigma ^{*}$ we have $f(x)=1$ if $x\in L$ and $f(x)=0$ if $x\not\in L$.\textcolor{white}{\index[subject]{$\UL$}}
\end{definition}
We now define $\coUL$, based on the definition of $\mbox{co-}\mathcal{C}$ from Definition \ref{compdefinition} where $\mathcal{C}$ is a complexity class.
\begin{definition}\label{coUL}
Let $\Sigma$ be the input alphabet. For a language $L\in\Sigma^*$, the complement of $L$ is $\overline{L}=\Sigma^*-L$. We define the complexity class $\coUL=\{ \overline{L}\subseteq \Sigma ^*|L\in\UL\}$.\textcolor{white}{\index[subject]{$\coUL$}}
\end{definition}
\begin{definition}
Let $\Sigma$ be the input alphabet. We say that $L\in\UL\cap\coUL$ if $L\in\UL$ and $L\in\coUL$.\textcolor{white}{\index[subject]{$\UL\cap\coUL$}}
\end{definition}

\begin{theorem}\label{min-unique-st-connectivity-UL}
Let $\mathcal{G}$ be a class of graphs, where each graph in $\mathcal{G}$ is given in terms of its adjacency matrix, and let $H=(V,E)\in\mathcal{G}$. If there is a polynomially bounded logarithmic space computable function $f$ that on input $H$ outputs a weighted graph $f(H)$ so that
\begin{enumerate}
\item $f(H)$ is min-unique, and
\item $H$ has an $st$-path if and only if $f(H)$ has an $st$-path
\end{enumerate}
then the $st$-connectivity problem for $\mathcal{G}$ is in $\UL\cap\coUL$.
\end{theorem}
\begin{proof}
It suffices to give a $\UL\cap\coUL$ algorithm for the reduced graph. For $H\in\mathcal{G}$, let $G=f(H)$ be a directed graph with a min-unique weight function $w$ on its edges. We first construct an unweighted graph $G'$ from $G$ by replacing every edge $e$ in $G$ with a directed path of length $w(e)$. It is easy to see that $st$-connectivity is preserved. That is, there is an $st$-path in $G$ if and only if there is one in $G'$. Since $G$ is a min-unique graph, it is straightforward to argue that the shortest path between any two vertices in $G'$ is unique. Let us call this directed graph $G'$ as {\bf\emph{unweighted min-unique graph}}\textcolor{white}{\index[subject]{unweighted min-unique graph}}.

Let $c_k$ and $\Sigma _k$ denote the number of vertices which are at a distance at most $k$ from $s$ and the sum of the lengths of the shortest path to each of them, respectively. Let $d(v)$ denote the length of the shortest path from $s$ to $v$. If no such path exists, then $d(v)=|V|+1$. We have,
\[
{\ssum}_k=\sum_{\begin{array}{c}{\scriptstyle\{{v\in V}}|{\scriptstyle{d(v)\leq k}\}}\end{array}}d(v).
\]
We first give an unambiguous routine (Algorithm \ref{algo:shortest-path}) to evaluate the predicate ``$d(v)\leq k$" when given the values of $c_k$ and $\ssum _k$. The algorithm will output the correct value of the predicate (True/False) on a unique path and outputs ``?" on the rest of the paths.
\begin{algorithm}[H]
\caption{Shortest-Path: Determining whether $d(v)\leq k$ or not. $\newline$ {\bf Input:} $(G,v,k,c_k,\ssum _k)$, where $G=(V,E)$ is a unweighted min-unique graph given in terms of its adjacency matrix, $v\in V$ and $k\in\N$.\newline{\bf Output:} True if $d(v)\leq k$. Otherwise False or ?. \newline {\bf Complexity:} $\UL$.}\label{algo:shortest-path}
\begin{algorithmic}[1]
\State Initialize $count\leftarrow 0$; $sum\leftarrow 0$; $path.to.v\leftarrow$False
\For {each $x\in V$}
	\State {Non-deterministically guess if $d(x)\leq k$}
	\If {\emph{guess is Yes}}
		\State {Non-deterministically guess a path of length $l\leq k$ from $s$ to $x$}
        \If {\emph{guess is correct}}
			\State{Set $count\leftarrow count+1$}
			\State{Set $sum\leftarrow sum+l$}
			\If {$x=v$}
                    \State{Set $path.to.v\leftarrow$True}
            \EndIf
		\Else            
            \State return ``?"
        \EndIf
	\EndIf
\EndFor
\If {$count=c_k$ and $sum=\ssum _k$}
	\State{return $path.to.v$}
\Else
	\State{return ``?"}
\EndIf    
\end{algorithmic}
\end{algorithm}
We will argue that Algorithm \ref{algo:shortest-path} is unambiguous.
\begin{enumerate}
\item If Algorithm \ref{algo:shortest-path} incorrectly guesses that $d(x)>k$ for some vertex $x$ then $count<c_k$ and so it returns ``?" in line 20. Thus, consider the computation paths that correctly guess the set $\{ x|d(x)\leq k\}$.
\item If at any point the algorithm incorrectly guesses the length $l$ of the shortest path to $x$, then one of the following two cases occur.
\begin{enumerate}
\item If $d(x)>l$ then no path from $s$ to $x$ would be found and the algorithm returns ``?" in line 13.
\item If $d(x)<l$ then the variable \emph{sum} would be incremented by a value greater than $d(x)$ and thus \emph{sum} would be greater than $\ssum _k$ causing the algorithm to return ``?" in line 20.
\end{enumerate}
\end{enumerate}
Thus there will remain only one computation path where all the guesses are correct and the algorithm will output the correct value of the predicate on this unique path. Finally, we note that Algorithm \ref{algo:shortest-path} is easily seen to be computable in logarithmic space.

Next, we describe an unambiguous procedure (Algorithm \ref{algo:computing-ck}) that computes $c_k$ and $\ssum _k$ given $c_{k-1}$ and $\ssum _{k-1}$.
\begin{algorithm}[H]
\caption{Computing $c_k$ and $\ssum _k.\newline${\bf Input:} $(G,k,c_{k-1},\ssum _{k-1})$, where $G=(V,E)$ is a unweighted min-unique graph given in terms of its adjacency matrix, and $k\in\N$.\newline {\bf Output:} $c_k,\ssum _k$. \newline {\bf Complexity:} $\UL$.}\label{algo:computing-ck}
\begin{algorithmic}[1]
\State Initialize $c_k\leftarrow c_{k-1}$ and $\ssum _k\leftarrow \ssum _{k-1}$
\For {each $v\in V$}
	\If {$\neg (d(v)\leq k-1)$}
		\For {each $x$ such that $(x,v)\in E$}    
			\If {$d(x)\leq k-1$}
				\State {Set $c_k\leftarrow c_k+1$}    
				\State {Set $\ssum _k\leftarrow \ssum _k+k$}
			\EndIf
        \EndFor        
    \EndIf
\EndFor    
\State{{\bf return} $c_k$ and $\ssum _k$}                
\end{algorithmic}
\end{algorithm}
Algorithm \ref{algo:computing-ck} uses Algorithm \ref{algo:shortest-path} as a subroutine. Other than making function calls to Algorithm \ref{algo:shortest-path}, this routine is deterministic, and so it follows that Algorithm \ref{algo:computing-ck} is also unambiguous.
We will argue that Algorithm \ref{algo:computing-ck} computes $c_k$ and $\ssum _k$. The subgraph consisting only of $s$ $(d(x)\leq 0)$ is trivially min-unique and $c_0=1$ and $\ssum _0=0$. Inductively, it is easy to see that
\[
\begin{array}{c}
c_k=c_{k-1}+|\{ v|d(v)=k\}|,\\
\ssum _k=\ssum _{k-1}+k\times|\{v|d(v)=k\}|.
\end{array}
\]
In addition, $d(v)=k$ if and only if there exists $(x,v)\in E$ such that $d(x)\leq k-1$ and $\neg (d(v)\leq k-1)$. Both of these predicates can be computed using Algorithm \ref{algo:shortest-path}. Combining these facts, we see that Algorithm \ref{algo:computing-ck} computes $c_k$ and $\ssum _k$ given $c_{k-1}$ and $\ssum _{k-1}$.
\begin{algorithm}[H]
\caption{Determining if there exists a path from $s$ to $t$ in a min-unique graph $G.\newline$ {\bf Input:} A min-unique graph $G=(V,E)$ given in terms of its adjacency matrix, and vertices $s,t\in V$.\newline {\bf Output:} True if there exists a directed path from $s$ to $t$ in $G$. Otherwise False. \newline {\bf Complexity:} $\UL$.}\label{algo:stpathmin-unique}
\begin{algorithmic}[1]
\State We obtain the unweighted min-unique graph $G'$ from $G$ as described in the beginning of this proof.
\State Initialize $c_0\leftarrow 1,\ssum _0\leftarrow 0,k\leftarrow 0$
\For {$k\leftarrow 1,\ldots n$}
	\State {Compute $c_k$ and $\ssum _k$ by invoking Algorithm \ref{algo:computing-ck} on $(G',k,c_{k-1},\ssum_{k-1})$}
\EndFor
\State {Invoke Algorithm \ref{algo:shortest-path} on $(G',t,n,c_n,\ssum _n)$ and return its value}
\end{algorithmic}
\end{algorithm}
As a final step, we give the main routine that invokes Algorithm \ref{algo:computing-ck} to check for $st$-connectivity in a min-unique graph. Since there is an $st$-path if and only if $d(t)\leq n$, it suffices to compute $c_n$ and $\ssum _n$ and invoke Algorithm \ref{algo:shortest-path} on the input $(G,t,n,c_n,\ssum _n)$. This procedure is presented as Algorithm \ref{algo:stpathmin-unique}. To ensure that the algorithm runs in logarithmic space, we do not store all intermediate values for $c_k,\ssum _k$. Instead, we only keep the most recently computed values and re-use space. Similar to Algorithm \ref{algo:computing-ck}, this procedure is deterministic and so the entire routine is unambiguous. Thus reachability in min-unique graphs can in fact be decided in $\UL\cap\coUL$.
\end{proof}

\begin{note}\label{min-unique-wrt-s}
We recall Definition \ref{minunique} and note that, in the proof of Theorem \ref{min-unique-st-connectivity-UL}, it is not necessary to have a weight function that is min-unique such that for every pair of vertices $u$ and $v$ there exists a directed path from $u$ to $v$ if and only if there exists a directed path from $u$ to $v$ which has unique minimum weight. Instead, it is sufficient that the min-uniqueness is with respect to only the vertex $s$. In other words, it is sufficient that the function $f$ outputs $f(H)$ such that there exists a directed path from $s$ to any vertex $v$ in $H$ if and only if there is a unique minimum weight directed path from $s$ to $v$ in $f(H)$.
\end{note}

\begin{definition}\label{fnldefn}
Let $\Sigma$ be the input alphabet and let $\Gamma$ be the output alphabet. We define $\FNL$ to be the complexity class of all functions $f:\Sigma^*\rightarrow\Gamma^*$ such that for any given input string $x\in\Sigma^*$ there exists a $\NL$-Turing machine $M$ which outputs $f(x)$ at the end of all of its accepting computation paths.\textcolor{white}{\index[subject]{$\FNL$}\index[subject]{$\FUL$}}
\end{definition}
\begin{definition}\label{fuldefn}
Let $\Sigma$ be the input alphabet and let $\Gamma$ be the output alphabet. We define $\FUL$ to be the complexity class of all functions $f:\Sigma^*\rightarrow\Gamma^*$ such that for any given input string $x\in\Sigma^*$ there exists a $\NL$-Turing machine $M$ which has at most one accepting computation path and $M$ outputs $f(x)$ at the end of its unique accepting computation path.
\end{definition}
Let $\Sigma$ be the input alphabet and $\Gamma$ be the output alphabet. We say that a function $f:\Sigma\rightarrow\Gamma^*$ is $\UL$ computable if $f\in\FUL$.
\begin{theorem}\label{min-unique-wrt-s-nlul}
$\NL =\UL$ if and only if there is a polynomially-bounded $\UL$ computable weight function $f$ so that for any directed acyclic graph $G$, we have $f(G)$ is min-unique with respect to $s$.
\end{theorem}
\begin{proof}
Assume $\NL =\UL$. Therefore, given a directed graph $G=(V,E)$ in terms of its adjacency matrix, and vertices $s,t\in V$ as input, the problem of determining if there exists a directed path from $s$ to $t$ in $G$ is in $\UL$. Also, we know from Theorem \ref{chap2-NLcomplement} that $\NL =\coNL$. So $\UL =\coUL$.

\emph{Claim}: The language $A=\{(G,s,t,k)|\exists$ a path from $s$ to $t$ of length $\leq k\}$ is in $\UL$.

\begin{cproof}
We reduce $(G,s,t)$ to an instance of $\LDAG$ in $O(\log n)$ space as in Theorem \ref{sldagstconcomplete}. Using a deterministic $O(\log n)$ space bounded Turing machine We then query the $\DSTCON$ oracle if there exists a directed path from $s$ to the copy of the vertex $t$ in every layer starting from the second layer of the instance of $\LDAG$. We accept the input if there is a directed path from $s$ to $t$ in layer $i$ where $i\leq k$. Otherwise we reject. Our claim now follows because of Theorem \ref{nlclosureunderTuring} and our assumption that $\NL =\UL$.
\end{cproof}

We now compute a subgraph of $G$ which is a directed tree rooted at $s$ such that there exists a directed path from $s$ to a vertex $v$ in $G$ if and only if there exists a directed path from $s$ to $v$ in the tree. We say that a vertex $v$ is in level $k$ if the minimum length of any directed path from $s$ to $v$ is of length $k>0$. A directed edge $(u,v)$ is in the tree if for some $k>0$,
\begin{enumerate}
\item $v$ is in level $k$, and
\item $u$ is the lexicographically first vertex in level $k-1$ so that $(u,v)$ is a directed edge.
\end{enumerate}
It is clear that this is indeed a well-defined tree and to decide if an edge $e=(u,v)$ is in this tree, we submit the following query to $A$: is it true that $\forall w<u,~l\leq (k-1)~[(\bigwedge (G,s,w,l)\not\in A)\wedge (w,v)\in A]$. If an edge is in the tree we assign the weight $1$ to it. For the rest of the edges we assign weight $n^2$. It is clear that the shortest path from $s$ to any vertex with respect to this weight function is min-unique. It is also easy to see that this weight function is computable in $\L ^{A}\subseteq\UL$.

Conversely, by following the proof of Theorem \ref{min-unique-st-connectivity-UL} under the assumption that the complexity of computing weighted graph $f(H)$ using the weight function $f$ is $\UL$ and the observation in Note \ref{min-unique-wrt-s} we get $\NL=\UL$.
\end{proof}

\subsection{Some more applications of Isolating Lemma and non-deterministic counting}\label{NLULpoly}
\begin{definition}\label{nonuniformC}
Let $\ssum =\{ 0,1\}$ be the input alphabet and let $\mathcal{C}$ be a complexity class. We define $\mathcal{C}/poly$ as the complexity class of all languages $L\subseteq\ssum^*$ for which there exists a sequence of ``advice strings" $\{\alpha (n)|n\in \N\}$, where $|\alpha (n)|\leq p(n)$ for some polynomial $p$, and a language $L'\in\mathcal{C}$ such that $x\in L$ if and only if $(x,\alpha (|x|))\in L'$, for every $x\in \ssum^*$.\textcolor{white}{\index[subject]{$\mathcal{C}/\poly$, non-uniform complexity class $\mathcal{C}$}}\textcolor{white}{\index[subject]{NL/poly, or non-uniform NL}}\textcolor{white}{\index[subject]{(UL$\cap$co-UL)/poly, or non-uniform (UL$\cap$co-UL)}}
\end{definition}

\begin{theorem}
$(\UL\cap\coUL){\rm /poly}\subseteq \NLpoly$.
\end{theorem}
\begin{proof}
It is easy to see from Definition \ref{UL} and Definition \ref{NLdefn} that $\UL\subseteq\NL$. Also we know from the Immerman-Szelepcsenyi Theorem (also see Theorem \ref{chap2-NLcomplement}) that $\NL=\coNL$ and so $(\UL\cap\coUL)\subseteq\NL$. Using Definition \ref{nonuniformC}, it is now trivial to show that $(\UL\cap\coUL){\rm /poly}\subseteq \NLpoly$.
\end{proof}

\begin{theorem}\label{chap2-thm-nlpolycontulpoly}
$\NLpoly\subseteq (\UL\cap\coUL){\rm /poly}$.
\end{theorem}
\begin{proof}
Using Definition \ref{nonuniformC}, we infer that it is sufficient to show that $\NL\subseteq (\UL\cap\coUL){\rm /poly}$. We know from Theorem \ref{DSTCONisNLcomplete} that, $\DSTCON$ is complete for $\NL$ under $\Lredn$. Therefore to prove our result, it is sufficient to show that $\DSTCON\in(\UL\cap\coUL){\rm /poly}$.

Let us now consider an input instance $(G,s,t)$ of $\DSTCON$, where $G=(V,E)$ is a directed graph and $n=|V|$. We assume that $G$ is given as input in terms of its $n\times n$ adjacency matrix $M$. Also vertices $s$ and $t$ of $G$ are marked using two special symbols of the input alphabet $\Sigma$. As a result the size of the input $(G,s,t)$ is $(n^2+2)$ symbols. We also assume the following ordering of edges in $G$: if $M(i,j)=1$ then it is the $((i-1)\times |V|+j)^{th}$ edge of $G$. Otherwise if $M(i,j)=0$ then the $((i-1)\times |V|+j)^{th}$ edge does not exist in $G$. In total there can be at most $n^2$ edges in $G$.

Let us define a family $\F _{u,v}$ of subsets of $E$ such that if $S\subseteq E$ and $S\in \F_{u,v}$ then edges in $S$ form a directed path from $u$ to $v$ in $G$. Number of such families $\F _{u,v}$ is at most $2\times {n\choose 2}$. Let $r>4n^4$ and $A=\{ 1,\ldots ,(n^2+2)\}$. It follows from Corollary \ref{chap3-coro-nlpolyulpolytailormade} that there is a collection of weight functions $f_1,\ldots ,f_{n^2+2}$ such that each $f_i:A\rightarrow \{ 1,\ldots ,r\}$, and for every collection $\F _{u,v}$ of families of non-empty subsets of $E$, $\exists$ some $f_i$ such that $f_i$ is good for all $\F_{u,v}$, where $1\leq i\leq (n^2+2)$ and $u,v\in V$. In other words, the set of weight functions $f_1,\ldots ,f_{n^2+2}$ which we obtain using Corollary \ref{chap3-coro-nlpolyulpolytailormade} is such that, there exists some $f_i$ using which there exists a directed path from $u$ to $v$ in $f_i(G)$ of unique minimum weight between every pair of vertices $u,v$ for which there exists a directed path from $u$ to $v$ in $G$, where the weight of a directed path is the sum of the weights of the edges in the directed path, $1\leq i\leq (n^2+2)$ and $f_i(G)$ is the weighted directed graph obtained after assigning weights to edges of $G$ according to $f_i$.

Given $f_i$, we replace every edge $(u,v)$ that exists in $G$ with a directed path of length $f_i((u,v))$, where $1\leq i\leq (n^2+2)$. Let the resulting directed graph be $G_i$. We assume that vertices in $G$ which are in $G_i$ are numbered from $1$ to $|V|$ and newly added vertices get numbering greater than $|V|$ in $G_i$. As a result we obtain $(n^2+2)$ directed graphs $G_1,\ldots G_{n^2+2}$ such that
\begin{itemize}
    \item the directed graph $G_i$ has a directed path from a vertex $u$ to another vertex $v$ if and only if there exists a directed path from $u$ to $v$ in $G_i$, and
    \item there exists some $1\leq i\leq (n^2+2)$ such that $G_i$ is min-unique.
\end{itemize}
It is easy to see that, given the directed graph $G=(V,E)$ as input, if we are given $(n^2+2)$ weight functions then we can output the $(n^2+2)$ directed graphs $G_1,\ldots ,G_{n^2+2}$ in $O(\log n)$ space, where $n=|V|$.

A useful observation is that if $G_i$ is min-unique, then we can use Algorithm \ref{algo:shortest-path} (Shortest-Path) on input $(G_i,v,k,c_k,\Sigma_k)$ to decide whether $d(v)\leq k$ or not in $\UL$. Here we assume that correct values of $c_k$ and $\Sigma_k$ are provided. We complete the proof of this theorem by giving two algorithms. The proof of correctness and the complexity analysis of following algorithms is similar to Theorem \ref{min-unique-st-connectivity-UL}.

The following algorithm computes $c_k$ and $\Sigma_k$ if the input directed graph is min-unique, or it outputs that the input directed graph is a $BAD.GRAPH$ and therefore it is not min-unique.
\begin{algorithm}[H]
\caption{Computing $c_k$ and $\ssum _k$ for directed graphs.$\newline${\bf Input:} $(G,k,c_{k-1},\ssum _{k-1})$, where $G=(V,E)$ is a directed graph given in terms of its adjacency matrix,  and $k\in\N$.\newline {\bf Output:} $c_k,\ssum _k$ and the flag $BAD.GRAPH$. \newline {\bf Complexity:} $\UL$.}\label{algo:computing-ck-gen-graphs}
\begin{algorithmic}[1]
\State Initialize $c_k\leftarrow c_{k-1}$ and $\ssum _k\leftarrow \ssum _{k-1}$
\For {each $v\in V$}
	\If {$\neg (d(v)\leq k-1)$}
\algstore{firstpart:ckBADGRAPH}
\end{algorithmic}
\end{algorithm}
\clearpage
\begin{algorithm}[H]
\setcounter{algorithm}{7}
\caption{Computing $c_k$ and $\ssum _k$ for directed graphs (continued).}\label{algo:computing-ck-contd}
\begin{algorithmic}[1]
\algrestore{firstpart:ckBADGRAPH}
		\For {each $x$ such that $(x,v)\in E$}    
            \If {$d(x)\leq k-1$}
				\State {Set $c_k\leftarrow c_k+1$}    
				\State {Set $\ssum _k\leftarrow \ssum _k+k$}
                \For {$x'\neq x$}
                    \If {($(x',v)$ is an edge and $d(x')\leq (k-1)$)}                
                        \State {$BAD.GRAPH\leftarrow True$}
                    \EndIf
                \EndFor
			\EndIf
        \EndFor        
    \EndIf
\EndFor    
\State{{\bf return} $c_k$ and $\ssum _k$ and $BAD.GRAPH$} \end{algorithmic}
\end{algorithm}

\begin{algorithm}[H]
\caption{Determining if there exists a path from $s$ to $t$ in a directed graph $G$ which is assumed to be min-unique.$\newline$ {\bf Input:} A directed graph $G=(V,E)$ given in terms of its adjacency matrix which is assumed to be min-unique, and vertices $s,t\in V$.\newline {\bf Output:} True if there exists a directed path from $s$ to $t$ in $G$. Otherwise False. Either $G$ is not min-unique or there does not exist any directed path from $s$ to $t$ in $G$.\newline {\bf Complexity:} $\UL$.}\label{algo:stpathmin-unique-general-graph}
\begin{algorithmic}[1]
\State Initialize $BAD.GRAPH=False, c_0\leftarrow 1,\ssum _0\leftarrow 0,k\leftarrow 0$
\Repeat
    \State {$k\leftarrow (k+1)$}
	\State {Compute $c_k$ and $\ssum _k$ by invoking Algorithm \ref{algo:computing-ck-gen-graphs} on $(G,k,c_{k-1},\ssum_{k-1})$}
\Until{($c_{k-1}=c_k$ or $BAD.GRAPH=True$)}
\If {$BAD.GRAPH=False$}
    \State {there is a directed path from $s$ to $t$ in $G$ if and only if $d(t)\leq k$}
\EndIf
\end{algorithmic}
\end{algorithm}

Assume that we are given an input instance $(G,s,t)$ of $\DSTCON$. Let functions $f_i:A\rightarrow \{ 1,\ldots ,r\}$ be as stated above, which we have obtained from Corollary \ref{chap3-coro-nlpolyulpolytailormade} be given as the advice string, where $1\leq i\leq (n^2+2)$. Each weight function $f_i$ is a vector of weights $\vec{\bf f_i}=(w_1,w_2,\ldots ,$ $w_l)$, where $l=(n^2+2)$ and $1\leq i\leq (n^2+2)$. Clearly the length of each $\vec{\rm\bf f_i}$ is $(n^2+2)^k$ (where $k>0$ is a constant), which is a polynomial in the size of the input $(G,s,t)$. Therefore the length of the advice string is also a polynomial in the size of the input.

Also it is possible to use this advice string to assign weights to edges of $G$ in $O(\log n)$ space to obtain $(n^2+2)$ weighted directed graphs $G_1,\ldots ,G_{n^2+2}$. It is clear that there exists $i$ such that $G_i$ is min-unique, where $1\leq i\leq (n^2+2)$. Using Algorithm \ref{algo:stpathmin-unique-general-graph} given above, it follows that we can decide if there exists a directed path from $s$ to $t$ in $G$ in $\UL {\rm /poly}$. Since we have shown in Theorem \ref{chap2-NLcomplement} that $\NL=\coNL$, it follows that we can in fact decide if there exists a directed path from $s$ to $t$ in an input instance of $\DSTCON$ in $(\UL\cap\coUL ){\rm \poly}$. As a result it follows that $\DSTCON\in(\UL\cap\coUL){\rm /poly}$ which proves our result.

\end{proof}
\begin{corollary}\label{NLULcoUL}
$\NLpoly=(\UL\cap\coUL){\rm /poly}$.
\end{corollary}

\section{A combinatorial property of $\boldsharpL$}
\subsection{Some more results on directed st-connectivity}\label{directedstconn}

\begin{lemma}\label{polynomialweightlemma}
Let $X=\{ 1,\ldots ,n\}$ be a set and let $\F\subseteq 2^X$ such that $|\F |\leq p(n)$ for a polynomial $p(n)$. Let $r>(n+1)^2p^2(n)$ be a prime number and for each $1\leq i\leq r$ and $j\in X$ define the weight function $w_i:{\rm [}n{\rm ]}\longrightarrow \Z _r$ as $w_i(j)=(i^j \mod r)$. Further for each subset $Y\subseteq X$ define
\[
w_i(Y)=\Sigma_{j\in Y}w_i(j)(\mod r).
\]
There exists a weight function $w_m$ such that $w_m(Y)\neq w_m(Y')(\mod r)$ for any two distinct $Y,Y'\in \F$.
\end{lemma}
\begin{proof}
For any $1\leq m\leq r$ and $Y\in \F$, we can interpret $w_m(Y)$ as the value of the polynomial $q_Y(z)=\Sigma _{j\in Y}z^j$ at the point $z=m$ over the field $\Z _r$. For $Y\neq Y'$, notice that the polynomials $q_Y(z)$ and $q_{Y'}(z)$ are distinct and their degrees are at most $n$. Hence, $q_Y(z)$ and $q_{Y'}(z)$ can be equal for at most $n$ values of $z$ in the field $\Z _r$. Equivalently, if $Y\neq Y'$ then $w_i(Y)=w_i(Y')$ for at most $n$ weight functions $w_i$. Since there are ${|\F |}\choose{2}$ pairs of distinct sets in $\F$, it follows that there are at most ${{|\F |}\choose{2}}\cdot n<n\cdot p^2(n)$ weight functions $w_i$ for which $w_i(Y)=w_i(Y')$ for some pair of sets $Y,Y'\in \F$. Since $r>n\cdot p^2(n)$, there is a weight function as claimed by the lemma.
\end{proof}

\begin{theorem}\label{countdirectedpaths}
Let $(G,s,t)$ be an input instance of $\LDAGSTCON$ and let the number of directed paths from $s$ to $t$ in $G$ be at most $p(n)$ for some polynomial $p(n)$, where $G=(V,E)\in \LDAG$ and $n=|E|$. Also let $E=\{ 1,\ldots ,n\}$ and let $r>(n+1)^2p^2(n)$ be a prime number and for each $1\leq i\leq r$ and $j\in E$ define the weight function $w_i:E\longrightarrow \Z _r$ as $w_i(j)=(i^j \mod r)$. Further for each subset $Y\subseteq E$ define
\[
w_i(Y)=\Sigma_{j\in Y}w_i(j)(\mod r).
\]
Now let $\F\subseteq 2^X$ such that if $X\in \F$ then edges in $X$ form a directed path from $s$ to $t$ in $G$. It is then possible to determine a weight function $w_m$ such that $w_m(X)\neq w_m(X')(\mod r)$ for any two distinct $X,X'\in \F$, and the number of directed paths from $s$ to $t$ in $G$ in $\FNL$.
\end{theorem}
\begin{proof}
We iteratively start with the first weight function and first replace each edge in $G$ with weight $w$ by a directed path of length $w$. Let the resulting directed graph obtained be $G'$. It is easy to note that the number of directed paths from $s$ to $t$ in $G$ is equal to the number of directed paths from $s$ to $t$ in $G'$. Since it is shown in Theorem \ref{sldagstconcomplete} that $\LDAGSTCON$ is $\NL$-complete under logspace many-one reductions it is possible to obtain an input instance $(G'',s'',t'')$ of $\LDAGSTCON$ when we are given $(G',s,t)$ as input. Upon following the proof of Theorem \ref{sldagstconcomplete} we infer that the number of directed paths from $s$ to $t$ in $G'$ is equal to the number of directed paths from $s''$ to $t''$ in $G''$. Let us initialize a counter $c$ to $0$. We consider the simple layered directed acyclic graph structure of $G''$ and vertices which are copies of $t$ in all the polynomially many layers of $G''$. We query the $\NL$ oracle if there exists a directed path in $G''$ from $s''$ to the copy of the vertex $t$ in each of the layers of $G''$. If the oracle returns ``yes" then we increment the counter $c$ by $1$. For a given weight function we can compute $c$ in $O(\log |G|)$ space. Using Lemma \ref{polynomialweightlemma}, it follows that there exists a weight function $w_m$ such that $w_m(X)\neq w_m(X')(\mod r)$ for any two distinct $X,X'\in \F$, where $1\leq m\leq r$. Clearly any such weight function will result in the maximum value for the counter $c$. As a result by storing $c$ for successive weight functions and updating it by comparison we can find the weight function $w_m$ also. Note that the maximum value of $c$ is the number of directed paths from $s$ to $t$ in $G$ itself. Since $|G''|$ is a polynomial in $|G|$ it follows that we can find the weight function $w_m$ in $\FL^{\mbox{\NL}}$ which is $\FNL$ due to Theorem \ref{nlclosureunderTuring}.\textcolor{white}{\index[subject]{$\FL^{\rm \mbox{\NL}}$}}
\end{proof}

\begin{theorem}\label{polynomialacceptingtheorem}
Let $G=(V,E)$ be a directed graph given in terms of its adjacency matrix as input, and let $s,t\in V$. Also let $p$ be a positive integer whose unary representation can be computed by a deterministic $O(\log |G|)$ space bounded Turing machine. Then we can determine if the number of directed paths from $s$ to $t$ in $G$ is at least $(p+1)$ in $\FNL$. Otherwise if the number of directed paths from $s$ to $t$ in $G$ is lesser than $(p+1)$ then the $\FNL$ machine outputs the number of directed paths from $s$ to $t$ in $G$.
\end{theorem}
\begin{proof}
Due to Theorem \ref{sldagstconcomplete} we know that $\LDAGSTCON$ is $\NL$-complete under logspace many-one reductions. We therefore follow Theorem \ref{sldagstconcomplete} and obtain $H\in\LDAG$ and vertices $s',t'\in V(H)$ from $G$. Upon following the proof of Theorem \ref{sldagstconcomplete} we infer that the number of directed paths from $s$ to $t$ in $G$ is equal to the number of directed paths from $s'$ to $t'$ in $H$. Let us consider the subgraph of $H$ induced by vertices that are in at least one directed path from $s'$ to $t'$ in $H$. Let this subgraph be $H'$. We use induction on the number of layers $\lambda\geq 2$ in $H'$ and consider subgraphs of $H'$ formed by vertices in the first $\lambda$ layers of $H'$ such that the number of directed paths from $s'$ to vertices in layer $\lambda$ of $H'$ is at most $(p+1)$ or a polynomial in $|G|$. Note that it is easy to compute this upper bound on the number of directed paths using $O(\log |G|)$ space. We now once again use Theorem \ref{sldagstconcomplete} and Theorem \ref{countdirectedpaths} to compute the number of directed paths from $s'$ to all the vertices in layer $\lambda$ in $H'$. If the number of directed paths is greater than or equal to $(p+1)$ then we move to the accepting configuration, output $(p+1)$ and stop. Otherwise finally we would have computed the number of directed paths from $s'$ to $t'$ in $H$ which is lesser than $(p+1)$. We then move to the accepting configuration, output this value and stop. In all stages of this proof, we use a deterministic $O(\log |G|)$ space bounded Turing machine with access to the $\FNL$ oracle. The deterministic $O(\log |G|)$ space bounded Turing machine submits queries deterministically to the $\FNL$ oracle. In the intermediary stages, the reductions are done by submitting queries deterministically to the $\FNL$ oracle and after reading the reply given by the oracle on the oracle tape. As a result given the input $G$ it is possible to determine if the number of directed paths from $s$ to $t$ is at least $(p+1)$ is in $\FL^{\rm\mbox{FNL}}$ which is $\FNL$ due to Theorem \ref{nlclosureunderTuring}.
\end{proof}

\begin{corollary}
Verifying if there are polynomially many accepting computation paths for a $\NL$-Turing machine on a given input is in $\FNL$.
\end{corollary}
\begin{proof}
Let $M$ be a $\NL$-Turing machine and let $g$ be the canonical logspace many-one reduction from $L(M)$ to $\LDAGSTCON$ obtained by using the seminal result that the $st$-connectivity problem for directed graphs is complete for $\NL$ under logspace many-one reductions and Theorem \ref{sldagstconcomplete}. We note that if $x$ is the input string then $g(x)=(G,s,t)$ and the number of accepting computation paths of $M$ on $x$ is equal to the number of directed paths from $s$ to $t$ in $G$. We now use Theorem \ref{polynomialacceptingtheorem} to complete the proof.\textcolor{white}{\index[subject]{FL$^{\rm \mbox{FNL}}$}}
\end{proof}

\begin{corollary}\label{ldag-st-subgraph-ul}
Assume that $\NL =\UL$. Let $G=(V,E)$ be a directed graph given in terms of its adjacency matrix as input, and let $s,t\in V$. Also let $p$ be a positive integer whose unary representation can be computed by a deterministic $O(\log |G|)$ space bounded Turing machine. Then we can determine if the number of directed paths from $s$ to $t$ in $G$ is at least $(p+1)$ in $\FUL$. Otherwise if the number of directed paths from $s$ to $t$ in $G$ is lesser than $(p+1)$ then the $\FUL$ machine outputs the number of directed paths from $s$ to $t$ in $G$.
\end{corollary}
\begin{note}\label{st-subgraph}
The number of accepting computation paths of a $\NL$-Turing machine is not altered if it simulates a $\NL$-Turing machine of a language $L\in \UL$ during intermediate stages to verify if some input string is in $L$.
\end{note}

\subsection{Choosing at most polynomially many number of computation paths from the computation tree of a $\NL$-Turing machine}
We refer to Definition \ref{appendix-defn-choosefunction} and Theorem \ref{appendix-thm-choosefunction} in the Appendix A for ${n\choose k}$, where $n,k\in\Z^+$. In Theorem \ref{sharplchoosefl}, after assuming $\NL =\UL$, we show that if $f\in\sharpL$ and $g\in\FL$ such that $g(x)$ is the unary representation of a positive integer $k$, where $x\in\Sigma^{*}$ is the input, then the number of ways of choosing exactly $k=|g(x)|$ distinct paths from amongst the $f(x)$ accepting computation paths of the $\NL$-Turing machine corresponding to $f$ is in $\sharpL$.
\begin{theorem}\label{sharplchoosefl}
  Assume that $\NL =\UL$. Let $\Sigma$ be the input alphabet, $f\in \sharpL$ and $g\in \FL$ such that, for the given input string $x\in\Sigma ^*$, $g(x)$ is a positive integer $k$ in the unary representation. Then the function
 ${{f(x)}\choose {k}}\in \sharpL$.
\end{theorem}
\begin{proof}
Let $x\in \Sigma^*$ be the input string. Given $x$, we obtain the number $k=|g(x)|$ in $\FL$. It is easy to note that $k$ is upper bounded by a polynomial in $|x|$. It follows from our assumption that $\NL =\UL$ and Definition \ref{UL} that there exists a $\NL$-Turing machine $M'$ to which if we give a directed graph $G'$ along with two vertices $s'$ and $t'$ in $G'$ as input then $M'$ outputs ``yes" at the end of the unique accepting computation path and ``no" at the end of all the other computation paths if there exists a directed path from $s'$ to $t'$ in $G'$. Otherwise if there does not exist any directed path from $s'$ to $t'$ in $G'$ then $M'$ outputs ``no" at the end of all of its computation paths. Now given the input $x$, we obtain an instance of $\LDAGSTCON$, say $(G,s,t)$, using a logspace many-one reduction from Theorem \ref{sldagstconcomplete}. It is easy to note that the adjacency matrix of the graph $G$ that we obtain is of size
 polynomial in $|x|$. Once again we assume that $s$ is in row $1$
  and $t$ is in row $n$.  Also any two paths are
  distinct if there exists a vertex in one of the paths that is not in
  the other path. It follows from Proposition \ref{parsicomplete} that $f(x)$ is equal to the number the directed
  paths from $s$ to $t$ in $G$.  We associate the label $(i,j)$ to a vertex of
  $G$ if it is the $j^{th}$ vertex in the $i^{th}$ row of $G$, where $1\leq i,j\leq n$. In the following algorithm the row number is denoted by $\lambda$ and a vertex in row $\lambda $ is denoted by $(\lambda ,\omega )$, where $1\leq \lambda ,\omega\leq n$. The number of distinct paths we have chosen till row $\lambda$ is denoted by $\varphi$ and $\eta$ denotes the number of vertices in row $\lambda$ that are in the $\varphi$ distinct paths. We use non-determinism to compute $\varphi$ and $\eta$. Let us consider the SHARPLCFL algorithm described below. Input to the SHARPLCFL algorithm is $(0^k, (G,s,t))$ where $(G,s,t)$ is an input instance of $\LDAGSTCON$. If using SHARPLCFL, we are able to non-deterministically choose exactly $k=|g(x)|$ distinct directed paths from $s$ to $t$ in $G$ then that computation path ends in the accepting configuration. Otherwise the computation path ends in the rejecting configuration.\textcolor{white}{\index[subject]{rejecting configuration}}
\begin{algorithm}[H]
\caption{SHARPLCFL \newline {\bf Input:}$(0^k, (G,s,t))$, where $k\in \N$, $G=(V,E)$ is an input instance of $\LDAGSTCON$ and $s,t\in V$.\newline {\bf Output:} \emph{accepts} if $k$ distinct directed paths from $s$ to $t$  have been chosen in $G$. Otherwise \emph{rejects}. \newline {\bf Complexity:} $\sharpL$.}\label{algo:sharplcfl}
\begin{algorithmic}[1]
\State $\lambda\leftarrow 1,~\varphi\leftarrow 1,~\eta\leftarrow 1$
\While{$\lambda\leq (n-1)$}
	\State $\varphi'\leftarrow 0,~\eta'\leftarrow 0,~\varphi ''\leftarrow 0,~\eta ''\leftarrow 0,~\omega\leftarrow 1$
	\While{$(\omega\leq n)$}    
		\State Nondeterministically either choose $(\lambda ,\omega)$ or skip $(\lambda ,\omega)$
		\If{$(\lambda,\omega)$ is chosen nondeterministically}
			\If{($M'(G,s,(\lambda ,\omega))$ returns``yes") and ($M'(G,(\lambda,\omega),t)$ returns ``yes")}        
				\State $\eta'\leftarrow\eta'+1$            
				\If{$\eta'>\eta$}
					\State reject the input
				\Else
					\State $\sigma\leftarrow 0,~\psi\leftarrow 0$
                    \While{$\exists$ a neighbour $(\lambda +1,\rho )$ of $(\lambda ,\omega )$ that is yet to be visited}
                        \State Nondeterministically either choose $(\lambda +1 ,\rho)$ or skip $(\lambda +1 ,\rho)$
\algstore{sharplcfl:firstpart}
\end{algorithmic}
\end{algorithm}
\begin{algorithm}[H]
\setcounter{algorithm}{7}
\renewcommand{\addcontentsline}[3]{}
\caption{SHARPLCFL (continued)}\label{algo:sharplcfl-continued1}
\begin{algorithmic}[1]
\algrestore{sharplcfl:firstpart}
						\If{$(\lambda +1,\rho)$ is chosen nondeterministically}
                            \State $\sigma\leftarrow \sigma +1$
							\If{$M'(G,(\lambda +1,\rho),t)$ returns ``no"}
								\State reject the input
                            \ElsIf{$(\varphi=k~or~\varphi''=k) ~and~\sigma\geq 2$}
								\State reject the input
							\EndIf
                        \EndIf
                    \EndWhile
                    \If{$(\sigma =0~or~\sigma>k)$}
						\State reject the input
					\EndIf
                    \If{$\#(directed~paths~from~s~to~(\lambda,\omega))>\varphi -\varphi '$}    
					\State $\psi \leftarrow {\rm max(}1,\varphi -\varphi '{\rm )}$
				    \Else
						\State $\psi\leftarrow\#(directed~paths~from~s ~to~(\lambda ,\omega))$
				    \EndIf
                    \State Nondeterministically choose a number $\alpha$ from  $1~to~\psi$
                    \State $\varphi '\leftarrow\varphi '+\alpha$
                    \State Nondeterministically choose a number $\beta$ from $0$ to $\sigma$
					\State $\varphi ''\leftarrow\alpha\beta+\varphi ''$
					\State $\eta ''\leftarrow\eta ''+\beta$
					\If{$(\eta''=0)$ or $((\varphi =k$ or $\varphi ''>k)$ and $\eta ''>\eta)$}
						\State reject the input
					\EndIf                        
                    \If{$\varphi ''>k$}	\State $\varphi ''\leftarrow k$	\EndIf
				\EndIf
			\Else
				\State reject the input
			\EndIf
		\EndIf
		\State $\omega\leftarrow\omega +1$
	\EndWhile
	\If{$\lambda +1<n$ and $(\eta '\neq\eta$ or $\varphi '<\varphi)$}
		\State reject the input
	\ElsIf{$\lambda +1=n$ and $(\eta '\neq\eta$ or $\varphi ''\neq k)$}
        \State reject the input
	\EndIf
	\State $\lambda\leftarrow\lambda +1,~\varphi\leftarrow\varphi '',~\eta\leftarrow\eta ''$
\EndWhile    
\State accept the input
\end{algorithmic}
\end{algorithm}

We note the following points about the SHARPLCFL algorithm.

Always $1\leq\lambda\leq n$ and $1\leq \omega\leq n$. In the {\rm SHARPLCFL} algorithm, if we choose a vertex non-deterministically we verify if it is in a directed path from $s$ to $t$ in lines 7 and 17.

The variable $\eta '$ is the number of vertices we are choosing non-deterministically in row $\lambda$ and it must be equal to $\eta$ after we have made non-deterministic choices on all the vertices in row $\lambda$ failing which we reject the input in lines 50-54. $\varphi '$ is used to verify if the number of distinct directed paths from $s$ to $t$ that pass through $\eta '$ vertices chosen non-deterministically in row $\lambda$ is atleast $\varphi$ failing which we reject the input in lines 50-54.

The variable $\sigma$ computed in line 16 inside the nested while loop from line 13 to 23 is the number of vertices chosen non-deterministically as the neighbours of $(\lambda ,\omega)$ in row $\lambda +1$ such that these vertices are in at least one directed path from $s$ to $t$ in $G$. Here $1\leq \sigma\leq n$.

We compute $\psi$ in lines 28 and 30 in $\FUL$ by Corollary \ref{ldag-st-subgraph-ul} without altering the number of accepting computation paths. Note that $\psi\geq 1$ since $(\lambda ,\omega)$ is in at least one directed path from $s$ to $t$ in $G$. After lines 28 and 30, $\psi$ is either ${\rm max(}1,\varphi -\varphi ')$ or {\rm \#(}directed paths from $s$ to $(\lambda ,\omega )${\rm )}. Therefore $1\leq\psi\leq\varphi$ always. From lines 1, 37-38 and 55 it follows that $1\leq \varphi\leq k$ always and so in line 35 we always have $\varphi ''\leq kn+\varphi ''\leq 2kn^2$. $1\leq \alpha\leq \psi$ and so $0\leq \varphi'\leq 2kn$ always. Also $0\leq \varphi ''\leq k$ in line 4 at the begining of the while loop. Therefore $0\leq\varphi ''\leq 2kn^2$ always.

Always $0\leq \beta\leq \sigma$, $0\leq \eta '\leq n$, $0\leq \eta ''\leq n^2$ and $1\leq \eta\leq n^2$. If $\eta >n$ then we reject the input due to the condition in lines 50 and 52.  Variables $\varphi ',~\eta '$ and $\eta ''$ are updated in the while loop from line 4 to 49 and these values do not decrease inside this loop.

Using non-determinism to increment $\eta ''$ in lines 34 and 36 by $\beta$ is to non-deterministically avoid the possibility of counting vertices in row $\lambda +1$ that are common neighbours of two distinct vertices in row $\lambda$ more than once.

In lines 32 and 34, assume that we are always non-deterministically choosing the correct value of $\alpha $ and $\beta$ respectively in the algorithm. Then our algorithm proceeds correctly and ends in an accepting configuration if and only if we have non-deterministically chosen exactly $k$ distinct directed paths from $s$ to $t$ in $G$. On the contrary if $\varphi ''$ is updated with an incorrect value of $\alpha$ or if $\eta ''$ is incremented by an incorrect value of $\beta$, then in those iterations of the while loop from line 4 to 49, the algorithm proceeds by assuming that an alternate set of non-deterministic choices have been made on vertices in row $\lambda +1$ which agree with $\eta ''$ and $\varphi ''$.

Any two directed paths formed by non-deterministically choosing two different vertices in row $\lambda$ in lines 5-6 are distinct irrespective of their neighbours non-deterministically chosen in the row $\lambda +1$ in lines 14-15. As a result if the number of directed paths from $s$ to $t$ in $G$ is lesser than $k$ then the value of $\varphi ''$ computed in line 30 is always lesser than $k$ and these inputs are rejected in lines 52-53.

At the end of the while loop in line 49, $\varphi ''$ is the number of directed paths from $s$ to $t$ computed non-deterministically, that we need for subsequent stages of our algorithm. Also $\eta ''$ is the number of distinct vertices that are in row $\lambda +1$ in $\varphi ''$ distinct directed paths from $s$ to $t$ in $G$ which is also computed non-deterministically.

Now assume that the number of directed paths from $s$ to $t$ in $G$ is at least $k$. The cases where we reject the input since the non-deterministic choices made on the neighbours of $(\lambda ,\omega)$ in row $\lambda +1$ results in increasing the number of distinct directed paths non-deterministically chosen to be greater than $k$ is in lines 19, 20, 37 and 38.

Lines 19 and 20 are pertaining to the case when we are in vertex $(\lambda ,\omega)$ in row $\lambda$ and we are visiting the neighbours of $(\lambda ,\omega)$ in row $\lambda +1$. In this case we have already non-deterministically chosen $k$ distinct directed paths from $s$ to $t$ and we have also chosen vertices in row $\lambda +1$ in excess that results in increasing the number of distinct directed paths chosen non-deterministically from $s$ to $t$ to be greater than $k$. Similarly in lines 37 and 38 we have the case when we have chosen a $(\lambda ,\omega )$ in row $\lambda$ and we have visited all the neighbours of $(\lambda ,\omega )$ in row $\lambda +1$. If {\rm (}$\varphi =k$ or $\varphi ''>k${\rm )} and $\eta ''>\eta$ then it implies that we have chosen more than $k$ distinct directed paths from $s$ to $t$ in $G$ that most recently includes directed paths that pass through $(\lambda ,\omega$) and $\eta ''$ distinct neighbours of $(\lambda ,\omega)$ in row $\lambda +1$ and so in lines 37 and 38 we reject the input.

The case when we reject the input since we have not chosen $\eta$ vertices in row $\lambda <n-1$ or at least $\varphi$ distinct directed paths till row $\lambda <n-1$ is in lines 50 and 51. The case when we reject the input since we have not chosen exactly $k$ distinct directed paths from $s$ to $t$, however we have moved till row $n-1$ is in lines 52 and 53.

In the SHARPLCFL algorithm, since we keep track of only a constant number of variables all of which take non-negative integer values and the values of these variables are upper bounded by a polynomial in the size of the graph $G$, which is once again polynomial in the input size $|x|$ we get a $\NL$-Turing machine that executes the SHARPLCFL algorithm. Since we have assumed that $\NL =\UL$ and we use $M'$ to verify if there exists a directed path from a vertex $s'$ to a vertex $t'$ in lines 13 and 27, it follows that the number of accepting computation paths of the $\NL$-Turing machine described by our SHARPLCFL algorithm does not get altered upon simulating $M'$ (also see Note \ref{st-subgraph}). As a result it follows that the $\NL$-Turing machine described by the SHARPLCFL algorithm stops in an accepting state if and only if we start from vertex $s$ in row $1$ and reach vertex $t$ in row $n$ via exactly $k$ distinct directed paths. Now as mentioned in Proposition \ref{parsicomplete}, since $f(x)$ is
  equal to the number of directed paths from $s$ to $t$ in $G$ and since we have assumed $\NL =\UL$, it
  follows that the number of accepting computation paths of this $\NL$-Turing machine is ${{f(x)}\choose{k}}$. When $f(x)<k$ this $\NL$-Turing machine has no
  accepting computation paths and so we get ${{f(x)}\choose {k}}=0$.
 This shows that ${{f(x)}\choose{k}}\in \sharpL$.
\end{proof}

\section*{Exercises}
\begin{enumerate}
\item A directed graph $G=(V,E)$ is called strongly connected if every pair of vertices $u,v\in V$ are connected by a directed path in each direction. Show that language of all directed graphs that are strongly connected is logspace many-one complete for $\NL$, where the directed graph is given as input in terms of its adjacency matrix.
\item Show that the language of all directed graphs that contain at least one directed cycle is logspace many-one complete for $\NL$ where the directed graph is given as input in terms of its adjacency matrix.
\item Show that $\NL$ is closed under union without using Theorem \ref{nlclosureunderTuring}.
\item Show that $\NL$ is closed under intersection without using Theorem \ref{nlclosureunderTuring}.
\item Show that $\NL$ is closed under star operation (Kleene closure).
\item Point out the error in the following deduction:
\begin{quote}
    Let $\Sigma$ be the input alphabet, $L\subseteq\Sigma^*$ and $L\in\NL$. Let $M'$ be the $\NL$ Turing machine that accepts $L$ such that on any input $x$ of length $n$ there are $2^{n^k}$ computation paths of $M'$ on $x$, where $k>0$ is a constant. For any given input $x\in\Sigma^*$, we know that $x\in L$ if and only if $acc_{M'}>0$. Let $f(x)=acc_{M'}(x)+2^{n^k}$, where $n=|x|$. Then $f\in\sharpL$. Also if $M$ is the $\NL$ Turing machine of $f$ then the number of computation paths of $M$ on any input $x$ of length $n$ is $2^{n^k+1}$. Clearly $L\in\NL$, and $x\in L$ if and only if $f(x)>2^{n^k}$. Now, if $L'\in\PL$ then there exists $g\in\GapL$ such that on any input $x$ we have $x\in L'$ if and only if $g(x)>0$. So $\PL\subseteq\NL$ which implies that $\NL =\PL$.
\end{quote}
\item ({\bf Polynomially bounded Isolating Lemma}) Let $p(x)$ be a polynomial and let $r>n^2p^2(n)$ be a prime, where $n>0$ and $n\in \Z$. Also let $0<m<r$ and $m\in \Z$. Define $e_m(\pi )=\Sigma _{i=1}^n(m^{ni+\pi (i)}\mod r)$, where $\pi\in S_n$ is a permutation of $n$ elements. Let $\pi_1,\pi_2\ldots ,\pi_t\in S_n$ for some $t\leq p(n)$. Show that there exists a $m<r$ such that $e_m(\pi _i)\neq e_m(\pi _j)$, for all $i\neq j$ and $1\leq i,j\leq t$. \label{chap1-exercise-polyIsoLemma}
\item\label{chap1-exercise-DGI} The Directed Graph Isomorphism problem consists of deciding whether there is a bijection between the set of nodes of the two input directed graphs $G$ and $H$ such that the bijection preserves edge relations. Consider the Colored Directed Graph Isomorphism problem, in which the nodes are colored by different colors, such that we have to decide whether there is a bijection between the set of nodes of the two input directed graphs $G$ and $H$ which preserves the colors of the nodes and edge relations, where $G$ and $H$ are given in terms of their adjacency matrix. In other words, the bijection maps vertices colored with the same color (say, red) in $G$ to vertices colored with the same color (say, red) in $H$ and it also preserves edge relations between every pair of vertices. The $k$-Colored Directed Graph Isomorphism problem is the Colored Directed Graph Isomorphism problem where the number of vertices colored with a color is at most a constant $k>0$.

The Directed Graph Automorphism problem consists of deciding if there exists an isomorphism from a directed graph to itself, where the input directed graph is given in terms of its adjacency matrix. Similarly, the Colored Directed Graph Automorphism problem consists of deciding if there exists an isomorphism from a directed graph, whose nodes are colored, to itself. The $k$-Colored Directed Graph Automorphism problem consists of deciding if there exists an automorphism from a colored directed graph to itself, where the number of vertices having a color is less than or equal to $k$, for some constant $k>0$.
\begin{itemize}
\item Show that if $k=2,3$ then $k$-Colored Directed Graph Isomorphism is logspace many-one complete for $\NL$.
\item Show that if $k=2,3$ then $k$-Colored Directed Graph Automorphism is in $\NL$.
\end{itemize}
\end{enumerate}
\section*{Open problems}
\begin{enumerate}
    \item Is $\sharpL$ closed under division by a function in $\FL$, which outputs a non-negative integer on a given input? In other words, is the quotient obtained by dividing the value of a $\sharpL$ function by a function in $\FL$, which outputs a non-negative integer on any given input, also in $\sharpL$?
    \item Is $\sharpL$ closed under the modulo operation by a function in $\FL$, which outputs a non-negative integer on a given input? In other words, is the remainder obtained by dividing the value of a $\sharpL$ function by a function in $\FL$, which outputs a non-negative integer on any given input, also in $\sharpL$?
    \item Is it possible to strengthen Theorem \ref{min-unique-wrt-s-nlul} by showing the following: ``$\NL =\UL$ if and only if there is a polynomially bounded $\UL$ computable weight function $f$ so that for any directed acyclic graph $G$, we have $f(G)$ is {\bf\emph{min-unique}}?"
    \item Is it possible to prove Theorem \ref{sharplchoosefl} unconditionally, that is without assuming $\NL =\UL$?
    \item Is $\GapL$ closed under division by a function in $\FL$, which outputs an integer on any given input? In other words, is the quotient obtained by dividing the value of a $\GapL$ function by a function in $\FL$ which outputs an integer on any given input, also in $\GapL$?
    \item Is $\GapL$ closed under the modulo operation by a function in $\FL$, which outputs an integer on all inputs? In other words, is the remainder obtained by dividing the value of a $\GapL$ function by a function in $\FL$ which outputs an integer on any given input, also in $\GapL$?
\end{enumerate}

\section*{Notes}
The Immerman-Szelepcsenyi Theorem (Section \ref{chap02-immermanszelepcsenyitheorem}, Theorem \ref{ISTheorem}) is a celebrated result in the Theory of Computational Complexity \cite{BDG1995, AB2009, DK2014, Sip2013} which generalizes the result that $\NL =\coNL$ shown in Theorem \ref{chap2-NLcomplement} for non-deterministic space bounded complexity classes above $\NL$.  Even though the algorithm in the proof of Theorem \ref{chap2-NLcomplement} is based on the algorithm given by Michael Sipser\textcolor{white}{\index[authors]{Sipser, Michael}} in \cite[Section 8.6, Theorem 8.27]{Sip2013}, our proof is slightly different from this proof since it depends on an input instance of the $\NL$-complete language SLDAGSTCON. However our proof has a shortcoming that it is not as much general as the proof in \cite[Section 8,6, Theorem 8.27]{Sip2013} which seems to work even for showing that the directed $st$-connectivity problem is closed under complement for restricted type of inputs such as directed planar graphs and so on. Also both these proofs use the counting method which is commonly called as the double inductive counting method\textcolor{white}{\index[subject]{double inductive counting}} (for example, see \cite[Appendix]{BDG1995}). However the term {\bf\em non-deterministic counting} seems to be more apt and akin to the method used to prove our theorem and so we refer to the method as the non-deterministic counting method in this manuscript.\textcolor{white}{\index[authors]{Alvarez, Carme}\index[authors]{Jenner, Birgit}\index[authors]{Allender, Eric}\index[authors]{Ogihara, Mitsunori}\index[authors]{Toran, Jacobo}}\textcolor{white}{\index[subject]{non-deterministic oracle Turing machine}}

Logarithmic space bounded counting classes were first introduced in analogy with the counting classes in the polynomial time setting such as $\sharpP$\textcolor{white}{\index[subject]{$\sharpP$, pronounced as sharpP}}. The logarithmic space bounded complexity class $\sharpL$ was defined by Carme Alvarez and Birgit Jenner in \cite{AJ1993}. Complexity classes $\GapL$ were defined by Eric Allender and Mitsunori Ogihara in \cite{AO1996}.

\textcolor{white}{\index[authors]{Reinhardt, Klaus}\index[authors]{Allender, Eric}\index[authors]{Mulmuley, Ketan}\index[authors]{Vazirani, Umesh}\index[authors]{Vazirani, Vijay}\index[authors]{Buntrock, Gerhard}}Results shown in  Section \ref{salientGapL} are from \cite{AO1996, BDHM1992} and the excellent textbook by Lane A. Hemaspaandra and Mitsunori Ogihara \cite[Chapter 9]{HO2001}\textcolor{white}{\index[authors]{Hemaspaandra, Lane A.}\index[authors]{Ogihara, Mitsunori}}.

The Isolating Lemma\textcolor{white}{\index[subject]{Perfect Matching}\index[subject]{UL/poly}} was invented by Ketan Mulmuley, Umesh Vazirani and Vijay Vazirani, see \cite[Chapter 12]{MR1995}. It plays an important role in designing randomized parallel algorithms for problems such as the perfect matching in undirected graphs. Our proof of the Isolating Lemma in Section \ref{Isolating-Lemma} is based on Lemma 4.1 in the textbook by Lane A. Hemaspaandra and Mitsunori Ogihara \cite{HO2001}. The complexity class $\UL$ was first defined and studied by Gerhard Buntrock, Birgit Jenner, Klaus Jorn Lange and Peter Rossmanith in \cite{BJLR1991}\textcolor{white}{\index[authors]{Jenner, Birgit}\index[authors]{Lange, Klaus Jorn.}\index[authors]{Rossmanith, Peter}}. The result that $\NLpoly =\ULpoly$ is based on the Isolating Lemma and it was proved by Klaus Reinhardt and Eric Allender in \cite{RA2000} (see also \cite[Section 4.3]{HO2001}). Our exposition of this result in Sections \ref{min-unique-graphs} and \ref{NLULpoly} (and especially our proof of Theorem \ref{min-unique-st-connectivity-UL}) is from results shown by Chris Bourke, Raghunath Tewari and N. V. Vinodchandran in \cite{BTV2009}\textcolor{white}{\index[authors]{Bourke, Chris}\index[authors]{Tewari, Raghunath}\index[authors]{Vinodchandran, N. Variyam}\index[authors]{Agrawal, Manindra}\index[authors]{Hoang, Thanh Minh}\index[authors]{Thierauf, Thomas}\index[authors]{Pavan, Aduri}}. Theorem \ref{min-unique-wrt-s-nlul} is due to Aduri Pavan, Raghunath Tewari and N. V. Vinodchandran and it is from \cite{PTV2012}.

We also note that Manindra Agrawal, Thanh Minh Hoang and Thomas Thierauf have shown a restricted version of the Isolating Lemma in \cite{AHT2007}, where the weight function is computable in deterministic logspace when the number of subsets of the set $A$ in the family $\F$ is at most a polynomial in the size of $A$. This result is stated in Exercise no.(\ref{chap1-exercise-polyIsoLemma}) of this chapter. They use this result to show that the problem of counting the number of perfect matchings in instances of the polynomially bounded perfect matching problem for bipartite and general undirected graphs is in $\CeqL$.\textcolor{white}{\index[subject]{Perfect Matching}} Exercise problem (\ref{chap1-exercise-DGI}) of this chapter on the directed graph isomorphism is adapted from the results proved on the undirected graph isomorphism by \cite{JKMT2003}.

Restricted versions of the directed $st$-connectivity problem, such as for planar graphs, grid graphs and graphs embedded on surfaces, each of which defines and uses deterministic logspace computable weight functions to isolate directed paths between a pair of vertices in the input graph, have been investigated and many interesting results on their computational complexity that use logarithmic space bounded counting classes are known due to Eric Allender, David A. Mix Barrington, Chris Bourke, Tanmoy Chakraborty, Samir Datta, Sambuddha Roy, Raghunath Tewari and N. V. Vinodchandran, from \cite{BTV2009, ABCDR2009}. We also note that Raghunath Tewari and N. V. Vinodchandran define a deterministic logspace computable weight function in \cite{TV2012} that can be used to isolate a directed path between any two vertices in a directed planar graph.\textcolor{white}{\index[authors]{Tewari, Raghunath}\index[authors]{Vinodchandran, N. Variyam}\index[authors]{Allender, Eric}\index[authors]{Barrington, David A. Mix}\index[authors]{Chakraborty, Tanmoy}\index[authors]{Bourke, Chris}\index[authors]{Datta, Samir}\index[authors]{Roy, Sambuddha}\index[subject]{computational complexity}}

We refer to \cite[Chapter 1, Section 1-3]{TM1997} for a nice introduction to Boolean logic and its normal forms such as the disjunctive normal form (DNF) and the conjunctive normal form (CNF) of propositional formulae. It is a standard result in Boolean logic on the satisfiability of Boolean formulae that a Boolean formula $\phi$ is equivalent to a Boolean formula $\phi '$ in the CNF such that $\phi$ is satisfiable if and only if $\phi '$ is satisfiable. We recall the definition of 3-CNF from \cite[Chapter 34, Section 34.4, pp.1076]{CLRS2022}. Also any Boolean formula $\phi '$ which is in CNF, is equivalent to a Boolean formula $\phi ''$ which is in the 3-CNF \cite[Chapter 34, Theorem 34.10]{CLRS2022} such that $\phi '$ is satisfiable if and only if $\phi ''$ is satisfiable. It is well known and fundamental result of computational complexity that the problem (denoted by 3SAT), of determining if a Boolean formula which is in 3-CNF has a satisfying assignment of its Boolean variables is NP-complete under polynomial time many-one reductions. Similar to 3-CNF and 3SAT, it is easy to define 2-CNF and 2SAT. In analogy with the NP-completeness of 3SAT we are able to show in Theorem \ref{2SATisNLcomplete} that 2SAT is logspace many-one complete for $\NL$ which is an interesting observation. Note that unlike Propositions \ref{sldagcountsharplcomplete} and \ref{parsicomplete}, we are unable to immediately conclude from Theorem \ref{2SATisNLcomplete} that $\sharp$2SAT is logspace many-one complete for $\sharpL$ and we leave this question as a conjecture.\textcolor{white}{\index[subject]{CNF, conjunctive normal form}\index[subject]{DNF, disjunctive normal form}\index[subject]{3-CNF}\index[subject]{3SAT}\index[subject]{computational complexity}\index[subject]{NP-complete}\index[subject]{polynomial time many-one reduction}}

Computer programs which solve decision problems or compute functions essentially map inputs to outputs. Descriptive Complexity is a branch of Theoretical Computer Science developed by N. Immerman \cite{Imm1999}, which uses first- and second-order mathematical logic to describe computer programs. In descriptive complexity, using first-order languages it is demonstrated that all measures of complexity can be mirrored in logic and most important complexity classes have very elegant and clean descriptive characterizations. In fact, N. Immerman gives a  descriptive complexity proof of the Immerman--Szelepcsenyi Theorem (Theorem \ref{ISTheorem}) in \cite{Imm1999}.\textcolor{white}{\index[authors]{Immerman, Neil}\index[subject]{descriptive complexity}\index[subject]{first-order mathematical logic}\index[subject]{second-order mathematical logic}\index[subject]{first-order languages}}

\chapter{Modulo-based Logarithmic space bounded counting classes}\label{chapter03-MOD}
In this chapter we introduce modulo-based logarithmic space bounded counting classes and prove many interesting properties about these complexity classes.
\begin{definition}\label{ModkL}
Let $\Sigma$ be the input alphabet. Also let $k\in\Z^{+}$ and let $k\geq 2$. We say that $L\subseteq\Sigma ^{*}$ is a language in the complexity class $\ModkL$ if there exists a function $f\in \sharpL$ such that for any input $x\in\Sigma ^{*}$ we have $x\in L$ if and only if $f(x)\not\equiv 0(\mod k)$.\textcolor{white}{\index[subject]{Mod$_p$L, where $p\geq 2$ is a prime}}
\end{definition}

\begin{proposition}\label{modulokeveryj}
Let $\Sigma$ be the input alphabet. Also let $k\in\Z^{+}$ and let $k\geq 2$. For every $j\in\Z$, there exists a function $g\in\sharpL$ such that on any input $x\in\Sigma^*$, we have $x\in L$ if and only if $g(x)\not\equiv j(\mod k)$.
\end{proposition}
\begin{proof}
Let $L\in\ModkL$ such that on any input $x\in\Sigma^*$, we have $x\in L$ if and only if $f(x)\not\equiv 0(\mod k)$, where $f\in\sharpL$. Without loss of generality, assume that $0\leq j<k$. Since $\sharpL$ is closed under addition (Proposition \ref{sharplfolklore}), it follows that $g(x)=(f(x)+j)\in\sharpL$. As a result, if $M_g$ is the $\NL$-Turing machine corresponding to $g$ then $x\in L$ if and only if $f(x)\not\equiv 0(\mod k)$ if and only if $g(x)\not\equiv j(\mod k)$.
\end{proof}
\begin{lemma}\label{modulok-funda}
Let $\Sigma$ be the input alphabet. Also let $p\in\N$ such that $p\geq 2$ and $p$ is a prime.
\begin{enumerate}
\item\label{modulok-fermat} There exists a function $g\in\sharpL$ such that on any input $x\in\Sigma^*$, we have
\begin{itemize}
\item if $x\in L$ then $g(x)\equiv 1(\mod p)$, and 
\item if $x\not\in L$ then $g(x)\equiv 0(\mod p)$.
\end{itemize}
\item\label{modulok-hapazhard} There exists a function $g\in\sharpL$ such that on any input $x\in\Sigma^*$, we have
\begin{itemize}
\item if $x\in L$ then $g(x)\equiv i(\mod p)$, and 
\item if $x\not\in L$ then $g(x)\equiv j(\mod p)$.
\end{itemize}
\end{enumerate}
\end{lemma}
\begin{proof}
\begin{enumerate}
\item Let $L\in\ModpL$ such that on any input $x\in\Sigma^*$, we have $x\in L$ if and only if $f(x)\not\equiv 0(\mod p)$, where $f\in\sharpL$. Now we define $g(x)=(f(x))^{p-1}$, where $x\in\Sigma^*$. It follows from Proposition \ref{sharplfolklore} that $g\in\sharpL$. Using the Fermat's Little Theorem, it follows that if $x\in L$ then \{$f(x)\not\equiv 0(\mod p)$ if and only if $g(x)\equiv 1(\mod p)$\}. On the other hand if $x\not\in L$ then \{$f(x)\equiv 0(\mod p)$ if and only if $g(x)\equiv 0(\mod p)$\}.
\item Let $L\in\ModpL$ such that on any input $x\in\Sigma^*$, there exists $f\in\sharpL$ such that if $x\in L$ then $f(x)\equiv 1(\mod p)$ and if $x\not\in L$ then $f(x)\equiv 0(\mod p)$. Now we define $g(x)=((i-j)f(x))+j$, where $x\in\Sigma^*$. It follows from Proposition \ref{sharplfolklore} that $g\in\sharpL$. Also if $x\in L$ then \{$f(x)\equiv 1(\mod p)$ if and only if $g(x)\equiv i(\mod p)$\}. On the other hand if $x\not\in L$ then \{$f(x)\equiv 0(\mod p)$ if and only if $g(x)\equiv j(\mod p)$\}.
\end{enumerate}
\end{proof}

\begin{theorem}\label{modulok-closureproperties}
Let $\Sigma$ be the input alphabet. Also let $p\in\N$ such that $p\geq 2$ and $p$ is a prime. Then,
\begin{enumerate}
\item\label{modulok-intersection} $\ModpL$ is closed under intersection,
\item\label{modulok-complement} $\ModpL$ is closed under complement, and
\item\label{modulok-union} $\ModpL$ is closed under union.
\end{enumerate}
\end{theorem}
\begin{proof}
\begin{enumerate}
\item  Let $L_1,L_2\in\ModpL$ and let $f_1,f_2\in\sharpL$ such that given any input $x\in\Sigma^*$, we have $x\in L_i$ if and only if $f_i(x)\not\equiv 0(\mod p)$, where $i=1,2$. Let us define $f=f_1f_2$. It follows from Proposition \ref{sharplfolklore} that $f\in\sharpL$. Since $p$ is a prime, on any input $x\in\Sigma^*$, we have $f(x)\not\equiv 0(\mod p)$ if and only if $f_1(x)\not\equiv 0(\mod p)$ and $f_2(x)\not\equiv 0(\mod p)$. Therefore $L_1\cap L_2\in\ModpL$.
\item Let $L\in\ModpL$ using $f\in\sharpL$. For any $j\in\N$, define $L_j=\{ x|f(x)\not\equiv j(\mod k)\}$. Then, $\overline{L}=\cap_{1\leq j<p}L_j$. However $L_j\in\ModpL$ due to Proposition \ref{modulokeveryj} and $\ModpL$ is closed under intersection from (\ref{modulok-intersection}). Therefore $\overline{L}\in\ModpL$.\textcolor{white}{\index[subject]{$\overline{L}$, complement of a language $L$}}
\item Follows from (\ref{modulok-intersection}) and (\ref{modulok-complement}).
\end{enumerate}
\end{proof}
\begin{definition}\label{join-operation}
Let $\Sigma$ be the input alphabet such that $\{ 0,1\}\subseteq\Sigma$ and let $L_1,L_2\subseteq\Sigma ^{*}$. We define the join of $L_1$ and $L_2$ to be $L_1\vee L_2=\{ x1|x\in L_1\}\cup \{ y0|y\in L_2\}$.\textcolor{white}{\index[subject]{$\vee$, join operator}}
\end{definition}
\begin{proposition}\label{modulok-join}
Let $\Sigma$ be the input alphabet. Also let $k\in\Z^+$ such that $k\geq 2$. $\ModkL$ is closed under join.
\end{proposition}
\begin{proof}
Let $L_i\in \ModkL$ and assume that we decide if an input string $x\in\Sigma^{*}$ is in $L_i$ using functions  $f_i\in\sharpL$, where $i=1,2$. We assume without loss of generality that the size of any  input string $x$ is at least $1$. Let $x=x_1x_2\cdots x_{n+1}$ and let $y=x_1\cdots x_n$ where $n\in\Z^{+}$. Also let 
\[ f(x)=\left \{ \begin{array}{ll}
	f_1(y) & \mbox{if $x_{n+1}=1$}\\
	f_2(y) & \mbox{otherwise if $x_{n+1}=0$.}
\end{array}
\right. \]
It is easy to note that $f\in\sharpL$. Now $f(x)\not\equiv 0(\mod k)$ for any input $x=yx_{n+1}$ if and only if exactly one of the following is true: ($x_{n+1}=1$ and $y\in L_1$) or ($x_{n+1}=0$ and $y\in L_2$). Clearly this shows that $L_1\vee L_2\in\ModkL$ which completes the proof.
\end{proof}
\begin{theorem}\label{modulok-logspaceTuringclosed}
Let $\Sigma$ be the input alphabet, $p\in\N$ and $p\geq 2$ such that $p$ is a prime.
$\L^{\mbox{\ModpL}}=\ModpL$.\textcolor{white}{\index[subject]{L$^{\mbox{\ModpL}}$, where $p\geq 2$ is a prime}}
\end{theorem}
\begin{proof}
Our proof of this result is similar to the proof of Theorem \ref{nlclosureunderTuring}. Let $L\in\L^{\mbox{\ModpL}}$. Then there exists a $O(\log n)$-space bounded deterministic Turing machine $M^A$, that has access to a language $A\in\ModpL$ as an oracle such that given any input $x\in\Sigma^*$, $M^A$ decides if $x\in L$ correctly. Since $M$ requires at most $O(\log n)$ space on any input of size $n$, we infer that the number of queries that $M$ submits to the oracle $A$ is a polynomial in the size of the input. Also it follows from Lemma \ref{modulok-funda}(\ref{modulok-fermat}) that, for the oracle $A$ there exists a $\NL$-Turing machine $M_A$ corresponding to which we have $g\in\sharpL$ such that on any input $x\in\Sigma^*$, we have if $x\in A$ then $g(x)\equiv 1(\mod p)$, and if $x\not\in A$ then $g(x)\equiv 0(\mod p)$. Since we know from Theorem \ref{modulok-closureproperties}(\ref{modulok-complement}) that $\ModpL$ is closed under complement, let $M_{\overline{A}}$ be the $\NL$-Turing machine corresponding to which we have $h\in\sharpL$ such that on any input $x\in\Sigma^*$, we have if $x\in \overline{A}$ then $h(x)\equiv 1(\mod p)$, and if $x\not\in \overline{A}$ then $h(x)\equiv 0(\mod p)$. Let us now consider the following algorithm implemented by a $\NL$-Turing machine $N$ given input $x\in\Sigma^*$.
\begin{algorithm}[H]
\setcounter{algorithm}{8}
\caption{Closure-Logspace-Turing-for-$\ModpL$  \newline {\bf Input:} $x\in \Sigma^*$.\newline {\bf Output:} \emph{accept} or \emph{reject}. The $\NL$-Turing machine that implements this algorithm obeys the property of $\ModpL$ in Lemma \ref{modulok-funda}(\ref{modulok-fermat}). \newline {\bf Complexity:} $\ModpL$.}\label{algo:modpllogspaceTuringclosed}
\begin{algorithmic}[1]
\While{$N$ has not reached any of its halting configurations}
	\State $N$ simulates $M$ on input $x$ until a query $y$ is generated.
	\State $N$ non-deterministically guesses if $y\in A$.
	\If{$N$ guesses that $y\in A$}
		\State $N$ simulates $M_A$ on input $y$.
		\If{$M_A$ rejects $y$}
			\State $N$ also rejects $x$ and stops.
		\EndIf
	\Else
		\State $N$ simulates $M_{\overline{A}}$ on $y$
		\If{$M_{\overline{A}}$ rejects $y$}
			\State $N$ also rejects $x$ and stops.
		\EndIf
	\EndIf
	\State $N$ continues to simulate $M$ on $x$.
\EndWhile
\end{algorithmic}
\end{algorithm}
Let $f\in\sharpL$ denote the number of accepting computation paths of $N$. Since $N$ rejects the input $x$ in step 4 of the above stated algorithm whenever its simuation of $M_A$ or $M_{\overline{A}}$ rejects their input $y$, it follows that $N$ accepts the input $x$ at the end of any computation path if and only if it guesses the oracle reply correctly which is the same as in the end of the computation path of $M_A$ or $M_{\overline{A}}$. As a result, on any given input $x\in\Sigma^*$, if $x\in L$ and if $N$ has made correct guesses for each of the query strings generated then $f(x)\equiv 1(\mod p)$, since $f$ is a polynomial length product of $g$ and $h$. However if $x\in L$ and $N$ has made at least one incorrect guess for the query string $y$ that is generated then $f(x)$ remains unaffected since $N$ rejects at the end of such computation paths. Conversely, if $x\not\in L$ and if $N$ has made the correct guesses for each of the query strings generated then $f(x)\equiv 0(\mod p)$, once again since $f$ is a polynomial length product of $g$ and $h$. Similar to the previous case, if $x\not\in L$ and $N$ has made at least one incorrect guess for the query string $y$ that is generated then $f(x)$ remains unaffected since $N$ rejects at the end of such computation paths. This shows that there exists $f\in\sharpL$ such that for any input $x\in\Sigma^*$, we have if $x\in L$ then $f(x)\equiv 1(\mod p)$ and if $x\not\in L$ then $f(x)\equiv 0(\mod p)$ which proves that $L\in\ModpL$.
\end{proof}
\begin{theorem}\label{modulok-modplTuringclosed}
Let $p\in\N$ and $p\geq 2$ such that $p$ is a prime. $\ModpL^{\mbox{\ModpL}}=\ModpL$\textcolor{white}{\index[subject]{Mod$_p$L$^{\mbox{\ModpL}}$, where $p\geq 2$ is a prime}}.
\end{theorem}
\begin{proof}
Let $L\in\ModpL^{\mbox{\ModpL}}$. Then there exists a $\NL$-Turing machine $M^A$, that has access to a language $A\in\ModpL$ as an oracle such that given any input $x\in\Sigma^*$, $M^A$ decides if $x\in L$ correctly. Since $M$ is a non-deterministic Turing machine, we assume that $M$ follows the Ruzzo-Simon-Tompa oracle access mechanism stated in Section \ref{chap1-NOTM} to access the oracle $A$. Also since it requires at most $O(\log n)$ space on any input of size $n$, we infer that the number of queries that $M$ submits to the oracle $A$ is a polynomial in the size of the input. Now remaining part of our proof of this theorem is is based on Theorem \ref{modulok-logspaceTuringclosed} and it is similar to Theorem \ref{chap2-NLhierarchycollapses}.
\end{proof}
\begin{proposition}\label{modulok-jdividesk}
If $j,k\in\Z^+$ such that $j,k\geq 2$ and $j|k$ then $\ModjL\subseteq\ModkL$.\textcolor{white}{\index[subject]{Mod$_j$L, where $j\geq 2$ is an integer}}
\end{proposition}
\begin{proof}
Let $k=cj$. If $L\in\ModjL$ using a $\NL$-Turing machine $N$ such that $f\in\sharpL$ is the number of accepting computation paths of $N$ on any input, then we can define $g=cf$ which is in $\sharpL$ due to Proposition \ref{sharplfolklore}. Clearly, given an input $x\in\Sigma^*$, we have $x\in L$ if and only if $f(x)\not\equiv 0(\mod j)$ if and only if $g(x)\not\equiv 0(\mod k)$ which proves the result.
\end{proof}

\begin{theorem}{\bf (Kummer)}\label{thm-kummertheorem}
Let $p$ be a prime, and let $n=a+b$. Then,
\[
{n\choose a}\equiv 0({\rm mod}~p^c)
\]
if and only if the number of carries when adding $a$ to $b$ in base-$p$ is at least $c$.
\end{theorem}
\begin{corollary}\label{cor-kummercorollary}
Let $p$ be a prime. Then,
\[
{n\choose p^k}\equiv 0({\rm mod}~p)
\]
if and only if the coefficient of $p^k$ in the base-$p$ expansion of $n$ is zero.
\end{corollary}
\begin{proof}
Suppose that the coefficient of $p^k$ in the base-$p$ expansion of $n$ is not zero. Using base-$p$ arithmetic, we have
\[
n=d_mp^m+\cdots +d_k p^k+d_{k-1}p^{k-1}+\cdots +d_0 p^0,
\]
\[
p^k=0+p^k+ 0+\cdots +0,\\
\]
\[
n-p^k=d_mp^m+\cdots +(d_k-1)p^k+d_{k-1}p^{k-1}+\cdots +d_0 p^0,
\]
so $p$ can be added to $n-p^k$ in base-$p$without carrying. On the other hand, if the coefficient of $p^k$ in the base-$p$ expansion of $n$ is zero then the coefficient of $p^k$ in the base-$p$ expansion of $n-p^k$ must be $p-1$; therefore, there must be a carry when adding $1$ to $p-1$ in the $p^k$'s position. The corollary, follows from Kummer's theorem with $c=1$.
\end{proof}

\begin{lemma}\label{modulok-primepowerprime}
Let $p,e\in\N$ and $p\geq 2$ such that $p$ is a prime. $\ModpeL =\ModpL$.
\end{lemma}
\begin{proof}
It follows from Proposition \ref{modulok-jdividesk} that $\ModpL\subseteq\ModpeL$\textcolor{white}{\index[subject]{Mod$_{p^e}$L, where $p\geq 2$ is a prime and $e\geq 1$}}. To prove the converse, we use induction on $e$. Assume that the converse result is true for some $e\geq 1$. Let $L\in\ModpeplusL$ using a $\NL$-Turing machine $N$. Let $g\in\sharpL$ denote the number of accepting computation paths. A number $g$ is divisible by $p^{e+1}$ if and only if

(1). $g$ is divisible by $p^e$, and

(2). the coefficient of $p^e$ in the base-$p$ expansion of $g$ is $0$.

Also using Corollary \ref{cor-kummercorollary}, we get that condition (2) is equivalent to ${{g}\choose{p^{e}}}\equiv 0(\mod p)$. Therefore, on a given input $x\in\Sigma^*$, we have $g(x)\not\equiv 0(\mod p^{e+1})$ if and only if $\{g(x)\not\equiv 0(\mod p^e)$ or ${{g(x)}\choose{p^{e}}}\not\equiv 0(\mod p)\}$. Now let $L_1=\{x\in\Sigma^*|g(x)\not\equiv 0(\mod p^e)\}$ and $L_2=\{x\in\Sigma^*|{{g(x)}\choose{p^{e}}}\not\equiv 0(\mod p)\}$. By the induction hypothesis $L_1\in\ModpL$. Also it follows from Theorem \ref{sharplchooseconstant} that  $L_2\in\ModpL$. As a result $L_1\cup L_2\in\ModpL$ which proves our result. 
\end{proof}
\begin{lemma}\label{modulok-jk}
Let $j$ and $k$ be relatively prime. Then $L\in\ModjkL$ if and only if there exists $L_j\in\ModjL$ and $L_k\in\ModkL$ such that $L=L_j\cup L_k$.\textcolor{white}{\index[subject]{Mod$_{jk}$L, where $j,k\geq 2$ are integers}}
\end{lemma}
\begin{proof}
Let $L\in\ModjkL$ using a $\NL$-Turing machine $M$ such that $f(x)\in\sharpL$ denotes the number of accepting computation paths of $M(x)$, where $x\in\Sigma^*$ and $\Sigma$ is the input alphabet. Also let $L_i=\{x|f(x)\not\equiv 0(\mod i)\}$ for $i=j,k$. Clearly $L=L_j\cup L_k$.

Conversely, let $L_i\in\ModiL$ using $\NL$-Turing machine $M_i$ such that $f_i(x)$ denotes the number of accepting computation paths of $M(x)$, where $x\in\Sigma^*$ and $i=j,k$. It follows from Proposition \ref{sharplfolklore} that $\sharpL$ is closed under addition and multiplication and so $f(x)=(kf_j(x)+jf_k(x))\in\sharpL$. Clearly, $x\in L_j\cup L_k$ if and only if $f(x)\not\equiv 0(\mod jk)$ which proves the result.
\end{proof}
\begin{theorem}\label{modulok-characterize}
Let $k=p_1^{e_1}p_2^{e_2}\cdots p_m^{e_m}$ be the prime factorization of $k\in\N$ and $k\geq 2$. $L\in\ModkL$ if and only if there are languages $L_i\in\ModpieiL$ such that $L=\cup_{i=1}^mL_i$, where $1\leq i\leq m$. Especially the complexity class $\ModkL =\ModponetopmL$.\textcolor{white}{\index[subject]{Mod$_{{p_i}^{e_i}}$L, where $p_i\geq 2$ is a prime and $e_i\geq 1$}\index[subject]{Mod$_{p_1p_2\cdots p_m}$L}}
\end{theorem}
\begin{proof}
By Lemma \ref{modulok-primepowerprime}, $\ModpeL =\ModpL$. Our theorem follows from Lemma \ref{modulok-jk} using induction on $m$.
\end{proof}
\begin{corollary}\label{modulok-union-corollary}
$\ModkL$ is closed under union, where $k\in\N$ and $k\geq 2$.
\end{corollary}
\begin{proof}
It follows from Theorem \ref{modulok-closureproperties}(\ref{modulok-union}) that $\ModpL$ is closed under union if $p\in\N$ is a prime. Now using Lemma \ref{modulok-primepowerprime} and applying Theorem \ref{modulok-characterize} we obtain our result.
\end{proof}

\section[$\boldModL$: an extension of modulo]{$\boldModL$: an extension of modulo}\label{modl-section}
\begin{tcolorbox}[colback=gray!35!white,colframe=white]
{\bf\textit{From now onwards, and the rest of this chapter, when we consider functions in $\GapL$, we do not necessarily assume that the computation tree of a function in $\GapL$ is a complete binary tree.}}
\end{tcolorbox}
For the purpose of the complexity class $\ModL$ which we study here, we first recollect Theorems \ref{chap1-crrtheorem} and \ref{chap1-booleancircuittheorem} and basics of logarithmic space bounded computation from Section \ref{chap1-sec-logspacecomputation}.
\begin{definition}\label{modldefn}
Let $\Sigma$ be the input alphabet. A language $L\subseteq\Sigma ^{*}$ is in the complexity class $\ModL$ if there is a function $f\in\GapL$ and a function $g\in\FL$ such that on any input $x\in\Sigma^{*}$,\textcolor{white}{\index[subject]{ModL}\index[subject]{$\FL^{\rm \mbox{ModL}}$}\index[subject]{$\FL^{\rm \mbox{GapL}}$}}
\begin{itemize}
\item $g(x)=1^{p^e}$ for some prime $p$ and a positive integer $e$, and
\item $x\in L\Leftrightarrow f(x)\not \equiv 0(\mod p^e)$.
\end{itemize}\textcolor{white}{\index[subject]{ModL}}
\end{definition}
\begin{definition}
Let $\Sigma$ be an input alphabet and let $\Gamma$ be an output alphabet. We define $\FLGapL$ to be the complexity class of all functions $f:\Sigma^*\rightarrow\Gamma^*$, which are logspace Turing reducible to the complexity class of functions in $\GapL$.
\end{definition}
\begin{definition}
Let $\Sigma$ be an input alphabet and let $\Gamma$ be an output alphabet. We define $\FLModL$ to be the complexity class of all functions $f:\Sigma^*\rightarrow\Gamma^*$, which are logspace Turing reducible to the complexity class of languages in $\ModL$.
\end{definition}
The {\bf\textit{prime distribution function}} $\pi (n)$ specifies the number of primes that are less than or equal to $n$.\textcolor{white}{\index[subject]{prime distribution function}\index[subject]{Prime Number Theorem}\index[subject]{Chinese Remainder Theorem}}
\begin{theorem}{\bf (Prime Number Theorem)}
\[
\lim_{n\rightarrow \infty}\frac{\pi (n)}{n/\ln n}=1.
\]
\end{theorem}
The approximation $n/\ln n$ gives reasonably accurate estimates of $\pi (n)$ even for small $n$.
\begin{theorem}{\bf (Chinese Remainder Theorem)}
Let $n=n_1n_2\cdots n_k$, where the $n_i$ are relatively prime. Consider the correspondence
\begin{eqnarray}\label{crt-eqn}
a & \leftrightarrow & (a_1,\ldots ,a_k),    
\end{eqnarray}
where $a\in\Z_n$, $a_i\in\Z_{n_i}$, and
\begin{eqnarray*}
a_i\equiv a({\rm mod}~n_i)
\end{eqnarray*}
for $i=1,2,\ldots k$. Then, mapping (\ref{crt-eqn}) is a one-to-one correspondence (bijection) between $\Z_n$ and the Cartesian product $\Z_{n_1}\times\Z_{n_2}\times\cdots\times\Z_{n_k}$. Operations performed on the elements of $\Z_n$ can be equivalently performed on the corresponding $k$-tuples by performing the operations independently in each coordinate position in the appropriate system. That is, if
\begin{eqnarray*}
a & \leftrightarrow & (a_1,\ldots ,a_k),\\
b & \leftrightarrow & (b_1,\ldots ,b_k),
\end{eqnarray*}
then
\begin{eqnarray*}
(a+b)({\rm mod}~n) & \leftrightarrow & ((a_1+b_1)({\rm mod}~n_1),\ldots ,(a_k+b_k)({\rm mod}~n_k)),\\
(a-b)({\rm mod}~n) & \leftrightarrow & ((a_1-b_1)({\rm mod}~n_1),\ldots ,(a_k-b_k)({\rm mod}~n_k)),\\
(ab)({\rm mod}~n) & \leftrightarrow & ((a_1b_1)({\rm mod}~n_1),\ldots ,(a_kb_k)({\rm mod}~n_k)).
\end{eqnarray*}
\end{theorem}

\begin{lemma}\label{AV2010-lemma}
$\FLModL =\FLGapL$.\textcolor{white}{\index[subject]{L$^{\rm \mbox{ModL}}$}}
\end{lemma}
\begin{proof}
For the forward inclusion it suffices to show that $\ModL\subseteq\LGapL$, since $\L^{\mbox{\LGapL}}$ $=\LGapL$ from Proposition \ref{lgaplequalsllgapl}. Suppose $L\in\ModL$ is witnessed by a function $f\in\GapL$ and a function $g\in\FL$ that computes prime powers in unary. Now, a logarithmic space bounded deterministic Turing machine can retrieve the value of $f(x)$ for any input $x$ to the $\GapL$ oracle as stated in Section \ref{functionoracleaccess2}. It is also easy to see that checking if $f(x)\not\equiv 0(\mod |g(x)|)$ is also computable in logarithmic space. As a result it follows that $L\in\LGapL$. Therefore $\LModL\subseteq\LGapL$.

For the reverse inclusion, let $L\in\LGapL$ be computed by a logarithmic space bounded deterministic Turing machine with access to a function $f\in\GapL$ as an oracle. (It follows from Theorem \ref{lgaplequalslsharpl} that $\LsharpL=\LGapL$. So we can assume that the function $f\in\sharpL$ and that it is non-negative on all inputs). For $x\in\Sigma^n$, we have size$(f(x))\leq p(n)$, for some polynomial $p(n)$. By the Prime Number Theorem, the number of primes between $2$ and $p^2(n)$ is $O(p^2(n)/\log n)>p(n)$, for sufficiently large $n$. Also the first $p(n)$ primes are of size $O(\log n)$. Furthermore, the product of the first $p(n)$ primes is greater than $f(x)$. Also it is easy to see using Corollary \ref{chap1-cor-crrtheorem}, that checking if a $O(\log n)$-bit integer is a prime can be done in logarithmic space. Furthermore, in logarithmic space, we can also compute the $i^{th}$ prime for $1\leq i\leq p(n)$. Let $p_i$ denote the $i^{th}$ prime. We define the function $g\in\FL$ as follows: $g(x,0^{p(|x|)},i)=0^{p_i}$ if $i\leq p(|x|)$, and it is $0^2$ otherwise.

We define the following language in $\ModL$
\[
L'=\{\langle x,0^{p(|x|)},i,k\rangle\mid f(x)\equiv k(\mod g(x,0^{p(|x|)},i)),i\leq p^2(|x|),k\leq p^2(|x|)\}.
\]
In order to show that $L\in\LModL$ we need to simulate the $\LGapL$ Turing machine for $L$ with a $\LModL$ computation. Clearly, it suffices to show that each $\GapL$ query $f(x)$ made by the base logarithmic space bounded deterministic Turing machine can be simulated in $\LModL$. For each $1\leq i\leq p(n)$, we can query $L'$ for $\langle x,0^{p(|x|)},i,k\rangle$ for different values of $k\leq p^2(|x|)$ to find $f(x)(\mod p_i)$.

Now, by the Chinese Remainder Theorem, $f(x)$ is uniquely determined by $f(x)(\mod p_i)$, for $1\leq i\leq p(n)$. Moreover, given these residues $f(x)(\mod p_i)$, for $1\leq i\leq p(n)$, it follows from Theorem \ref{chap1-crrtheorem}, that it is possible to compute $f(x)$ in logarithmic space (in fact, given the residues of $f(x)(\mod p_i)$, for $1\leq i\leq p(n)$, it is possible to compute $f(x)$ in DLOGTIME-uniform $\TCzero$ which is a very restricted form of uniformity of the uniform $\TCzero$ complexity class). Hence a logarithmic space bounded deterministic oracle Turing machine with access to the $\ModL$ oracle $L'$ can recover $f(x)$ for each query $x$. As a result it follows easily that $L$ is in $\LModL$.\textcolor{white}{\index[subject]{$\L^{\rm \mbox{ModL}}$}}
\end{proof}

\subsection{Characterization of \textrm{\boldModL}}\label{ModLcharacterize}
In this section we study the complexity class $\ModL$ and obtain a characterization of $\ModL$ that shows that a $\sharpL$ function $f$ and a $\FL$ function $g$ that outputs a prime number in unary representation are
sufficient to decide if a given input $x$ is in a language $L\in
\ModL$, however under the assumption that $\NL =\UL$. More precisely we are able to show a characterization of $\ModL$ in Theorem \ref{modlprime} that if we assume $\NL =\UL$ and we have a language $L\subseteq\Sigma^{*}$ with $L\in\ModL$ then we can decide whether an input $x\in\Sigma^{*}$ is in $L$ using $f\in\sharpL$ and $g\in\FL$ where $g(x)$ is a prime number $p$ that is output by $g$ in the unary representation such that if $x\in L$ then $f(x)\equiv 1(\mod p)$ and if $x\not\in L$ then $f(x)\equiv 0(\mod p)$. As an immediate consequence of Theorem \ref{modlprime}, 
we show that $\ModL$ is closed under complement assuming $\NL =\UL$ in Corollary \ref{modlcomplement}.
\begin{lemma}\label{gapltosharpl}
  Let $L\in \ModL$. There exists a function $f\in \sharpL$ and a
  function $g\in \FL$ such that on any input string $x$,
  \begin{itemize}
  \item $g(x)=0^{p^e}$ for some prime $p$ and a positive integer $e$,
    and
  \item $x\in L$ if and only if $f(x)\not \equiv 0(\mod p^e)$.
  \end{itemize}
\end{lemma}
\begin{proof}
  Let $L\in \ModL$ be witnessed by functions $f'\in \GapL$ and $g\in
  \FL$ as in Definition \ref{modldefn}. Now given $f'\in \GapL$ there exists
  $f_1,f_2\in \sharpL$ such that on any input string $x$ we have
  $f'(x)=f_1(x)-f_2(x)$. Consider $f(x)=f_1(x)+(p^e-1)f_2(x)$. Since
  $\sharpL$ is closed under multiplication by a $\FL$ function that
  outputs a positive integer, and also under addition, we have $f(x)\in \sharpL$. Moreover on a given
  input $x$, we have $f'(x)\not \equiv 0(\mod p^e)$ if and only if
  $f'(x)=f_1(x)-f_2(x)\not \equiv 0(\mod p^e)$ if and only if
  $f(x)=f_1(x)+(p^e-1)f_2(x)\not \equiv 0(\mod p^e)$. As a result we
  can replace the $\GapL$ function $f'$ by the $\sharpL$ function $f$
  to decide if any given input string $x$ is in $L$.
\end{proof}

\begin{theorem}\label{modlprime}
  Assume that $\NL =\UL$ and let $L\in \ModL$. There exists a
  function $f\in \sharpL$ and a function $g\in \FL$ such that on any
  input string $x$,
  \begin{itemize}
  \item $g(x)=0^p$ for some prime $p>0$, and,
  \item if $x\in L$ then $f(x)\equiv 1(\mod p)$,
  \item if $x\not\in L$ then $f(x)\equiv 0(\mod p)$.
  \end{itemize}
\end{theorem}

\begin{proof}
  Let $L\in \ModL$. It follows from Lemma \ref{gapltosharpl} that
  there exists $f'\in \sharpL$ and $g'\in \FL$ such that on any input
  string $x$ we have $g'(x)=0^{p^e}$ for some prime $p$ and a positive
  integer $e$, and $x\in L$ if and only if $f'(x)\not \equiv 0(\mod
  p^e)$. Let $g$ be a $\FL$ function that outputs the prime $p$ in unary when
  given the input $x$. Now assume that there exists a $f''\in \sharpL$
  such that $f'(x)\not \equiv 0(\mod p^e)$ if and only if $f''(x)\not
  \equiv 0(\mod p)$. Then $x\in L$ if and only if $f''(x)\not \equiv
  0(\mod p)$. Define $f(x)=(f''(x))^{(p-1)}$. Using Fermat's Little
  Theorem we have, if $x\in L$ then $f(x)\equiv 1(\mod
  p)$. Otherwise if $x\not \in L$ then $f(x)\equiv 0(\mod p)$. We
  therefore prove the theorem statement if we define the function
  $f''$ such that $f'(x)\not \equiv 0(\mod p^e)$ if and only if
  $f''(x)\not \equiv 0(\mod p)$.

  It is easy to see that we can compute the largest power of $p$ that divides $p^e=|g'(x)|$ in $\FL$ (for this, we simply iteratively compute $\lfloor\frac{|g'(x)|}{p}\rfloor$ until this value is $0$, which is in ${\rm U_L}$-$\NCone$ as stated in Theorem \ref{chap1-booleancircuittheorem}).
  
  If $e=1$ then we define $f''=f'$ where $f'$ is the $\sharpL$ function in Lemma \ref{gapltosharpl}. It is clear that on an input string $x$ we have $g(x)=g'(x)=0^p$ for
  some prime $p$ and $x\in L$ if and only if $f'(x)\not \equiv 0(\mod p^e)$ which is true if and only if $f''(x)\not \equiv 0(\mod p)$.
  
  Otherwise $e\geq 2$. Here we have $g'(x)=0^{p^e}$ and $g(x)=0^p$. Now we use induction on
  $(e-1)$ to define $f''$. Inductively assume that for $1\leq i\leq
  (e-1)$ we have functions $f_i\in \sharpL$ such that $f'(x)\not
  \equiv 0(\mod p^i)$ if and only if $f_i(x)\not \equiv 0(\mod
  p)$. For the case when $(e-1)=1$ we can have $f_{e-1}=f'$. Then it
  is clear that given an input string $x$, we have $f'(x)$ is
  divisible by $p^e$ if and only if
  \begin{enumerate}
  \item $f'(x)$ is divisible by $p^{e-1}$, and
  \item \label{condition2}the coefficient of $p^{e-1}$ in the base-$p$
    expansion of $f'(x)$ is zero.
  \end{enumerate}
  Here, using Corollary \ref{cor-kummercorollary} we get that, condition (\ref{condition2}) is equivalent to
  ${{f'(x)}\choose{p^{e-1}}}\equiv 0(\mod p)$. Therefore $f'(x)\not \equiv 0(\mod p^e)$ if and only if
  $\{ f'(x)\not \equiv 0(\mod p^{e-1})$ or
  ${{f'(x)}\choose{p^{e-1}}}\not \equiv 0(\mod p)\}$ which is true if
  and only if $\{ f_{e-1}(x)\not \equiv 0(\mod p)$ or
  ${{f'(x)}\choose{p^{e-1}}}\not \equiv 0(\mod p)\}$. Let $f_e'(x)=(f_{e-1}(x))^{p-1}$ ${{f'(x)}\choose{p^{e-1}}}^{p-1}+((f_{e-1}(x))^{p-1}+p-1)$ ${{f'(x)}\choose{p^{e-1}}}^{p-1}+(f_{e-1}(x))^{p-1}$ $({{f'(x)}\choose{p^{e-1}}}^{p-1}+(p-1))$.
  
  Now define $f_e(x)=(f_e'(x))^{p-1}$. Using
  Theorem \ref{sharplchoosefl} and closure properties of $\sharpL$ it
  follows that $f_e\in \sharpL$. Moreover on an input $x$, if $f'(x)\equiv 0(\mod p^e)$ then $f_e(x)\equiv 0(\mod p)$. On the other hand if $f'(x)\not\equiv 0(\mod p^e)$ then we consider the following two cases.\newline  
case 1: $f'(x)\equiv 0(\mod p^{e-1})$: Then $f_{e-1}(x)\equiv 0(\mod p)$ and we also have ${{f'(x)}\choose{p^{e-1}}}\not\equiv 0(\mod p)$.\textcolor{white}{\index[subject]{${f'(x)}\choose{p^{e-1}}$}} As a result from the definition of $f_e(x)$ we get $f_e(x)\equiv 1(\mod p)$.\newline
case 2: $f'(x)\not\equiv 0(\mod p^{e-1})$: Then $f_{e-1}(x)\not \equiv 0(\mod p)$ and we can have either ${{f'(x)}\choose{p^{e-1}}}\not\equiv 0(\mod p)$ or ${{f'(x)}\choose{p^{e-1}}}\equiv 0(\mod p)$. However in both of these cases we get $f_e(x)\equiv 1(\mod p)$. As a result defining $f''=f_e$ we complete the proof.
\end{proof}


\begin{corollary}\label{modlcomplement}
Assume that $\NL =\UL$. $\ModL$ is closed under complement.
\end{corollary}

\begin{proof}
  Let $L\in \ModL$. Then by Theorem \ref{modlprime} there exists $f
  \in \sharpL$ and $g\in \FL$ such that on any input $x$, we have
  $g(x)=0^p$ for some prime $p$ and if $x\in L$ then $f(x)\equiv
  1(\mod p)$. Otherwise if $x\not\in L$ then $f(x)\equiv 0(\mod p)$.

  Let $h(x)=(f(x)+(p-1))^{(p-1)}$. Using closure properties of $\sharpL$ it follows that $h(x)\in \sharpL$. Using Fermat's Little Theorem we have, if $x\in L$ then $h(x)\equiv 0(\mod
  p)$ and if $x\not\in L$ then $h(x)\equiv 1(\mod p)$. Clearly this
  shows $\overline{L}\in \ModL$ or that $\ModL$ is closed under
  complement.
\end{proof}

\section{Closure properties of $\boldModL$}\label{chap04-closureModL}
As a consequence of the characterization of $\ModL$ that we have shown in Theorem \ref{modlprime}, it is shown in Corollary \ref{modlcomplement} that $\ModL$ is closed under complement under the assumption that $\NL =\UL$. We continue to study the complexity class $\ModL$ and obtain its closure properties which follow due to the above stated characterization of $\ModL$.
\subsection{An unconditional closure property of \textrm{ModL}}
We recall the definition of the $\vee$ (join) operation from Definition \ref{join-operation}.
\begin{theorem}\label{modljoin}
If $\{ 0,1\}\subseteq\Sigma$ and $L_1\in\ModL$ then $L_1\vee L_2\in\ModL$.
\end{theorem}
\begin{proof}
Let $L_i\in \ModL$ and assume that we decide if an input string $x\in\Sigma^{*}$ is in $L_i$ using functions  $f_i\in\GapL$ and $g_i\in \FL$ where $1\leq i\leq 2$ respectively. We assume without loss of generality that the size of any  input string $x$ is at least $1$. Let $x=x_1x_2\cdots x_{n+1}$ and let $y=x_1\cdots x_n$ where $n\in\Z^{+}$. Also let 
\[ f(x)=\left \{ \begin{array}{ll}
	f_1(y) & \mbox{if $x_{n+1}=1$}\\
	f_2(y) & \mbox{otherwise if $x_{n+1}=0$,}
\end{array}
\right. \]
and
\[ g(x)=\left \{ \begin{array}{ll}
	g_1(y) & \mbox{if $x_{n+1}=1$}\\
	g_2(y) & \mbox{otherwise if $x_{n+1}=0$.}
\end{array}
\right. \]
It is easy to note that $f\in\GapL$ and $g\in\FL$. Since $|g_1(x)|$ or $|g_2(x)|$ is always a prime power on any input string $x\in\Sigma ^{*}$ it follows that $|g(x)|$ is also a prime power. Now $f(x)\not\equiv 0(\mod |g(x)|)$ for any input $x=yx_{n+1}$ if and only if exactly one of the following is true:
($x_{n+1}=1$ and $y\in L_1$) or ($x_{n+1}=0$ and $y\in L_2$). Clearly this shows that $L_1\vee L_2\in\ModL$ and that we can decide if any input string $x\in\Sigma ^{*}$ is in $L_1\vee L_2$ using $f\in\GapL$ and $g\in\FL$ which completes the proof.
\end{proof}
In Theorem \ref{modlunconclos}(\ref{modlmany}) we show that $\ModL$ is closed under logspace many-one reductions ($\leq ^{\L}_m$)  and similarly in Theorem \ref{modlunconclos}(\ref{modlulmany}) we show that $\ModL$ is closed under unambiguous logspace many-one reductions ($\leq ^{\ULredn}_m$). Using Theorem \ref{modlprime} and Corollary \ref{modlcomplement} we show in Theorem \ref{modlconclos}(\ref{modl1trt}) that $\ModL$ is closed under logspace truth-table reductions that make one query to the $\ModL$ oracle ($\leq ^{\L}_{1-\trt}$) assuming $\NL =\UL$.  We also show that if $\NL =\UL$ then $\ModL$ is closed under unambiguous logspace many-one reductions that make one query to the oracle ($\leq ^{\ULredn}_{1-\trt}$) in Theorem \ref{modlconclos}(\ref{modlul1trt}). The following table lists the closure properties of $\ModL$ that are known.
\begin{table}[h!]
\centering
\begin{tabular}{||c|c|l||}\hline\hline
\emph{Closure property} & \emph{Assumptions} & \emph{Reference} \\ \hline\hline
$\vee$ (join) & unconditional & Theorem \ref{modljoin} \\ \hline
complement & $\NL =\UL$ & Corollary \ref{modlcomplement} \\ \hline
${\Lredn}$ & unconditional & Theorem \ref{modlunconclos}(\ref{modlmany}) \\ \hline
$\leq^{\ULredn}_m$ & unconditional & Theorem \ref{modlunconclos}(\ref{modlulmany}) \\ \hline
$\leq_{1-\trt}^{\L}$ & $\NL =\UL$ & Theorem \ref{modlconclos}(\ref{modl1trt}) \\ \hline
$\leq^{\ULredn}_{1-\trt}$ & $\NL =\UL$ & Theorem \ref{modlconclos}(\ref{modlul1trt}) \\ \hline\hline
\end{tabular}
\caption{Closure properties of $\ModL$}\label{table:closuremodl}
\end{table}

As a consequence of these results we show in Corollary \ref{modlulmodl} that $\UL_{1-\trt}^{\ModLredn}=\ModL$ assuming $\NL =\UL$.

\subsection{Reducibilities of $\boldModL$}\label{modl-closure-properties}
We recall the Definitions \ref{L-Turing-reduction}, \ref{L-oracle-class} and \ref{NL-oracle-class} stated in Section \ref{chap02-immermanszelepcsenyitheorem} of Chapter 2.
\begin{definition}\label{logspace1ttdefn}\textcolor{white}{\index[subject]{$\leq ^{\L}_{1-\trt}$, logspace truth-table reduction that makes one query to the oracle}}
Let $\Sigma$ be the input alphabet and let $L_1,L_2\subseteq\Sigma ^{*}$. We say that $L_1$ is logspace $1$-truth-table reducible to $L_2$ and denoted by $L_1\leq^{\L}_{1-\trt}L_2$ if there exists a $O(\log n)$ space bounded Turing machine $M^{L_2}$ that has access to $L_2$ as an oracle such that $M^{L_2}$ decides if any given input $x\in\Sigma^{*}$ is in $L_1$ by making exactly one query to the oracle $L_2$, where $n=|x|$.
\end{definition}
\begin{definition}\label{ulmanyonedefn}
Let $\Sigma$ be the input alphabet and let $L_1,L_2\subseteq\Sigma ^{*}$. We say that $L_1\leq ^{\ULredn}_mL_2$ if there exists a $\FNL$ Turing machine $M$ such that on any input $x\in\Sigma ^{*}$ the number of accepting computation paths of $M(x)$ is at most $1$. Also if $x\in L_1$ then there exists an accepting computation path of $M$ on input $x$ such that if we obtain $y\in\Sigma^{*}$ as the output at the end of the accepting computation path then $y\in L_2$. If $x\not\in L_1$ then either $M$ does not have any accepting computation path on input $x$ or if there exists an accepting computation path of $M$ on input $x$ and $y\in\Sigma^{*}$ is the output at the end of this computation path then $y\not\in L_2$.\textcolor{white}{\index[subject]{$\leq^{{\rm UL}}_m$, unambiguous logspace many-one reducible}}
\end{definition}
\begin{definition}\label{ul1ttdefn}
Let $\Sigma$ be the input alphabet and let $L_1,L_2\subseteq\Sigma ^{*}$. We say\textcolor{white}{\index[subject]{$\leq ^{\UL}_{1-\trt}$, unambiguous logspace truth-table reduction that makes one query to the oracle}} that $L_1\leq ^{\ULredn}_{1-\trt}L_2$ if there exists a $\NL$-Turing machine $M^{L_2}$ with access to the oracle $L_2$ such that on any input $x\in\Sigma ^{*}$ the number of accepting computation paths of $M^{L_2}(x)$ is at most $1$. Also $M^{L_2}$ submits exactly one query $q_x\in\Sigma^{*}$ to the $L_2$ oracle on any input $x\in\Sigma^{*}$ to decide if $x\in L_1$.  We assume that the query is submitted by the $\NL$-Turing machine $M^{L_2}$ in a deterministic manner to the $L_2$ oracle according to the Ruzzo-Simon-Tompa oracle access mechanism. Here if $x\in L_1$ then there exists exactly one accepting computation path of $M^{L_2}$ on input $x$. Otherwise if $x\not\in L_1$ then $M^{L_2}$ rejects the input $x$ on all of its computation paths.\textcolor{white}{\index[subject]{$\leq ^{\UL}_{1-\trt}$, unambiguous logspace truth-table reduction that makes one query to the oracle}}
\end{definition}
We recall the definiton of $\ModL$ stated in Definition \ref{modldefn}.
\begin{definition}
Let $\Sigma$ be the input alphabet and let $L_1,L_2\in \Sigma ^*$. We define the complexity class $\UL^{\ModL}_{1-\trt}=\{L_1|L_1\leq ^{\ULredn}_{1-\trt}L_2$, where $L_2\in\ModL\}$.\textcolor{white}{\index[subject]{UL$^{\rm\mbox{ModL}}_{\mbox{1-tt}}$}}
\end{definition}

\begin{lemma}\label{fillacceptpaths}
Let $\Sigma$ be the input alphabet and let $g\in \FL$ such that on any input $x\in \Sigma ^*$, we have $g(x)$ to be a positive integer $p$ in the unary notation that depends on the input $x$. There exists a $\NL$-Turing machine $M$ such that the lexicographically least $|g(x)|$ computation paths are exactly the only accepting computation paths in the computation tree of $M$.
\end{lemma}
\begin{proof}
Let us consider the following algorithm of a non-deterministic Turing machine given an input $x\in\Sigma^*$.
\begin{algorithm}[H]
\caption{Lex-Least-accept-$\FL _g$ \newline {\bf Input:} $x\in \Sigma^*$.\newline {\bf Output:} \emph{accept} or \emph{reject.} The $\NL$-Turing machine $M$ that implements this algorithm is such that the lexicographically least $|g(x)|$ computation paths are the only accepting computation paths of $M(x)$ and so $\acc _M(x)=|g(x)|$.\newline {\bf Complexity:} $\NL$.}\label{algo:fillacceptpaths}
\begin{algorithmic}[1]
\State Compute $p=|g(x)|$.
\State Compute $k$ such that $2^{k-1}\leq p<2^k$.
\State Non-deterministically choose $k$ bits from $\{ 0,1\}$ that describes a computation path of length $k$ in the computation tree of depth $k$. Let $m$ be the number describing the computation path that is chosen.
\If{$m\leq p$}
    \State accept and stop.
\Else
    \State reject and stop.
\EndIf
\end{algorithmic}
\end{algorithm}
Since $g(x)=0^p$ for some positive integer $p$ and $k\in\Z^+$ such that $2^{k-1}\leq p<2^k$, it is easy to see that $k,p\in O(\log n)$ As a result it is easy to note that the above algorithm can be implemented by a non-deterministic Turing machine that uses at most $O(\log n)$ space where $n=|x|$ and $x\in\Sigma^*$ is the input. Let $M$ be the $\NL$-Turing machine that simulates the above stated algorithm on input $x$. Clearly, the $\NL$-Turing machine $M$ non-deterministically makes $k$ choices and accepts in a computation path if the non-deterministic choices form a binary string less than or equal to $p$. $M$ rejects at the end of all other computation paths which proves our result.
\end{proof}

\begin{theorem}\label{modlunconclos}
Let $\Sigma$ be the input alphabet and let $L_1,L_2\subseteq \Sigma ^{*}$. Also let $L_2\in\ModL$. Now
\begin{enumerate}
\item\label{modlmany} if $L_1{\Lredn} L_2$ then $L_1\in\ModL$, and
\item\label{modlulmany} if $L_1\leq^{\ULredn}_mL_2$ then $L_1\in\ModL$.
\end{enumerate}
\end{theorem}
\begin{proof}
\begin{enumerate}
\item Let $L_2\in\ModL$ and let $f_2\in\GapL$ and $g_2\in\FL$ be functions using which we decide if an input $x\in\Sigma ^{*}$ is in $L_2$. Also let $L_1\Lredn L_2$ using $f\in\FL$. If $x\in\Sigma^{*}$ is the input then $x\in L_1$ if and only if $f(x)\in L_2$. However $L_2\in\ModL$ and therefore $f(x)\in L_2$ if and only if $f_2(f(x))\not\equiv 0(\mod p^e)$ where $p^e=|g_2(f(x))|\in\Z^{+}$ where $p,e\in\Z^{+}$ with $p$ being a prime and $e\geq 1$. Now it follows from Lemma \ref{composeFLsharpLGapL} that $(f_2\circ f)\in\GapL$ and $(g_2\circ f)\in\FL$. Clearly $g_2(f(x))$ is a prime power in the unary representation for any $x\in\Sigma ^{*}$. As a result if $x\in\Sigma ^{*}$ then $x\in L_1$ if and only if $f_2(f(x))\not\equiv 0(\mod p^e)$ where $p^e=|g_2(f(x))|$ which shows $L_1\in\ModL$.

\item We recall the definition $\leq^{\ULredn}_m$ from Definition \ref{ulmanyonedefn}. The proof of this assertion is similar to Theorem \ref{modlunconclos}(\ref{modlmany}). We use Lemma \ref{gapltosharpl} for $L_2\in \ModL$ and assume that there exists $f\in\sharpL$ and $g\in\FL$ such that on any input $x\in\Sigma^{*}$ we have $g(x)=1^{p^e}$ where $p,e\in\Z^{+}$, $p$ is a prime, $e\geq 1$ and $x\in L_2$ if and only if $f(x)\not\equiv 0(\mod p^e)$. Let us denote the $\leq^{\ULredn}_m$ reduction by $h$. Clearly, $h$ has at most one accepting computation path. Let us now consider the following algorithm implemented by a non-deterministic Turing machine given an input $x\in\Sigma^*$.
\begin{algorithm}[H]
\caption{$\UL$-many-one-$\ModL$ \newline {\bf Input:} $x\in \Sigma^*$.\newline {\bf Output:}  \emph{accept} or \emph{reject.} The $\NL$-Turing machine that implements this algorithm obeys the property of $\ModL$ in Lemma \ref{gapltosharpl}. \newline {\bf Complexity:} $\ModL$.}\label{algo:UL-many-one-ModL}
\begin{algorithmic}[1]
\State Simulate the $\NL$-Turing machine $M_h$ corresponding to the $\FUL$ function $h$ on the input $x$.
\If{the computation path of $M_h(x)$ ends in an accepting configuration}
    \State Let $y=M_h(x)$.
\Else
    \State Lex-Least-accept-FL$_g(x)$ and stop.
\EndIf
\State Simulate $M(y)$ and stop.
\end{algorithmic}
\end{algorithm}
It is easy to see that this algorithm can be implemented by a non-deterministic Turing machine that uses $O(\log n)$ space, where $n=|x|$. In the above algorithm, upon simulating $M_h$ on input $x$, if the computation path that we obtain is an accepting computation path and $y\in\Sigma^{*}$ is the output at the end of the accepting computation path, then simulating the $\NL$-Turing machine for $L_2\in\ModL$ with $y$ as the input shows that the congruence relation based on the $\sharpL$ function $f$ and the $\FL$ function $g$ that is used to decide if any input $x\in\Sigma ^{*}$ is in $L_2$ can also be used for deciding if $x\in L_1$. Otherwise if the computation path of the $\leq^{\ULredn}_m$ reduction that we obtain is a rejecting computation path then using the non-deterministic algorithm stated in Lemma \ref{fillacceptpaths}, it follows that our $\NL$-Turing machine can make sufficiently many non-deterministic choices so that the number of accepting computation paths that we obtain at the end of each of these rejecting computation paths is divisible by $|g(x)|$. This shows $L_1\in\ModL$.
\end{enumerate}
\end{proof}

\begin{theorem}\label{modlconclos}
Let $\Sigma$ be the input alphabet and let $L_1,L_2\subseteq \Sigma ^{*}$. Also let $L_2\in\ModL$ and assume that $\NL =\UL$.
\begin{enumerate}
\item \label{modl1trt} If $L_1\leq ^{\L}_{1-\trt}L_2$ then $L_1\in\ModL$, and 
\item \label{modlul1trt} If $L_1\leq^{\ULredn}_{1-\trt}L_2$ then $L_1\in\ModL$.
\end{enumerate}
\end{theorem}
\begin{proof}
\begin{enumerate}
\item We recall the definition of $\leq^{\L}_{1-\trt}$ from Definition \ref{logspace1ttdefn}. Let $L_1\leq ^{\L}_{1-\trt}L_2$ using $f\in\FL$ that makes exactly one query to the oracle $L_2$. In other words, $f$ correctly decides if an input $x\in\Sigma^{*}$ is in $L_1$ by making exactly one query to the oracle $L_2$. Also let  $q_x\in\Sigma ^{*}$ be the query string that is generated by $f$ for the input $x\in\Sigma ^{*}$. Clearly $f$ decides if $x\in L_1$ based on $x$ and the reply of the oracle $L_2$ when the oracle $L_2$ is given the query $q_x$ as input. Therefore let $f'\in\FL$ be the function that generates and outputs the query $q_x$ generated by $f$ when it is given the input $x$. Also let $g'\in\FL$ be such that $g'(x,\chi _{L_2}(q_x))=f(x)$ where $\chi _{L_2}$ is the characteristic function of $L_2$. It is clear that $g'$ correctly decides whether any input $x\in L_1$ if the reply of the oracle $L_2$ is also given to it. Since $L_2\in\ModL$ and we have assumed that $\NL =\UL$, it follows from Theorem \ref{modlprime} that there exists $f_2\in\sharpL$ and $g_2\in\FL$ such that on any input $x\in\Sigma^{*}$ we have if $x\in L_2$ then $f_2(x)\equiv 1(\mod p)$ and if $x\not \in L_2$ then $f_2(x)\equiv 0(\mod p)$, where $g_2(x)=1^p$, $p\in\Z^{+}$ and $p$ is a prime. Let us consider the function $(g_2\circ f')$. Given an input $x\in \Sigma^*$, $g_2(f(x))=1^p$, for some prime $p\in\Z^+$ which depends on the query $f'(x)$ generated from the input $x$. It is clear that $(g_2\circ f')\in\FL$. Consider the function $(g'\circ f_2\circ f')$ and let $M$ denote the $\NL$-Turing machine corresponding to this function. The $\NL$-Turing machine $M$ considers the output of a computation path of the $\NL$-Turing machine corresponding to $(f_2\circ f')(x)$ as $\chi _{L_2}(q_x)$. Since $g'\in\FL$ it follows that the output of $M$ is the same at the end of all the accepting computation paths of the $\NL$-Turing machine corresponding to $(f_2\circ f')$. In other words, either $M$ accepts at the end of all the accepting computation paths of the $\NL$-Turing machine corresponding to $(f_2\circ f')$ or $M$ rejects at the end of all the accepting computation paths of the $\NL$-Turing machine corresponding to $(f_2\circ f')$. Similarly either $M$ accepts at the end of all the rejecting computation paths of the $\NL$-Turing machine corresponding to $(f_2\circ f')$ or $M$ rejects at the end of all the rejecting computation paths of the $\NL$-Turing machine corresponding to $(f_2\circ f')$. Let us now consider the following algorithm.
\begin{algorithm}[H]
\caption{$\L$-1-truth-table-$\ModL$ \newline {\bf Input:} $x\in \Sigma^*$.\newline {\bf Output:} \emph{accept} or \emph{reject.} The $\NL$-Turing machine $M'$ that implements this algorithm obeys the property of $\ModL$ in Theorem \ref{modlprime}.\newline {\bf Complexity:} $\ModL$.}\label{algo:L-1-truth-table-ModL}
\begin{algorithmic}[1]
\State Compute $p=|g_2(f'(x))|$.
\State $counter=0$.
\State $lex{_-}least{_-}flag={\mbox{True}}$.
\While{$counter<p-1$}
    \State Simulate the $\NL$-Turing machine $M$ on input $x$ as follows.
    \If{$M(x)$ rejects at the end of its computation path}
        \State reject and stop.
    \ElsIf{$M(x)$ accepts and $(f_2\circ f')(x)$ rejects}
        \If{the computation path of $M(x)$ is not the lexicographically least}
\algstore{L1-tt-firstpart}
\end{algorithmic}
\end{algorithm}
\clearpage
\begin{algorithm}[H]
\setcounter{algorithm}{11}
\caption{$\L$-1-truth-table-$\ModL$ (continued)}\label{algo:L-1-truth-table-ModL-continued}
\begin{algorithmic}[1]
\algrestore{L1-tt-firstpart}    
            \State $lex{_-}least{_-}flag={\mbox{False}}$.
        \EndIf
    \EndIf
    \State $counter=counter+1$.
\EndWhile
\If{$lex{_-}least{_-}flag={\mbox{True}}$}
    \State Lex-Least-accept-FL$_{g_2}(f'(x))$.    
    \State $M'$ non-deterministically branches sufficiently many times on its computation path as in Corollary \ref{modlcomplement}.
\EndIf
\end{algorithmic}
\end{algorithm}

Let $x\in\Sigma^{*}$ be the input and let $g_2(f'(x))=1^p$. In the above algorithm, if the output of $M$ at the end of any of its computation paths is the same as the output of the computation path of the $\NL$-Turing machine corresponding to $(f_2\circ f')$ on the input $x$ then $M'$ makes $(p-1)$ times the simulation of $M$ on input $x$ and stops. In this case, for a given input $x\in\Sigma^{*}$, if $L_1\leq ^{\L}_{1-\trt}L_2$ then at the end of every computation path the reduction is similar to a logspace many-one reduction from $L_1$ to $L_2$. In this case using Fermat's Little Theorem, it follows from our observations that if $x\in L_1$ then $\acc _{M'}(x)\equiv 1(\mod p)$. However if $x\not \in L_1$ then $\acc _{M'}(x)\equiv 0(\mod p)$. This shows $L_1\in\ModL$. On the other hand if we have the output of $M$ along any of its computation paths is the complement of the output of the computation path of the $\NL$-Turing machine corresponding to $(f_2\circ f')$ on a given input $x\in\Sigma^{*}$ then also $M'$ makes $(p-1)$ simulations of $M$ on input $x$. However in these $(p-1)$ simulations of $M$ by $M'$, along any of the computation paths of $M'$ that is not the lexicographically least computation path we assume that $M'$ stops with the output of $M$ on input $x$. Along the lexicographically least computation path that we obtain in these $(p-1)$ simulations of $M$ by $M'$ we assume that $M'$ makes sufficiently many non-deterministic choices at the end of this computation path as in Lemma \ref{fillacceptpaths} so that we obtain $\acc _{M'}(x)=\acc _M(x)^{p-1}+(p-1)$. In this case using Fermat's Little Theorem, it follows from our observations that if $x\in L_1$ then $\acc _{M'}(x)\equiv 0(\mod p)$ and if $x\not \in L_1$ then $\acc _{M'}(x)\not\equiv 0(\mod p)$ which implies $\overline{L_1}\in \ModL$. We now use our assumption that $\NL=\UL$ and so it follows from Corollary \ref{modlcomplement} that $L_1\in\ModL$ and this completes the proof.

\item We recall the definition of $\leq^{\ULredn}_{1-\trt}$ from Definition \ref{ul1ttdefn}. We follow the Ruzzo-Simon-Tompa oracle access mechanism stated in Section \ref{chap1-NOTM} and generate the query string $q_x\in\Sigma^*$ given the input $x\in\Sigma^*$ in a deterministic manner using $O(\log n)$ space. Our proof follows from our assumption that $\NL=\UL$ and a case analysis similar to the proof of Theorem \ref{modlconclos}(\ref{modl1trt}) depending on whether the $\leq^{\ULredn}_{1-\trt}$ reduction is similar to a $\leq^{\ULredn}_m$ reduction or if the output of the computation paths of the $\NL$-Turing machine that is used in the reduction is the complement of the output of the computation paths of the $\NL$-Turing machine for the oracle $L_2$ on the query string $q_x$.
\end{enumerate}
\end{proof}
\begin{corollary}\label{modlulmodl}
Assume that $\NL =\UL$. $\UL ^{\ModLredn}_{1-\trt}=\ModL$.
\end{corollary}
\begin{proof}
Since we know that $\NL$ is closed under complement from Theorem \ref{chap2-NLcomplement}, our assumption that $\NL =\UL$ implies $\UL =\coUL$. Proof of our result now follows from Theorem \ref{modlconclos}(\ref{modlul1trt}).
\end{proof}

\section[Relations among Modulo-based]{Relations among Modulo-based}
\begin{definition}
Let $\Sigma$ be the input alphabet and let $L_1,L_2\subseteq\Sigma^{*}$. We say that $L_1$ is logspace conjunctive truth-table reducible to $L_2$, denoted by $L_1\leq ^{\L}_{\ctt}L_2$, if there exists $f\in\FL$ such that given an input $x\in\Sigma ^{*}$ of length $n$ we have $f\in\FL$ such that $f(x)=\{ y_1\ldots ,y_{p(n)}\}$ and $x\in L_1$ if and only if $y_i\in L_2$ for every $1\leq i\leq p(n)$ where $p(n)$ is a polynomial in $n$.\textcolor{white}{\index[subject]{$\leq ^{\L}_{\ctt}$, logspace conjunctive truth-table reduction}}
\end{definition}
\begin{definition}
Let $\Sigma$ be the input alphabet and let $L_1,L_2\subseteq\Sigma^{*}$. We say that $L_1$ is logspace $k$-conjunctive truth-table reducible to $L_2$, denoted by $L_1\leq ^{\L}_{k-\ctt}L_2$, if there exists $f\in\FL$ such that given an input $x\in\Sigma ^{*}$ of length $n$ $f(x)=\{ y_1\ldots ,y_k\}$ and $x\in L_1$ if and only if $y_i\in L_2$ for every $1\leq i\leq k$.\textcolor{white}{\index[subject]{$\leq ^{\L}_{k-\ctt}$, logspace conjuctive truth-table reduction that makes exactly $k$ queries to the oracle}}
\end{definition}
\begin{definition}
Let $\Sigma$ be the input alphabet and let $L_1,L_2\subseteq\Sigma^{*}$. We say that $L_1$ is logspace disjunctive truth-table reducible to $L_2$, denoted by $L_1\leq ^{\L}_{\dtt}L_2$, if there exists $f\in\FL$ such that given an input $x\in\Sigma ^{*}$ of length $n$ we have $f(x)=\{ y_1,\ldots ,y_{p(n)}\}$ and $x\in L_1$ if and only if $y_i\in L_2$ for at least one $1\leq i\leq p(n)$ where $p(n)$ is a polynomial in $n$.\textcolor{white}{\index[subject]{$\leq ^{\L}_{\dtt}$, logspace disjunctive truth-table reduction}}
\end{definition}
\begin{definition}
Let $\Sigma$ be the input alphabet and let $L_1,L_2\subseteq\Sigma^{*}$. We say that $L_1$ is logspace $k$-disjunctive truth-table reducible to $L_2$ for some $k\in\Z^{+}$, $k\geq 1$, denoted by $L_1\leq ^{\L}_{k-\dtt}L_2$, if there exists  $f\in\FL$ such that given an input $x\in\Sigma ^{*}$ of length $n$ we have $f(x)=\{ y_1,\ldots ,y_k\}$ and $x\in L_1$ if and only if $y_i\in L_2$ for at least one $1\leq i\leq k$.\textcolor{white}{\index[subject]{$\leq ^{\L}_{k-\dtt}$, logspace disjunctive truth-table reduction that makes exactly $k$ queries to the oracle}}
\end{definition}
It follows from the definition of $\ModL$ that if $p\in\Z^{+}$ is a prime then $\ModpL\subseteq\ModL$. However if $k\in\Z^{+}$ such that $k\geq 6$ is a composite number that has more than one distinct prime divisor then it is not known if $\ModkL\subseteq\ModL$. We show that if $\ModL$ is closed under $\leq ^{\L}_{l-\dtt}$ reductions for some $l\in\Z^{+}$ such that $l\geq 2$ then $\ModkL\subseteq\ModL$ for all $k\in\Z^{+}$ such that $k\geq 6$ is a composite number that has at most $l$ distinct prime divisors.
\begin{theorem}\label{modklmodl}
If $\ModL$ is closed under $\leq ^{\L}_{l-\dtt}$ reductions where $l\in\Z^{+}$ and $l\geq 2$ then $\ModkL\subseteq\ModL$ for all $k\in\Z^{+}$ such that $k\geq 6$ is a composite number that has at least $2$ and at most $l$ distinct prime divisors.
\end{theorem}
\begin{proof}
Let us assume that $\ModL$ is closed under $\leq ^{\L}_{l-\dtt}$ reductions for some $l\in\Z^{+}$ such that $l\geq 2$. Let $\Sigma$ be the input alphabet and let $\sharp\not\in\Sigma$. Also let $L\in\Sigma ^{*}$ and assume that $L\in\ModkL$, where $k=p_1^{e_1}\cdots p_m^{e_m}$ is a composite number such that $p_i$ is a prime, $e_i\in\Z^{+}$ with $e_i\geq 1$ and $p_i\neq p_j$ for all $1\leq i<j\leq m\leq l$.

Using Lemma \ref{gapltosharpl}, it follows from the definition of $\ModkL$, $\GapL$, $\sharpL$ and Proposition \ref{sldagcountsharplcomplete} that determining if the number of $st$-paths in an instance $(G,s,t)$ of SLDAG is $\not\equiv 0(\mod k)$ is logspace many-one complete for $\ModkL$. Let us denote this problem by Mod$_kst$Path. Also it is easy to infer that for any function  $f\in\sharpL$, we have $f(x)\not\equiv 0(\mod k)$ if and only if $f(x)\not\equiv 0(\mod p_i)$ for at least one of the primes $p_i|k$, where $1\leq i\leq m$. Based on these observations let us define $L_k=\{\langle x\sharp 1^{p_i}\rangle |f(x)\not\equiv 0(\mod |g(\langle x\sharp 1^{p_i}\rangle )|)~\forall ~p_i|k$ where $g\in\FL$ and $g(\langle x\sharp 1^{p_i}\rangle )=1^{p_i}\}$. As a result, it follows from Definition \ref{modldefn} that $L_k\in\ModL$.

Therefore if $x\in\Sigma^*$ is the input then $x\in L$ if and only if $\langle x\sharp 1^{p_i}\rangle \in L_k$ for at least one prime $p_i$ where $1\leq i\leq m$. However it is easy to note that given $x$ as input, a $O(\log |x|)$ space bounded Turing machine $M$ can obtain the prime factorization of $k$ and also output $\langle x\sharp 1^{p_i}\rangle$ for all primes $p_i|k$ where $1\leq i\leq m$. If $m<l$ then we assume that $M$ outputs the last query $\langle x\sharp 1^{p_{j}}\rangle$ that it generates with repetition sufficiently many times to output $l$ strings where $1\leq j\leq m\leq l$. Now $x\in L$ if and only if $\langle x\sharp 1^{p_j}\rangle\in L_k$ for at least one $1\leq j\leq l$. However we have assumed that $\ModL$ is closed under $\leq ^{\L}_{l-\dtt}$ reductions which implies  $L\in\ModL$. This shows that $\ModkL\subseteq\ModL$ whenever $k\geq 6$ is a composite number such that the number of distinct prime divisors of $k$ is at least $2$ and at most $l$.
\end{proof}

\begin{theorem}\label{coceqlmodl}
If $\ModL$ is closed under $\leq ^{\L}_{\dtt}$ reductions then $\coCeqL\subseteq\ModL$.
\end{theorem}
\begin{proof}
Let us assume that $\ModL$ is closed under $\leq ^{\L}_{\dtt}$ reductions. Recalling the definition of Mod$_kst$Path from the proof of Theorem \ref{modklmodl}, we know that Mod$_kst$Path is a canonical complete language for $\ModL$ under logspace many-one reductions. Also given $x\in\Sigma^*$ as an input, the problem of determining if $f(x)\neq 0$ is complete for $\coCeqL$ under logspace many-one reductions, where $f$ is a logspace many-one complete function for $\sharpL$. Using the Chinese Remainder Theorem we know that the $\det (A)$ is uniquely determined by its residues modulo all the primes from $2$ and $n^4$. Therefore $f(x)\neq 0$ if and only if $f(x)\not\equiv 0(\mod p)$ for some prime $2\leq p\leq n^4$.

Similar to the proof of Lemma \ref{modklmodl} it is easy to note that there exists a $O(\log n)$ space bounded Turing machine $M$ that when given the input string $x$ as input computes all the primes from $2$ to $n^4$ and also outputs the pairs $\langle x\sharp 1^{p_i}\rangle$ for all of the primes from $2$ to $n^4$. Clearly $f(x)\neq 0$ if and only if $\langle x\sharp 1^{p_i}\rangle\in\Mod_kst$Path for at least one of the primes $p\in\{ 2,\ldots ,n^4\}$. However we have assumed that $\ModL$ is closed under $\leq ^{\L}_{\dtt}$ reductions from which it follows that we can determine if $f(x)\neq 0$ in $\ModL$ and this implies $\coCeqL\subseteq\ModL$.
\end{proof}

\begin{corollary}
Assume that $\NL =\UL$. If $\ModL$ is closed under $\leq^{\L}_{\dtt}$ reductions then $\CeqL\subseteq\ModL$.
\end{corollary}
\begin{proof}
Proof follows from Theorem \ref{coceqlmodl} and \ref{modlcomplement}.
\end{proof}

\begin{lemma}\label{bddlogspacectt}
Assume that $\NL =\UL$. If $\ModL$ is closed under $\leq ^{\L}_{l-\dtt}$ reductions where $l\in\Z^{+}$ and $l\geq 2$ then $\ModL$ is closed under $\leq ^{\L}_{l-\ctt}$ reductions.
\end{lemma}
\begin{proof}
Let $\Sigma$ be the input alphabet and let $\sharp\not\in\Sigma$. Also let $L_1,L_2\subseteq\Sigma ^{*}$ and let $L_2\in\ModL$ be such that $L_1\leq ^{\L}_{l-\ctt}L_2$ using a function $f\in\FL$. Therefore if $x\in\Sigma ^{*}$ is the input then we have $f(x)=\langle y_1\sharp\cdots\sharp y_l\rangle$ such that $x\in L_1$ if and only if $y_i\in L_2$ for all $1\leq i\leq l$.

However this is equivalent to $x\not\in L_1$ if and only if $y_i\in\overline{L_2}$ for at least one $1\leq i\leq l$ where $\overline{L_2}$ denotes the complement of $L_2$. However since we have assumed $\NL =\UL$ it follows from Corollary \ref{modlcomplement} that $\overline{L_2}\in\ModL$. These observations show that $\overline{L_1}\leq ^{\L}_{l-\dtt}\overline{L_2}$ and since we have assumed $\ModL$ is closed under $\leq ^{\L}_{l-\dtt}$ we have $\overline{L_1}\in\ModL$. Once again from our assumption that $\NL =\UL$ using the result that $\ModL$ is closed under complement shown in Corollary \ref{modlcomplement} it follows that $L_1\in\ModL$.
\end{proof}

\begin{theorem}\label{logspacectt}
Assume that $\NL =\UL$. If $\ModL$ is closed under $\leq^{\L}_{\dtt}$ reductions then $\ModL$ is closed under $\leq ^{\L}_{\ctt}$ reductions.
\end{theorem}
\begin{proof}
Proof of this result is similar to the proof of Lemma \ref{bddlogspacectt}. We need to observe that the number of instances of $L_2$ that can be output by a $\leq^{\L}_{\dtt}$ reduction is at most a polynomial in the size of the input. Since we have assumed $\NL =\UL$ it follows from Corollary \ref{modlcomplement} that $\ModL$ is closed under complement and by using this property of $\ModL$ we obtain this result.
\end{proof}

\section*{Exercises}
\begin{enumerate}
\item Show that $\NL$ is closed under $\leq ^{\L}_{\ctt}$ reductions without using Theorems \ref{nlclosureunderTuring} and \ref{chap2-NLcomplement} .
\item Show that $\NL$ is closed under $\leq ^{\L}_{\dtt}$ reductions without using Theorems \ref{nlclosureunderTuring} and \ref{chap2-NLcomplement}.
\item Let $\Sigma$ be the input alphabet and let $A\subseteq\Sigma^*$. Define the complexity class $\ModL^A$.
\item Define the complexity class $\ModL^{\ULredn}$.
\item Show that $\ModL^{\ULredn}=\ModL$ assuming $\UL =\coUL$.
\end{enumerate}
\section*{Open problems}
\begin{enumerate}
\item Does there exist any containment relation between $\parityL$ and $\NL$? Is $(\parityL \Delta$ $\NL )$ non-empty? 
\item Is $\ModkL\subseteq\ModL$ if $k>0$ is a composite number?
\item Is $\ModkLH =\ModkL$, where $k\in\N$ and $k$ is a composite?
\end{enumerate}

\section*{Notes}
Modulo-based logarithmic space bounded counting classes $\ModkL$, where $k\in\Z$ and $k\geq 2$, were by defined Gerhard Buntrock, Carsten Damm, Ulrich Hertrampf and Christoph Meinel in \cite{BDHM1992}\textcolor{white}{\index[authors]{Buntrock, Gerhard}\index[authors]{Damm, Carsten}\index[authors]{Hertrampf, Ulrich}\index[authors]{Meinel, Christoph}}. All the results in this chapter prior to Section \ref{modl-section} have been shown in the polynomial-time setting by Richard Beigel and John Gill in \cite{BG1992} for complexity classes $\ModkP$ and $\ModpP$, where $k,p\in\N$, $k,p\geq 2$ and $p$ is a prime. Theorems \ref{modulok-logspaceTuringclosed} and \ref{modulok-modplTuringclosed} are due to Ulrich Hertrampf, Steffan Reith and Heribert Vollmer \cite{HRV2000}\textcolor{white}{\index[authors]{Reith, Steffan}\index[authors]{Vollmer, Heribert}}. The remaining results starting from Proposition \ref{modulok-jdividesk} to Corollary \ref{modulok-union-corollary} are from \cite{BDHM1992}.

The modulo-based logarithmic space bounded counting class $\ModL$ was defined by V. Arvind and T. C. Vijayaraghavan in \cite{AV2010,Vij2008} to
tightly classify the complexity of the problem called LCON which is to solve a system of linear equations modulo a composite number $k$, where $k$ is given in terms of its prime factorization such that every distinct prime power divisor that occurs in the prime factorization of $k$ is given in the unary representation. In \cite{AV2010}, it is shown that LCON\textcolor{white}{\index[subject]{LCON}}$\in\LModLpoly$\textcolor{white}{\index[subject]{L$^{\rm \mbox{ModL}}$/poly}\index[subject]{BP.NC$^2$}\index[subject]{Mod$_p$P}} and in $\BPNCtwo$. Along with LCON, some more problems on linear congruences modulo a composite number $k$ which is given as a part of the input in terms of its prime factorization, where every prime power is given in unary representation, and a host of problems on Abelian permutation groups have been shown to be in $\LModLpoly$ and in $\BPNCtwo$\textcolor{white}{\index[subject]{L$^{\rm \mbox{ModL}}$/poly}\index[subject]{BP.NC$^2$}}. \textcolor{white}{\index[authors]{Beigel, Richard}\index[authors]{Gill, John}\index[authors]{Arvind, V.}\index[authors]{Vijayaraghavan, T. C.}}The complexity class $\ModL$ generalizes $\ModkL$, for any $k\in \Z$ and $k\geq 2$ since the moduli varies with the input. A very useful observation on $\ModL$ which uses the Chinese Remainder Theorem is Lemma \ref{AV2010-lemma} and it is from \cite[Lemma 3.3]{AV2010}. Given a number $n$ as residues modulo polynomially many primes, the result that it is possible to compute or find out $n$ from its residues in logspace-uniform $\NCone$ is due to Andrew Chiu, George Davida and Bruce Litow \cite{CDL2001}. This result was improved by W. Hesse, David A. Mix Barrington and Eric Allender in \cite{HAB2002}\textcolor{white}{\index[authors]{Chiu, Andrew}\index[authors]{Davida, George}\index[authors]{Litow, Bruce}\index[authors]{Hesse, William}\index[authors]{Allender, Eric}\index[authors]{Barrington, David A. Mix}} wherein it is shown that the same operation of computing the number $n$ from its residues modulo polynomially many primes is in DLOGTIME-$\TCzero$.

Results on $\ModL$ in Section \ref{ModLcharacterize} are from \cite{Vij2022}. An important consequence of the characterization of $\ModL$ shown in Theorem \ref{modlprime} is that, assuming $\NL =\UL$, we are able to show that $\ModL$ is the logspace analogue of the polynomial time counting complexity class ModP defined by J. K$\ddot{\rm o}$bler and Seinosuke Toda in \cite{KT1996}. Closure properties of $\ModL$ shown in Section \ref{chap04-closureModL} and results on reducibilities of $\ModL$ shown in Section \ref{modl-closure-properties} is from \cite{Vij2010}\textcolor{white}{\index[authors]{K$\ddot{\rm o}$bler, Johannes}\index[authors]{Toda, Seinosuke}\index[subject]{ModP}}.

\chapter{Probabilistic Logarithmic space bounded counting class: $\boldPL$}\label{chapter04-PL}
\begin{tcolorbox}[colback=gray!35!white,colframe=white]
{\bf\textit{From now onwards for the rest of this chapter, when we consider functions in $\GapL$, we do not necessarily assume that the computation tree of a function in $\GapL$ is a complete binary tree.}}
\end{tcolorbox}
\section{Closure properties of $\boldPL$}\label{PL-properties}
We recall the definition of $\PL$ from Definition \ref{pldefinition}.
\begin{proposition}\label{characterizePL}
Let $\Sigma$ be the input alphabet. The class $\PL$ consists of those languages $L\subseteq\Sigma^*$ such that for some $\GapL$ function $f$ and all $x\in\Sigma^*$
\begin{itemize}
\item if $x\in L$ then $f(x)>0$
\item if $x\not\in L$ then $f(x)<0$.
\end{itemize}
\end{proposition}
\begin{proof}
For a $\GapL$ function $f$, the $\GapL$ function $2f(x)-1$ has the same sign as $f(x)$ for positive and negative values of $f(x)$ and takes value $-1$ for $f(x)=0$.
\end{proof}

\begin{corollary}\label{plclosurecomplement}
$\PL$ is closed under complement.
\end{corollary}

\begin{lemma}\label{gaplfunctionsdefine}
Define
\[\begin{array}{c}
P_n(x) = (x-1)\prod_{i=1}^n(x-2^i)^2. \\
A_n(x)=P_n(-x))^{\frac{n}{2}+1}-(P_n(x))^{\frac{n}{2}+1}. \\
B_n(x)=P_n(-x))^{\frac{n}{2}+1}+(P_n(x))^{\frac{n}{2}+1}. \\
S_n(x)=\frac{A_n}{B_n}.
\end{array}\]
Then, for $n\geq 1$,
\begin{enumerate}
\item If $1\leq x\leq 2^n$ then $0\leq 4P_n(x)<-P_n(-x)$.
\item If $1\leq x\leq 2^n$ then $1\leq S_n<\frac{5}{3}$.
\item If $-2^n\leq x\leq -1$ then $-\frac{5}{3}<S_n(x)\leq -1$.
\end{enumerate}
\end{lemma}
\begin{proof}
\begin{enumerate}
\item Since $x\geq 1$, we have that $P_n(x)\geq 0$. Clearly, $x-1<x+1$, and $(x-2^i)^2<(-x-2^i)^2$ for $i=1,\ldots ,n$. Also, if $2^{k-1}\leq x<2^k$ then $4(x-2^k)^2\leq 4.2^{2k-2}=2^{2k}<(-x-2^k)^2$. Together these imply that $4P_n(x)<-P_n(-x)$.
\item If $P_n(x)=0$ then $S_n(x)=1$. If $P_n(x)\neq 0$, then we can write
\[
S_n(x)=1+\frac{2}{\left(\left(\frac{-P_n(-x)}{P_n(x)}\right)-1\right)}.
\]
Simple algebra and part (1) yield the desired result.
\item This follows from (2) and the fact that $S_n(x)$ is an odd function, i.e., $S_n(-x)=-S_n(x)$.
\end{enumerate}
\end{proof}

\begin{theorem}\label{plclosureunion}
$\PL$ is closed under union.
\end{theorem}
\begin{proof}
Fix two languages $D$ and $E$ in $\PL$. We will show that $D\cup E$ is also in $\PL$.

Let $f_D$ and $f_E$ be the functions given by Proposition \ref{characterizePL} for $D$ and $E$, respectively. By Lemma \ref{computationpathsupperbound}, let $n=n(|x|)$ be a polynomial such that $2^n$ bounds the maximum absolute value of $f_D(x)$ and $f_E(x)$. Define $A_D(x)=A_n(f_D(x))$. Similarly, define $B_D(x), S_D(x), A_E(x), B_E(x)$, and $S_E(x)$. Note that $A_D(x), B_D(x)$, $A_E(x)$, and $B_E(x)$ are $\GapL$ functions.

Let $H(x)=S_D(x)+S_E(x)+1$. By Lemma \ref{gaplfunctionsdefine}, we have
\begin{enumerate}
\item If $x\in D$ and $x\in E$ then $H(x)\geq 3$.
\item If $x\in D$ and $x\not\in E$ then $H(x)\geq \frac{1}{3}$.
\item If $x\not\in D$ and $x\in E$ then $H(x)\geq \frac{1}{3}$.
\item If $x\not\in D$ and $x\not\in E$ then $H(x)\leq -1$.
\end{enumerate}
Thus we have that $x\in D\cup E$ if and only if $H(x)>0$.

We would be finished if $H$ were a $\GapL$ function. However unfortunately it may even take nonintegral values. We do have 
\[
H(x)=\frac{A_D(x)}{B_D(x)}+\frac{A_E(x)}{B_E(x)}+1=\frac{A_D(x)B_E(x)+A_E(x)B_D(x)+B_D(x)B_E(x)}{B_D(x)B_E(x)}.
\]
Note that for nonzero integers $p$ and $q$ we have $p/q>0$ if and only if $pq>0$. Then we define 
\[
H'(x)=(A_D(x)B_E(x)+A_E(x)B_D(x)+B_D(x)B_E(x))(B_D(x)B_E(x)).
\]
We have that 
\begin{enumerate}
\item $H'(x)$ is $\GapL$ function.
\item For all $x\in\Sigma^*$, $H(x)>0$ if and only if $H'(x)>0$.
\item $x\in D\cup E$ if and only if $H'(x)>0$.
\end{enumerate}
Finally applying Proposition \ref{characterizePL} to $H'$, we have $D\cup E\in\PL$.
\end{proof}

\begin{corollary}\label{plclosureintersection}
$\PL$ is closed under intersection.
\end{corollary}
\section{$\boldPL$ is closed under $\boldPL$-Turing reductions}
Similar to Definition \ref{NL-oracle-class}, we define the complexity class $\PL^{\PL}$ as follows.
\begin{definition}\label{PL-oracle-class}
Let $\Sigma $ be the input alphabet. \textcolor{white}{\index[subject]{$\PL^{\rm \mbox{PL}}$}}We define $\PL^{\PL}$ to be the complexity class of all languages $L\subseteq\Sigma^*$ for which there exists a $O(\log n)$ space bounded non-deterministic Turing machine $M$ that has oracle access to a language $A\in\PL$ such that for any $x\in\Sigma^*$, the number of accepting computation paths of $M(x)\geq 2^{p(n)-1}$ if and only if $x\in L$, where $n=|x|$, $A\subseteq\Sigma^*$ and $2^{p(n)}$ is the number of accepting computation paths of $M$ on any input of size $n$ for some polynomial $p(n)$. Here we assume that $M$ submits queries to the oracle $A$ according to the Ruzzo-Simon-Tompa oracle access mechanism.
\end{definition}
In Theorem \ref{plTuringclosure} of this section, we show that $\PL$ is closed under $\PL$-Turing reductions\textcolor{white}{\index[subject]{PL-Turing reduction}}. In other words, we show that $\PL^{\PL}=\PL$.
\begin{definition}{\bf (Low-Degree Polynomials to Approximate the Sign Function)}
Let $m$ and $r$ be positive integers. Define:
\begin{eqnarray}
P_m(z) & = & (z-1)\prod_{1\leq i\leq m} (z-2^i)^2.\\
Q_m(z) & = & -P_m(z)-P_m(-z).\\
A_{m,r}(z) & = & (Q_m(z))^{2r}.\\
B_{m,r}(z) & = & (Q_m(z))^{2r}+(2P_m(z))^{2r}.\\
R_m,r(z) & = & \left(\frac{2P_m(z)}{Q_m(z)}\right)^{2r}.\\
S_{m,r}(z) & = & (1+R_{m,r}(z))^{-1}.
\end{eqnarray}
$R_{m,r}(z)$ and $S_{m,r}(z)$ are two auxiliary functions that will help us understand the properties of $A_{m,r}(z)$ and $B_{m,r}(z)$.
\end{definition}
\begin{lemma}\label{polynomiallemma1}
\begin{enumerate}
\item For all positive integers $m$ and $r$, $S_{m,r}(z)=\frac{A_{m,r}(z)}{B_{m,r}(z)}$.
\item For every positive integers $m$ and $r$, both $A_{m,r}(z)$ and $B_{m,r}(z)$ are polynomials in $z$ of degree $O(rm)$.
\item For all integers $m,r\geq 1$ and every integer $z$,
\begin{enumerate}
\item if $1\leq z\leq 2^m$, then $1-2^{-r}\leq S_{m,r}(z)\leq 1$, and 
\item if $-2^m\leq z\leq -1$, then $0\leq S_{m,r}(z)\leq 2^{-r}$.
\end{enumerate}
\end{enumerate}
\end{lemma}
\begin{proof}
The proofs of (1) and (2) are by routine calculation. We leave them to the reader. To prove (3), let $m$ and $r$ be positive integers. First consider the case when $1\leq z\leq 2^m$. In this case $P_m(z)\geq 0$ and $P_m(-z)<0$. We prove the following claim.
\begin{claim}\label{plclaim1}
If $1\leq z\leq 2^m$, then  $0\leq P_m(z)<-\frac{P_m(-z)}{9}$.
\end{claim}
\begin{cproof}
The claim clearly holds for $z=1$. So suppose that $1\leq z\leq 2^m$. There is a unique $i$, $1\leq i\leq m$, such that $2^i\leq z<2^{i+1}$. Let $t$ be that $i$. Then (i) $2^t\leq z$ and (ii) $z/2<2^t$. By combining (i) and (ii) we get $0\leq (z-2^t)<\frac{z}{2}$, and from (ii) we get $\frac{z}{2}<|-z-2^t|/3$. By combining the two inequalities, we have $(z-2^t)^2<\frac{(-z-2^t)^2}{9}$. Note that $z-1<z+1$ and $|z-2^i|\leq z+2^i$ for every $i$, $1\leq i\leq m$. Thus, in light of the definition of $P_m$, $P_m(z)<-\frac{P_m(-z)}{9}$.
\end{cproof}

Now by the above claim $0\leq P_m(z)<-\frac{P_m(-z)}{9}$. Combining this with $P_m(-z)\leq 0$ yields $Q_m(z)>-\frac{8P_m(-z)}{9}>0$. Thus
\[
0\leq R_{m,r}(z)<\left(\frac{2\left(\frac{-1}{9}\right)P_m(-z)}{-\frac{8}{9}P_m(-z)}\right)^{2r}=\left(\frac{1}{4}\right)^{2r}<2^{-r}.
\]
Since $R_{m,r}(z)\geq 0$ and since for ever $\delta\geq 0$, $(1+\delta )>0$ and $(1+\delta)(1\delta )=1-\delta^2\leq 1$, we have
\[
1\geq S_{m,r}(z)=\frac{1}{1+R_{m,r}(z)}>1-R_{m,r}(z)>1-2^{-r}.
\]
Hence (3a) holds.

Next consider the case when $-2^m\leq z\leq -1$. For this range of values of $z$, in light of Claim \ref{plclaim1} we have $0\leq P_m(-z)<-\frac{P_m(z)}{9}$. This implies $0<Q_{m,r}(z)<-P_m(z)$. Thus
\[
R_{m,r}(z)\geq \left(\frac{2P_m(z)}{-P_m(z)}\right)^{2r}=4^r>2^r.
\]
Hence (3b) holds.
\end{proof}

\begin{lemma}\label{polynomiallemma2}
For each $L\in\PL$ and each polynomial $r$, there exists $\GapL$ functions $g:\Sigma^*\rightarrow \N$ and $h:\Sigma^*\rightarrow\N$ such that, for all $x\in\Sigma^*$,
\begin{enumerate}
\item if $\chi_L(x)=b$, then $1-2^{-r(|x|)}\leq \frac{g(\langle x,b\rangle )}{h(x)}\leq 1$, and
\item if $\chi_L(x)\neq b$, then $0\leq \frac{g(\langle x,b\rangle )}{h(x)}\leq 2^{-r(|x|)}$.
\end{enumerate} 
\end{lemma}
\begin{proof}
Let $L\in\PL$ and $f$ be a $\GapL$ function witnessing, in the sense of Proposition \ref{characterizePL}, the membership of $L$ in $\PL$. Then, for every $x\in\Sigma^*$, the absolute value of $f(x)$ is at least one. Let $m$ be a polynomial such that, for every $x\in\Sigma^*$, the absolute value of $f(x)$ is at most $2^{m(|x|)}$. Let $r$ be an arbitrary polynomial. Define:
\begin{eqnarray*}
h(x) & = & B_{m(|x|),r(|x|)}(f(x)),\\
g(\langle x,1\rangle ) & = & A_{m(|x|),r(|x|)}(f(x)), and\\
g(\langle x,0\rangle ) & = & B_{m(|x|,r(|x|)}(f(x))-A_{m(|x|),r(|x|)}(f(x)).
\end{eqnarray*}
Then, for every $x\in\Sigma^*$, $g(\langle x,0\rangle )+g(\langle x,1\rangle )=h(x)$. For every $x\in\Sigma^*$, by Part 1 of Lemma \ref{polynomiallemma1}, we have $S_{m(|x|),r(|x|)}(f(x))=\frac{g(\langle x,1\rangle )}{h(x)}$ and since $g(\langle x,0\rangle )+g(\langle x,1\rangle )=h(x)$, $1-S_{m(|x|),r(|x|)}(f(x))=\frac{g(\langle x,0\rangle )}{h(x)}$. So, by Lemma \ref{polynomiallemma1}, it follows that $1-2^{-r(|x|)}\leq \frac{g(\langle x,1\rangle )}{h(x)}\leq 1$ if $f(x)>0$, and $2^{-r(|x|)}\geq \frac{g(\langle x,1\rangle )}{h(x)}\geq 0$ if $f(x)<0$. Since $\frac{g(\langle x,0\rangle )}{h(x)}=1-\frac{g(\langle x,1\rangle )}{h(x)}$, we have $1-2^{-r(|x|)}\leq \frac{g(\langle x,0\rangle )}{h(x)}\leq 1$ if $f(x)<0$, and $2^{-r(|x|)}\geq \frac{g(\langle x,0\rangle )}{h(x)}\geq 0$ if $f(x)>0$. Now it remains to prove that both $g$ and $h$ are in $\GapL$, which is easy because of the closure properties of $\GapL$ under addition and multiplication of up to polynomial length. We leave that verification as an exercise for the reader.
\end{proof}

\begin{theorem}\label{plTuringclosure}
$\PL^{\PL}=\PL$.\textcolor{white}{\index[subject]{PL$^{\rm \mbox{PL}}$}}
\end{theorem}
\begin{proof}
Let $L\in \PL^{\PL}$ such that $N$ is a non-deterministic $O(\log n)$-space bounded oracle Turing machine which obeys the Ruzzo-Simon-Tompa oracle access mechanism as described in Section \ref{chap1-NOTM} in Chapter \ref{chap01-introduction} and which makes oracle queries to a language $A\in\PL$ to correctly decide if an input string $x\in\Sigma^*$ is in $L$ or not, where $\Sigma$ is the input alphabet. We use $\Gamma_N^A(x)$ to denote the answer sequence that the oracle $A$ provides to $N$ on input $x$. Let $p$ be a polynomial bounding the running time of $N$. Let $q$ be a polynomial such that, for every $x\in\Sigma^*$,
\begin{itemize}
\item $acc_{N^A}(x)\leq 2^{q(|x|)}$, and
\item $x\in L\Leftrightarrow acc_{N^A}(x)\geq 2^{q(|x|)-1}$.
\end{itemize}
Let $m$ be a polynomially bounded function, as defined above, that maps each integer $n$ to the number of queries that $N$ makes on each input of length $n$. Then $m(n)\leq p(n)$ for all $n$. For each $x\in\Sigma^*$ and $w$, $|w|=m(|x|)$, let $\alpha (x,w)$ denote the number of accepting computation paths that $M$ on input $x$ would have if for every $i$, $1\leq i\leq m(|x|)$, the oracle answer to the $i^{th}$ query of $N$ on input $x$ is taken to be yes if the $i^{th}$ bit of $w$ is a $1$, and it is no if the $i^{th}$ bit of $w$ is a $0$. Then, for every $x\in\Sigma^*$, $\alpha (x,\Gamma_N^A(x))=\acc_{N^A}(x)$, and for every $w\in\Sigma^{m(|x|)}$, $\alpha (x,w)\leq 2^{q(|x|)}$. We now define a $\GapL$ function $d$ such that on any given input string $x\in\Sigma^*$, we have $x\in L$ if and only if $d(x)\geq 0$.

By Lemma \ref{polynomiallemma2}, there exists non-negative function $g\in\GapL$ and a strictly positive function $h\in\GapL$ such that, for all $x\in\Sigma^*$ and $b\in\{ 0,1\}$,
\begin{eqnarray*}
1-2^{-r(|x|)}\leq \frac{g(\langle x,b\rangle )}{h(x)}&\leq&1 {\rm ~if~}\chi_A(x)=b,~{\rm and}\\
0\leq \frac{g(\langle x,b\rangle )}{h(x)}&\leq&2^{-r(|x|)} {\rm ,~otherwise.}
\end{eqnarray*}

Define, for each $x\in\Sigma^*$,
\[
s(x)=\sum_{|w|=m(|x|)}\alpha (x,w)\prod_{1\leq i\leq m(|x|)}g(\langle y_{x,i},w_i\rangle )
\]
and
\[
t(x)=\prod_{1\leq i\leq m(|x|)}h(y_{x,i}).
\]
We claim that for every $x\in\Sigma^*$,
\[
x\in L\Leftrightarrow \frac{s(x)}{t(x)}\geq 2^{q(|x|)-1}-\frac{1}{4}.
\]
To prove the claim let $x\in\Sigma^*$. First suppose that $x\in L$. For $w=\Gamma_N^A$, the fraction
\[
\kappa (x,w)=\frac{\prod_{1\leq i\leq m(|x|)}g(\langle y_{x,i},w_i\rangle )}{t(x)}
\]
is at least
\[
1-m(|x|)2^{-min\{r(|y_{x,1}|),\ldots ,r(|y_{x,m(|x|)}|)\}}.
\]
Because $m$ is bounded by $p$, because $r$ is a natural polynomial, and because the machine $N$ is a length-increasing query generator, the above amount is at least
\[
1-p(|x|)2^{-p(|x|)-q(|x|)-1}>1-2^{-q(|x|)-1}.
\]
Note that in the summation expression of $s(x)$, there exists a computation path which corresponds to some $w$ such that the bits of $w$ are consistent with the sequence of oracle replies $\Gamma_N^A(x)$. For other every other $w$, we always add non-negative numbers to $s(x)$ irrespective of whether $w=\Gamma_N^A(x)$ or not. Moreover, $w$ is a non-deterministically chosen string of polynomial length, and so we get that $s(x)\in\GapL$. Thus the fraction $\frac{s(x)}{t(x)}$ is at least
\[
2^{q(|x|)-1}(1-2^{-q(|x|)-1})=2^{q(|x|)-1}-\frac{1}{4}.
\]
Next suppose that $x\not\in L$. For $w=\Gamma_N^A$, $\kappa(x,w)\leq 1$ and $\alpha (x,w)=\acc_{N^A}(x)\leq 2^{q(|x|)}-1$. For other $w$ of length $m(|x|)$, $\alpha (x,w)\leq 2^{q(|x|)}$ and 
\[
\kappa (x,w)\leq 2^{-min\{r(y_{x,1}),\ldots ,r(|y_{x,m(|x|)}|)\}}.
\]
Because $m$ is bounded by $p$, because $r$ is a natural polynomial, and because the machine $N$ is a length-increasing query generator, the above number is at most  $2^{-p(|x|)-q(|x|)-1}$. Since the number of $w$, $|w|=m(|x|)$, such that $w\neq \Gamma_N^A(x)$ is $2^{m(|x|)}-1<2^{p(|x|)}$, $\frac{s(x)}{t(x)}$ is less than
\[
2^{q(|x|)-1}-1+2^{p(|x|)}2^{-p(|x|)-q(|x|)-1}\\
=2^{q(|x|)-1}-1+2^{-q(|x|)-1}\\
\leq 2^{q(|x|)-1}-\frac{1}{2}.
\]
Thus the claim holds.

Now define
\[
d(x)=4s(x)-(2^{q(|x|+1}-1)t(x).
\]
Then, for every $x\in\Sigma^*$, $x\in L$ if and only if $d(x)\geq 0$. Clearly $d\in\GapL$ which proves our theorem.
\end{proof}

\subsection{An alternate proof of $\boldNL\subseteq\boldPL$}
\begin{theorem}\label{ceqlcontainedinpl}
$\CeqL\subseteq\PL$.
\end{theorem}
\begin{proof}
Let $L\subseteq\Sigma^*$ and let $L\in\CeqL$. Let $f\in\GapL$ be such that for any input $x\in\Sigma^*$ we have $x\in L$ if and only if $f(x)=0$. Let $L_1=\{x\in\Sigma^*|f(x)>0\}$ and let $L_2=\{x\in\Sigma^*|f(x)<0\}$. It is easy to note that $x\in L$ if and only if both of the following two conditions are false:

1. is $f(x)>0$, equivalently $x\in L_1$,

2. is $f(x)<0$, equivalently $x\in L_2$.

Clearly $L_1\in\PL$. Determining if $f(x)>0$ is therefore possible using a logarithmic space bounded deterministic Turing machine that has access to $L_1$ as an oracle. Next to determine if $f(x)<0$ let us consider the function $h(x)=-(2f(x)+1)$. Using Theorem \ref{plTuringclosure} it follows that $h\in\GapL$. We now observe that if $x\in L_2$ then $f(x)<0$ and so $h(x)>0$. Conversely, if $x\not\in L_2$ then $f(x)\geq 0$ and so $h(x)<0$. As a result we get $L_2\in\coPL$\textcolor{white}{\index[subject]{$\coPL$}}. Using Corollary \ref{plclosurecomplement} it follows that $L_2\in\PL$. Consider $L'=L_1\cup L_2$. Then it follows from Theorem \ref{plclosureunion} that $L'\in\PL$. Now a logarithmic space bounded deterministic Turing machine that has access to $L'$ as an oracle can therefore determine if the input $x\in\Sigma^*$ is such that $f(x)=0$. This shows that $\CeqL\subseteq\L^{\PL}$. However $\L^{\PL}\subseteq\PL^{\PL}=\PL$ by Theorem \ref{plTuringclosure} from which we get $\CeqL\subseteq\PL$.\textcolor{white}{\index[subject]{$\L^{\rm \mbox{PL}}$}}
\end{proof}

\section*{Notes}
The complexity class $\PL$ was defined by John Gill in \cite{Gil1977}. The definition of $\PL$ which we have stated in Definition \ref{pldefinition} is drastically different from the definition given by John Gill in \cite{Gil1977}. We state a definition of $\PL$ from \cite{AO1996} as follows: $\PL$ is defined to be a complexity class of languages $A$ for which there exists a probabilistic Turing machine (in the sense of \cite{Gil1977}; that is, a Turing machine with access to a source of unbiased random bits), such that on input $x$ the machine never uses more than $\log |x|$ space, and $x\in A$ if and only if the probability that the machine reaches an accepting configuration is greater than one half. As mentioned in \cite{AO1996}, the source of the difficulty in analyzing $\PL$ is because probabilistic logspace machines can perform useful work after exponentially many computation steps. However H. Jung in \cite{Jun1985} has shown that at least in the unbounded error model (which defined the class $\PL$), the polynomial time restriction causes no loss of power. In other words, H. Jung has shown that the probabilistic Turing machine associated with any language $A\in\PL$ can be assumed to be running in time which is polynomial in the size of the input. Eric Allender and Mitsunori Ogihara in \cite{AO1996} have given a simpler proof of this Jung's Theorem in \cite[Theorem 6]{AO1996}. Also our definition of $\PL$ stated in Definition \ref{pldefinition} is shown to be equivalent to the conventional definition of $\PL$ using Jung's Theorem by Eric Allender and Mitsunori Ogihara in \cite[Proposition 3]{AO1996}. As a result we use their proposition to define the complexity class $\PL$. \cite{AO1996} also show some more interesting closure properties of $\PL$ such as the closure of $\PL$ under logspace conjunctive truth-table reductions and logspace disjunctive truth-table reductions which use low degree polynomials defined by \cite{BRS1995}.

Results shown in Section \ref{PL-properties} that $\PL$\textcolor{white}{\index[subject]{PLH, Probabilistic Logarithmic Space Hierarchy}} is closed under complement, union and intersection are derived from the results shown in the polynomial time setting for the complexity class PP\textcolor{white}{\index[subject]{PP}} in \cite[Section 4.4]{For1997} by Lance Fortnow. The framework of low-degree polynomials to approximate the sign of a $\GapL$ function used by Mitsunori Ogihara in \cite{Ogi1998} which leads us to prove in Theorem \ref{plTuringclosure} that the complexity class $\PL$ is closed under $\PL$-Turing reductions\textcolor{white}{\index[subject]{$\PL$-Turing reduction}} is obtained from the polynomial time counting classes setting and it is due to Richard Beigel, Nick Reingold and Daniel Spielman\textcolor{white}{\index[authors]{Beigel, Richard}\index[authors]{Reingold, Nick}\index[authors]{Spielman, Daniel}\index[authors]{Allender, Eric}} in \cite{BRS1995}.\textcolor{white}{\index[authors]{Ogihara, Mitsunori}}\textcolor{white}{\index[subject]{C$_=$LH, Exact Counting Logspace Hierarchy}\index[authors]{Fortnow, Lance}\index[authors]{Jung, Hermann}\index[authors]{Gill, John}}

\chapter[Complete problems and Hierarchies]{Complete problems and Hierarchies}
In this chapter we list the set of problems that are logspace many-one complete for various logarithmic space bounded counting classes we have studied in Chapters \ref{chap01-introduction} to \ref{chapter04-PL}.
\section{Problems logspace many-one complete for $\boldNL$}
\noindent{\bf 2SAT (Satisfiability of 2-CNF Boolean formulae)}\newline
INSTANCE: Boolean formula $\phi$ in the 2-CNF.\newline
QUESTION: Is there an assignment of Boolean values for the variables of $\phi$ that is a satisfying assignment for $\phi$?\newline
REFERENCE: Theorem \ref{2SATisNLcomplete}\textcolor{white}{\index[subject]{2SAT}\index[subject]{2-CNF}}.\newline

\noindent{\bf DSTCON (Directed $st$-connectivity)}\newline
INSTANCE: Directed graph $G=(V,E)$ which is given in terms of its adjacency matrix, $s,t\in V$.\newline
QUESTION: Is there a directed path from $s$ to $t$ in $G$?\newline
REFERENCE: Theorem \ref{DSTCONisNLcomplete}.\newline

\noindent{\bf SLDAGSTCON (Simple Layered Directed Acyclic Graph $st$-connectivity)}\newline
INSTANCE: Directed graph $G=(V,E)$ which is given in terms of its adjacency matrix, $s,t\in V$ where the vertices in $G$ are arranged as a square matrix such that there are $n$ rows and every row has $n$ vertices. Any edge in this graph is from a vertex in the $i^{th}$ row to a vertex in the $(i+1)^{st}$ row, where $1\leq i\leq (n-1)$ and $n\geq 2$. Also $s$ is a vertex in the first row and $t$ is a vertex in the last row of $G$.\newline
QUESTION: Is there a directed path from $s$ to $t$ in $G$?\textcolor{white}{\index[subject]{SLDAGSTCON}}\newline
REFERENCE: Theorem \ref{sldagstconcomplete}.

\section{Problems logspace many-one complete for $\boldCeqL$}
\noindent{\bf ExactDSTCON}\newline
INSTANCE:\textcolor{white}{\index[subject]{ExactDSTCON}} Directed graph $G=(V,E)$ which is given in terms of its adjacency matrix, $s_1,t_1,s_2,t_2\in V$, where at least three vertices in $\{ s_1,s_2,t_1,t_2\}$ should be necessarily distinct.\newline
QUESTION: Is the number of directed paths from $s_1$ to $t_1$ equal to the number of directed paths from $s_2$ to $t_2$ in $G$?\newline
REFERENCE: Follows from Definition \ref{ceqldefinition} and the definition of GapDSTCON given below.\newline

\noindent{\bf ExactSLDAGSTCON}\newline
INSTANCE:\textcolor{white}{\index[subject]{ExactSLDAGSTCON}} Directed graph $G=(V,E)$ which is given in terms of its adjacency matrix, $s_1,t_1,s_2,t_2\in V$, where the vertices in $G$ are arranged as a square matrix such that there are $n$ rows and every row has $n$ vertices. Any edge in this graph is from a vertex in the $i^{th}$ row to a vertex in the $(i+1)^{st}$ row, where $1\leq i\leq (n-1)$ and $n\geq 2$. Also $s_1,s_2$ are vertices in the first row and $t_1,t_2$ are vertices in the last row of $G$ such that at least three vertices in $\{ s_1,s_2,t_1,t_2\}$ should be necessarily distinct.\textcolor{white}{\index[subject]{ExactSLDAGSTCON}}\newline
QUESTION: Is the number of directed paths from $s_1$ to $t_1$ equal to the number of directed paths from $s_2$ to $t_2$ in $G$?\newline
REFERENCE:  Follows from Definition \ref{ceqldefinition} and the definition of GapSLDAGSTCON given below.\newline

\section{Problems logspace many-one complete for $\boldsharpL$}
\noindent{\bf $\sharp$DSTCON (sharp Directed $st$-connectivity)}\newline
INSTANCE:\textcolor{white}{\index[subject]{$\sharp$DSTCON}} Directed graph $G=(V,E)$ which is given in terms of its adjacency matrix, $s,t\in V$.\textcolor{white}{\index[subject]{$\sharp$DSTCON}}\newline
QUESTION: Count the number of directed paths from $s$ to $t$ in $G$?\newline
REFERENCE: Proposition \ref{dstconsharplcomplete}.\newline

\noindent{\bf $\sharp$SLDAGSTCON (sharp Simple Layered Directed Acyclic Graph $st$-connectivity)}\newline
INSTANCE:\textcolor{white}{\index[subject]{$\sharp$SLDAGSTCON}} Directed graph $G=(V,E)$ which is given in terms of its adjacency matrix, $s,t\in V$, where the vertices in $G$ are arranged as a square matrix such that there are $n$ rows and every row has $n$ vertices. Any edge in this graph is from a vertex in the $i^{th}$ row to a vertex in the $(i+1)^{st}$ row, where $1\leq i\leq (n-1)$ and $n\geq 2$. Also $s$ is a vertex in the first row and $t$ is a vertex in the last row of $G$.\textcolor{white}{\index[subject]{$\sharp$SLDAGSTCON}}\newline
QUESTION: Count the number of directed paths from $s$ to $t$ in $G$?\newline
REFERENCE: Proposition \ref{sldagcountsharplcomplete}.

\section{Problems logspace many-one complete for $\boldGapL$}
\noindent{\bf GapDSTCON}\newline
INSTANCE:\textcolor{white}{\index[subject]{GapDSTCON}} Directed graph $G=(V,E)$ which is given in terms of its adjacency matrix, $s_1,t_1,s_2,t_2\in V$, where at least three vertices in $\{ s_1,s_2,t_1,t_2\}$ should be necessarily distinct.\textcolor{white}{\index[subject]{GapDSTCON}}\newline
QUESTION: Find the number of directed paths from $s_1$ to $t_1$ minus the number of directed paths from $s_2$ to $t_2$ in $G$?\newline
REFERENCE: Follows from $\sharp$DSTCON defined above and Lemma \ref{sharpLGapLconnect}.\newline

\noindent{\bf GapSLDAGSTCON}\newline
INSTANCE:\textcolor{white}{\index[subject]{GapSLDAGSTCON}} Directed graph $G=(V,E)$ which is given in terms of its adjacency matrix, $s_1,t_1,s_2,t_2\in V$, where the vertices in $G$ are arranged as a square matrix such that there are $n$ rows and every row has $n$ vertices. Any edge in this graph is from a vertex in the $i^{th}$ row to a vertex in the $(i+1)^{st}$ row, where $1\leq i\leq (n-1)$ and $n\geq 2$. Also $s_1,s_2$ are vertices in the first row and $t_1,t_2$ are vertices in the last row of $G$ such that at least three vertices in $\{ s_1,s_2,t_1,t_2\}$ should be necessarily distinct.\newline
QUESTION: Find the number of directed paths from $s_1$ to $t_1$ minus the number of directed paths from $s_2$ to $t_2$ in $G$?\newline
REFERENCE: Follows from $\sharp$SLDAGSTCON defined above and Lemma \ref{sharpLGapLconnect}.

\section{Problems logspace many-one complete for {\bf Mod$_{\bf\it k}$L}}
\noindent{\bf Mod$_kst$GapDSTCON}\textcolor{white}{\index[subject]{Mod$_kst$GapDSTCON}}\newline
INSTANCE:\textcolor{white}{\index[subject]{Mod$_kst$GapDSTCON}} Directed graph $G=(V,E)$ which is given in terms of its adjacency matrix, $s_1,t_1,s_2,t_2\in V$, where  at least three vertices in $\{ s_1,s_2,t_1,t_2\}$ should be necessarily distinct.\newline
QUESTION: Is the (number of directed paths from $s_1$ to $t_1$ minus the number of directed paths from $s_2$ to $t_2$ in $G$) not divisible by $k$, where $k\in\N$ and $k\geq 2$?\newline
REFERENCE: Follows from Definition \ref{ModkL}, $\sharp$DSTCON, Proposition \ref{sharplfolklore} and using elementary modulo arithmetic as in Lemma \ref{gapltosharpl}.\newline

\noindent{\bf Mod$_kst$GapSLDAGSTCON}\textcolor{white}{\index[subject]{Mod$_kst$GapSLDAGSTCON}}\newline
INSTANCE:\textcolor{white}{\index[subject]{Mod$_kst$GapSLDAGSTCON}} Directed graph $G=(V,E)$ which is given in terms of its adjacency matrix, $s_1,t_1,s_2,t_2\in V$, where the vertices in $G$ are arranged as a square matrix such that there are $n$ rows and every row has $n$ vertices. Any edge in this graph is from a vertex in the $i^{th}$ row to a vertex in the $(i+1)^{st}$ row, where $1\leq i\leq (n-1)$ and $n\geq 2$. Also $s_1,s_2$ are vertices in the first row and $t_1,t_2$ are vertices in the last row of $G$ such that at least three vertices in $\{ s_1,s_2,t_1,t_2\}$ should be necessarily distinct\newline
QUESTION: Is the (number of directed paths from $s_1$ to $t_1$ minus the number of directed paths from $s_2$ to $t_2$ in $G$) not divisible by $k$, where $k\in\N$ and $k\geq 2$?\newline
REFERENCE: Follows from Definition \ref{ModkL}, $\sharp$SLDAGSTCON, Proposition \ref{sharplfolklore} and using elementary modulo arithmetic as in Lemma \ref{gapltosharpl}.\newline

\noindent{\bf Mod$_kst$Path}\textcolor{white}{\index[subject]{Mod$_kst$Path}}\newline
INSTANCE\textcolor{white}{\index[subject]{Mod$_kst$Path}}: Directed graph $G=(V,E)$ which is given in terms of its adjacency matrix, $s,t\in V$.\newline
QUESTION: Is the number of directed paths from $s$ to $t$ in $G$ not divisible by $k$, where $k\in\N$ and $k\geq 2$?\newline
REFERENCE: Follows from Definition \ref{ModkL}, $\sharp$DSTCON and the proof of Theorem \ref{modklmodl}.\newline

\noindent{\bf Mod$_kst$SLDAGSTCON}\newline
INSTANCE:\textcolor{white}{\index[subject]{Mod$_kst$SLDAGSTCON}} Directed graph $G=(V,E)$ which is given in terms of its adjacency matrix, $s,t\in V$, where the vertices in $G$ are arranged as a square matrix such that there are $n$ rows and every row has $n$ vertices. Any edge in this graph is from a vertex in the $i^{th}$ row to a vertex in the $(i+1)^{st}$ row, where $1\leq i\leq (n-1)$ and $n\geq 2$. Also $s$ is a vertex in the first row and $t$ is a vertex in the last row of $G$.\newline
QUESTION:  Is the number of directed paths from $s$ to $t$ in $G$ not divisible by $k$, where $k\in\N$ and $k\geq 2$?\newline
REFERENCE: Follows from Definition \ref{ModkL}, $\sharp$SLDAGSTCON. Similar to the logspace many-one completeness of Mod$_kst$Path for $\ModkL$.

\section{Problems logspace many-one complete for $\boldModL$}
\noindent{\bf ModDSTCON}\newline
INSTANCE:\textcolor{white}{\index[subject]{ModDSTCON}} Directed graph $G=(V,E)$ which is given in terms of its adjacency matrix, $s,t\in V$.\newline
QUESTION: Is the number of directed paths from $s$ to $t$ in $G$ not divisible by $k$, where $k=|g(x)|$ for some $g\in\FL$ and $x=(G,s,t)$?\newline
REFERENCE: Follows from Definition \ref{modldefn}, $\sharp$DSTCON and Lemma \ref{gapltosharpl}.\newline

\noindent{\bf ModGapDSTCON}\newline
INSTANCE:\textcolor{white}{\index[subject]{ModGapDSTCON}} Directed graph $G=(V,E)$ which is given in terms of its adjacency matrix, $s_1,t_1,s_2,t_2\in V$, where at least three vertices in $\{ s_1,s_2,t_1,t_2\}$ should be necessarily distinct.\newline
QUESTION: Is the (number of directed paths from $s_1$ to $t_1$ minus the number of directed paths from $s_2$ to $t_2$ in $G$) not divisible by $k$, where $k=|g(x)|$ for some $g\in\FL$ and $x=(G,s,t)$?\newline
REFERENCE: Follows from Definition \ref{modldefn} and $\sharp$DSTCON.\newline

\noindent{\bf ModGapSLDAGSTCON}\newline
INSTANCE:\textcolor{white}{\index[subject]{ModGapSLDAGSTCON}} Directed graph $G=(V,E)$ which is given in terms of its adjacency matrix, $s_1,t_1,s_2,t_2\in V$, where the vertices in $G$ are arranged as a square matrix such that there are $n$ rows and every row has $n$ vertices. Any edge in this graph is from a vertex in the $i^{th}$ row to a vertex in the $(i+1)^{st}$ row, where $1\leq i\leq (n-1)$ and $n\geq 2$. Also $s_1,s_2$ are vertices in the first row and $t_1,t_2$ are vertices in the last row of $G$ such that at least three vertices in $\{ s_1,s_2,t_1,t_2\}$ should be necessarily distinct.\newline
QUESTION: Is the (number of directed paths from $s_1$ to $t_1$ minus the number of directed paths from $s_2$ to $t_2$ in $G$) not divisible by $k$, where $k=|g(x)|$ for some $g\in\FL$ and $x=(G,s,t)$?\newline
REFERENCE: Follows from Definition \ref{modldefn} and $\sharp$DSTCON.\newline

\noindent{\bf ModSLDAGSTCON}\newline
INSTANCE:\textcolor{white}{\index[subject]{ModSLDAGSTCON}} Directed graph $G=(V,E)$ which is given in terms of its adjacency matrix, $s,t\in V$, where the vertices in $G$ are arranged as a square matrix such that there are $n$ rows and every row has $n$ vertices. Any edge in this graph is from a vertex in the $i^{th}$ row to a vertex in the $(i+1)^{st}$ row, where $1\leq i\leq (n-1)$ and $n\geq 2$. Also $s$ is a vertex in the first row and $t$ is a vertex in the last row of $G$.\newline
QUESTION:  Is the number of directed paths from $s$ to $t$ in $G$ not divisible by $k$, where $k=|g(x)|$ for some $g\in\FL$ and $x=(G,s,t)$?\newline
REFERENCE: Follows from Definition \ref{modldefn}, $\sharp$SLDAGSTCON and Lemma \ref{gapltosharpl}.

\section{Problems logspace many-one complete for $\boldPL$}
\noindent{\bf ProbDSTCON}\newline
INSTANCE:\textcolor{white}{\index[subject]{ProbDSTCON}} Directed graph $G=(V,E)$ which is given in terms of its adjacency matrix, $s_1,t_1,s_2,t_2\in V$, where at least three vertices in $\{ s_1,s_2,t_1,t_2\}$ should be necessarily distinct.\newline
QUESTION: Is the number of directed paths from $s_1$ to $t_1$ minus the number of directed paths from $s_2$ to $t_2$ in $G$ greater than $0$?\newline
REFERENCE: Follows from Definition \ref{pldefinition}.\newline

\noindent{\bf ProbSLDAGSTCON}\newline
INSTANCE:\textcolor{white}{\index[subject]{ProbSLDAGSTCON}} Directed graph $G=(V,E)$ which is given in terms of its adjacency matrix, $s_1,t_1,s_2,t_2\in V$, where the vertices in $G$ are arranged as a square matrix such that there are $n$ rows and every row has $n$ vertices. Any edge in this graph is from a vertex in the $i^{th}$ row to a vertex in the $(i+1)^{st}$ row, where $1\leq i\leq (n-1)$ and $n\geq 2$. Also $s_1,s_2$ are vertices in the first row and $t_1,t_2$ are vertices in the last row of $G$ such that at least three vertices in $\{ s_1,s_2,t_1,t_2\}$ should be necessarily distinct.\newline
QUESTION: Is the number of directed paths from $s_1$ to $t_1$ minus the number of directed paths from $s_2$ to $t_2$ in $G$ greater than $0$?\newline
REFERENCE: Follows from Definition \ref{pldefinition}.

\section{Closure properties of logarithmic space bounded counting classes}
We summarize the most important closure properties of language based logarithmic space bounded counting classes shown in Chapters 2 to 4 as Table 5.1.
\begin{center}
\begin{table}[h!]
\begin{tabular}{||c||c|c|c|c|c||}\hline\hline
\emph{Complexity} &\multicolumn{5}{c||}{Closure property}\\\cline{2-6}
\emph{class of}& & & &logspace&$\mathcal{C}$-\\
\emph{languages}&union&complement&intersection&Turing&Turing\\
$\mathcal{C}$&$\cup$&$\overline{L}$&$\cap$&reduction&reduction\\\hline\hline
$\NL$ & \checkmark & \checkmark & \checkmark & \checkmark & \checkmark\\\hline
$\CeqL$ & \checkmark & unknown & unknown & unknown & unknown\\\hline
$\ModpL$, & & & & &\\
$p\in\N$ is a& \checkmark & \checkmark & \checkmark & \checkmark & \checkmark\\
prime.& & & & &\\\hline
$\ModkL$, & & & & &\\
$k\in\N$ is a& \checkmark & unknown & unknown & unknown & unknown\\
composite.& & & & &\\\hline
 & & \checkmark & & & \\
$\ModL$ & unknown & assuming & unknown & unknown & unknown\\
 & & $\NL =\UL$ & & &\\\hline
$\PL$ & \checkmark & \checkmark & \checkmark & \checkmark & \checkmark\\\hline\hline
\end{tabular}
\caption{Fundamental closure properties of logarithmic space bounded counting classes. Here, \checkmark $~$denotes that we have shown that property for the corresponding complexity class in Chapters 1 to 4.}\label{table:closureLBLSBCC}
\end{table}
\end{center}
\section[Logarithmic space bounded counting]{Logarithmic space bounded counting class hierarchies}\label{sec:defnLSBCCH}
In exploring the power and limitations of logarithmic space bounded counting classes, it is a practice to define logarithmic space bounded counting class hierarchies for every logarithmic space bounded counting class.
\begin{definition}\label{nlh-defn}
Let $\Sigma$ be the input alphabet and let $\NLH_1=\NL$. For $i\geq 2$, we define $\NLH_i$ as the class of all languages $L\subseteq \Sigma^*$ such that there exists a $O(\log n)$ space bounded non-deterministic oracle Turing machine $M^A$ that has access to a language $A\in\NLH_{i-1}$ as an oracle and we have $x\in L$ if and only if $\acc_{M^A}(x)>0$, where $n$ denotes the size of the input. Here we assume that $M^A$ submits queries to the oracle $A$ according to the Ruzzo-Simon-Tompa oracle access mechanism. The non-deterministic logspace hierarchy, denoted by $\NLH$, is defined as\textcolor{white}{\index[subject]{$\NLH$}}
\[
\NLH=\cup_{i\geq 1}\NLH_i
\]
\end{definition}
\begin{definition}\label{modplh-defn}
Let $\Sigma$ be the input alphabet and let $\ModpLH_1=\ModpL$, where $p\in\N$ and $p$ is a prime. For $i\geq 2$, we define $\ModpLH_i$ as the class of all languages $L\subseteq \Sigma^*$ such that there exists a $O(\log n)$ space bounded non-deterministic oracle Turing machine $M^A$ that has access to a language $A\in\ModpL$ as an oracle with $A\in\ModpLH_{i-1}$ and we have $x\in L$ if and only if $\acc_{M^A}(x)\not\equiv 0(\mod p)$, where $p\in\N$, $p$ is a prime and $n$ denotes the size of the input. Here we assume that $M^A$ submits queries to the oracle $A$ according to the Ruzzo-Simon-Tompa oracle access mechanism. The modulo-prime logspace hierarchy, denoted by $\ModpLH$ is defined as,\textcolor{white}{\index[subject]{$\ModpLH$}}
\[
\ModpLH=\cup_{i\geq 1}\ModpLH_i,
\]
where $p\in\N$ and $p$ is a prime.
\end{definition}
\begin{definition}\label{plh-defn}
Let $\Sigma$ be the input alphabet and let $\PLH_1=\PL$. For $i\geq 2$, we define $\PLH_i$ as the class of all languages $L\subseteq \Sigma^*$ such that there exists a $O(\log n)$ space bounded non-deterministic oracle Turing machine $M^A$ that has access to a language $A\in\PL$ as an oracle with $A\in\PLH_{i-1}$ and we have $x\in L$ if and only if $\acc_{M^A}(x)>0$, where $n$ denotes the size of the input. Here we assume that $M^A$ submits queries to the oracle $A$ according to the Ruzzo-Simon-Tompa oracle access mechanism. The probabilistic logspace hierarchy, denoted by $\PLH$ is defined as,\textcolor{white}{\index[subject]{$\PLH$}}
\[
\PLH=\cup_{i\geq 1}\PLH_i
\]
\end{definition}

\begin{definition}\label{modklh-defn}
Let $\Sigma$ be the input alphabet and let $\ModkLH_1=\ModkL$, where $k\in\N$ and $k\geq 2$. For $i\geq 2$, we define $\ModkLH_i$ as the class of all languages $L\subseteq\Sigma^*$ such that there exists a $O(\log n)$ space bounded non-deterministic oracle Turing machine $M^A$ that has access to a language $A\in\ModkLH_{i-1}$ and we have $\acc_{M^A}(x)\not\equiv 0(\mod k)$, where $k\in\N$, $k\geq 2$ and $n$ denotes the size of the input. Here we assume that $M^A$ submits queries to the oracle $A$ according to the Ruzzo-Simon-Tompa oracle access mechanism. The modulo-k logspace hierarchy, denoted by $\ModkLH$ is defined as,\textcolor{white}{\index[subject]{$\ModkLH$}}
\[
\ModkLH=\cup_{i\geq 1}\ModkLH_i,
\]
where $k\in\N$ and $k\geq 2$.
\end{definition}

\begin{theorem}\label{collapse-nlceqlmodplpl}
Let $\mathcal{C}$ be any of the following logarithmic space bounded counting classes: $\NL$ or $\ModpL$, where $p\in\N$ and $p$ is a prime, or $\PL$.
\[
\mathcal{C}{\rm H}=\cup_{i\geq 1}\mathcal{C}{\rm H}_i=\mathcal{C}.
\]
\end{theorem}
\begin{proof}
Let $\mathcal{C}$ be $\NL$ or $\CeqL$ or $\ModpL$, where $p\in\N$ and $p$ is a prime, or $\PL$. We know from the results shown in Chapters 1 to 3 which is summarized in Table \ref{table:closureLBLSBCC} that each of these complexity classes denoted by $\mathcal{C}$ is closed under $\mathcal{C}$-Turing reductions. It is trivial from the definition of $\mathcal{C}{\rm H}$ that $\mathcal{C}{\rm H}_1$=$\mathcal{C}$. We now use induction on $i\geq 1$ and assume that $\mathcal{C}{\rm H}_i=\mathcal{C}$ and consider $\mathcal{C}{\rm H}_{i+1}$. The remaining part of this proof easily follows from the definition of $\mathcal{C}{\rm H}$ using the result that $\mathcal{C}$ is closed under $\mathcal{C}$-Turing reductions and from our inductive assumption that $\mathcal{C}{\rm H}_i$=$\mathcal{C}$.
\end{proof}

Since it is not known whether $\ModkL$ is closed under the $\ModkL$-Turing reductions, where $k\in\N$ and $k\geq 2$ is a composite number, we cannot show that $\ModkLH =\ModkL$ in Theorem \ref{collapse-nlceqlmodplpl}.

\section[Hierarchies and Boolean circuits with oracle gates]{Hierarchies and Boolean circuits with oracle gates}\label{sec:defnBooleancktsLSBCC}
We recall the definition of standard unbounded fan-in basis from Definition \ref{chap1-unbfaninbasis-defn-booleanbasis}, admissible encoding scheme from Definition \ref{chap1-admissibleencoding}, and family of $\L$-uniform circuits from Definition \ref{chap1-luniformcircuitfamily}.
\begin{definition}\label{booleanoraclebasis}
We define $\mathcal{B}_1(\mathcal{C})=\{\neg ,(\wedge ^n)_{n\in\N},(\vee ^n)_{n\in\N},\mathcal{C}\}$ as the standard unbounded fan-in basis with oracle gates for $\mathcal{C}$, a complexity class of languages.\textcolor{white}{\index[subject]{$\mathcal{B}_1(\mathcal{C})$}}
\end{definition}
\begin{definition}\label{sdunbfaninoracle-defn}
Let $\mathcal{B}_1(\mathcal{C})$ be a standard unbounded fan-in basis with oracle gates for the complexity class of languages $\mathcal{C}$, and let $s,d:\N\rightarrow\N$. We define the complexity class, $\SD_{\mathcal{B}_1(\mathcal{C})}(s,d)$ as the class of all sets $A\subseteq \{ 0,1\}^*$ for which there is a circuit family $(C_n)_{n\in\N}$ over the basis $\mathcal{B}_1(\mathcal{C})$ of size $O(s)$ and depth $O(d)$ that accepts $A$.\textcolor{white}{\index[subject]{$\SD_{\mathcal{B}_1(\mathcal{C})}(s,d)$}}
\end{definition}
We recall the definition of ${\rm U_L}$-$\ACzero$ from Definition \ref{chap1-aczero-defn}.
\begin{definition}\label{aczeroC-defn}
Let $\mathcal{C}$ be any of the following logarithmic space bounded counting classes: $\NL$ or $\ModpL$, where $p\in\N$ and $p$ is a prime, or $\PL$. We define $\L$-uniform $\ACzero(\mathcal{C},i)$, denoted by ${\rm U_L}$-$\ACzero (\mathcal{C},i)$, as the class of all languages computable by circuits, $\L$-uniform $\SD_{\mathcal{B}_1(\mathcal{C})} (q(n),O(1))$ over the basis $\mathcal{B}_1(C)$ such that there are no more than $i$ oracle gates for $\mathcal{C}$ in any directed path from an input gate to an output gate in any circuit, where $q(n)$ is a polynomial in $n$ and $n$ is the size of the input. We define ${\rm U_L}$-$\ACzero (\mathcal{C})=\cup_{i\geq 0}{\rm U_L}$-$\ACzero(\mathcal{C},i)$.\textcolor{white}{\index[subject]{${\rm U_L}$-$\ACzero (\mathcal{C},i)$}\index[subject]{${\rm U_L}$-$\ACzero (\mathcal{C})$}}
\end{definition}
It is easy to note that ${\rm U_L}$-$\ACzero\subseteq$ ${\rm U_L}$-$\ACzero(\mathcal{C},0)\subseteq{\rm U_L}$-$\ACzero(\mathcal{C})$ and $\mathcal{C}\subseteq{\rm U_L}$-$\ACzero(\mathcal{C},1)\subseteq{\rm U_L}$-$\ACzero(\mathcal{C})$ follow from Definitions \ref{chap1-aczero-defn} and \ref{aczeroC-defn}.

\begin{theorem}\label{booleancktshierarchiescollapse}
Let $\mathcal{C}$ be any of the following logarithmic space bounded counting classes: $\NL$ or $\ModpL$, where $p\in\N$ and $p$ is a prime, or $\PL$. ${\rm U_L}$-$\ACzero (\mathcal{C})=\mathcal{C}{\rm H}$. In particular, for $i\geq 1$ we have ${\rm U_L}$-$\ACzero(\mathcal{C},i)=\mathcal{C}{\rm H}_i$.
\end{theorem}
\begin{proof}
We know from Theorem \ref{collapse-nlceqlmodplpl} that $\mathcal{C}{\rm H}=\mathcal{C}$. Also, it follows from Definition \ref{aczeroC-defn} that $\mathcal{C}\subseteq{\rm U_L}$-$\ACzero (\mathcal{C})$ which proves that $\mathcal{C}{\rm H}\subseteq {\rm U_L}$-$\ACzero (\mathcal{C})$.

Conversely, we know that ${\rm U_L}$-$\ACzero\subseteq\mathcal{C}$. Now, consider the class ${\rm U_L}$-$\ACzero(\mathcal{C},1)$. We want to show that ${\rm U_L}$-$\ACzero(\mathcal{C},1)\subseteq\mathcal{C}{\rm H}_1$.

Let $L\in{\rm U_L}$-$\ACzero(\mathcal{C},1)$ and let $(C_n)_{n\in\N}$ be a $\L$-uniform circuit family accepting $L$. Therefore the admissible encoding scheme $\overline{C_n}$ of $C_n$ is computable in space $O(\log n)$ when $1^n$ is given as an input. Clearly it follows from the Definition \ref{aczeroC-defn} that there is at most $1$ oracle gate $\mathcal{C}$ on any directed path from an input gate to an output gate in $C_n$, where $n\geq 1$.

It is easy to note that given an admissible encoding scheme $\overline{C_n}$ of $C_n$ we can assume without loss of generality that any two sub-circuits of $C_n$ with the $\mathcal{C}$ oracle gate as the root do not have any gate in common. This is easy to observe since $C_n$ is a constant depth circuit and a $O(\log n)$ space bounded deterministic Turing machine can obtain $\overline{C_n}$ when given $1^n$ as input and also make sufficiently many copies of Boolean gates and assign predecessors from the successive levels using $\overline{C_n}$ such that this property is preserved for all sub-circuits with the $\mathcal{C}$ oracle gate as the root. Using this assumption it follows that a $O(\log n)$ space bounded deterministic Turing machine can also compute inputs to the oracle gate and it writes the input on the oracle tape of a $\mathcal{C}$ oracle. We assume that the deterministic $O(\log n)$ Turing machine similarly submits all the queries to the $\mathcal{C}$ oracle that are possible and it also gets the membership replies from $\mathcal{C}$ oracle. Once again based on the $\L$-uniform circuit description, and using oracle replies it is possible for a $O(\log n)$ space bounded deterministic Turing machine to compute the output of $C_n$ on the given input. Since the entire computation of evaluating a ${\rm U_L}$-$\ACzero(\mathcal{C},1)$ circuit that decides if an input string of length $n$ is in $L$ is in $\L^{\mathcal{C}}$ and $\mathcal{C}$ is closed under logspace Turing reductions (Table \ref{table:closureLBLSBCC}), it follows that ${\rm U_L}$-$\ACzero (\mathcal{C},1)\subseteq\C$.

We now use induction on $i$ and assume for $i\geq 1$ that ${\rm U_L}$-$\ACzero (\mathcal{C},i)\subseteq\mathcal{C}{\rm H}_i$ and consider the class ${\rm U_L}$-$\ACzero (\mathcal{C},i+1)$. We want to show that ${\rm U_L}$-$\ACzero (\mathcal{C},i+1)\subseteq\mathcal{C}{\rm H}_{i+1}$. The proof for this step is similar to the base case. It uses the inductive assumption and the property of $\mathcal{C}$ that $\mathcal{C}{\rm H}_{i}=\mathcal{C}$ which follows from the collapse of $\mathcal{C}{\rm H}$ to $\mathcal{C}$ shown in Theorem \ref{collapse-nlceqlmodplpl}, which in turn is implied by the property of $\mathcal{C}$ that it is closed under $\mathcal{C}$-Turing reductions (Table \ref{table:closureLBLSBCC}). Thus we get the proof.
\end{proof}
\begin{corollary}
Let $\mathcal{C}$ be any of the following logarithmic space bounded counting classes: $\NL$ or $\ModpL$, where $p\in\N$ and $p$ is a prime, or $\PL$. ${\rm U_L}$-$\ACzero (\mathcal{C})=\mathcal{C}$.
\end{corollary}

\section*{Exercises}
\begin{enumerate}
\item Define the $\ModL$ hierarchy, denoted by $\ModLH$.
\item Define the $\sharpL$ hierarchy, denoted by $\sharpLH$ and show that $\ModLH =\sharpLH$.
\end{enumerate}

\section*{Notes}
\begin{figure}[h!]
\begin{center}
\includegraphics[scale=0.97]{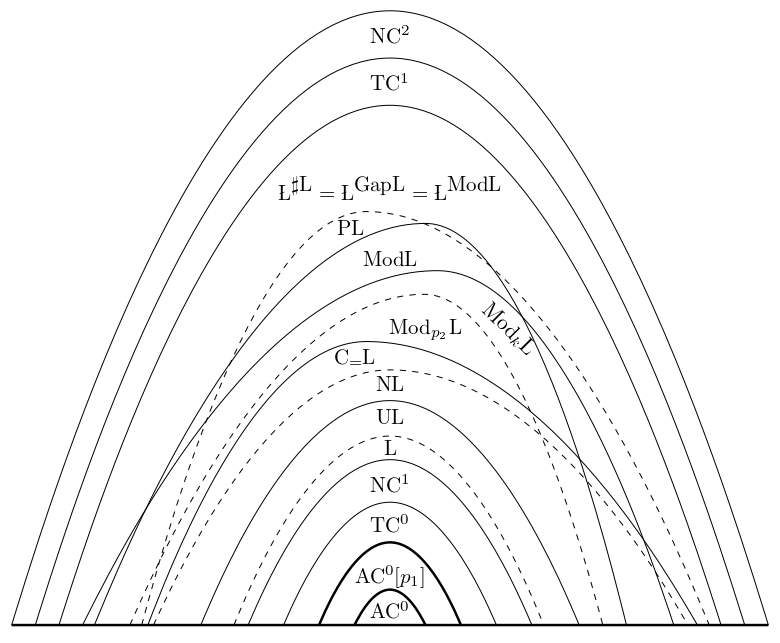}
\end{center}
\caption{\it The landscape of complexity classes contained in $\NCtwo$. In this figure, thick line between two complexity classes $\mathcal{C}_1$ and $\mathcal{C}_2$ denotes that $\mathcal{C}_1\subset\mathcal{C}_2$. Dashed line above a complexity class $\mathcal{C}$ denotes that $\mathcal{C}=$co-$\mathcal{C}$ (in fact $\mathcal{C}$ is closed under $\mathcal{C}$-Turing reductions). Here $p_1,p_2\geq 2$ are primes and $k\geq 6$ is a composite number that has more than one distinct prime divisor and $p_2|k$. We assume that the circuit complexity classes are $\L$-uniform circuit complexity classes.}\textcolor{white}{\index[subject]{NC$^2$}\index[subject]{TC$^1$}\index[subject]{L$^{\rm\mbox{ModL}}$}\index[subject]{L$^{\rm\mbox{GapL}}$}\index[subject]{$\PL$}\index[subject]{ModL}\index[subject]{Mod$_k$L, where $k\geq 2$ is an integer}\index[subject]{Mod$_{p_2}$L, where $p_2$ is a prime},\index[subject]{L, Logarithmic Space}\index[subject]{NC$^1$}\index[subject]{TC$^0$, threshold circuits of constant depth}\index[subject]{AC$^0$}\index[subject]{AC$^0$[$p_1$]}\index[subject]{L$^{\sharpL{\mbox{\rm}}}$}}
\end{figure}

In exploring the power and limitations of logarithmic space bounded counting classes, it is a routine exercise to define logarithmic space bounded counting class hierarchies for every logarithmic space bounded counting class that has been defined. These hierarchies are similar to the well known Polynomial Hierarchy (PH) which is defined based on the complexity class $\NP$. Our definitions and collapse results on hierarchies for $\CeqL$, $\PL$ and $\ModpL$ in Section \ref{sec:defnLSBCCH} is from \cite{AO1996,Ogi1998,ABO1999,HRV2000}. Theorem \ref{booleancktshierarchiescollapse} was shown for $\mathcal{C}=\PL$ and $\mathcal{C}=\CeqL$ in \cite{AO1996}. We note that using \cite[Theorems 2.18 and 2.32]{Voll1999} it is possible to show that we can evaluate ${\rm U_L}$-$\NCone$ circuits in $\L$.

\subsection*{Pause to ponder}
Now what happens if instead, we define the $\sharpL$ hierarchy or a complexity class of logspace uniform constant depth Boolean circuits that also contains oracle gates for $\sharpL$ functions? This has also been well studied by Eric Allender and Mitsunori Ogihara in \cite{AO1996}. Let us now define $\sharpLH$ which is based on functions in $\sharpL$ \cite{AO1996}:

Let $\Sigma$ be the input alphabet and let $\sharpLH_1=\sharpL$. For $i\geq 2$, we define $\sharpLH_i$ to be the class of all functions $f:\Sigma^*\rightarrow\Z ^+$ such that there exists a $O(\log n)$-space bounded non-deterministic oracle Turing machine $M^g$ that has access to a function $g:\Sigma^*\rightarrow\Z^+$ as an oracle with $g\in\sharpLH_{i-1}$ and $f(x)=\acc_{M^g}(x)$, where $n$ denotes the size of the input. Here we assume that $M^g$ submits queries to the oracle function $g$ according to the Ruzzo-Simon-Tompa oracle access mechanism. The counting logspace hierarchy, denoted by $\sharpLH$ is defined as,
\[
\sharpLH=\cup_{i\geq 1}\sharpLH_i
\]

An interesting question is, does the $\sharpLH$ collapse? Few points with regard to this question are as follows:
\begin{enumerate}
    \item In \cite{Coo1985}, Stephen A. Cook proves that the following four linear algebraic problems of computing the determinant of an integer matrix, computing the entries of the powers of an integer matrix, computing the entries of the iterated product of a set of integer matrices, and computing the entries of the inverse of an integer matrix are all $\L$-uniform $\NCone$-Turing equivalent (refer to \cite{Coo1985} for the definition of $\L$-uniform $\NCone$-Turing reduction).
    \item Eric Allender and Mitsunori Ogihara have shown that $\L$-uniform $\ACzero$ $(\sharpL)$ $=\sharpLH$ in \cite{AO1996}.
    \item Let us now assume that $\L$-uniform $\NCone (\sharpL)=\L$-uniform $\ACzero (\sharpL)$.
    \item As we will see in Corollary \ref{determinantgaplcomplete} in Chapter 6 of this monograph, it is well known that computing the determinant of integer matrices is logspace many-one complete for $\GapL$.\textcolor{white}{\index[authors]{Mahajan, Meena}\index[authors]{Vinay, V.}\index[authors]{Allender, Eric}\index[authors]{Ogihara, Mitsunori}\index[authors]{Cook, Stephen A.}}
    \item So shall we expect all the four problems introduced by S. A. Cook in \cite{Coo1985} to be logspace many-one equivalent which will imply that these four problems are also logspace many-one complete for $\GapL$? If so, then as a consequence of results shown by Eric Allender, it implies that $\L$-uniform $\NCone (\sharpL)=\GapL$ which will mean that $\sharpLH$ collapses to $\GapL$.
\end{enumerate}

\chapter[The complexity of computing the determinant]{The complexity of computing the determinant}
\section{Permutations}\label{basic-facts-permutations-matrices}
Let $\Omega =\{ 1,\ldots ,n\}$. A permutation on $\Omega$ is a one-to-one and onto mapping from $\Omega$ to itself. The set of all permutations of $\Omega$ is denoted by $S_n$, and it is easy to see that $|S_n|=n!$.\textcolor{white}{\index[subject]{permutation}}
\begin{proposition}
$S_n$ forms a group under the operation of composition of mappings, which we call as multiplication of permutations, and we say that $S_n$ is the symmetric group of degree $n$ on $\Omega$.
\end{proposition}
Given $\theta\in S_n$ and $s\in\Omega$, let $s\theta$ denote an element $s'\in\Omega$ such that $\theta (s)=s'$. Similarly, let $s\theta^i$ denote the element $s''\in\Omega$ such that $\theta^i (s)=s''$, where $i\geq 1$.
\begin{proposition}
If $\theta\in S_n$ and $s\in\Omega$, then there is the smallest positive integer $t$ depending on $\theta$ and $s$ such that $s\theta^t=s$.
\end{proposition}
\begin{definition}
Given $\theta$ and $s\in\Omega$ such that $s\theta^t=s$, let $x_i=s\theta^i$, where $1\leq i\leq t$. The set $\{ x_1,\ldots ,x_t\}$ is called the orbit of $s$ under $\theta$.\textcolor{white}{\index[subject]{orbit}}
\end{definition}
\begin{proposition}\label{prop-orbitsequaldisjoint}
If $\theta\in S_n$, $s,s'\in\Omega$ and $A$ is the orbit of $s$ under $\theta$ and $A'$ is the orbit of $s'$ under $\theta$, then either $A=A'$ or $A\cap A'=\emptyset$.
\end{proposition}
\begin{proof}
$s'\in A$ if and only if $s\in A'$. Therefore, we have either $A=A'$ or $A\cap A'=\emptyset$.
\end{proof}
\begin{definition}
Given $\theta$ and $s\in\Omega$ such that $s\theta^t=s$, let us consider the orbit $A=\{ x_1,x_2,\ldots ,x_t\}$ of $s$ under $\theta$, where $x_i=s\theta^i$. Let $\{ y_1,y_2,\ldots ,y_t\}$ be elements of $A$ arranged in the increasing order. We define the {\bf\textit{orbicycle}} containing $s$ under the permutation $\theta$ to be $( y_1\theta ~y_2\theta~\cdot~ \cdot ~\cdot ~y_t\theta )$.\textcolor{white}{\index[subject]{orbicycle}}
\end{definition}
\begin{lemma}\label{lemm-permutationuniorbicycles}
Every permutation $\theta\in S_n$ can be uniquely expressed as a product of its orbicycles.
\end{lemma}
\begin{proof}
Since orbicycles are formed from orbits, it is clear from Proposition \ref{prop-orbitsequaldisjoint} that given two orbicycles, either they are equal or there does not exist any $x\in\Omega$ which is in two different orbicycles.

If we consider an orbit $A=\{ x_1,x_2,\ldots ,x_t\}$ of $s$ under $\theta$, where $x_i=s\theta^i$, $1\leq i\leq t$, and its orbicycle $( y_1\theta ~y_2\theta~\cdot~ \cdot ~\cdot ~y_t\theta )$ as a permutation in $S_n$ then it permutes elements of $A$ and point-wise fixes every element of $\Omega$ which is not in $A$. As a result, considering orbicycles of $\theta$ as individual permutations, if we multiply all orbicycles of $\theta$ we get $\theta$. By re-ordering orbicycles of $\theta$ based on the least element in each orbicycle we get that every permutation $\theta\in S_n$ can be uniquely expressed as a product of its orbicycles.
\end{proof}
\begin{example}(Even permutation)\label{chap6-exampleorbicycleeven}
Let $\Omega=\{ 1,\ldots ,9\}$. Consider the permutation,
\[
\theta = \left (
\begin{array}{ccccccccc}
1 & 2 & 3 & 4 & 5 & 6 & 7 & 8 & 9\\
5 & 4 & 8 & 7 & 6 & 1 & 3 & 2 & 9
\end{array}
\right ).
\]
Orbit of $1$ under $\theta=\{ 5,6,1\}$, orbit of $2$ under $\theta=\{ 4,7,3,8,2\}$ and orbit of $9$ under $\theta=\{ 9\}$. Orbicycle containing $1$ under $\theta=(5~6~1)$, orbicycle containing $2$ under $\theta=(4~8~7~3~2)$ and orbicycle containing $9$ under $\theta=(9)$. It is easy to see that $\theta =(5~6~1)(4~8~7~3~2)(9)$.
\end{example}
\begin{example}(Odd permutation)\label{chap6-exampleorbicycleodd}
Let $\Omega=\{ 1,\ldots ,9\}$. Consider the permutation,
\[
\theta '= \left (
\begin{array}{ccccccccc}
1 & 2 & 3 & 4 & 5 & 6 & 7 & 8 & 9\\
1 & 8 & 7 & 4 & 6 & 2 & 5 & 3 & 9
\end{array}
\right ).
\]
Orbit of $1$ under $\theta=\{ 1\}$, orbit of $2$ under $\theta=\{ 8,3,7,5,6,2\}$, orbit of $4$ under $\theta=\{ 4\}$ and orbit of $9$ under $\theta=\{ 9\}$. Orbicycle containing $1$ under $\theta=(1)$, orbicycle containing $2$ under $\theta=(8~7~6~2~5~3)$, orbicycle containing $4$ under $\theta=(4)$ and orbicycle containing $9$ under $\theta=(9)$. It is easy to see that $\theta '=(1)(8~7~6~2~5~3)(4)(9)$.
\end{example}

\begin{definition}\label{defn-2orbicycle}
We say that an orbicycle is a {\bf$\mathit{\mathbf{2}}$\textit{-orbicycle}} if $s\theta\neq s$, $s\theta^2=s$ for some $s\in\Omega$, and $s'\theta =s'$, for all $s'\in\Omega$ such that $s\neq s'$ and $s\theta\neq s'$.\textcolor{white}{\index[subject]{$2$-orbicycle}}
\end{definition}
\begin{theorem}\label{thm-permutation2orbicycles}
Every orbicycle can be uniquely expressed as a product of $2$-orbicycles. Every permutation can be uniquely expressed as a product of $2$-orbicycles.
\end{theorem}
\begin{proof}
Consider an orbit $A=\{ x_1,x_2,\ldots ,x_t\}$ of $s$ under $\theta$, where $x_i=s\theta^i$, $1\leq i\leq t$, and its orbicycle $( y_1\theta ~y_2\theta~\cdot~ \cdot ~\cdot ~y_t\theta )$. Then, $\theta$ is the product of $2$-orbicycles $(y_1\theta ~y_1)$ $(y_1\theta^2 ~y_1)~\cdot ~\cdot ~\cdot ~(y_1\theta^{t-1} ~y_1)$ in this order.

Since orbits of any two elements in $\Omega$ is either the same or both are disjoint, using Lemma \ref{lemm-permutationuniorbicycles}, it is easy to see that every permutation can be uniquely expressed as a product of $2$-orbicycles, maintaining their order.
\end{proof}
\begin{corollary}\label{cor-permutation2orbicycles}
Every orbicycle can be uniquely expressed as a product of $(l-1)$ $2$-orbicycles, where $l$ is the number of elements of $\Omega$ in the orbit of the orbicycle. Every permutation $\theta$ can be uniquely expressed as a product of $\sum_{i=1}^m(l_i-1)$ $2$-orbicycles, where $l_i$ denotes the number of elements in the orbit of the $i^{th}$ orbicycle, and $m$ denotes the number of orbicycles of $\theta$.
\end{corollary}

\begin{definition}
A permutation is said to be an even permutation if it is a product of an even number of $2$-orbicycles.\textcolor{white}{\index[subject]{even permutation}}
\end{definition}
\begin{definition}
A permutation is an odd permutation if it is not an even permutation.\textcolor{white}{\index[subject]{odd permutation}}
\end{definition}
\begin{lemma}\label{lemm-eveniseven}
A permutation can be expressed as the product of an even number of $2$-orbicycles if and only if it cannot be expressed as the product of an odd number of $2$-orbicycles. 
\end{lemma}
\begin{proof}
Consider the polynomial in $n$ variables
\[
p(x_1,\ldots ,x_n)=\prod_{i<j}(x_i-x_j).
\]
If $\theta\in S_n$, let $\theta$ act on the polynomial $p(x_1,\ldots ,x_n)$ by
\[
\theta :p(x_1,\ldots ,x_n)=\prod_{i<j}(x_i-x_j)\rightarrow\prod_{i<j}(x_{\theta (i)}-x_{\theta (j)}).
\]
It is clear that $\theta :p(x_1,\ldots ,x_n)\rightarrow\pm p(x_1,\ldots ,x_n)$. For instance, in $S_5$, if $\theta =(3~4~1)(5~2)$, then $\theta$ takes
\begin{eqnarray*}
p(x_1,\ldots ,x_5) & = & (x_1-x_2)(x_1-x_3)(x_1-x_4)(x_1-x_5)\\
 & & \times (x_2-x_3)(x_2-x_4)(x_2-x_5)\\
 & & \times (x_3-x_4)(x_3-x_5)\\
 & & \times (x_4-x_5)
\end{eqnarray*}
into 
\begin{eqnarray*}
(x_3-x_5)(x_3-x_4)(x_3-x_1)(x_3-x_2)(x_5-x_4)(x_5-x_1)(x_5-x_2)(x_4-x_1)\\
\times (x_4-x_2)(x_1-x_2),
\end{eqnarray*}
which can be easily verified to be $-p(x_1,\ldots ,x_n)$.

In particular, if $\theta$ is a $2$-orbicycle, then $\theta :p(x_1,\ldots ,x_n)\rightarrow -p(x_1,\ldots ,x_n)$. Extending this observation, since a permutation which is a $2$-orbicycle is an odd permutation, if $\theta$ is an odd permutation, then $\theta :p(x_1,\ldots ,x_n)\rightarrow -p(x_1,\ldots ,x_n)$.

Thus, if a permutation $\theta$ can be written as a product of an even number of $2$-orbicycles, then $\theta$ must leave $p(x_1,\ldots ,x_n)$ fixed. So any other way of writing $\theta$ as a product of $2$-orbicycles must be such that it leaves $p(x_1,\ldots ,x_n)$ fixed; which implies that $\theta$ is always a product of an even number of $2$-orbicycles.
\end{proof}

Based on Lemma \ref{lemm-eveniseven} we obtain the following.
\begin{lemma}\label{lemma-evenoddrules}
\begin{enumerate}
\item The product of two even permutations is an even permutation.
\item The product of an even permutation and an odd permutation is an odd permutation.
\item The product of two odd permutations is an even permutation.
\end{enumerate}
\end{lemma}
\begin{example}\label{chap6-exampleorbicycleevencont1}
Let $\Omega=\{ 1,\ldots ,9\}$ and let $\theta$ be the permutation in Example \ref{chap6-exampleorbicycleeven}. It is easy to see that the orbicycle containing $1$ under $\theta=(5~6~1)=(5~1)(6~1)$ and therefore this orbicycle is an even permutation. Similarly, the orbicycle containing $2$ under $\theta=(4~8~7~3~2)=(4~2)(7~2)(3~2)(8~2)$ and therefore this orbicycle is also an even permutation. The orbicycle containing $9$ under $\theta$ has zero $2$-orbicycles and therefore this orbicycle is also an even permutation. It is easy to see that $\theta =(5~1)(6~1)(4~2)(7~2)(3~2)(8~2)$. Since $\theta$ is a product of only even permutations it follows that $\theta$ is also an even permutation.
\end{example}
\begin{example}\label{chap6-exampleorbicycleoddcont1}
Let $\Omega=\{ 1,\ldots ,9\}$ and let $\theta '$ be the permutation in Example \ref{chap6-exampleorbicycleodd}. It is easy to see that the orbicycle containing $1$ under $\theta$ has zero $2$-orbicycles and so this orbicycle is an even permutation. Similarly the orbicycle containing $4$ is an even permutation and the orbicycle containing $9$ is also an even permutation. However the orbicycle containing $2$ under $\theta '=(8~7~6~2~5~3)=(8~2)(3~2)(7~2)(5~2)(6~2)$ which means that this orbicycle is an odd permutation. It is easy to see that $\theta '=(8~2)(3~2)(7~2)(5~2)(6~2)$. Since in writing $\theta '$ as a product of its orbicycles, we have odd number of odd permutations it follows that, $\theta '$ is an odd permutation.
\end{example}

\begin{theorem}
Let $A_n$ denote the set of all even permutations in $S_n$. $A_n$ is a normal subgroup of $S_n$ of index $2$ under multiplication of permutations, and it is called as the \emph{alternating group} of degree $n$.\textcolor{white}{\index[subject]{alternating group of degree $n$}}
\end{theorem}
\begin{proof}
$A_n$ is a group since the product of two even permutations is an even permutation.

Let $W$ be the group of real numbers $-1$ and $1$ under multiplication. Define $\psi :S_n\rightarrow W$ by $\psi (\theta )=1$ if $\theta$ is an even permutation, $\psi (\theta )=-1$ if $\theta$ is an odd permutation. Using Lemma \ref{lemma-evenoddrules} it is easy to see that $\psi$ is a homomorphism onto $W$. The kernel of $\psi$ is precisely $A_n$. Therefore $A_n$ is a normal subgroup of $S_n$. As a consequence, $\left (\frac{S_n}{A_n}\right )\approx W$ which implies, $o(W)=2=o\left (\frac{S_n}{A_n}\right )=\frac{o(S_n)}{o(A_n)}$. This completes the proof.
\end{proof}
\begin{definition}\label{defn-signpermutation}
Given a permutation $\theta\in S_n$, let $p$ denote the number of $2$-orbicycles in $\theta$. We define the sign of $\theta$, denoted by $sgn(\theta )$, as $(-1)^p$.\textcolor{white}{\index[subject]{sign of a permutation, denoted by $sgn(\theta)$}}
\end{definition}
\begin{proposition}
Let $\theta$ be an even permutation. $sgn(\theta )=1$. Let $\sigma$ be an odd permutation. $sgn(\sigma )=-1$.
\end{proposition}

\begin{definition}\label{defn-orbicycleinversion}
Given an orbicycle $( y_1\theta ~y_2\theta~\cdot~ \cdot ~\cdot ~y_t\theta )$ of a permutation $\theta$ on $\Omega =\{1,\ldots ,n\}$, we say that a pair $(y_i\theta, y_j\theta )$ is an inversion if $y_i\theta >y_j\theta$ whenever $y_i<y_j$. The number of inversions in $\theta$ is equal to the sum of the number of inversions in its orbicycles.\textcolor{white}{\index[subject]{inversion in an orbicycle}}
\end{definition}

\begin{theorem}
\begin{enumerate}
\item\label{sign-inversion} Given a permutation $\theta\in S_n$, let $i$ denote the number of inversions in $\theta$. $sgn(\theta )=(-1)^i$.
\item\label{sign-determinant} Given a permutation $\theta\in S_n$, $sgn(\theta )$ is equal to the determinant of the permutation matrix of $\theta$. 
\end{enumerate}
\end{theorem}
\begin{proof}
(\ref{sign-inversion}) Consider the proof of Lemma \ref{lemm-eveniseven} and $p(x_1,\ldots ,x_n)$ defined therein. Clearly, if a permutation $\theta$ is an even permutation, since it is a product of an even number of $2$-orbicycles it leaves $p(x_1,\ldots ,x_n)$ fixed. Otherwise if $\theta$ is an odd permutation, then $\theta :p(x_1,\ldots ,x_n)\rightarrow -p(x_1,\ldots ,x_n)$. In other words, if $\theta$ is an even permutation, then $\theta$ contains even number of inversions among elements of $\Omega$ when it is written as a product of its orbicycles due to which $\theta$ maps $p(x_1,\ldots ,x_n)$ to itself. On the other hand, if $\theta$ is an odd permutation, then $\theta$ contains odd number of inversions among elements of $\Omega$ when it is written as a product of its orbicycles due to which it maps $p(x_1,\ldots ,x_n)$ to $-p(x_1,\ldots ,x_n)$. This shows that $sgn(\theta )=(-1)^{i({\rm mod}~2)}=(-1)^i$, where $i$ is the number of inversions in $\theta$.

(\ref{sign-determinant}) Follows from the definition of the determinant of a matrix in equation (\ref{determinant-definition}) given below, and since the matrix is only a permutation matrix whose entries are only $0$ and $1$.
\end{proof}
\begin{example}
In continuation with Example \ref{chap6-exampleorbicycleevencont1}, $sgn(\theta )=1$. Also the number of inversions in $\theta =(2+8)=10$, which is an even number. It is easy to see that $sgn(\theta )=(-1)^{10}=1$. If we consider the permutation matrix of $\theta$, it is

\[
A=\left(
\begin{array}{ccccccccc}
0 & 0 & 0 & 0 & 1 & 0 & 0 & 0 & 0\\
0 & 0 & 0 & 1 & 0 & 0 & 0 & 0 & 0\\
0 & 0 & 0 & 0 & 0 & 0 & 0 & 1 & 0\\
0 & 0 & 0 & 0 & 0 & 0 & 1 & 0 & 0\\
0 & 0 & 0 & 0 & 0 & 1 & 0 & 0 & 0\\
1 & 0 & 0 & 0 & 0 & 0 & 0 & 0 & 0\\
0 & 0 & 1 & 0 & 0 & 0 & 0 & 0 & 0\\
0 & 1 & 0 & 0 & 0 & 0 & 0 & 0 & 0\\
0 & 0 & 0 & 0 & 0 & 0 & 0 & 0 & 1\\
\end{array}
\right ).
\]
Clearly, $\det (A)=1$.
\end{example}
\begin{example}
In continuation with Example \ref{chap6-exampleorbicycleoddcont1}, $sgn(\theta ')=-1$. Also the number of inversions in $\theta '=13$, which is an odd number. It is easy to see that $sgn(\theta ')=(-1)^{13}=-1$. If we consider the permutation matrix of $\theta '$, it is

\[
A'=\left(
\begin{array}{ccccccccc}
1 & 0 & 0 & 0 & 0 & 0 & 0 & 0 & 0\\
0 & 0 & 0 & 0 & 0 & 0 & 0 & 1 & 0\\
0 & 0 & 0 & 0 & 0 & 0 & 1 & 0 & 0\\
0 & 0 & 0 & 1 & 0 & 0 & 0 & 0 & 0\\
0 & 0 & 0 & 0 & 0 & 1 & 0 & 0 & 0\\
0 & 1 & 0 & 0 & 0 & 0 & 0 & 0 & 0\\
0 & 0 & 0 & 0 & 1 & 0 & 0 & 0 & 0\\
0 & 0 & 1 & 0 & 0 & 0 & 0 & 0 & 0\\
0 & 0 & 0 & 0 & 0 & 0 & 0 & 0 & 1\\
\end{array}
\right ).
\]
Clearly, $\det (A')=-1$.
\end{example}

\section{Mahajan-Vinay's Theorems on the Determinant}\label{mahajan-vinay1997}
We recall the definition of the determinant of a matrix. Let $A\in\R ^{n\times n}$. We define the \emph{determinant} of $A$ as:
\begin{eqnarray}
\det(A) & = & \sum_{\sigma\in S_n}sgn(\sigma )\prod_{i=1}^na_{i,\sigma (i)}.\textcolor{white}{\index[subject]{determinant}}\label{determinant-definition}
\end{eqnarray}
\begin{tcolorbox}[colback=gray!35!white,colframe=white]
{\bf\emph{In this chapter, given a matrix $A\in\R^{n\times n}$, we view $A$ as a directed graph such that the weight of the directed edge $(i,j)$ is equal to the entry $A(i,j)$, where $1\leq i,j\leq n$. If $A(i,j)=0$, then the directed edge $(i,j)$ does not exist in the directed graph.}}
\end{tcolorbox}

\begin{proposition}\label{removeselfloops}
Let $A\in\R^{n\times n}$ and let
\[
B=\left [\begin{array}{cc}
0 & -A\\
I_n & 0
\end{array}\right ].
\]
Then, $\det (A)=\det (B)$.
\end{proposition}

\begin{figure}[h!]
\begin{center}
\includegraphics[scale=0.9]{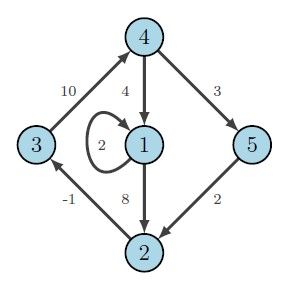}
\caption{This figure is a directed graph $G_1=(V_1,E_1)$. The following are clow sequences of $G_1$: $\mathcal{C}_1=((1,2,3,4),(2,3,4,5))$, $\mathcal{C}_2=(1,2,3,4,5,2,3,4)$, $\mathcal{C}_3=((1),(2,3,4,5))$. Note that $\mathcal{C}_3$ is a cycle cover of $G_1$. It is easy to note the following facts about clows of these clow sequences in $G_1$: $l((1,2,3,4))=l((2,3,4,5))=4$, $l((1,2,3,4,5,2,3,4))=9$ and $l((1))=1$. Here $w(\mathcal{C}_1)=w(\mathcal{C}_2)=19200$ and $w(\mathcal{C}_3)=-120$. Since $\mathcal{C}_3$ is a cycle cover we can obtain from $\mathcal{C}_3$, the following orbicycle cover permutation denoted by $\sigma$, of the vertices of $G_1$ based on the traversal of vertices of the individual cycles of $\mathcal{C}_3$: $(1)(3~4~5~2)$. It is easy to note that if $\sigma$ is written as a product of $2$-orbicycles then $\sigma =(3~2)(4~2)(5~2)$. Since there are an odd number of $2$-orbicycles it follows that $sgn(\mathcal{C}_3)=-1$. Clearly there are $3$ inversions in $\sigma$ $(3>2, 4>2, 5>2)$. So we once again infer that $sgn(\sigma )=-1$. Also we note the permutation matrix corresponding to $\sigma$ has determinant is $-1$. We get $clowsgn(\mathcal{C}_1)=1$ while $clowsgn(\mathcal{C}_2)=-1$ and $clowsgn(\mathcal{C}_3)=-1$. The determinant of the adjacency matrix of $G_1$=120. Note that $G_1$ has infinitely many clow sequences, each of which can be obtained by starting from the vertex $1$ and iteratively including edges from the cycle $(2,3,4,5)$ without ending a clow using the edge $(4,1)$.}\label{fig_chap6_1}
\end{center}
\end{figure}

Note that if $B\in \R^{2n\times 2n}$ in Proposition \ref{removeselfloops} is viewed as a directed graph, then there does not exist self-loop on any vertex in the directed graph represented by $B$. We need this important property to prove our results.
\begin{definition}\label{defn-cycle}
Let $G$ be a directed graph on $n$ vertices. A cycle is a sequence of $k$ distinct vertices $C=(v_1,v_2,\ldots ,v_k)$ such that $(v_i,v_{i+1}), (v_k,v_1)\in E(G)$, where $1\leq i\leq k-1$ and $k\leq n$.\textcolor{white}{\index[subject]{cycle}}
\end{definition}
\begin{definition}\label{defn-cyclelength}
Let $G$ be a directed graph on $n$ vertices. Given a cycle $C=(v_1,v_2,\ldots ,v_k)$ in $G$, we define the length of $C$ as $k$ and it is denoted by $l(C)$.\textcolor{white}{\index[subject]{length of a cycle}}
\end{definition}
\begin{definition}
Let $G$ be a weighted directed graph on $n$ vertices such that $w:E(G)\rightarrow \R$ is a weight function on the edges of $G$. Given a cycle $C=(v_1,v_2,\ldots ,v_k)$ in $G$, we define the weight of the cycle as $w(C)=\prod_{i=1}^k w(e_i)$, where $e_i=(v_i,v_{i+1})$, for $1\leq i\leq k-1$ and $e_k=(v_k,v_1)$.\textcolor{white}{\index[subject]{weight of a cycle}}
\end{definition}
\begin{definition}\label{defn-cyclehead}
Let $G$ be a directed graph on $n$ vertices. Given a cycle $C=(v_1,v_2,\ldots ,v_k)$ in $G$ such that $v_1\leq v_i$ for all $1\leq i\leq k$, we say that $v_1$ is the head of the cycle $C$ and it is denoted by $h(C)$.\textcolor{white}{\index[subject]{head of a cycle}}
\end{definition}
\begin{definition}\label{defn-clow}
Let $G=(V,E)$ be a directed graph on $n$ vertices. A walk $W=(v_1,v_2,\ldots ,v_l)$ in $G$ is a clow (abbreviation for ``closed-walk") starting from the least numbered vertex $v_1$ and ending at the same vertex such that there is exactly one incoming edge $(v_l,v_1)\in E(G)$ for $v_1$ in the clow and exactly one outgoing edge $(v_1,v_2)\in E(G)$ for $v_1$ in the clow, and $(v_i,v_{i+1})\in E(G)$, where $1\leq i\leq k-1$.\textcolor{white}{\index[subject]{clow}}
\end{definition}
\begin{definition}
Let $G=(V,E)$ be a directed graph on $n$ vertices and let $W=(v_1,v_2,\ldots ,v_l)$ be a clow in $G$. We define the length of the clow $W$ as the number of edges in the clow, and it is denoted by $l(W)$.\textcolor{white}{\index[subject]{length of a clow}}
\end{definition}
\begin{definition}\label{defn-clowweight}
Let $G$ be a weighted directed graph on $n$ vertices such that $w:E(G)\rightarrow \R$ is a weight function on the edges of $G$. Given a clow $W=(v_1,v_2,\ldots ,v_k)$ in $G$, we define the weight of the clow as $w(W)=\prod_{i=1}^k w(e_i)$, where $e_i=(v_i,v_{i+1})$, for $1\leq i\leq k-1$ and $e_k=(v_k,v_1)$.\textcolor{white}{\index[subject]{weight of a clow}}
\end{definition}
\begin{definition}
Let $G=(V,E)$ be a directed graph on $n$ vertices and let $W=(v_1,v_2,\ldots ,v_l)$ be a clow in $G$ such that the vertex $v_1$ is the least numbered vertex in $W$. We define $v_1$ as the head of the clow $W$ and it is denoted by $h(W)$.\textcolor{white}{\index[subject]{head of a clow}}
\end{definition}
In a directed graph, it is easy to see that every cycle is a clow. 
\begin{proposition}
Let $G=(V,E)$ be a directed graph on $n$ vertices and let $W=(v_1,v_2,\ldots ,v_l)$ be a clow in $G$. Then $v_1$ occurs exactly once in the clow.
\end{proposition}
\begin{definition}\label{defn-clowsequence}
Let $G=(V,E)$ be a directed graph. A clow sequence $\mathcal{W}$ of $G$ is a sequence of clows $\mathcal{W}=(C_1,\ldots ,C_k)$ such that the sequence is ordered based on heads of clows, that is head($C_1)<$ head($C_2)<\cdots<$ head($C_k)$, and every vertex of $G$ is in at least one $C_i$, where $1\leq i\leq k$.\textcolor{white}{\index[subject]{clow sequence}}
\end{definition}
A directed graph can have infinitely many clow sequences; for example, see Figure \ref{fig_chap6_1}.
\begin{definition}
Let $G=(V,E)$ be a directed graph and let $\mathcal{W}=(C_1,\ldots $ $,C_k)$ be a clow sequence of $G$. We define the length of the clow sequence $\mathcal{W}$ as the sum of the lengths of the clows in $\mathcal{W}$ and it is denoted by $l(\mathcal{W})$.\textcolor{white}{\index[subject]{length of a clow sequence}}
\end{definition}
\begin{definition}
Let $G=(V,E)$ be a weighted directed graph such that $w:E\rightarrow \R$ is a weight function on the edges of $G$. The weight of a clow sequence $\mathcal{W}=(C_1,\ldots, C_m)$ is defined as $w(\mathcal{W})=\prod_{i=1}^mw(C_i)$.\textcolor{white}{\index[subject]{weight of a clow sequence}}
\end{definition}

\begin{definition}\label{defn-cyclecover}
Let $G=(V,E)$ be a directed graph. A clow sequence $\mathcal{W}=(C_1,\ldots $ $,C_k)$ is a cycle cover of $G$ if every $C_i$ is a cycle and there does not exist any vertex which is common to $C_i$ and $C_j$, where $1\leq i,j\leq k$ and $i\neq j$.\textcolor{white}{\index[subject]{cycle cover}}
\end{definition}

\begin{definition}
Let $G$ be a directed graph on $n$ vertices. Let $x,y_1,\ldots ,y_t\in\{ 1,\ldots ,n\}$ and $\sigma\in S_n$ such that $( y_1\sigma  ~y_2\sigma~\cdot~ \cdot ~\cdot ~y_t\sigma )$ is the orbicycle of $x$ under $\sigma$. We say that the orbicycle of $x$ under $\sigma$ satisfies the graph $G$ if $(y_i,y_i\sigma )\in E(G)$ for all $1\leq i\leq t$.
\end{definition}

\begin{definition}\label{defn-orbicyclecover}
Let $G$ be a directed graph on $n$ vertices. A permutation $\sigma$ in $S_n$ is called an orbicycle cover of $G$ if every orbicycle of $\sigma$ satisfies $G$.\textcolor{white}{\index[subject]{orbicycle cover}}
\end{definition}

\begin{lemma}\label{lemm-ccocc}
Let $G$ be a directed graph on $n$ vertices. If a clow sequence $\mathcal{W}=(C_1,\ldots $ $,C_k)$ is a cycle cover of $G$ then we can obtain $\sigma\in S_n$ from $\mathcal{W}$ which is an orbicycle cover of $G$. Conversely, if $\sigma\in S_n$ is an orbicycle cover of $G$ then we can obtain a clow sequence $\mathcal{W}$ from $\sigma$ which is also a cycle cover of $G$.

There exists a one-to-one and onto mapping between cycle covers of $G$ and permutations in $S_n$ which are orbicycle covers of $G$.
\end{lemma}
\begin{proof}
Let $\mathcal{W}=(C_1,\ldots $ $,C_k)$ be a clow sequence which is also a cycle cover of $G$. $\mathcal{W}$ is a collection of cycles such that head($C_1)<$ head($C_2)<\cdots<$ head($C_k)$. Let $i_l$ denote $l(C_i)$. $\sum_{l=1}^k i_l=n$ and $1\leq i_l\leq n$, where $1\leq i\leq k$.

Let $C_i=(u_{i,1},\ldots ,u_{i,i_l})$, where $u_{i,1}={\rm head}(C_i)$, $u_{i,j}\in V$, $1\leq u_{i,j}\leq n$, $1\leq i\leq k$ and $1\leq j\leq i_l\leq n$, $1\leq l\leq k$. 

Define $\sigma:V\rightarrow V$ such that for $1\leq i\leq k$, $\sigma(u_{i,j})=u_{i,j+1}$ if $1\leq j< i_l$ and $\sigma(u_{i,i_l})=u_{i,1}$. Since $\mathcal{W}$ is a cycle cover, individual clows of $\mathcal{W}$ are cycles and each vertex of $G$ is in exactly one cycle in $\mathcal{W}$. Therefore, $\sigma$ is a permutation of $V(G)$. It is easy to see from the definition of $\sigma$ that if we write $\sigma$ as a product of orbicycles then every orbicycle satisfies $G$. This implies that $\sigma\in S_n$ is an orbicycle cover of $G$ which we have obtained from $\mathcal{W}$.

Conversely if $\sigma$ is an orbicycle cover of $G$, then it follows from the definition of an orbicycle cover (Definition \ref{defn-orbicyclecover}) that every orbicycle of $\sigma$ satisfies $G$. We therefore obtain cycles in $G$ corresponding to each orbicycle. By arranging vertices in each cycle with the head of the cycle as the least numbered vertex, and ordering the collection of cycles based on their heads, we in turn obtain a clow sequence which is also a cycle cover of $G$.

It is easy to see that this mapping between cycle covers of $G$ and permutations in $S_n$ which are orbicycle covers of $G$ is in fact a one-to-one and onto mapping.
\end{proof}
\begin{definition}\label{defn-signclowsequence}
Let $G$ be a directed graph on $n$ vertices.
\begin{enumerate}
\item Let $\mathcal{W}=(C_1,\ldots ,C_k)$ be a cycle cover in $G$ and let $\sigma\in S_n$ be the permutation of $V(G)$ which we obtain from $\mathcal{W}$ such that $\sigma$ is an orbicycle cover of $G$. We define the sign of the cycle cover $\mathcal{W}$ as $sgn(\sigma)$.\textcolor{white}{\index[subject]{sign of a cycle cover}}
\item We define the sign of a clow sequence $\mathcal{W}=(C_1,\ldots ,C_k)$ which is not a cycle cover, called the clowsign of $\mathcal{W}$ and denoted by $clowsgn(\mathcal{W})$, as $(-1)^{m}$ where $0\leq m\leq k$ is the number of components of $\mathcal{W}$ which does not include fixed point cycles that are clows in $\mathcal{W}$. We say that $\mathcal{W}$ is a clow sequence that has positive sign if $m$ is even; otherwise $m$ is odd and $\mathcal{W}$ is called as a clow sequence that has negative sign.\textcolor{white}{\index[subject]{sign of a clow sequence, denoted by $clowsgn(\mathcal{W})$}}
\end{enumerate}
\end{definition}
If $\mathcal{W}$ is a cycle cover then $\mathcal{W}$ is also a clow sequence. Therefore, without loss of generality we denote the sign of a cycle cover $\mathcal{W}$ also as $clowsgn(\mathcal{W})$.

\begin{figure}[h!]
\begin{center}
\includegraphics[scale=0.9]{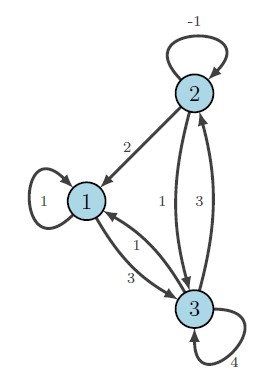}
\caption{This figure is a directed graph $G_2=(V_2,E_2)$. The following are clow sequences, which are cycle covers of $G_2$: $\mathcal{C}_1=((1),(2),(3))$, $\mathcal{C}_2=((1),(2,3))$, $\mathcal{C}_3=((1,3,2))$, and $\mathcal{C}_4=((1,3),(2))$. Examples of clow sequences which are not cycle covers of $G_2$: $\mathcal{C}_5=(1,3,2,3,2)$ and $\mathcal{C}_6=(1,3,2)(2,3)$. Here $l(\mathcal{C}_1)=l(\mathcal{C}_2)=l(\mathcal{C}_3)=l(\mathcal{C}_4)=3$ and $l(\mathcal{C}_5)=l(\mathcal{C}_6)=5$. Also $w(\mathcal{C}_1)=-4$, $w(\mathcal{C}_2)=3$, $w(\mathcal{C}_3)=18$, $w(\mathcal{C}_4)=-3$, and $w(\mathcal{C}_5)=w(\mathcal{C}_6)=54$. Permutations that we obtain from $\mathcal{C}_1$, $\mathcal{C}_2$, $\mathcal{C}_3$, and $\mathcal{C}_4$ are $(1)(2)(3)$, $(1)(3~2)$, $(3~1~2)$, and $(3~2~1)$ respectively. Therefore, $clowsgn(\mathcal{C}_1)=1$, $clowsgn(\mathcal{C}_2)=-1$, $clowsgn(\mathcal{C}_3)=1$ and $clowsgn(\mathcal{C}_4)=-1$. Also, $clowsgn(\mathcal{C}_5)=-1$ and $clowsgn(\mathcal{C}_6)=1$. The determinant of the adjacency matrix of $G_2$ is $14$ and it satisfies the equation \ref{determinant-cyclecover-defn} in Theorem \ref{determinant-cyclecover}. Also note that equation \ref{determinant-clowsequence-eqn} in Theorem \ref{determinant-theorem1} is also satisfied.}\label{fig_chap6_2}
\end{center}
\end{figure}
\begin{theorem}\label{determinant-cyclecover}
Let $A\in\R^{n\times n}$ be the adjacency matrix of a weighted directed graph $G$. Then,
\begin{eqnarray}
\det (A) & = & \sum_{\mathcal{C}:a~cycle~cover}clowsgn(\mathcal{C})w(\mathcal{C}).\label{determinant-cyclecover-defn}
\end{eqnarray}
\end{theorem}
\begin{proof}
It follows from the definition of determinant of $A$ given in equation \ref{determinant-definition}, the definition of the weight of a cycle cover (Definition \ref{defn-clowweight}) and the definition of the clowsign of a cycle cover as the sign of the permutation of the vertices of $G$ obtained from the cycle cover (Definition \ref{defn-signclowsequence}).
\end{proof}

\begin{theorem}\label{determinant-theorem1}{\rm(\bf Mahajan-Vinay, Theorem A)}
Let $A\in\R^{n\times n}$ be the adjacency matrix of a weighted directed graph $G$ such that $A(i,i)=0$ for all $1\leq i\leq n$. Now,
\begin{eqnarray}
\det (A) & = & \sum_{\mathcal{W}:a~clow~sequence}clowsgn(\mathcal{W})w(\mathcal{W}).\label{determinant-clowsequence-eqn}
\end{eqnarray}
\end{theorem}
\begin{proof}
Given the matrix $A\in\R ^{n\times n}$, our assumption in the statement of this theorem that $A(i,i)=0$, for all $1\leq i\leq n$, is very important for its proof which follows below. We prove this theorem by showing that the contribution of clow sequences that are not cycle covers is zero on the right-hand side of the above equation. Consequently, only cycle covers contribute to the summation yielding exactly the determinant of $A$ using Theorem \ref{determinant-cyclecover}.

We define a mapping called an involution (denoted by $\varphi$) on the set of clow sequences of $G$. An involution $\varphi$ on a set is a bijection with the property that $\varphi ^2$ is the identity map on the set. The involution that we define has the domain to be the set of all clow sequences. Also $\varphi^2$ is the identity mapping on the set of all clow sequences, $\varphi$ maps a cycle cover to itself and otherwise maps a clow sequence $\mathcal{W}$ to another clow sequence $\varphi (W)$ such that $\mathcal{W}$ and $\varphi(\mathcal{W})$ have the same weight but opposite clowsigns as follows.

Let $\mathcal{W}=(C_1,\ldots ,C_m)$ be a clow sequence. Now, choose the smallest $1\leq i\leq m$ such that in $\mathcal{W}$, we have $(C_{i+1},\ldots ,C_m)$ is a set of vertex disjoint cycles in $G$. If $i=0$ then any such clow sequence $\mathcal{W}$ is a cycle cover and the involution $\varphi$ maps $\mathcal{W}$ to itself. Otherwise, if such an $1\leq i\leq m$ exists, then we traverse the clow $C_i$ starting from the head until at least one of the following two possibilities occur:
\begin{enumerate}
    \item we visit a vertex that touches one of the cycles in $(C_{i+1},\ldots ,C_m)$,
    \item we visit a vertex that is a part of a cycle within $C_i$.
\end{enumerate}
Without loss of generality, we consider the first such vertex and let it be $v$. It is not difficult to note that some vertex $v$ might satisfy both of these above stated possibilities simultaneously. If the clow $C_i$ has such a vertex $v$ we employ case 1 that follows when considering any such vertex $v$. We now consider these two possibilities:

{\bf Case 1: (merging of clows)} Suppose $v$ touches $C_j$. We map $\mathcal{W}$ to a clow sequence
\[
\mathcal{W}'=(C_1,\ldots ,C_{i-1},C_i',C_{i+1},\ldots,C_{j-1},C_{j+1},\ldots,C_m).
\]
The modified clow sequence $C_i'$ is obtained by merging $C_i$ and $C_j$ as follows: insert the cycle $C_j$ into $C_i$ at the first occurrence (from the head) of $v$. For example, let $C_i=(8,11,10,14)$ and $C_j=(9,10,12)$. Then the new clow is $(8,11,10,12,9,10,14)$. Figure \ref{clow_seq_opp_signs} illustrates the mapping.
\begin{figure}[h!]
\begin{center}
\includegraphics[scale=0.7]{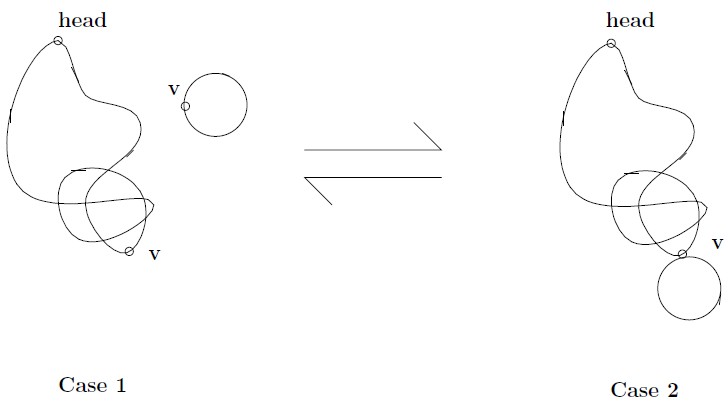}
\caption{Pairing clow sequences of opposite signs}\label{clow_seq_opp_signs}
\end{center}
\end{figure}

The head of $C_i'$ is clearly the head of $C_i$. The new sequence has the same multiset of edges and hence the same weight as the original sequence. It also has one component less than the original sequence $\mathcal{W}$.

{\bf Case 2: (splitting of clows)} Suppose $v$ completes a cycle $C$ in $C_i$. 

We now modify the sequence $\mathcal{W}$ by deleting $C$ from $C_i$ and introducing $C$ as a new clow in an appropriate position, depending on the minimum labeled vertex in $C$, which we make the head of $C$. For example, let $C_i=(8,11,10,12,9,10,14)$. Then $C_i$ changes to $(8,11,10,14)$ and the new cycle $(9,10,12)$ is inserted in the clow sequence.

To show that the modified sequence continues to be a clow sequence, note that the head of $C$ is greater than the head of $C_i$; hence $C$ occurs after $C_i$. Also the head of $C$ is distinct from the heads of $C_j$ ($i<j\leq m$) since $v$ does not touch any of the clows $C_j$ $(i<j\leq m)$ for otherwise we would have merged clows as mentioned in case 1. In fact, $C$ is disjoint from all cycles $C_j$ ($i<j\leq m$) . Further the new sequence has the same multiset of edges and hence the same weight as the original sequence. It also has exactly one component more than the original sequence. Figure \ref{clow_seq_opp_signs} illustrates the mapping. In the new clow sequence, vertex $v$ in cycle $C_i'$ would have been chosen by our traversal, and it satisfies case 1.

In both of the above cases, the new clow sequence constructed maps back to the original clow sequence. Therefore the mapping is a weight preserving involution. Furthermore, the number of clows in the original sequence and in the new sequence in both these cases differ by one and there are no length 1 cycles in $G$. So they have opposite clowsigns. As a result the contribution of the weight of clow sequences which are not cycle covers on the right hand side in the statement of this theorem get canceled. This completes the proof.
\end{proof}
\subsection{Constructing a special weighted layered directed acyclic graph ${\bf H}$}\label{constructwldag}
Given a $n\times n$ matrix $A$, we define a weighted layered directed acyclic graph $\rm H$ with three special vertices $s,t_+$ and $t_-$ that will satisfy the following property:
\begin{eqnarray}
\det (A) & = & \sum_{\rho :s\leadsto t_+}w(\rho )-\sum_{\eta :s\leadsto t_-}w(\eta).\label{determinant-graphtheory-eqn}
\end{eqnarray}
Here the weight of a walk is simply the product of the weights of the edges appearing in it. The idea is that there exists a one-to-one mapping from the set of all cycle covers of positive sign of the directed graph $G$ represented by the matrix $A$ to the set of all walks $s\leadsto t_+$ in $\rm H$. Similarly there exists a one-to-one mapping from the set of all cycle covers of negative sign of the directed graph $G$ represented by the matrix $A$ to the set of all walks $s\leadsto t_-$ in $\rm H$. To prove the above statement, given the matrix $A\in\R^{n\times n}$, we first obtain the matrix $B\in\R^{2n\times 2n}$ using Proposition \ref{removeselfloops} such that $B(i,i)= 0$, where $1\leq i\leq 2n$. Therefore, without loss of generality for the rest of this section, we assume that the input matrix $A\in\R ^{n\times n}$ and $A(i,i)=0$, for all $1\leq i\leq n$. In other words, if $G$ denotes the directed graph represented by $A$ then there are no self-loops on vertices in $G$.

We can view the directed graph $\rm H$ in 3-dimensions as follows: there exists a vertex $s$ above and in between the two 3-dimensional cubes of dimension $n\times n\times n$ each, and two other vertices $t_-$ and $t_+$ below the two cubes. The cube above the vertex $t_+$ is called as the $0$-cube and the cube above the vertex $t_-$ is called as the $1$-cube. Every vertex in the two 3-dimensional cubes of this graph $H$ is represented by a $4$-tuple: $(p,i,h,u)$. The $1^{st}$ component $p$ of the $4$-tuple $(p,i,h,u)$ of a vertex denotes if a path which starts from $s$ is in the $0$-cube or the $1$-cube. The $2^{nd}$ component $i$ of the $4$-tuple $(p,i,h,u)$ of a vertex denotes the layer in which the path that is traversed starting from $s$ is present. The $3^{rd}$ component $h$ of the $4$-tuple $(p,i,h,u)$ of a vertex denotes the head of the clow that is traversed in $G$. The $4^{th}$ component $u$ of the $4$-tuple $(p,i,h,u)$ of a vertex denotes the vertex of the clow in $G$ with head $h$ that is traversed. The set of vertices of $\rm H$ is therefore $\{ s,t_+,t_-\}\cup\{[p,i,h,u]|p\in\{0,1\}, h,u,i\in\{1,\ldots ,n\}\}$. 

We will traverse walks that are clow sequences in the directed graph $G$ represented by $A$. This is mapped to directed paths in $\rm H$ starting from $s$. If a clow sequence in $G$ has a clow with head $h$ and the clow we are traversing is in vertex $u$ then it is mapped to a vertex $(p,i,h,u)$ in $\rm H$. The first $n$ edges of any such clow sequence is mapped to paths from $s$ in $\rm H$ and these paths have either $t_-$ or $t_+$ as the destination vertex.

The edge set of $\rm H$ consists of the following edges:
\begin{enumerate}
    \item $(s,[0,1,h,h])$ for $h\in\{1,\ldots ,n\}$; this edge has weight $1$,
    \item $([p,i,h,u],[\overline{p},i+1,h,v])$ if $i<n$ and $v>h$; this edge has weight $A(u,v)$,
    \item $([p,i,h,u],[p,i+1,h',h'])$ if $i<n$ and $h'>h$; this edge has weight $A(u,h)$,
    \item $([1,n,h,u],t_-)$. Here, if $u>h$ and the directed edge $(u,h)$ exists in the directed graph $G$ represented by $A$ then this edge exists and it has weight $A(u,h)$, and
    \item $([0,n,h,u],t_+)$. Here, if $u>h$ and the directed edge $(u,h)$ exists in the directed graph $G$ represented by $A$ then this edge exists and it has weight $A(u,h)$.
\end{enumerate}
In this definition of $\rm H$, if we have not created a directed edge from a vertex $a$ to a vertex $b$ or that we have not mentioned the weight of the directed edge $(a,b)$ then the weight of the directed edge $(a,b)$ in $\rm H$ is assumed to be $0$. Given the original matrix $A\in\R^{n\times n}$, after reducing $A$ to $B$ as in Proposition \ref{removeselfloops} and obtaining $\rm H$, we have the number of vertices in $\rm H$ is $2(2n)^3+3=16n^3+3$ and the number of edges in $H$ is at most $4n^4$. Without loss of generality, in this weighted layered directed acyclic graph $\rm H$, we say that there are $(n+2)$ layers, vertex $s$ is in the layer $0$, vertices $t_+$ and $t_-$ are in the layer $n+1$, and the vertex $[p,i,h,u]$ is in the layer $i$, where $1\leq i\leq n$. Since $\rm H$ is a weighted layered directed acyclic graph, it follows that any walk from $s$ to $t_+$ in $H$ is actually a directed path. Similarly any walk from $s$ to $t_-$ in $\rm H$ is actually a directed path.
\subsection{Facts about directed paths from $s$ in the special directed graph $\bf H$}\label{factsaboutspecialgraph}
\begin{enumerate}
\item Any directed path from $s$ always starts from a vertex (represented by the $4$-tuple $[0,1,h,h]$) in layer $1$ of the $0$-cube. In the starting vertex of the cubes, $(h,h)$ denotes that we are starting to traverse a walk in $G$ which is in fact a clow sequence and that $h$ is the head of the clow and we are at present in $h$.
\item The second component of a vertex $[p,i,h,u]$ also denotes the layer of the cube we are at present in. It also denotes ($1+$the number of edges that we have traversed in the clow sequence of $G$).
\item If we are traversing a clow sequence of $G$, then it is first mapped to a directed path in $\rm H$ starting from $s$. Also we traverse the edges of the directed graph $\rm H$ for only the first $n$ edges of the clow sequence of $G$. This is important since every cycle cover of $G$ has exactly $n$ edges and such a mapping of the first $n$ edges of clow sequences to directed paths from $s$ suffices for us to prove Theorem \ref{determinant-theorem2}.
\item In the mapping from traversing a clow sequence in $G$ to a directed path from $s$ in $\rm H$, we trace a directed path in $\rm H$ connecting vertices along edges which alternate between the vertices of the $0$-cube and the $1$-cube of $\rm H$. This is due to premise $(2)$ in the definition of $\rm H$ in Section \ref{constructwldag}.
\item If a clow of a clow sequence ends then we start the next clow of the clow sequence in the next layer of the same cube of the directed graph $\rm H$ as defined in the premise $(3)$ in the definition of $\rm H$ in Section \ref{constructwldag}.
\item Premises $\{(4)$ and $(5)\}$ deal with the layer $n$ of the cubes and state those edges that have non-zero weight which connect from the $n^{th}$ layer of the cubes to vertices $t_+$ and $t_-$.
\end{enumerate}
\subsection{Determinant, the special directed graph ${\bf H}$ and $\boldGapL$}
\begin{theorem}\label{determinant-theorem2}{\rm(\bf Mahajan-Vinay, Theorem B)}
Let $A\in\R^{n\times n}$ and let $\rm H$ be the weighted layered directed acyclic graph described above. Then,
\begin{eqnarray}
\det (A) & = & \sum_{\rho :s\leadsto t_+}w(\rho )-\sum_{\eta :s\leadsto t_-}w(\eta).\label{determinant-graphtheory-eqn2}
\end{eqnarray}
\end{theorem}
\begin{proof}
To prove this theorem, as mentioned in Section \ref{constructwldag}, given the matrix $A\in\R^{n\times n}$, we first obtain the matrix $B\in\R^{2n\times 2n}$ using Proposition \ref{removeselfloops} such that $B(i,i)=0$, where $1\leq i\leq 2n$. Therefore, without loss of generality for the rest of this proof, we assume that the input matrix $A\in\R ^{n\times n}$ and $A(i,i)=0$, for all $1\leq i\leq n$. In other words, if $G$ denotes the directed graph represented by $A$ then there are no self-loops on vertices in $G$.

{\bf Case 1:} We first show that there exists a one-to-one mapping from the set of all cycle covers of positive clowsign in the directed graph $G$ represented by the matrix $A$ into the set of all paths $s\leadsto t_+$ in $\rm H$. Similarly there exists a one-to-one mapping from the set of all cycle covers of negative clowsign in the directed graph $G$ represented by the matrix $A$ into the set of all paths $s\leadsto t_-$ in $\rm H$.

Let $\mathcal{W}=(C_1,\ldots ,C_k)$ be a cycle cover of $G$. Clearly, $C_i$ are a collection of vertex disjoint cycles such that $h(C_1)<h(C_2)<\cdots <h(C_k)$. Using Lemma \ref{lemm-ccocc}, it follows that cycle covers of $G$ whose weight is non-zero yield permutations of vertices of $G$ whose weight is non-zero and the same weight respectively. Clearly the clowsign of a cycle cover is the $sgn$ of the permutation which we obtain from the cycle cover. It follows from the definition of $\rm H$ that from $s$ we move to the vertex $[0,1,h_1,h_1]$ and the weight of this edge is $1$. In the sequence of vertices visited along this path, the first component denotes the cube we are at present in. The second component is the layer of the cube we are present in. It also denotes $1+$the number of edges that we have traversed in the clow sequence of $G$ and it is monotonically increasing. Also the third component is the head of the clow that is traversed in $G$ and this component is also monotonically non-decreasing and takes, say $k$ distinct values $h_1,\ldots ,h_k$. Consider the maximal segment of the directed path in this clow with the third component $h_i$. The fourth component on this maximal segment is a vertex of the clow with the head $h_i$ in $G$. When this clow is completely traversed, a new clow with a larger head must be started. Depending on whether we have the clowsign of the cycle cover is positive or negative, we end the directed path which started from $s$ in $t_+$ or in $t_-$ respectively. This is precisely modeled by the edges of $\rm H$. Clearly, the weight of the path from $s$ to $t_+$ or to $t_-$ is equal to the weight of the cycle cover which is the product of the weights of the edges in the cycle cover. Also no two distinct cycle covers of the same clowsign will yield the same directed path from $s$ to vertices $t_-$ or $t_+$. Since it is not possible to either merge clows (which are in fact pair of individual disjoint cycles) in cycle covers or to split clows (individual cycles) in cycle covers, we get this mapping to be a one-to-one mapping.

{\bf Case 2:} We consider the case of a clow sequence $\mathcal{W}=(C_1,\ldots ,C_k)$ which is not a cycle cover. Such clow sequences are not fully traversed by the directed path that starts from $s$ since any directed path in $\rm H$ is of length only $n+1$ and its first edge from $s$ to a vertex in layer $1$ of the $0$-cube is not a part of the clow sequence. As a result only the first $n$ edges of the clow sequence are traversed using edges of, say a $s\leadsto t_+$ path in $\rm H$. We observe here also using the same argument as in the case of cycle covers that no two partial clow sequences, which are different and form only the first $n$ edges of the actual clow sequences of $G$, will result in the same directed path from $s$ to $t_+$ or from $s$ to $t_-$.

We note that if a clow of a clow sequence which has length $>n$ does not end a clow at the $n^{th}$ edge, then the weight of the clow sequence is $0$. This property is preserved due to the definition of $\rm H$ and its edges from the layer $n$ to vertices $t_+$ or $t_-$. Based on if $k$ is even or odd, and hence the sign of $\mathcal{W}$ is positive or negative respectively, we will demonstrate that there exists a directed path from $s$ to $t_+$ or from $s$ to $t_-$, respectively in $\rm H$.

Now, without loss of generality, let us consider a clow sequence $\mathcal{W}$ of negative clowsign which is not a cycle cover and the partial clow sequence $\mathcal{W'}$ obtained using the first $n$ edges of this clow sequence $\mathcal{W}$ such that the $n^{th}$ edge of $\mathcal{W}$ ends a clow in $\mathcal{W}$. Using the merge operation on a pair of clows or the split operation on a clow of $\mathcal{W'}$, we can therefore obtain another partial clow sequence $\mathcal{W''}$ such that $w(\mathcal{W'})=w(\mathcal{W''})$ and $clowsgn(\mathcal{W'})$ and $clowsgn(\mathcal{W''})$ are opposite to each other.

{\bf\textit{Claim:}} We claim that the number of negative partial clow sequences that we can obtain starting from any such partial clow sequence $\mathcal{W'}$ containing exactly $n$ edges of a clow sequence $\mathcal{W}$ is equal to the number of positive partial clow sequences that we can obtain starting from the same partial clow sequence $\mathcal{W'}$ of the clow sequence $\mathcal{W}$. 

Assuming that the above claim is true, it is easy to see that all these partial clow sequences, irrespective of their sign, have the same weight. As a result an equal number of positive partial clow sequences and negative partial clow sequences will cancel each other in the right-hand side of the equation in the theorem statement as in the proof of Theorem \ref{determinant-theorem1}, yielding the fact that any non-zero contribution to the equation is always because of the sign and weight of cycle covers only. Our theorem therefore follows from Theorem \ref{determinant-cyclecover}.

{\bf\textit{Proof of claim:}} To prove our statement regarding partial clow sequences that contains exactly $n$ edges, we first define a relation $\sim$ on such partial clow sequences such that any two partial clow sequences $\mathcal{W}_1$ and $\mathcal{W}_2$ are related if and only if we can obtain $\mathcal{W}_2$ from $\mathcal{W}_1$ using a merge operation on a pair of clows of $\mathcal{W}_1$ or a split operation on a clow of $\mathcal{W}_1$. It is clear that $\sim$ is symmetric. Also $\mathcal{W}_1$ and $\mathcal{W}_2$ have the same number of edges $n$ and the same weight. However their signs are opposite to each other.

We now define a bipartite graph $G'$ whose set of vertices is the set of all such partial clow sequences that are related by the relation $\sim$ such that partial clow sequences of positive clowsign are in one partition $U$ and partial clow sequences of negative clowsign are in the other partition $V$. Also we put a directed edge from a vertex $u\in U$ to a vertex $v\in V$ or vice-versa if the two partial clow sequences represented by these two vertices $u$ and $v$ are related according to $\sim$. It is clear that since there are no self-loops in the directed graph represented by the input matrix $A$, we do not obtain a partial clow sequence of the same clowsign following a merge operation of a pair of clows of a partial clow sequence or a split operation of a clow of a partial clow sequence. As a result any directed edge in $G'$ is always from a vertex in one partition to a vertex in the other partition. So our graph $G'$ is bipartite. Note that $\sim$ is symmetric and so whenever there exists a directed edge between two vertices in $G'$, there exists a back edge also. Therefore we can treat this bipartite graph $G'$ as an undirected bipartite graph. Also this undirected bipartite graph $G'$ is connected since any partial clow sequence represented by a vertex is always related to any other partial clow sequence by a sequence of merge or split operations and so there exists an undirected path connecting any two vertices in $G'$.

Now, for any vertex $u\in U$, let $m$ denote the number of pairs of clows in the clow sequence represented by $u$ that can be merged. Similarly let $s$ denote the number of clows in the clow sequence represented by $u$ that can be split. Then degree of $u$ is $m+s$. Let $p=m+s$. It is easy to observe that if $v\in V$ is a neighbour of $u$ in $G'$ then the degree of $v$ is also $p$. In fact the undirected bipartite graph $G'$ is $(p,p)$-regular. Using a simple counting argument on the number of edges in $G'$ along its partitions $U$ and $V$, we can show that any such $(p,p)$-regular undirected bipartite graph $G'$ has the same number of vertices in both the partitions $U$ and $V$. This shows that the number of partial clow sequences of positive sign and the number of partial clow sequences of negative sign that can be obtained from any such partial clow sequence $\mathcal{W'}$ are equal from which the claim follows.
\end{proof}

\begin{theorem}\label{determinant-computeingapl}
Let $A\in\Z^{n\times n}$. Computing $\det (A)$ is in $\GapL$.
\end{theorem}
\begin{proof}
It is easy to see that given matrix $A$ as input, we can obtain the adjacency matrix of the weighted layered directed acyclic graph $\rm H$ defined in Section \ref{constructwldag} in using space which is logarithmic in the size of $A$. As a result it follows that the proof of Theorem \ref{determinant-theorem2} actually shows that the problem of computing $\det (A)$ is logspace many-one reducible to the problem GapDSTCON which is shown to be complete for $\GapL$ in Chapter 4. This completes the proof.
\end{proof}

\begin{theorem}\label{determinant-hardforgapl}
Let $A\in\Z^{n\times n}$. Computing $\det (A)$ is logspace many-one hard for $\GapL$.
\end{theorem}
\begin{proof}
In Chapter 4, we have shown that the problem GapSLDAGSTCON is logspace many-one complete for $\GapL$. Let $(G,s,t_+,t_-)$ be an input instance of GapSLDAGSTCON. Let $n$ denote the number of layers in $G$. We assume without loss of generality that $G$ is not an empty graph. We have to compute the quantity $D$, which is number of directed paths from $s$ to $t_+$ minus the number of directed paths from $s$ to $t_-$ in $G$. We show that it is possible to obtain a matrix $A$ in $O(\log N)$ space such that $\det (A)=D$, where $N$ is the size of the input instance.

For this, first we obtain a new graph $G'$ from $G$ as follows. We first replace each directed edge in $G$ which exists between vertices of adjacent layers in $G$ by a directed path of length $2$ such that both these edges have weight $1$. We then add a vertex $x$ and the following edges of weight $1$: $\{ (t_+,s),(t_-,x),(x,s)\}$, and self-loops having weight $1$ on all vertices of $G'$ except $s$. Clearly $G'$ contains $(2n-1)$ layers, where the $(2n-1)^{th}$ layer contains $(n+1)$ vertices including $x$. $G'$ contains $(n^2+|E(G)|+1)$ vertices. Also $G'$ contains $(2|E(G)|+3)$ ordinary edges, $(n^2+|E(G)|)$ self-loops. Therefore $G'$ contains $(n^2+3|E(G)|+3)$ edges. Since the graph $G'$ is still a weighted layered directed graph, we assume that the vertices in $G'$ are labeled by a pair of indices $(i,j)$ such that the vertex $(i,j)$ is the $j^{th}$ vertex in layer $i$, where $1\leq i\leq 2(n-1)$ and $1\leq j\leq n$. The vertex $s$ is in the first layer of $G'$, and vertices $t_+$ and $t_-$ are in the last layer of $G'$. We also assume an ordering of the vertices first based on the layers and then from the left-most vertex to the right-most vertex in each layer.

It follows from the definition of $G'$ that each $s\leadsto t_+$ path in $G$ corresponds to a cycle of odd length in $G'$ which contains $s$ and $t_+$. Due to self-loops on all vertices of $G'$ except $s$, using self-loop on any vertex which is not in this directed cycle, we therefore get a cycle cover in $G'$. Using Lemma \ref{lemm-ccocc} it follows that every such cycle cover in turn yields an even permutation of vertices of $G'$. As a result the sign of every such permutation is positive. Similarly, each $s\leadsto t_-$ path in $G$ corresponds to a cycle cover of even length in $G'$ which in turn yields a permutation of the vertices of $G'$ of negative sign in $G'$. Conversely, we note that any cycle cover in $G'$ should have a cycle containing $s$ and since there is no self-loop on $s$, any cycle which contains $s$ should contain the vertex $t_+$ and the directed edge $(t_+,s)$, or the cycle should contain both vertices $x$ and $t_-$ and directed edges $(t_-,x)$ and $(x,s)$. We have therefore shown that there exists a directed path from $s$ to $t_+$ in $G$ if and only if there exists a cycle cover in $G'$ containing $s$ and $t_+$. Similarly, there exists a directed path from $s$ to $t_-$ in $G$ if and only if there exists a cycle cover in $G'$ containing $s$ and vertices $x$ and $t_-$.

Let $A$ denote $adj(G')$, the adjacency matrix of $G'$. It is also easy to note that the weight of the cycle cover in $G'$ is equal to the product of edge weights which is the product of entries of $A$. Since any newly added edge to $G$ used to obtain $G'$ has weight $1$, it follows from the definition of the determinant (equation \ref{determinant-definition}), Theorem \ref{determinant-cyclecover} and Theorem \ref{determinant-theorem2} that $\det (A)$ equals $D$. Since we can obtain the matrix $A$ from the input instance of GapSLDAGSTCON using at most $O(\log n)$ space, where $n$ is the size of the input instance, we have proved our theorem.
\end{proof}
\begin{corollary}\label{determinantgaplcomplete}
Let $A\in\Z^{n\times n}$. Computing $\det (A)$ is logspace many-one complete for $\GapL$.
\end{corollary}
\begin{corollary}\label{determinantsharplhard}
Let $A\in\Z^{n\times n}$. Computing $\det (A)$ is logspace many-one hard for $\sharpL$.
\end{corollary}
\begin{proof}
The proof of this result is similar to Theorem \ref{determinant-hardforgapl}. In Chapter 4, we have shown that the problem $\sharp\LDAGSTCON$ is logspace many-one complete for $\sharpL$. We show that this problem is logspace many-one reducible to the problem of computing the determinant of an integer matrix.

Let $(G,s,t)$ be an input instance of $\sharp\LDAGSTCON$. We show that it is possible to obtain a matrix $A$ in $O(\log n)$ space such that $\det (A)=N$, where $n$ is the size of the input instance and $N$ is the number of directed paths from $s$ to $t$ in $G$. We first replace each directed edge in $G$ which exists between vertices of adjacent layers in $G$ by a directed path of length $2$ such that both these edges have weight $1$. We then add the following edges of weight $1$: $(t,s)$ and self-loops having weight $1$ on all vertices except $s$. Let the resulting directed graph be $G'$. Now any $s\leadsto t$ path in $G$ yields a cycle cover in this new graph which further yields a permutation of the vertices of $G'$. As in Theorem \ref{determinant-hardforgapl}, here we note that permutations which are obtained from cycle covers have positive sign. Also there are no permutations which we obtain that have negative sign. Therefore, as in Theorem \ref{determinant-hardforgapl}, it follows from the definition of the determinant (equation \ref{determinant-definition}), Theorem \ref{determinant-cyclecover} and Theorem \ref{determinant-theorem2} that the number of $s\leadsto t$ paths in $G$ is equal to the $\det (adj(G'))$, where $adj(G')$ denotes the adjacency matrix of $G'$. Since we can obtain $A$ from the input instance of $\sharp\LDAGSTCON$ using at most $O(\log n)$ space, where $n$ is the size of the input instance, our theorem follows.
\end{proof}

\section{Applications of computing the Determinant}\label{applicationDeterminant}
In this section, we assume familiarity with basic notions of vector spaces over a field such as linear independence and linear dependence of vectors. We get the following results on linear algebraic problems.
\begin{theorem}\label{LA1}
\begin{enumerate}
\item\label{singularCeqL} Let $A\in\Z^{n\times n}$. Determining if $\det (A)=0$ is logspace many-one complete for $\CeqL$.
\item\label{lex-leastcolumns} Given a set of vectors $S=\{\vec{\rm\bf v}_1,\ldots ,\vec{\rm\bf v}_k\}$ with integer entries over the field of rationals $\Q$, determining if a column in $S$ is in the lexicographically least set of linearly independent vectors in $S$ over $\Q$ is in $\LCeqL$.
\item \label{ranktheorem} Let $A\in\Z^{n\times n}$. Computing the $rank(A)$ is in $\LCeqL$.
\item Given $(A,\vec{\rm\bf b})$, where $A\in\Z^{m\times n}$ and $\vec{\rm\bf b}\in\Z^{m\times 1}$, determining if there exists $\vec{\rm\bf x}\in\Q^{n\times 1}$ such that $A\vec{\rm\bf x}=\vec{\rm\bf b}$ is in $\LCeqL$ and it is logspace many-one hard for $\CeqL$.
\end{enumerate}
\end{theorem}
\begin{proof}
\begin{enumerate}
\item It follows from Corollary \ref{determinantgaplcomplete} and Definition \ref{ceqldefinition}.
\item We use the greedy method of iteratively picking columns from $S$ in a lexicographically least manner such that the rank of the sub-matrix formed continues to increase. We simultaneously determine if the column that we want is in the lexicographically least subset of linearly independent columns $S$. Clearly, using a $O(\log n)$-space bounded deterministic Turing machine with access to a $\CeqL$ oracle, we can therefore obtain the lexicographically least subset of columns in $S$.
\item  Since we know that the $rank(A)$ is the cardinality of the lexicographically least subset of linearly independent columns of $A\in\Z^{n\times n}$ over $\Q$ and finding the size of this subset of columns of $A$ is a $\leq_{\rm T}^{\L}$ reduction to $\CeqL$.
\item Since the system of linear equations $A\vec{\rm\bf x}=\vec{\bf\rm b}$ has a solution if and only if matrices $A$ and $[A;\vec{\rm\bf b}]$ have the same rank, it follows from Theorem \ref{LA1}(\ref{ranktheorem}) that this problem is in $\LCeqL$.

It follows from Theorem \ref{LA1}(\ref{singularCeqL}) that given $A\in\Z^{n\times n}$, the problem of determining if $\det (A)=0$ is logspace many-one complete for $\CeqL$. This is equivalent to determining if the system of linear equations $AX=0$ does not have any solution $X\in\Z^{n\times n}$, where $0$ denotes the $n\times n$ matrix containing only zeros. As a result the problem of determining if a system of linear equations does not have any solution is logspace many-one hard for $\CeqL$.

We now show that the problem of determining if a system of linear equations has a solution is logspace many-one reducible to its complement problem of determining if a system of linear equations does not have a solution. We claim that the system $A\vec{\rm\bf x}=\vec{\rm\bf b}$ is feasible if and only if there exists a $\vec{\rm\bf y}$ such that $A^T\vec{\rm\bf y}=0$ and $\vec{\rm\bf b}^T\vec{\rm\bf y}=1$. Let $W$ be the subspace spanned by the columns of $A$. The system is feasible if and only if $\vec{\rm\bf b}\in W$. From elementary linear algebra we know that $\vec{\rm\bf b}$ can be uniquely written as $\vec{\rm\bf b}=\vec{\bf\rm v}+\vec{\rm\bf w}$, where $\vec{\rm\bf v}$ is perpendicular to $W$ (i.e., $\vec{\rm\bf v}^TA=0$) and $\vec{\rm\bf w}\in W$. If $\vec{\rm\bf v}\neq 0$, then since $\vec{\rm\bf v}^T\vec{\rm\bf w}=0$, we have $\vec{\rm\bf v}^T\vec{\rm\bf b}=\vec{\rm\bf v}^T\vec{\rm\bf v}>0$, and we may let $\vec{\rm\bf y}=\frac{1}{\vec{\rm\bf v}^T\vec{\rm\bf v}}\vec{\rm\bf v}$. Thus if $A\vec{\rm\bf x}=\vec{\rm\bf b}$ is infeasible, then there exists $\vec{\rm\bf y}$ such that $A^T\vec{\rm\bf y}=\vec{\rm\bf 0}$ and $\vec{\rm\bf b}^T\vec{\rm\bf y}=1$. Conversely, if such a $\vec{\rm\bf y}$ exists then $A\vec{\rm\bf x}=\vec{\rm\bf b}$ is infeasible. This shows that the problem of determining if a system of linear equations has a solution is logspace many-one hard for $\CeqL$.
\end{enumerate}
\end{proof}

\begin{corollary}
Let $p\in\N$ such that $p$ is a prime.
\begin{enumerate}
\item Let $A\in\Z^{n\times n}$. Determining is $\det (A)\not\equiv 0(\mod p)$ is logspace many-one complete for $\ModpL$.
\item Let $A\in\Z^{n\times n}$. Computing the $rank(A)$ over $\Z_p$ is in $\ModpL$.
\item Given a set of vectors $S=\{\vec{\rm\bf v}_1,\ldots ,\vec{\rm\bf v}_k\}$ with integer entries over the field $\Z_p$, determining if a column in $S$ is in the lexicographically least set of linearly independent vectors in $S$ over $\Z_p$ is in $\ModpL$.
\item Given $(A,\vec{\rm\bf b})$, where $A\in\Z^{m\times n}$ and $\vec{\rm\bf b}\in\Z^{m\times 1}$, determining if there exists $x\in\Q^{n\times 1}$ such that $A\vec{\rm x}=\vec{\rm\bf b}(\mod p)$ is logspace many-one complete for $\ModpL$.
\end{enumerate}
\end{corollary}

It is unknown if $\GapL$ is closed under division. However it is possible to use Theorem \ref{LA1}(\ref{ranktheorem}) and show the following result concerning division of two functions in $\GapL$ and on $rank (A)$, where $A\in\Z^{n\times n}$.
\begin{proposition}
Let $A\in\Z^{n\times n}$. Then there exists functions $g,h\in\GapL$ such that $rank(A)=\frac{g(A)}{h(A)}$.
\end{proposition}
\begin{proof}
It follows from Theorem \ref{LA1}(\ref{ranktheorem}) that determining if $rank(A)=r$ is in $\CeqL$. As a result there exists a function $f\in\GapL$ such that
\[
rank(A)=r \Longleftrightarrow f(A,r)=0.
\]
Define functions
\begin{eqnarray*}
g(A) & = & \sum_{r=0}^nrf(A,r),\\
h(A) & = & \sum_{r=0}^nf(A,r).
\end{eqnarray*}
Then we have $g,h\in\GapL$ and $rank(A)=\frac{g(A)}{h(A)}$.
\end{proof}

\subsection{Matrix problems reducible to computing the determinant}

Let us consider the following problem.\newline\newline
\noindent{\bf POWERELEMENT}\newline
INPUT: $A\in\R^{n\times n}$, and $i,j,m\in\N$ such that $1\leq i,j,m\leq n$.\newline
OUTPUT: $(A^m)_{i,j}$, the $(i,j)^{th}$ entry of $A^m$.
\begin{proposition}
{\rm POWERELEMENT}$\Lredn${\rm Determinant}.\textcolor{white}{\index[subject]{POWERELEMENT}}
\end{proposition}
\begin{proof}
Given an input instance $(A,i,j,m)$ of POWERELEMENT, we can without loss of generality assume that $i=1$ and $j=n$ since it is possible to use elementary row and column transformations to obtain a $n\times n$ matrix $M$ from the matrix $A$ such that the $(i,j)^{th}$ element of $A$ is the $(1,n)^{th}$ element of $M$, and $M^m_{1,n}=A^m_{i,j}$. So let $i=1$ and $j=n$.

We now construct a matrix $B$ such that $A^m_{i,j}=\det (B)$. We interpret $A$ as representing a directed bipartite graph on $2n$ nodes. That is, the nodes of the bipartite graph are arranged in two columns of $n$ nodes each. In both columns, nodes are numbered from $1$ to $n$. If entry $a_{k,l}$ of $A$ is not zero, then there is an edge labeled $a_{k,l}$ from node $k$ in the first column to node $l$ in the second column. Now, take $m$ copies of this graph, put them in a sequence and identify each second column of nodes with the first column of the next graph in the sequence. Call the resulting graph as $G'$. The directed graph $G'$ has $m+1$ columns of nodes. The weight of a path in $G'$ is the product of all labels on the edges of the path. The crucial observation now is that the entry at position $(1,n)$ in $A^m$ is the sum of the weights of all paths in $G'$ from node $1$ in the first column to node $n$ in the last column. Call these two nodes as $s$ and $t$ respectively.

The graph $G'$ is further modified: for each edge $(k,l)$ with label $a_{k,l}$, introduce a new node $u$ and replace the edge by two edges $(k,u)$ with label $1$ and $(u,l)$ with label $a_{k,l}$. Now all paths from $s$ to $t$ have even length, but still the same weight. Add an edge labeled $1$ from $t$ to $s$. Finally, add self-loops labeled $1$ to all nodes, except $t$. Call the resulting graph $G$.

Let $B$ be the adjacency matrix of $G$. The determinant of $B$ can be expressed as the sum over all weighted cycle covers of $G$. However, every cycle cover of $G$ consists of a directed path from $s$ to $t$, (due to the extra edge from $t$ to $s$) and self-loops for the remaining nodes. The single nontrivial cycle in each cover has odd length, and thus corresponds to an even permutation. Therefore, $\det (B)$ is precisely the sum over all weighted directed paths from $s$ to $t$ in $G'$. We conclude that $\det (B)=(A^m)_{1,n}$ as desired.
\end{proof}

Let us consider the following problem.\newline\newline
\noindent{\bf ITMATPROD}\textcolor{white}{\index[subject]{ITMATPROD}}\newline
INPUT: a set of $n$ matrices $A_1,\ldots ,A_n$ of dimension $n\times n$ each, indexes $1\leq i,j\leq n$.\newline
OUTPUT: the $(i,j)^{th}$ element of the product $\prod_{1\leq i\leq n}A_i$.
\begin{proposition}
We have the following.
\begin{enumerate}
\item {\rm POWERELEMENT}$\leq^{\rm L}_m$ {\rm ITMATPROD}, and
\item {\rm ITMATPROD}$\leq^{\rm L}_m$ {\rm POWERELEMENT}.
\end{enumerate}
In other words, {\rm POWERELEMENT} and {\rm ITMATPROD} are logspace many-one equivalent, denoted by {\rm POWERELEMENT}$\equiv^{\rm L}_m$ {\rm ITMATPROD}.
\end{proposition}
\begin{proof}
It is obvious that POWERELEMENT$\Lredn$ITMATPROD. Conversely, let $A_1\ldots ,A_n$ be $n\times n$ matrices, and let $B$ be the $(n^2+n)\times (n^2+n)$ matrix consisting of $n\times n$ blocks which are all zero except for $A_1,\ldots ,A_n$ appearing above the diagonal of zero blocks. Then $B^n$ has the product $A_1A_2\cdots A_n$ in the upper right corner.
\end{proof}

Let us consider the following problem.\newline\newline
\noindent{\bf MATINV}\textcolor{white}{\index[subject]{MATINV}}\newline
INPUT: $A\in\R^{n\times n}$ and $i,j\in\N$ such that $1\leq i,j\leq n$.\newline
OUTPUT: the $(i,j)^{th}$ element of $A^{-1}$ in the form $(numerator,denominator)$ which is $((-1)^{i+j}$ $\det(A(j|i)),\det (A))$, where $(-1)^{i+j}\det(A(j|i))$ denotes the $(i,j)^{th}$ entry in the co-factor matrix of $A$.
\begin{proposition}
{\rm POWERELEMENT}$\Lredn${\rm MATINV}
\end{proposition}
\begin{proof}
Let $N$ be the $n^2\times n^2$ matrix consisting of $n\times n$ blocks which are all zero except for $n-1$ copies of $A$ above the diagonal of zero blocks. Then $N^n=0$ and $(I_n-N)^{-1}=I_n+N+N^2+\cdots N^{n-1}=$
\[
\left [\begin{array}{c c c c c}
                    I_n & A & A^2 & \cdots & A^{n-1}\\
                    0 & I_n & A & \cdots & A^{n-2}\\
                    \vdots & & & & \vdots\\
                    0 & & & \cdots & I_n
       \end{array}\right ],
\] where $I_n$ denotes the $n\times n$ identity matrix.
\end{proof}
\begin{proposition}
{\rm MATINV}$\Lredn${\rm Determinant}
\end{proposition}
\begin{proof}
All the entries of the co-factor matrix of the input matrix $A$ are determinants of minors of $A$ with appropriate sign.
\end{proof}

\begin{theorem}
{\rm POWERELEMENT, ITMATPROD, MATINV}$~\in\GapL$.
\end{theorem}
\begin{theorem}
{\rm POWERELEMENT, ITMATPROD, MATINV} are logspace many-one hard for $\sharpL$.
\end{theorem}

\section[Logarithmic space bounded counting classes and Boolean circuits]{Logarithmic space bounded counting classes and Boolean circuits}\label{sharpLHTCone-sec}
We recall Definitions \ref{chap1-TCzero-defn} and \ref{chap1-TCone-defn} from Chapter 1. It can be shown using Theorem \ref{determinant-computeingapl} that computing the determinant of an integer matrix is in the circuit-based complexity class ${\rm U_L}$-$\TCone$ as follows. Observing the proof of Theorem \ref{determinant-computeingapl}, it follows that given a matrix $A\in\Z^{n\times n}$, the graph $H_A$ can be output in ${\rm U_L}$-$\ACzero$. To complete the proof of the above assertion, we show that POWERELEMENT$\in{\rm U_L}$-$\TCone$ first. 

We need the following standard and basic functions involving integers:\newline
\noindent{\bf ADD}\textcolor{white}{\index[subject]{ADD}}\newline
INPUT: two $n$ bitnumbers $a=a_{n-1}\cdots a_0$ and $b=b_{n-1}\cdots b_0$.\newline
OUTPUT: $s=s_n\cdots s_0$, where $s=_{\rm def}a+b$.

\noindent{\bf ITADD}\textcolor{white}{\index[subject]{ITADD}}\newline
INPUT: $n$ numbers in binary with $n$ bits each.\newline
OUTPUT: the sum of the input numbers in binary.

\noindent{\bf MULT}\textcolor{white}{\index[subject]{MULT}}\newline
INPUT: two $n$ bitnumbers $a=a_{n-1}\cdots a_0$ and $b=b_{n-1}\cdots b_0$.\newline
OUTPUT: the product of the input numbers in binary.

\noindent{\bf ITMULT}\textcolor{white}{\index[subject]{ITMULT}}\newline
INPUT: $n$ numbers in binary with $n$ bits each.\newline
OUTPUT: the product of the input numbers in binary.

We can without loss of generality assume that the functions that we have defined are length respecting, which means the following. Let $f:\{ 0,1\}^*\rightarrow\{ 0,1\}^*$. We say that $f$ is length-respecting if whenever $|x|=|y|$ then also $|f(x)|=|f(y)|$. Also corresponding to any length-respecting function $f:\{ 0,1\}^*\rightarrow\{ 0,1\}^*$, the notion of computing the bits of the output of the function $f$ on any given input in $\{ 0,1\}^*$ is defined as follows.
\begin{definition}
Let $f:\{ 0,1\}^*\rightarrow\{ 0,1\}^*$, $f=(f^n)_{n\in\N}$, be length-respecting. Let for every $n$ and $|x|=n$ the length of $f^n(x)$ be $r(n)$. Let $f_i^n:\{ 0,1\}^n\rightarrow \{ 0,1\}$ be the Boolean function that computes the $i^{th}$ bit of $f^n$, i.e., if $f^n(x)=a_1a_2\cdots a_{r(n)}$ then $f_i^n(x)=a_i$, where $1\leq i\leq r(n)$. Then, ${\rm bits}(f)$ denotes the class of all those functions, i.e., ${\rm bits}(f)=_{\rm def}\{f_i^n|n,i\in\N, i\leq r(n)\}$.\textcolor{white}{\index[subject]{${\rm bits}(f)$}}
\end{definition}
Let $\FSD_{\mathcal{B}_1\cup {\rm bits (g)}}$ $(n^{O(1)},1)$\textcolor{white}{\index[subject]{$\FSD_{\mathcal{B}_1\cup {\rm bits (g)}}$ $(n^{O(1)},1)$}} denote the complexity class of all length-respecting functions $f:\{ 0,1\}^*\rightarrow \{ 0,1\}^*$ such that computing ${\rm bits}(f)$ is in the complexity class $\SD_{\mathcal{B}_1\cup {\rm bits (f)}}(n^{O(1)},1)$. We also need the definition of {\em constant-depth reductions}.
\begin{definition}
Let $f,g:\{ 0,1\}^*\rightarrow \{ 0,1\}^*$ be length-respecting. We say that $f$ is constant-depth reducible to $g$ if $f\in \FSD_{\mathcal{B}_1\cup {\rm bits (g)}}(n^{O(1)},1)$ and it is denoted by $f\leq_{\rm cd}g$. Here, a gate $v$ for a function in $\rm {bits}(g)$ contributes to the size of its circuit with $k$, where $k$ is the fan-in of $v$. We say that $f$ is constant-depth equivalent to $g$ and we write $f\equiv_{\rm cd} g$ if $f\leq _{\rm cd}g$ and $g\leq _{\rm cd} f$.
\end{definition}
It is a standard result in circuit complexity theory that MAJ$\equiv_{\rm cd}$ ITADD$\equiv_{\rm cd}$ MULT$\equiv_{\rm cd}$ ITMULT. As a result MULT, ITADD$\in{\rm U_L}$-$\TCzero$. Therefore it follows that given $\vec{\rm\bf u},\vec{\rm\bf v}\in\Z^{n\times n}$, the inner product of $\vec{\rm\bf u}$ and $\vec{\rm\bf v}$, denoted by $\langle \vec{\rm\bf u},\vec{\rm\bf v}\rangle$ is computable in ${\rm U_L}$-$\TCzero$. As a result it is clear that it is possible to compute the $(i,j)^{th}$ entry in the product of an input pair of square integer matrices in ${\rm U_L}$-$\TCzero$. As a consequence if we are given a set of $n$ square integer matrices as input where every entry in each of these matrices is of size $n$ then we can compute the entries in product of these matrices by combining these $\TCzero$ circuits in a pairwise manner to obtain a Boolean circuit that contains $\neg$ gates, unbounded fan-in $\vee$,$\wedge$ and MAJ gates and whose size is a polynomial in $n$ and depth $O(\log n)$. Clearly this shows that the problem of computing the $(i,j)^{th}$ entry of the powers of an input matrix that has entries in $\Z$ is in ${\rm U_L}$-$\TCone$ which implies POWERELEMENT$\in{\rm U_L}$-$\TCone$. It is also a standard fact that ${\rm U_L}$-$\ACzero\subseteq$${\rm U_L}$-$\TCone$.
As a result, combining the ${\rm U_L}$-$\ACzero$ circuit that many-one reduces the problem of computing the determinant of an integer matrix to computing the difference of the $(i,j)^{th}$ entries of powers of the adjacency matrix of the layered directed acyclic graph $H_A$, it follows that we can compute the $\det (A)$ in $\subseteq{\rm U_L}$-$\TCone$. We therefore get the following theorem.
\begin{theorem}\label{determinantTCone}
Let $A\in\Z^{n\times n}$. Computing $\det (A)\in{\rm U_L}$-$\TCone$.
\end{theorem}
This result also shows the following.
\begin{corollary}\label{gaplTCone}
$\GapL\subseteq{\rm U_L}$-$\TCone$.
\end{corollary}
\begin{corollary}\label{sharpLHTCone}
$\sharpLH\subseteq{\rm U_L}$-$\TCone$.
\end{corollary}
\begin{proof}
It follows from Corollary \ref{gaplTCone} that $\sharpL\subseteq{\rm U_L}$-$\TCone$. We therefore get $\sharpLH_1=\sharpL\subseteq{\rm U_L}$-$\TCone$. We can now prove our result by showing that for $i\geq 2$, $\sharpLH_i\subseteq{\rm U_L}$-$\TCone$ using induction on the number of levels of $\sharpLH$.
\end{proof}
\section*{Exercises}
\begin{enumerate}
\item Show that computing the permanent of a square integer matrix modulo $2$ is logspace many-one complete for $\parityL$.
\item Show that determining if any two linear representations $M_1$ and $M_2$ of linearly representable matroids over $\Z_2$ represent the same matroid is logspace many-one complete for $\parityL$.
\item Let $A\in \Q^{n\times m}$ be the input rational matrix. Show that computing a maximal set of linearly independent columns of $A$ over $\Q$ is in ${\rm FL}^{\rm\CeqL}$.
\item Let $A\in \Q^{n\times m}$ be the input rational matrix. Show that determining whether a column of $A$ is in the lexicographically least set of linearly independent columns of $A$ over $\Q$ is in $\LCeqL$.
\item Define the $\GapL$ hierarchy, denoted by $\GapLH$. Show that $\GapLH =\sharpLH$. Also show that $\GapLH ={\rm U_L}$-$\ACzero (\GapL )$.
\item Let $A\in\Q^{m\times n}$ and $\b \in \Q ^n$. Assume that the system of linear equations $A\x =\b$ has a solution. Show that it is possible to compute a solution $\x$ to $A\x =\b$ in $\GapLH_3$. More precisely, show that a logspace machine with access to a function in  $\GapLH_2$ as an oracle can obtain a solution $\x$ to $A\x=\b$.
\item Let $f(x),g(x)\in \Q [x]$ be monic polynomials given as input in terms of a vector of its coefficients. Show that the $\deg (\gcd (f(x),g(x)))$ can be computed in $\PL$.
\item Let $f(x),g(x),h(x)\in \Q [x]$ be monic polynomials given as input in terms of a vector of its coefficients. Show that we can test if $h(x)=\gcd (f(x),g(x))$ in $\LCeqL$.
\item Let $f(x),g(x)\in \Q [x]$ be monic polynomials given as input in terms of a vector of its coefficients. Show that the coefficients of the $\gcd (f(x),g(x))$ can be computed in $\GapLH_5$. More precisely, show that a logspace machine with access to a function in $\GapLH_4$ as an oracle can compute the coefficients of the $\gcd (f(x),g(x))$.
\item Let $A\in \Q^{n\times m}$ be the input rational matrix. Show that it is possible to compute the coefficients of the minimal polynomial of $A$ in $\GapLH$. What is the precise complexity level of $\GapLH$ to compute the coefficients of the minimal polynomial of $A$?
\end{enumerate}

\section*{Notes}
Basic facts about permutations included in Section \ref{basic-facts-permutations-matrices} of this chapter is based on \cite[2.10]{Her1975}. We refer to \cite{DM1996,HJ2013,Str2006} for supplementary material on permutations and matrices.

The notation and terminology for all the definitions and results shown in Section \ref{mahajan-vinay1997} are based on the exposition and results due to Meena Mahajan and V. Vinay in \cite{MV1997}.\textcolor{white}{\index[authors]{Mahajan, Meena}\index[authors]{Vinay, V.}} The necessary background on Combinatorial Matrix Theory needed to follow results in Section \ref{mahajan-vinay1997} can be found in \cite[Chapter 9]{BR1991}. We have brought about significant modifications to the definition of cycle cover, clow, clow sequence, weight of a clow, weight of a clow sequence and the sign of a clow sequence in \cite{MV1997}. Our proof of Theorem \ref{determinant-theorem1} is based on \cite{MV1997}. Theorem \ref{determinant-cyclecover} has not been stated explicitly in \cite{MV1997} even though it is very useful to complete the proof of Theorem \ref{determinant-theorem2}. Theorem \ref{determinant-theorem2} is stated in \cite{MV1997}. Our proof of Theorem \ref{determinant-theorem2} is substantially different from \cite{MV1997} in view of the use of Proposition \ref{removeselfloops}. The $4$-tuple used to represent a vertex in the definition of the special weighted layered directed acyclic graph $\bf H$ and the definition of $\bf H$ is slightly different from the vertex definition in \cite{MV1997}. Once again Theorems \ref{determinant-computeingapl} and \ref{determinant-hardforgapl}, Corollaries \ref{determinantgaplcomplete} and \ref{determinantsharplhard} are not stated in \cite{MV1997} explicitly even though the proof of these results have been explained.

Results shown in Section \ref{applicationDeterminant} on applications of computing the determinant are based on problems defined and studied in \cite{ABO1999,BDHM1992,HT2005,Coo1985}.

The results on linear algebraic problems shown in this monograph is not exhaustive. Many more results on classifying the complexity of computing the coefficients of the minimal polynomial of an integer matrix \cite{HT2003}, computing the inertia of an integer matrix \cite{HT2005,HT2010}, computing the gcd of the coefficients of any two uni-variate polynomials with integer coefficients \cite{Vij2008,HT2010,AV2011} have been shown to be contained in logarithmic space bounded counting classes, especially the $\sharpLH$.

We show in Section \ref{sharpLHTCone-sec} that essentially all the logarithmic space bounded counting classes are contained in the Boolean circuit complexity class $\TCone$. The notion of length-respecting functions is from \cite[pp. 11]{Voll1999}. The definition of computing the bits of a function $f:\{ 0,1\}^*\rightarrow\{ 0,1\}^*$ is from \cite[pp. 18]{Voll1999}. The standard and basic functions defined in this section involving integers such as ADD, ITADD, MULT and ITMULT have been shown to be equivalent under constant-depth reductions in \cite[Corollary 1.45]{Voll1999}. Exercise problems $3$, $4$ and $6$ to $10$ of this chapter are used in \cite{AV2011} to tightly classify the complexity of a linear algebraic problem called the orbit problem using logarithmic space bounded counting classes.

Yet another very interesting problem in computational complexity is the problem of testing if two given undirected graphs are isomorphic. In other words, this problem known as the Graph Isomorphism problem, is about determining if there exists a bijection $\phi$ between the set of vertices of a pair of undirected graphs $(G_1,G_2)$ given as input such that given any two vertices $v_1,v_2\in V(G_1)$ we have $(v_1,v_2)\in E(G_!)$ if and only if $(\phi (v_1),\phi (v_2))\in E(G_2)$. There are many results known that classify the computational complexity of the Graph Isomorphism problem and its restricted variants using logarithmic space bounded counting classes. To mention a few references that contain these results, see  \cite{JKMT2003,Tor2004,AKV2005,JKMT2003b,Wag2007,Tor2008}.

\appendix
\chapter{Mathematical prerequisites}
\section{Number Theory}
\begin{theorem}{\bf (Fundamental Theorem of Arithmetic)}
Every integer $n>1$ can be uniquely represented as a product of prime powers.\textcolor{white}{\index[subject]{Fundamental Theorem of Arithmetic}}
\end{theorem}
\begin{definition}
Given integers $a_1,a_2,\ldots ,a_n$ all different from $0$, the least of the positive common multiples is called the {\rm least common multiple}, and it is denoted by ${\rm [}a_1,a_2,\ldots ,a_n{\rm ]}$.\textcolor{white}{\index[subject]{${\rm [}a_1,a_2,\ldots ,a_n{\rm ]}$}}
\end{definition}
\begin{definition}\label{appendix-defn-choosefunction}
Let $\alpha$ be any real number, and let $k$ be a non-negative integer. Then the binomial coefficient ${\alpha}\choose{k}$ is given by the formula\textcolor{white}{\index[subject]{${n}\choose {k}$}}
\[
{{\alpha}\choose{k}} = \frac{\alpha (\alpha -1)\cdots (\alpha -k+1)}{k!}.
\]
\end{definition}
Suppose that $n$ and $k$ are both integers. From the formula we see that if $0\leq k\leq n$ then ${{n}\choose{k}}=\frac{n!}{k!(n-k)!}$, whereas if $0\leq n<k$, then ${{n}\choose{k}}=0$. Here we employ the convention that $0!=1$.
\begin{theorem}\label{appendix-thm-choosefunction}
Let $\mathcal{S}$ be a set containing exactly $n$ elements. For any non-negative integer $k$, the number of subsets of $\mathcal{S}$ containing precisely $k$ elements is ${{n}\choose{k}}$.
\end{theorem}
\begin{theorem}
Let $f$ denote a polynomial with integral coefficients. If $a\equiv b({\rm mod~} m)$ then $f(a)\equiv f(b)({\rm mod~} m)$
\end{theorem}
\begin{theorem}
Let $m_1,m_2,\ldots ,m_r$ be non-zero integers. $x\equiv y({\rm mod~} m_i)$ for $i=1,2,\ldots ,r$ if and only if $x\equiv y({\rm mod ~[}m_1,\ldots ,m_r{\rm ]})$.
\end{theorem}
\begin{theorem}
If $b\equiv c({\rm mod~} m)$ then ${\rm gcd(}b,m{\rm )}={\rm gcd(}c,m{\rm )}$
\end{theorem}

\begin{theorem}{\rm (}{\bf Fermat's Little Theorem}{\rm )}
Let $p$ denote a prime. If $p$ does not divide $a$ then $a^{p-1}\equiv 1({\rm mod~}p)$. For every integer $a$, $a^p\equiv a({\rm mod~}p)$.\textcolor{white}{\index[subject]{Fermat's Little Theorem}}
\end{theorem}
\begin{theorem}{\rm (}{\bf Euler-Fermat Theorem}{\rm )}
If ${\rm gcd}(a,m)=1$, then
\[
a^{\phi (m)}\equiv 1({\rm mod~}m),
\]
where $\phi (m)$ denotes the number of positive integers less than or equal to $m$ that are relatively prime to $m$.\textcolor{white}{\index[subject]{Euler-Fermat Theorem}}
\end{theorem}
\noindent $\Z$ denotes the set of integers. $\Z^+$ denotes the set of non-negative integers. $\N$ denotes the set of natural numbers. $\Q$ denotes the set of rationals. $\R$ denotes the set of real numbers. $\mathbb{C}$ denotes the set of complex numbers.\textcolor{white}{\index[subject]{$\Z$}\index[subject]{$\N$}\index[subject]{$\Q$}\index[subject]{$\R$}\index[subject]{$\mathbb{C}$}}

\section{Asymptotic notation}
For any real number $x$, we denote the greatest integer less than or equal to $x$ by $\lfloor x\rfloor$ (read ``the floor of $x$") and the least integer greater than or equal to $x$ by $\lceil x\rceil$ read ``the ceiling of $x$"). Let $f,g:\N\rightarrow \Z^+$.
\begin{enumerate}
    \item $\Theta (g(n))=\{f(n):$ there exists positive constants $c_1,c_2$ and $n_0$ such that $0\leq c_1g(n)\leq f(n)\leq c_2g(n)$ for all $n\geq n_0\}$.\textcolor{white}{\index[subject]{$\Theta (g(n))$}}
    \item $O(g(n))=\{f(n):$ there exists positive constants $c$ and $n_0$ such that $0\leq f(n)\leq cg(n)$ for all $n\geq n_0\}$.\textcolor{white}{\index[subject]{$O(g(n))$}}
    \item $\Omega (g(n))=\{f(n):$ there exists positive constants $c$ and $n_0$ such that $0\leq cg(n)\leq f(n)$ for all $n\geq n_0\}$.\textcolor{white}{\index[subject]{$\Omega (g(n))$}}
\end{enumerate}

\section{Basics of Algebra \& notation}
\begin{itemize}
\item Order of a group $G$ is the cardinality of $G$, and it is denoted by $o(G)$.\textcolor{white}{\index[subject]{order of a group, denoted by $o(G)$}}
\item Let $G$ be a finite group and let $H$ be a subgroup of $G$. The index of $H$ in $G$, denoted by $[G:H]$, is the number of distinct left (or right) cosets of $H$ in $G$. In particular, the index of $H$ in $G$ is,
\[
[G:H]=\frac{o(G)}{o(H)}.
\]\textcolor{white}{\index[subject]{$[G:H]$}}
\item Let $\phi$ be a homomorphism of $G$ onto $\overline{G}$ with kernel $K$. Then, $K$ is a normal subgroup of $G$, and $G/K$ and $\overline{G}$ are isomorphic and it is denoted by $\left (\frac{G}{K}\right )\approx \overline{G}$.
\item If $G$ is a finite group and $N$ is a normal subgroup of $G$, then $[G:N]$ denotes the order of the quotient group $G/N$.
\item For $n\in\Z$ and $n\geq 1$, $(\Z_n,+_n)$ denotes the finite additive abelian group of integers modulo $n$.\textcolor{white}{\index[subject]{$\Z_n$}}
\item For $n\in\Z$ and $n\geq 1$, $(\Z_n^*,._n)$ denotes the finite multiplicative abelian group of integers modulo $n$. Elements of this group are elements of $\Z_n$ that are relatively prime to $n$.\textcolor{white}{\index[subject]{$\Z_n^*$}}
\item If $r$ is a prime, then $(\Z_r,+_r,._r)$ is a finite field under addition and multiplication of integers modulo $r$.
\item $\Q$ is a field under addition and multiplication of rationals and $\R$ is a field under addition and multiplication of reals.
\item For $n\geq 1$, $\Q^n$ denotes the set of all column vectors with $n$ components, each containing a rational. We can add two vectors in $\Q^n$ and multiply a vector in $\Q^n$ by a scalar from $\Q$. In other words, we can take linear combinations of vectors in $\Q^n$ with coefficients from $\Q$.\textcolor{white}{\index[subject]{$\Q^n$}}
\item A set of columns of a vector space $V$ over a field $\mathbb{F}$ is linearly dependent if there exists a linear combination of vectors with non-zero coefficients in $\mathbb{F}$ which is equal to the zero vector. Otherwise the set of columns is linearly independent.\textcolor{white}{\index[subject]{linearly independent}\index[subject]{linearly dependent}}
\item A set of vectors $B$, in a vector space $V$ over a field $\mathbb{F}$, spans $V$ if every vector in $V$ can be expressed as a linear combination of vectors in $B$ with coefficients in $\mathbb{F}$.
\item A basis for a vector space $V$ over a field $\mathbb{F}$ is a set of vectors $B$ such that vectors in $B$ are linearly independent, and $B$ spans $V$.\textcolor{white}{\index[subject]{basis of a vector space}}
\item Let $A\in\R ^{n\times n}$. We define the \emph{permanent} of $A$ as:
\begin{eqnarray}
{\rm perm}(A) & = & \sum_{\sigma\in S_n}\prod_{i=1}^na_{i,\sigma (i)}.\textcolor{white}{\index[subject]{permanent}}\label{permanent-definition-appendix}
\end{eqnarray}
\item Let $A\in\R ^{n\times n}$. We define the \emph{determinant} of $A$ as:
\begin{eqnarray}
\det(A) & = & \sum_{\sigma\in S_n}sgn(\sigma )\prod_{i=1}^na_{i,\sigma (i)}.\textcolor{white}{\index[subject]{determinant}}\label{determinant-definition-appendix}
\end{eqnarray}
\item Let $A\in\R^{n\times n}$. The \emph{characteristic polynomial} of $A$ is defined as $\det (\lambda I_n-A)$, where $\lambda$ is a variable and $I_n$ is the $n\times n$ identity matrix.\textcolor{white}{\index[subject]{characteristic polynomial}}
\item The Cayley-Hamilton Theorem states that \emph{every matrix $A$ satisfies its own characteristic equation}. In other words, if $\chi(x)$ denotes the characteristic polynomial of $A$, then $\chi(A)=0$.\textcolor{white}{\index[subject]{Cayley-Hamilton Theorem}}
\item Let $A\in\R^{n\times n}$. Roots of the characteristic polynomial of $A$ are called as \emph{eigenvalues} of $A$.\textcolor{white}{\index[subject]{eigenvalues}} Clearly, the roots of $A$ can be real or complex numbers.
\item Let $A\in\R^{n\times n}$. $rank (A)$ is defined as the maximum number of linearly independent columns of $A$ over $\R$.\textcolor{white}{\index[subject]{rank of a matrix $A$, denoted by $rank (A)$}}
\item Let $A\in\R^{n\times n}$. The \emph{minimal polynomial} of $A$ is the unique monic polynomial of the smallest degree satisfied by $A$.\textcolor{white}{\index[subject]{minimal polynomial}}
\item Let $A\in\R^{n\times n}$. The minimal polynomial of $A$ is a factor of the characteristic polynomial of $A$.
\item Let $A\in\R^{n\times n}$. Inertia of $A$, denoted by $i(A)$, is defined as the triple $i(A)=(i_+(A),i_-(A),i_0(A))$, where $i_+(A),i_-(A)$ and $i_0(A)$ are the number of eigenvalues of $A$, counting multiplicities, with positive, negative and zero real part.\textcolor{white}{\index[subject]{inertia of a matrix $A$, denoted by $i(A)$}}
\end{itemize}

\backmatter
\bibliographystyle{amsalpha}

\printindex[authors]
\printindex[subject]

\end{document}